\documentclass[a4paper,11pt,twoside]{memoir}
\chapterstyle{veelo}

\usepackage[table,dvipsnames]{xcolor}
\usepackage{TUINFDISS}
\usepackage{pdfpages}

\usepackage{amsmath}
\usepackage{url}
\usepackage{hyperref}
\usepackage{subfig}
\usepackage{graphicx}

\usepackage{tabularx}

\usepackage{gensymb}

\usepackage{verbatim}
\usepackage[lined,linesnumbered,algochapter]{algorithm2e} 

\usepackage{enumitem}

\usepackage{pgf}					

\usepackage{array}
\usepackage{multirow}

\usepackage{epigraph}

\usepackage{environ}
\usepackage[tikz]{bclogo}
\usepackage{tikz}

\usepackage{tikzpagenodes}

\usepackage{ngerman}
\usepackage[ngerman]{babel}
\usepackage{bibgerm,cite}
\usepackage[ngerman]{varioref}

\usepackage{subfig}

\usepackage{framed}

\DeclareUnicodeCharacter{00A0}{~}

\definecolor{mylightgray}{HTML}{EEEEEE}
\usetikzlibrary{calc,snakes,arrows}

\NewEnviron{pubinfo}[1]
  { \noindent\par\medskip\medskip\medskip\noindent
  
  \begin{tikzpicture}

    \node[inner sep=0pt,anchor=south] at (current page text area.south) (box) {\parbox[t]{0.95\textwidth}{%
	
	  \fcolorbox{mylightgray}{mylightgray}{%

      \begin{minipage}{.2\textwidth}
      \centering\tikz[scale=3]\node[scale=2,rotate=0]{\bcplume};
      \end{minipage}%
      \begin{minipage}{.73\textwidth}
      \medskip\textbf{#1}\par\smallskip
      \BODY
      \medskip
      \end{minipage} 
      }
      }%
    };

    \draw[yellow!100!black,line width=3pt] 
      ( $ (box.north west) + (0pt,0pt) $ ) -- ( $ (box.north west) + (0,0pt) $ ) -- ( $ (box.south west) + (0,-0pt) $ ) -- + (0pt,0);
  \end{tikzpicture} 
}

\NewEnviron{infobox}[1]
  { \noindent\par\medskip\medskip\medskip\noindent
  
  \begin{tikzpicture}

    \node[inner sep=0pt,anchor=south] at (current page text area.south) (box) {\parbox[t]{0.95\textwidth}{%
	
	  \fcolorbox{mylightgray}{mylightgray}{%

      \begin{minipage}{.2\textwidth}
      \centering\tikz[scale=3]\node[scale=2,rotate=0]{\bcinfo};
      \end{minipage}%
      \begin{minipage}{.73\textwidth}
      \medskip\textbf{#1}\par\smallskip
      \BODY
      \medskip
      \end{minipage} 
      }
      }%
    };

    \draw[yellow!100!black,line width=3pt] 
      ( $ (box.north west) + (0pt,0pt) $ ) -- ( $ (box.north west) + (0,0pt) $ ) -- ( $ (box.south west) + (0,-0pt) $ ) -- + (0pt,0);
  \end{tikzpicture} 
}

\newcommand\descleftmargin{0.8cm}
\newcommand\descrightmargin{1.5cm}
\epigraphfontsize{\small\itshape}
\newcommand{\x}{$\times$}

\thesistitle{Multi-Modal Music Information Retrieval: Augmenting Audio-Analysis with Visual Computing for Improved Music Video Analysis} 
\thesissubtitle{} 
\thesisdate{01.10.2019}

\thesisdegree           {Doktor der technischen Wissenschaften}
                        {Doktor der technischen Wissenschaften}
             
\thesisverfassung       {Verfasser}
\thesisauthor           {Alexander Schindler}
\thesisauthoraddress    {Beckmanngasse 4/12, 1140 Wien}
\thesismatrikelno       {9926045}
                        
\thesisbetreins         {Ao.univ.Prof. Dr. Andreas Rauber}
                        
\thesisbegutachtungeins {Univ. Prof. Mag. Dipl.-Ing. Dr. Markus Schedl}
\thesisbegutachtungzwei {Univ. Prof. Dr. Allan Hanbury}
                        
\makepagestyle          {numberCorner}
\makeevenfoot           {numberCorner}{\thepage}{}{}
\makeoddfoot            {numberCorner}{}{}{\thepage}

\begin{document}

\captionnamefont{\bfseries}

\frontmatter
\pagenumbering{roman}

%
%
%
%

\selectlanguage{ngerman}

\newgeometry{left=2.4cm,right=2.4cm,bottom=2.5cm,top=2cm}

\newlength{\tmpbaselineskip}
\setlength{\tmpbaselineskip}{\baselineskip}
\setlength{\baselineskip}{13.6pt}
\newlength{\tmpparindent}
\setlength{\tmpparindent}{\parindent}
\setlength{\parindent}{17pt}

\thispagestyle{tuinftitlepage}

%
%
\pagenumbering{Alph}

\begin{center}
	{\ \vspace{3.4cm}}
	
	\begin{minipage}[t][2.8cm][s]{\textwidth}%
		\centering
		\thesistitlefontHUGE\sffamily\bfseries\tuinfthesistitle\\
		\bigskip
		{\thesistitlefonthuge\sffamily\bfseries\tuinfthesissubtitle}
	\end{minipage}
	
	\vspace{1.3cm}
	
	{\thesistitlefontLARGE\sffamily \tuinfthesistype}
	
	\vspace{6mm}
	
	{\thesistitlefontlarge\sffamily zur Erlangung des akademischen Grades}
	
	\vspace{6mm}
	
	{\thesistitlefontLARGE\sffamily\bfseries \tuinfthesisdegree}
	
	\vspace{6.5mm}
	
	{\thesistitlefontlarge\sffamily eingereicht von}
	
	\vspace{6mm}
	
	{\thesistitlefontLarge\sffamily\bfseries \tuinfthesisauthor}
	
	\vspace{1.5mm}
	
	{\thesistitlefontlarge\sffamily Matrikelnummer \tuinfthesismatrikelno} 
	
	\vspace{1.4cm}
	
	\begin{minipage}[t][1.6cm][t]{\textwidth}%
		\vspace{0pt}\raggedright\thesistitlefontnormalsize\sffamily
		an der
		
		Fakult\"{a}t f\"{u}r Informatik der Technischen Universit\"{a}t Wien
	\end{minipage}
	
	\begin{minipage}[t][1.6cm][t]{\textwidth}%
		\vspace{-10pt}\thesistitlefontnormalsize\sffamily
		\tuinfthesisbetreuung: \tuinfthesisbetreins\\
	\end{minipage}
	
	\vspace{1cm}
	
	\begin{minipage}[t][1.5cm][t]{\textwidth}%
		\vspace{-30pt}\sffamily\thesistitlefontnormalsize
		\tuinfthesisbegutachtung
		\begin{tabbing}%
			\hspace{45mm} \= \hspace{63mm} \= \hspace{51mm} \kill
			\> \> \\
			\> \> \\
			\> \> \\
			\> {\raggedright\rule{51mm}{0.5pt}} \> {\raggedright\rule{51mm}{0.5pt}} \\
			\> \begin{minipage}[t][0.5cm][t]{51mm}\centering (\tuinfthesisbegutachtungeins)\end{minipage}
			\> \begin{minipage}[t][0.5cm][t]{51mm}\centering (\tuinfthesisbegutachtungzwei)\end{minipage}
		\end{tabbing}
	\end{minipage}
	
	\begin{minipage}[t][1.5cm][t]{\textwidth}%
		\vspace{75pt}\sffamily\thesistitlefontnormalsize
		\begin{tabbing}%
			\hspace{45mm} \= \hspace{63mm} \= \hspace{51mm} \kill
			Wien, \tuinfthesisdate \> {\raggedright\rule{51mm}{0pt}} \> {\raggedright\rule{51mm}{0.5pt}} \\
			\> \begin{minipage}[t][0.5cm][t]{51mm}\centering\end{minipage}
			\> \begin{minipage}[t][0.5cm][t]{51mm}\centering (\tuinfthesisauthor)\end{minipage}
			
		\end{tabbing}
	\end{minipage}
	
\end{center}

\pagestyle{empty}
\cleardoublepage

\pagenumbering{roman}

\setlength{\baselineskip}{\tmpbaselineskip}
\setlength{\parindent}{\tmpparindent}

\restoregeometry

\selectlanguage{english}


%
%
%
%

\newgeometry{left=2.4cm,right=2.4cm,bottom=2.5cm,top=2cm}


\thispagestyle{tuinftitlepage}

%
%
\pagenumbering{Roman}

\begin{center}
	{\ \vspace{3.4cm}}
	
	\begin{minipage}[t][2.8cm][s]{\textwidth}%
		\centering
		\thesistitlefontHUGE\sffamily\bfseries\tuinfthesistitle\\
		\bigskip
		{\thesistitlefonthuge\sffamily\bfseries\tuinfthesissubtitle}
	\end{minipage}
	
	\vspace{1.3cm}
	
	{\thesistitlefontLARGE\sffamily \tuinfthesistypeen}
	
	\vspace{6mm}
	
	{\thesistitlefontlarge\sffamily submitted in partial fulfillment of the requirements for the degree of}
	
	\vspace{6mm}
	
	{\thesistitlefontLARGE\sffamily\bfseries \tuinfthesisdegreeen}
	
	\vspace{6.5mm}
	
	{\thesistitlefontlarge\sffamily by}
	
	\vspace{6mm}
	
	{\thesistitlefontLarge\sffamily\bfseries \tuinfthesisauthor}
	
	\vspace{1.5mm}
	
	{\thesistitlefontlarge\sffamily Registration Number \tuinfthesismatrikelno} 
	
	\vspace{1.4cm}
	
	\begin{minipage}[t][1.6cm][t]{\textwidth}%
		\vspace{0pt}\raggedright\thesistitlefontnormalsize\sffamily
		to the Faculty of Informatics 
		
		at the Vienna University of Technology
	\end{minipage}
	
	\begin{minipage}[t][1.6cm][t]{\textwidth}%
		\vspace{-10pt}\thesistitlefontnormalsize\sffamily
		Advisor: \tuinfthesisbetreins\\
	\end{minipage}
	
	\vspace{1cm}
	
	\begin{minipage}[t][1.5cm][t]{\textwidth}%
		\vspace{-30pt}\sffamily\thesistitlefontnormalsize
		The dissertation has been reviewed by:
		\begin{tabbing}%
			\hspace{45mm} \= \hspace{63mm} \= \hspace{51mm} \kill
			\> \> \\
			\> \> \\
			\> \> \\
			\> {\raggedright\rule{51mm}{0.5pt}} \> {\raggedright\rule{51mm}{0.5pt}} \\
			\> \begin{minipage}[t][0.5cm][t]{51mm}\centering (\tuinfthesisbegutachtungeins)\end{minipage}
			\> \begin{minipage}[t][0.5cm][t]{51mm}\centering (\tuinfthesisbegutachtungzwei)\end{minipage}
		\end{tabbing}
	\end{minipage}
	
	\begin{minipage}[t][1.5cm][t]{\textwidth}%
		\vspace{75pt}\sffamily\thesistitlefontnormalsize
		\begin{tabbing}%
			\hspace{45mm} \= \hspace{63mm} \= \hspace{51mm} \kill
			Wien, \tuinfthesisdate \> {\raggedright\rule{51mm}{0pt}} \> {\raggedright\rule{51mm}{0.5pt}} \\
			\> \begin{minipage}[t][0.5cm][t]{51mm}\centering\end{minipage}
			\> \begin{minipage}[t][0.5cm][t]{51mm}\centering (\tuinfthesisauthor)\end{minipage}
			
		\end{tabbing}
	\end{minipage}
	
\end{center}

\pagestyle{empty}
\cleardoublepage

\pagenumbering{roman}

\setlength{\baselineskip}{\tmpbaselineskip}
\setlength{\parindent}{\tmpparindent}

\restoregeometry


\cleardoublepage
\selectlanguage{ngerman}
\chapter*{Erklaerung zur Verfassung der Arbeit}

\tuinfthesisauthor\\
\tuinfthesisauthoraddress

\vspace*{1.2cm}

Hiermit erklaere ich, dass ich diese Arbeit selbstqendig verfasst habe, 
dass ich die verwendeten Quellen und Hilfsmittel vollstaendig angegeben 
habe und dass ich die Stellen der Arbeit - einschliesslich Tabellen, 
Karten und Abbildungen -, die anderen Werken oder dem Internet im 
Wortlaut oder dem Sinn nach entnommen sind, auf jeden Fall unter Angabe 
der Quelle als Entlehnung kenntlich gemacht habe.\\

\vspace*{2cm}
\begin{tabbing}%
	\hspace{58mm} \= \hspace{28mm} \= \hspace{58mm} \kill
	{\raggedright\rule{58mm}{0.5pt}} \> \> {\raggedright\rule{58mm}{0.5pt}} \\
	\begin{minipage}[t][0.5cm][t]{58mm}
		\vspace{0pt}\sffamily\thesistitlefontnormalsize
		\centering (Ort, Datum)
	\end{minipage}
	\> \>
	\begin{minipage}[t][0.5cm][t]{58mm}
		\vspace{0pt}\sffamily\thesistitlefontnormalsize
		\centering (Unterschrift \tuinfthesisverfassung)
	\end{minipage}
\end{tabbing}

\selectlanguage{english}



\chapter*{Abstract}

The context of this thesis is embedded in the interdisciplinary research field of Music Information Retrieval (MIR) and in particular in the subsection which extracts information from the audio signal by means of digital signal analysis. Because music is in itself multi-modal, many approaches harness multiple input modalities such as audio, lyrics or music notes to solve MIR research tasks.
This thesis focuses on the information provided by the visual layer of music videos and how it can be harnessed to augment and improve tasks of the MIR research domain. The main hypothesis of this work is based on the observation that certain expressive categories such as genre or theme can be recognized on the basis of the visual content alone, without the sound being heard. This leads to the hypothesis that there exists a visual language that is used to express mood or genre. In a further consequence it can be concluded that this visual information is music related and thus should be beneficial for the corresponding MIR tasks such as music genre classification or mood recognition.

The validation of these hypotheses is approached analytically and experimentally. 
The analytical approach conducts literature search in the Musicology and Music Psychology research domain to identify studies on or documentations of production processes of music videos or visual branding in the music business. The history of the utilization of visual attribution is investigated beginning with illustrations on sheet music, album cover arts to music video production. This elaborates the importance of visual design and how the music industry harnesses it to promote new acts, increase direct sales or market values. In the pre-streaming era to attract more customers, album covers had to be as appealing and recognizable as possible to stand out in record shelves. Especially new artists whose style was yet unknown were visually branded and outfitted by music labels to be immediately identifiable in terms of style and music genre in magazines or on TV.

The experimental approach conducts a series of comprehensive experiments and evaluations which are focused on the extraction of visual information and its application in different MIR tasks. Due to the absence of appropriate datasets, a custom set is created, suitable to develop and test visual features which are able to represent music related information. This dataset facilitates the experiments presented in this thesis. The experiments include evaluations of visual features concerning their ability to describe music related information. This evaluation is performed bottom-up from low-level visual features to high-level concepts retrieved by means of Deep Convolutional Neural Networks. Additionally, new visual features are introduced capturing rhythmic visual patterns. In all of these experiments the audio-based results serve as benchmark for the visual and audio-visual approaches. For all experiments at least one audio-visual approach showed results improving over this benchmark. The experiments are conducted for three prominent MIR tasks \textit{Artist Identification}, \textit{Music Genre Classification} and \textit{Cross-Genre Classification}.

Subsequently, the results of the two approaches are compared with each other to establish relationships between the described production processes and the quantitative analysis. Thus, known and documented visual stereotypes such as the cowboy hat for American country music can be confirmed. Finally, it is shown that an audio-visual approach harnessing high-level semantic information gained from visual concept detection, leads to an improvement of up to 16.43\% over the audio-only baseline.

\cleardoublepage
\selectlanguage{ngerman}

\chapter*{Kurzfassung}

Der Kontext dieser Arbeit liegt im interdisziplinaeren Forschungsgebiet Music Information Retrieval (MIR) und insbesondere im Teilbereich, welcher Informationen aus dem Audiosignal mittels digitaler Signalanalyse extrahiert. Da Musik in sich multi-modal ist, nutzen viele Ansaetze mehrere Eingabemodalitaeten wie Audio, Liedtexte oder Musiknoten, um MIR-Forschungs\-aufgaben zu loesen.
Diese Dissertation untersucht den Informationsgehalt des visuellen Kanals von Musikvideos und wie dieser genutzt werden kann, um Aufgaben des Forschungsbereichs MIR zu erweitern und zu verbessern. Die grundsaetzliche Hypothese dieser Arbeit basiert auf der Beobachtung, dass bestimmte Ausdruckskategorien wie Stimmung oder Genre allein aufgrund des visuellen Inhalts erkannt werden koennen, ohne dass die dazugehoerige Tonspur gehoert wird. Davon leitet sich die Annahme der Existenz einer visuellen Sprache ab, die verwendet wird, um Emotionen oder Genre auszudruecken. In weiterer Konsequenz kann gefolgert werden, dass diese visuelle Information musikbezogen ist und daher fuer die entsprechenden MIR-Aufgaben wie die Klassifikation von Musik nach Genre oder Genre-uebergreifenden Themen wie Weihnachten oder Anti-Kriegs-Lied von Vorteil sein sollte.

Die Validierung dieser Hypothesen gliedert sich in einen analytischen und einen experimentellen Teil. Der analytische Ansatz basiert auf einer Literaturrecherche in den Forschungsbereichen Musikwissenschaft und Musikpsychologie, um Studien oder Dokumentationen von Produktionsprozessen von Musikvideos oder Visual Branding im Musikgeschaeft zu identifizieren. Dabei wird die Geschichte der Nutzung visueller Attributierung untersucht, beginnend mit Illustrationen von Noten bis zur Musikvideo-Produktion. Dies verdeutlicht die Bedeutung des visuellen Designs und dessen gezielte Nutzung durch die Musikindustrie, um neue Kuenstler zu foerdern und Direktverkaeufe oder Marktwerte zu erhoehen. Vor allem in der Zeit vor Online-Streaming-Diensten mussten Albumcover so ansprechend und erkennbar wie moeglich sein, um in den Regalen der Plattenlaeden aufzufallen. Besonders neue Kuenstler, deren Stil noch nicht bekannt war, wurden von Musiklabels optisch gebrandet und ausgestattet, um in Zeitschriften oder im Fernsehen sofort durch ihren Look bezueglich Stil und Musikrichtung eingeordnet werden zu koennen.

Der experimentelle Ansatz fuehrt eine Reihe umfassender Experimente und Evaluierungen durch, die sich auf die Extraktion von visueller Information und deren Anwendung in verschiedenen MIR-Aufgaben konzentrieren. 
Aufgrund des Fehlens geeigneter Datensets wird zuerst ein Set erstellt, welches geeignet ist, visuelle Merkmalebeschreibungen zu entwickeln und diese daraufhin zu testen, ob sie in der Lage sind, musikbezogene Informationen darzustellen. Dieses Datenset dient als Basis fuer die Experimente. Diese beinhalten Auswertungen visueller Eigenschaften hinsichtlich ihrer Faehigkeit, musikbezogene Informationen zu erfassen. Diese Evaluierungen erstrecken sich von einfachen visuellen Features zu komplexen Konzepten, welche unter Verwendung von Deep Convolutional Neural Networks extrahiert werden. Darueber hinaus werden neue Features eingefuehrt, die rhythmische visuelle Muster erfassen. In all diesen Experimenten dienen die audio-basierten Ergebnisse als Benchmark fuer die visuellen und audiovisuellen Ansaetze. In allen Experimenten zeigen die Ergebnisse der audio-visuellen Ansaetze Verbesserungen gegenueber dieser Benchmark. Die Experimente werden fuer die MIR-Aufgaben \textit {Kuenstler-Identifikation}, \textit{Musik Genren Klassifikation} und \textit{Genre-uebergreifende Klassifikation} durchgefuehrt.

Anschliessend werden die Ergebnisse der beiden Ansaetze miteinander verglichen, um Zusammenhaenge zwischen den beschriebenen Produktionsprozessen und der quantitativen Analyse herzustellen. So koennen bekannte und dokumentierte visuelle Stereotypen wie der Cowboyhut fuer amerikanische Country-Musik bestaetigt werden. Schliesslich wird gezeigt, dass ein audio-visueller Ansatz, der semantische Information auf hohem Niveau nutzt, die aus der visuellen Konzepterkennung gewonnen wird, zu einer Verbesserung von bis zu 16,43\% gegenueber der Nur-Audio-Basislinie fuehrt.

\selectlanguage{english}

\setsecnumdepth{subsubsection}
\setcounter{maxsecnumdepth}{2}
\setcounter{tocdepth}{1}

\cleardoublepage
\pagestyle{numberCorner}

\tableofcontents*


\mainmatter
\pagenumbering{arabic}
\pagestyle{numberCorner}


\chapter{Introduction}
\label{ch1:intro}

\epigraph{
	``An album is a whole universe, and the recording studio is a three-dimensional kind of art space that I can fill with sound. Just as the album art and videos are ways of adding more dimensions to the words and music. I like to be involved in all of it because it's all of a piece.''} 
{--- Natasha Khan, Bat for Lashes, 2013}

\begin{figure}
	\centering 
	\includegraphics[width=1.00\textwidth]{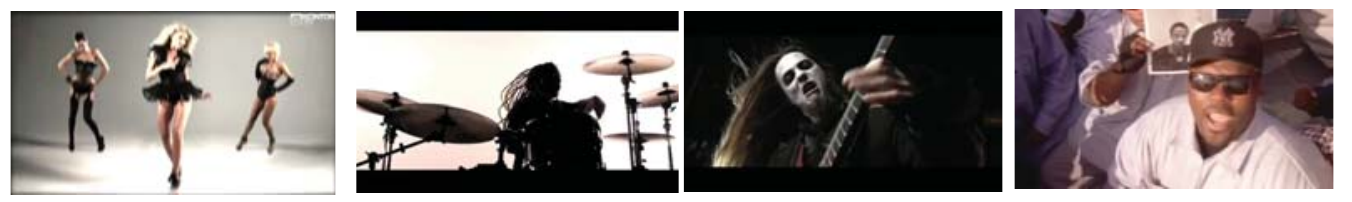}
	\caption{Examples of music genres that are easy to identify in images - a) Dance, b) Rock, c) Heavy Metal, d) Rap}
	\label{fig:ch1:genre_examples}
\end{figure}

%
In the second part of the last century the visual representation has become a vital part of music. Album covers grew out of their basic role of packaging to become visual mnemonics to the music enclosed \cite{jones1999steve}. Stylistic elements emerged into prototypical visual descriptions of genre specific music properties. Initially intended to aide or sway customers in their decision of buying a record, these artworks became an influential part of modern pop culture. The ``look of music'' became an important factor in people's appreciation of music. In that period the structure of the prevalent music business changed completely by shifting the focus from selling whole records to hit singles \cite{frith2005sound}. Their accentuation and the new accompanied intensification of celebrity around pop music played an important role in the development of the \textit{pinup culture}. The rise of music videos in the early 80s provided further momentum to this development. This emphasis on visual aspects of music transcended from topics directly connected with music production into aspects of our daily life. The relationship to fashion tightened, yielding to different styles that discriminated the various genres. An early study on music video consumption among young adolescents revealed that one of the primary intentions was social learning such as how to dress, act and relate to others \cite{sun1986adolescent}. Over the past decades music videos emerged from a promotional support medium into an art form of their own which distinctively influenced our pop-culture and became a significant part of it.  Still intended to promote new hit-singles and artists, contemporary music video productions typically make use of of film making techniques and roles such as screen-play, directors, producers, director of photography, etc. How much influence visual music had on us over half a century is hard to evaluate, but we grew accustomed to a visual vocabulary that is specific for a music style in a way that the genre of a music video can often be predicted despite the absence of sound (see Figure \ref{fig:ch1:genre_examples}).

The research field concerned with the analysis and processing of music related information is called \textit{Music Information Retrieval (MIR)}. Chapter \ref{ch2:soa_mir} will provide a detailed introduction to types and sources of music information as well as a broad overview of the challenges and tasks the field is facing.
One of the central research topics within MIR is the analysis of the audio content of music recordings.
As in many other Information Retrieval (IR) fields there is a discrepancy between the ambitious tasks a discipline is aiming to solve and the technical abilities available to approach them. This is commonly referred to as the \textit{Semantic Gap}. 
For audio content based music analysis this discrepancy derives from the inability to accurately describe the music content by means of a set of expressive numbers. These so called \textit{Features} are the basis for a wide range of further processing steps which are necessary for tasks such as music classification \cite{fu2011survey}, similarity estimation or music recommendation \cite{celma2006foafing}.
The methods applied draw from a wide range of well researched digital signal processing techniques. Through various transformations and aggregations the recorded audio signal is converted into numerical interpretations of timbre, rhythm, music theoretic properties and various abstract descriptions of spectral distribution and fluctuations \cite{Lidy2010}.
Finally, these features are used to train machine learning algorithms which are intended to label an unheard song by its genre, mood or estimate its similarity to other songs in a collection.
Although content based music feature development has made huge progress over the past two decades, the results for many tasks are still unsatisfactory.
Take music similarity estimation as an example. Music features already perform well in describing properties such as timbre and rhythm, and although ``\textit{sounding similar}'' is an obvious qualification for music similarity, there is a wide range of subliminal properties that cannot be captured or at least not without huge efforts. 
Such properties often fall into the domain of human perception and are generally difficult to model computationally. The interpretation if a song is happy or sad is highly subjective and culturally dependent \cite{balkwill1999cross}.  
But also more obvious attributes such as temporal relationships are relevant. A contemporary song may sound like a song by the Beatles, but explaining this relationship the other way around is an impossible causality. 
Nevertheless, also the supposedly objective measure of acoustic similarity is problematic to asses with state-of-the-art audio features. 
Completely similar sounding songs are usually rare due to implications about plagiarism. More frequent are tracks that share similar sounding parts such as verse or chorus. 
Often subjectively stated similarity only refers to the melody of the singing voice. Common machine learning based approaches to acoustic similarity usually work on a track based granularity and thus accumulate the similarity for an entire song. Analyzing the similarities based on sub-segments or structural elements of a composition would require to automatically detect these segments, which is a different unsolved MIR task. 


A well researched  \cite{fu2011survey,SCHINDLER_2012amr,schindler2012,lidy08_MIREX,lidy2005evaluation,lidy2011report} and extensively discussed \cite{sturm2012survey,sturm2013classification,sturm2014simple} task in MIR is automatic genre classification.  
%
Music genre is an ambiguous mean which is supposed to aide us in quickly finding new and relevant music. Back in the days when buying music was a solely physical act, big genre labels on the record shelves made buying music less cumbersome.
%
An album does not reveal its content right away which is physically encoded onto synthetic media. If you wanted to pre-listen a record before buying, you had to queue up in front of one of the few record players provided by the store. As customers usually are not blessed with unlimited time to iterate through this process - choose records, queue up, have a listen - over and over again, music genres are a convenient aide to short-cut customers to their preferred music. They are synonyms which describe a wide range of music properties such as rhythm, expression and instrumentation in a single term. Although, nowadays we can virtually browse through on-line music catalogs without leaving the house and we can pre-listen to songs just by the click of a mouse or by the tap on a smart-phone screen, music genres still remain relevant for describing and categorizing music. 
%
The major problem with genres is that on one side they are not well defined and on the other side artists do not comply with the existing definitions. Playing with genre boundaries and mixing different styles is a common mean of artistic expression.
%
This complicates MIR tasks such as automatic genre estimation but also music similarity retrieval. The difficulty lies in the proper definition of audio features so that machine learning algorithms are able to effectively discriminate the various abstractions of spectro-temporal distributions of music data into different classes such as music genres. This is aggravated by the nature of digital audio. The recorded audio signal is a representation of sound pressure levels sampled at sequential intervals at a virtually fixed location. These sampled values can be transformed into a spectral representation, but this only provides insight in which frequencies are dominant at the given moment. The problem is that the information about the source of the signal, such as guitars, violins or drums, is not available. Its estimation, also referred to as \textit{source separation} is a highly complicated research field, especially in polyphonic environments.

\begin{figure}[!ht]
	\centering
	\subfloat[Country Music]{{
			\includegraphics[width=.35\linewidth]{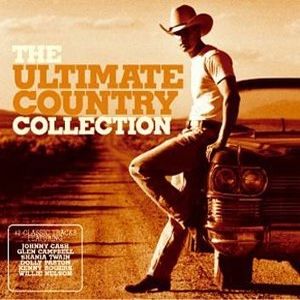}
	}}
	\qquad
	\subfloat[Dance Music]{{
			\includegraphics[width=.35\linewidth]{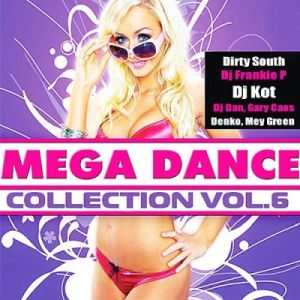}
	}}
	\hspace{0mm}
	\subfloat[Heavy Metal]{{
			\includegraphics[width=.35\linewidth]{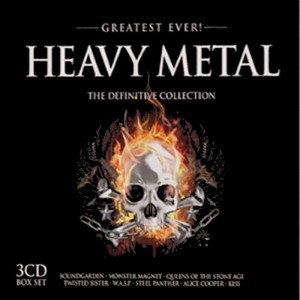}
	}}
	\qquad
	\subfloat[Classical Music]{{
			\includegraphics[width=.35\linewidth]{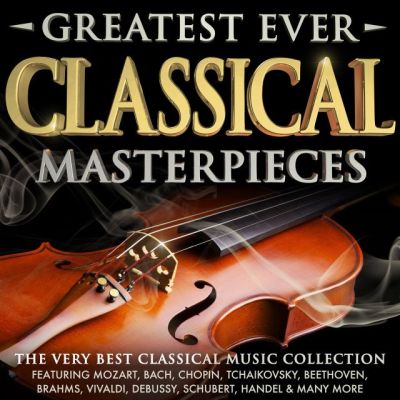}
	}}
	\caption{Examples of visual stereotypes. Genre specific compilations and 'Best of' albums often make use of visual stereotypes to advertise their content. Typical stereotypes such as the Cowboy hat for Country Music (A), over-sexualization in Dance Music (B), death and fire for Heavy Metal (C) and the display of classical instruments for classical music (D).} 
	\label{fig:ch1:stereotypes}
\end{figure}

Besides music genres music labels courted the customers attention by making their records as visually appealing and descriptive as possible. As more extensively elaborated in Chapter \ref{ch:visualmusic} the music industry created and made use of a distinct visual language to promote new artists and records more efficiently. They made and make extensive use of visual artists, fashion and visual media and developed well known visual stereotypes for the various genres (see Figure \ref{fig:ch1:stereotypes}).
%
This visual accentuation represents music related information which can be used to augment and improve MIR tasks such as automatic genre classification and artist identification as well as to realize new approaches to existing or newly identified tasks.

Approaches to harness this music related visual information are currently scarce in literature. This thesis represents a step towards the facilitation of audio-visual MIR approaches. First, a set of datasets are provided to facilitate experimentation. Then, a set of extensive experimentations and evaluations are presented which form the basis for the provided conclusions on the type and quality of information provided by music related artworks such as album cover arts and music videos and how to harness this information for MIR tasks.

\section{Opportunities of audio-visual analysis}
\label{ch1:mvir_opportunities}

The influence and outreach of music videos is on a constant rise as recently reported. In 2011 we conducted a survey among decision makers and stakeholders in the music industry which showed that YouTube\footnote{http://www.youtube.com/} is considered to be the leading online music service \cite{lidy2011report}. A different survey by Nielsen \cite{nielsen} involving 26,644 online consumers in 53 international markets revealed that '\textit{watching music videos on computer}' was the most mentioned consuming activity, practiced by 57\% of these consumers in the 3 months preceding the survey. 27\% mentioned to also have '\textit{watched music videos on mobile devices}' in the same period. In a large scale user-survey on requirements for music information services by Lee et al. \cite{LeeEtAl_2012_UndeUserRequFor} YouTube was listed as the second most preferred music service. Although the main usage scenario is identified as listening to music, music videos are mentioned to be highly valued by users. Stowell and Dixon \cite{StowellEtAl_2011_MirInSchoLess} identified YouTube as one of the most important technologies used in secondary school music classes. The authors suggest to offer MIR analysis for videos to enable open-ended exploration. Cunningham et al. \cite{CunninghamEtAl_2007_FindNewMusiA} identified music videos as a passive trigger to active music search and state that visual aspects of music videos can strongly influence the amount of attention paid to a song. Liem et al. \cite{liem2011need} strongly argue that there is a need for multimodal strategies for music information retrieval. The MIRES roadmap for music information research \cite{MiresRoadmap2013} identifies music videos as a potential rich source of additional information. As an open research problem it requires more attention of the audio/visual communities as well as appropriate data-sets.

Harnessing the visual layer of music videos provides a wide range of opportunities and possible scenarios. At first hand it presents a new way to approach existing MIR problems such as classification (e.g. genre, mood, artist, etc.), structural analysis (e.g. music segmentation, tempo, rhythm, etc.) or similarity retrieval. On another hand it may provide new innovative approaches to search for music. One intention is to use visual stimuli to query for music. This would enhance the traditional search spaces that are either based on audio content, textual or social data or a mixture of those. An investigation about the usage of music in films and advertising \cite{InskipEtAl_2008_MusiMoviAndMean} suggests that a search function matching video features to music features would have potential applications in film making, advertising or similar domains. But to use the visual domain for searching music, it has to be linked to the acoustic domain. This section outlines objectives and obstacles of an integrated audio-visual approach to music. Challenges for a selected set of music research disciplines are discussed.

The following descriptions of opportunities for research tasks of various research domains are based on observations made during the accumulation and definition of the Music Video Dataset which will be introduced in Chapter \ref{ch4.1:intro}.

\subsection{Music Information Retrieval}

At firsthand, harnessing music related visual information represents a valuable source for the MIR domain. Various research tasks could profit from such an audio-visual approach.

\begin{description}[leftmargin=!,itemsep=1pt,labelwidth=\widthof{\bfseries 12345},labelindent=\descleftmargin,rightmargin=0cm] 
	
	\item[Music Genre Recognition:] 
	Automatic genre recognition will be extensively discussed in Chapter \ref{ch2:mir_tasks}. It is generally concerned with assigning genre labels to recorded songs, which is mainly achieved through the application of signal processing and machine learning methods to the audio signal. The pros and cons concerning these approaches will be discussed in the next Chapter, but one of the major drawbacks is, that some music related information cannot be derived from the audio signal alone. Meta-Genres such as birthday, Christmas or road-trip songs do not share specific harmonics or instrumentation which produces a significant pattern in the audio signal. Still, they are often easy to categorize accordingly by human listeners.
	Music marketing uses visual cues that are commonly associated with terms of such meta-genres. Christmas albums and samplers are often held in the traditional colors pine green and heart red or decorated with typical items such as Christmas holly, a Christmas tree or Santa Clause. Such visual cues are intentionally easy to identify and mark the concerning album as Christmas music. Similarly, cakes and balloons are shown on covers for birthday party albums and roads and convertibles for road-trip music. Further examples are hugging couples or sunsets for romantic music, sunny beaches and women in bathing suits for summer-hit compilations, etc. (see Figure \ref{fig:ch1:stereotypes}). As a consequence the genre of music videos can often be recognized without hearing the corresponding song. Recent cross-modal approaches reported improved accuracy in detecting similar visual concepts (e.g. sky, people, plant life, clouds, sunset, etc.) \cite{huiskes2010new}. It has to be evaluated which cues are discriminative for each genre and which state-of-the-art technology from the image processing domain can be applied to reliably extract this information. 
	
	\item[Artist Identification:] 
	Artist recognition is an important task for music indexing, browsing and retrieval. Using audio content based descriptors only is problematic due to the wide range of different styles on either a single album or more severe throughout the complete works of an artist. Current image processing technologies (e.g. object detection, face recognition) could be used to identify the performing artist within a video. 
	
	\item[Music Emotion Recognition:] 
	Visual arts have developed a wide range of techniques to express emotions or moods. For example, some album covers for aggressive music show images of tortured or rioting people, Buddhist symbols and people performing yoga exercises are often seen on relaxing albums. 
	Another example is color which is a popular mean of expressing emotions. We have incorporated colors in our daily language for expressing emotions (e.g. feeling blue), gravity of events (e.g. black Friday), political parties, etc. Movie directors often pick warm colors for romantic scenes while using cold ones for fear and anxiety. Songwriters use similar effects to express emotions. Major chords are perceived happy, minor chords sad, dissonance creates an impression of dis-comfort, etc. 
	Such techniques have to be identified, evaluated and modeled to harness their information for effective emotion recognition. It has to be evaluated to which extent the visual techniques of expressing emotions and those of music correlate. Such correlations may be used to combine the acoustic with the visual domain.
	
	\item[Instrument Detection:]
	Instrument detection is an ambitious task currently suffering from poor performance of source separation approaches. Many music videos show the performing artists playing their instruments. This information can be used to augment audio based classification approaches.
	
	\item[Music Segmentation and Summarization:] 
	Music segmentation and summarization tries to identify coherent blocks within a composition such as verse or chorus. Typical approaches are based on self-similarities of sequentially extracted audio features. If no timbral or rhythmical differences can be detected, segmentation fails. The structure of a music video is often aligned to the compositional or lyrical properties of a track. Scenery or color changes depict the transition to another segment. Such transitions can be detected \cite{cotsaces2006video} and analyzed. Repetitions of the same images through the video clip for example may suggest repetitions of sections such as the chorus.
	
	\item[Tempo and Rhythmic Analysis:] 
	Current beat tracking approaches \cite{mckinney2007evaluation} rely on strong contrasting musical events to correctly estimate the track's tempo. Shot editing of music video can serve as valuable information in cases where audio based approaches fail. Further detecting and classifying dance motions can provide additional data for rhythmic analysis.
	
	\item[Music Video Similarity:] 
	Definitions of similarity of music videos are still missing. It is even undecided if similarity should be based on acoustic or visual properties or on a combination of both. Descriptions of music videos intuitively refer to visual cues. An extensive discussion including results from user evaluations should precede attempts to measure similarity between music videos.
	
\end{description}

\subsection{General Objectives}

This section envisions application scenarios that could be accomplished through an audio-visual approach which harnesses music related visual media.

\begin{figure}
	\centering
	\includegraphics[width=1.00\columnwidth]{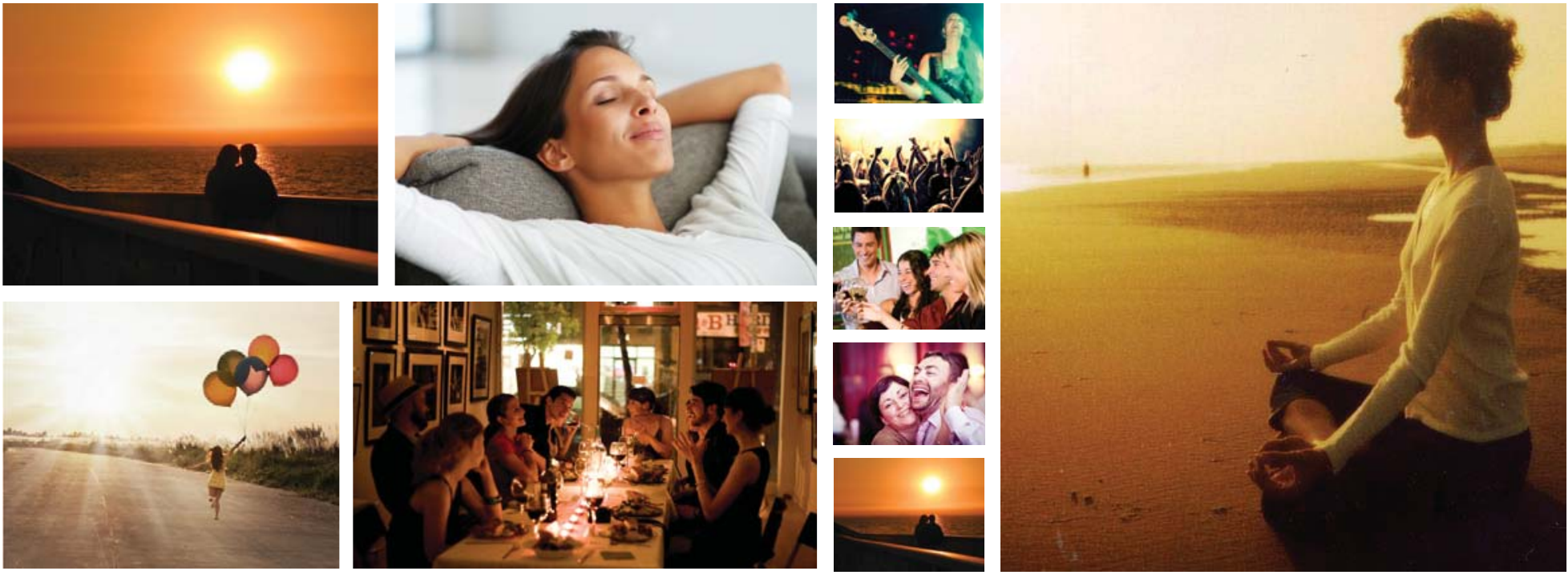}
	\caption{Example images as input for music search algorithms.} 
	\label{fig:ch1:search_by_image}
\end{figure}

\begin{description}[leftmargin=!,itemsep=1pt,labelwidth=\widthof{\bfseries 12345},labelindent=\descleftmargin,rightmargin=0cm] 
	
	\item[New Innovative Music Search:] Visual clues identified and extracted from music videos, such as correlations between color and music, may promote the development of new search methods for music or sounds. Traditionally, textual or audio content descriptions of seed songs are used to query for music. Although not applicable to all kinds of music, the intention for a search is often better described through a picture. Based on insights from the suggested research, a music query using a picture of a glass of wine and burning candles might return a collection of romantic music.
	
	\item[Music Visualization:] Identified audio-visual correlations can be used to improve music visualization or automatic lighting systems. LED-based color-changing lights are currently gaining popularity. An add-on to such systems could change the color according to musical transitions and mood.

	\item[Music for Movies or Advertisements:] An investigation about the usage of music in films and advertising \cite{InskipEtAl_2008_MusiMoviAndMean} suggests that a search function matching video features to music features would have potential applications in film making, advertising or similar domains. Understanding audio-visual correlations, especially in their cultural context, facilitates recommender systems suggesting music for individual scenes in a movie.
	
	\item[Adaptive Music Recommendation:] Similar to recommending music for videos, cameras can be used to observe the ambiance of an area. Based on the visual input from the camera, high level features such as number and motion of people, clothing, color of the room, weather conditions, sunrise, sunset, etc. can be used to select appropriate music.
	
\end{description}

\noindent
\textit{\smaller Parts of this introduction to Music Video Information Retrieval (MVIR) and its opportunities as presented in this and the previous section was published and presented in Alexander Schindler. ``A picture is worth a thousand songs: Exploring visual aspects of music'' at the 1st International Digital Libraries for Musicology workshop (DLfM 2014), London, UK, September 12 2014 \cite{Schindler2014dlfm}. }

\clearpage

\section{Research Questions}
\label{ch1:research_questoins}

\begin{figure}
	\centering
	\includegraphics[width=0.7\linewidth]{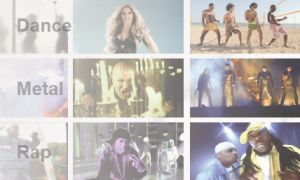}
	\caption{Examples of genre specific music videos.}
	\label{fig:musicvideoexamples}
\end{figure}

This thesis is based on the observation that many music videos have distinct visual properties which facilitate the derivation of further music characteristics from them such as tempo, rhythm, genre, mood, etc. (see Figure \ref{fig:musicvideoexamples}). This observation lead to the hypothesis on which this thesis is based on - that there are specific visual patterns which are connected with certain music characteristics. 
The overarching question of this thesis is, \textit{to which extent does visual information contribute to the performance of MIR tasks?} Because there exist recognizable repeating visual patterns the question is, does this information contribute to the performance of MIR research tasks? It has been shown, that adding further modalities such as text \cite{mayer2008rhyme,laurier2008multimodal} or symbolic music \cite{lidy08_MIREX} to audio signal analysis, does improve the performance of these tasks. Except for a few examples on small data-sets \cite{mayer2011} this has not yet been shown for visual information on a broad and generalized scale. This hypothesis lead to the development of the following research questions:\\

\noindent\textbf{Research Question 1 (RQ1): }

\hfill\begin{minipage}{\dimexpr\textwidth-1cm}
	\textit{Which visual features are able to capture task related information?}\\
	
	The main motivation of this research question is the identification of appropriate information retrieval approaches to extract music and task related information from visual sources. This will require to research state-of-the-art image processing, video processing and visual computing methods, as well as to develop custom features and to evaluate their effectiveness in capturing the relevant information. This research question will be addressed in Chapters \ref{ch6:intro}, \ref{ch7:intro} and \ref{ch8:intro}.
	
	Feature development and evaluation will be preceded by a literature search to identify documented processes in music production. The aim is to identify described correlations between visual and musical concepts which can be modeled by existing or to be developed features. This research will be adressed in Chapter \ref{ch:visualmusic}.\\
\end{minipage}

\noindent\textbf{Research Question 2 (RQ2): }

\hfill\begin{minipage}{\dimexpr\textwidth-1cm}
	\textit{How can the different modalities be combined to improve performance?}\\
	
	Having identified appropriate visual features to extract music related visual information, the next goal is to research approapriate methods to combine this information with those extracted from the acoustic layer. This research question will be addressed in Chapters \ref{ch6:intro}, \ref{ch7:intro} and \ref{ch8:intro} by evalting different approaches to fuse the different modalities in feature space as well as by ensemble methods.\\
\end{minipage}

\pagebreak

\noindent\textbf{Research Question 3 (RQ3): }

\hfill\begin{minipage}{\dimexpr\textwidth-1cm}
	\textit{To which extent can these visual patterns be used to derive further music characteristics?}\\
	
	The effectiveness of an information retrieval system to capture the semantically relevant information is always dependent on the underlying task it will be evaluated on. Thus, to properly answer the previous question it is also necessary to assess to which extent the extracted information can be used to predict further music related properties such as the corresponding genre or mood. To evaluate if the exploitation of the visual domain provides enough additional information to show significantly improvements over audio-only based approaches, the identified or new defined visual features will be evaluated on well known Music Information Retrieval tasks. This research question will also be addressed in Chapters \ref{ch6:intro}, \ref{ch7:intro} and \ref{ch8:intro}.\\
\end{minipage}




\noindent\textbf{Research Question 4 (RQ4): } 

\hfill\begin{minipage}{\dimexpr\textwidth-1cm}
	\textit{Is it possible to verify concepts within the production process?}\\
	

	Having identified appropriate ways to extract semantically meaningful information from the visual layer and how to consecutively apply further methods to predict certain music properties, it should be analyzed if identified visual patterns reported in literature (see RQ1) can be verified through automated analysis. This research question will also be addressed in Chapter \ref{ch8:intro}.\\
\end{minipage}

\pagebreak

\section{Structure of the Work}

\medskip\medskip

\begin{description}[leftmargin=!,itemsep=10pt,labelwidth=\widthof{\bfseries 1234567890},labelindent=\descleftmargin,rightmargin=0cm] 
	
	\item[Chapter \ref{ch1:intro}] provides a brief introduction to the problem domain, including the research field of music information retrieval, image processing, visual computing and music marketing. This chapter also introduces the problem statement and corresponding research questions and outlines the opportunities of harnessing music related visual information.
	
	\item[Chapter \ref{ch2:intro}] provides an extensive overview of the state-of-the-art in the concerned research domains, including music information retrieval, image processing and visual computing.
	
	\item[Chapter \ref{ch:visualmusic}] elaborates on audio-visual relationships in music by providing introductions to the history of album-cover-art and music videos. Further, common practices in music marketing are explained and the importance of visual accentuation is discussed. Finally, a detailed introduction into music video theory is provided as basis for the experiments in the consecutive chapters.
	
	\item[Chapter \ref{ch4:intro}] introduces the datasets that will be used in the experiments and evaluations. Two new datasets are defined which can be used to develop and evaluate methods to harness the visual layer of music videos and album-cover-arts. Further, ground-truth assignments to existing datasets are introduced.
	
	\item[Chapter \ref{ch5:intro}] investigates charachteristics of the newly introduced datasets MVS and MSD, as well as provided feature-sets and ground-truth assignments. Supporting experimental results are provided which are essential for the interpretation of the main experimental results and serve as baselines as well. 
	
	\item[Chapter \ref{ch6:intro}] presents an audio-visual approach to the music information retrieval task \textit{Artist Identification}. State-of-the-art music and audio features are combined with face recognition to an ensemble classifier. 
	
	\item[Chapter \ref{ch9:intro}] addresses the unsolved problem of shot boundary detection in music videos. It identifies an extended list and examples of transition types frequently found in music videos and sets them in contrast to state-of-the-art approaches to shot boundary detection.
	
	\item[Chapter \ref{ch7:intro}] provides an in-depth evaluation of primary, low-level and affective features towards their effectiveness in music genre classification tasks. This bottom up evaluation provides valuable insights into the complexity of the music related visual information. 
	
	\item[Chapter \ref{ch8:intro}] investigates the capability of high-level visual concept detection approaches to capture music related visual information. Aggregations of accumulated semantic concepts, extracted with deep convolutional neural networks, are combined with music features to improve the accuracy of music genre classification.
	
	
	
	\item[Chapter \ref{ch8:intro}] discusses results and insights of the conducted experiments, summarizes results and draws conclusions. It further points to open issues, research possibilities and future work.
	
	
\end{description}

\newpage

\section{Key Contributions}

The scientific contributions provide by this thesis are manifold but can be divided into the three categories \textit{Data}, \textit{Methods} and \textit{Code}:\\

\noindent\textbf{\larger Data:}

\hfill\begin{minipage}{\dimexpr\textwidth-1cm}
	
	\hfill
	\begin{enumerate}
		
		\item The \textit{Music Video Dataset (MVD)}: The MVD (see Chapter \ref{ch4.1:intro}) is the first dataset in the context of Music Information Retrieval which facilitates audio-visual research on a broad range of open MIR challenges such as music genre classification, cross-genre classification, artist recognition and similarity estimation.
		
		\item Additional gound-truth assignments and an expanded set of audio- and music-features are provided for the \textit{Million Song Dataset (MSD)} to facilitate comprehensive benchmarking experiments on a large scale (see Chapter \ref{ch4.2:intro}).\\
		
		\newcounter{enumTemp}
		\setcounter{enumTemp}{\theenumi}
	\end{enumerate}

\end{minipage}

\noindent\textbf{\larger Methods:}

\hfill\begin{minipage}{\dimexpr\textwidth-1cm}
	
	\hfill
	\begin{enumerate}
		\setcounter{enumi}{\theenumTemp}

		\item Methods to select and aggregate features provided with the Million Song Dataset (MSD) based on extensive evaluations of their performance in classification experiments (see Chapter \ref{ch5.2:intro}). This provides benefits for facilitating large scale experiments on the MSD.
		
		\item Methods using Deep Neural Networks (DNN) to improve music classification performance (see Chapter \ref{ch5:intro}). This model was subsequently used in the \textit{Detection and Classification of Acoustic Scenes and Events (DCASE)} 2016 evaluation campaign and was the winning contribution to the \textit{Domestic audio tagging} task \cite{lidy2016cqt}.
		
		\item Methods to harness audio-visual information to address the MIR research task of \textit{Artist Recognition} (see Chapter \ref{ch6:intro}).
		
		\item Methods to harness information provided by low-level image processing features in an audio-visual approach to music genre classification (see Chapter \ref{ch7:intro}).
		
		\item Methods to harness high-level visual concepts and music features in an audio-visual approach to music genre and cross-genre classification (see Chapter \ref{ch8:intro}).
		
		\item Surveys and positions on opportunities of audio-visual approaches to MIR (see Chapter \ref{ch1:mvir_opportunities}) as well as on open research problems in \textit{Shot-boundary Detection} for music videos (see Chapter \ref{ch9:intro}).\\

		\setcounter{enumTemp}{\theenumi}	
	\end{enumerate}
	
\end{minipage}

\pagebreak

\noindent\textbf{\larger Code:}

\hfill\begin{minipage}{\dimexpr\textwidth-1cm}

	\hfill
	\begin{enumerate}
		\setcounter{enumi}{\theenumTemp}
		
		\item The \textit{Music Video Information Retrieval - Toolset}\footnote{\url{https://github.com/slychief/mvir}}: a Python framework to extract and merge acoustic and visual features from music videos. It implements all the features introduced and used in this thesis.

		\item A conversion and optimization of the Matlab based \textit{Rhythm Patterns} feature-set extractor to the Python programming language\footnote{\url{https://github.com/tuwien-musicir/rp_extract}}.\\
		
	\end{enumerate}

\end{minipage}



\chapter{State-of-the-Art}
\label{ch2:intro}

In the course of this dissertation, the state-of-the-art of the involved research domains has undergone a significant transition from traditional feature-engineering based to data-driven \textit{deep learning} based approaches. This transition shows its most outstanding impacts in the visual computing domain where Top-5 error rates in well established international evaluation campaigns suddenly dropped from 26.2\% to 15.4\% through the application of Deep Convolutional Neural Networks in 2012 \cite{krizhevsky2012imagenet} and sequentially then dropped to 4.9\% \cite{SzegedyIV16}. This astounding improvement initiated the hype around \textit{Deep Learning}. The reason why this hype is still ongoing is, that many neural network based approaches live up to their expectations. Huge performance improvements are reported from various application domains. Different network architectures are reported regularly and adaption to different modalities such as text or speech also provide outstanding results. Neural networks have also gained the attention of the Music Information Retrieval (MIR) community. Several attempts are reported to define appropriate network architectures to efficiently learn music representations.

The research focus of this thesis is embedded in the Music Information Retrieval (MIR) domain. Thus, this chapter first provides a brief introduction into the MIR domain and then discusses more detailed related work and the state-of-the-art on multi-modal approaches for general and MIR specific problems.

\section{Music Information Retrieval}
\label{ch2:soa_mir}

Music Information Retrieval (MIR), a subfield of multimedia retrieval, is a multidisciplinary research field drawing upon the expertise from information retrieval, library science, musicology, audio engineering, digital signal processing, psychology, law and business \cite{downie2003music}. This relatively new discipline gained momentum in the early 2000s and is supported by an eager research community including the International Society for Music Information Retrieval (ISMIR) with its homonymous annual conference.

A general concern of the MIR research field is the extraction and interpretation of meaningful information from music. That this is a challenging endeavor can be seen by the different music representations such as audio recordings, music scores, lyrics, playlists, music videos, album cover arts, music blogs, etc. Music sheets contain formal descriptions of music compositions.
Audio recordings capture the information about the actual interpretation of a composition. 
Lyrics and music videos add further layers of artistic and semantic expression to a music track. Playlists set different tracks into relation and blog posts contain opinions about music, artists and performances. All this relevant information comes in different modalities.
Recordings are provided as digital audio requiring digital signal processing to extract the information. Lyrics and blog posts require text and natural language analysis. Music videos and album cover art analysis require image processing and computer vision techniques. Mining playlists and social media data requires knowledge of collaborative filtering methods. Finally, it has proven to be advantageous to combine two or more modalities to harness the information provided by the different layers. Given that wealth of music information one of the main tasks is to reduce as much information and to retain the most relevant notion of it to provide fast and accurate methods of indexing music and to provide new and intuitive ways to search and retrieve music.

Compared to image processing, MIR is a relatively young research domain. It started in the 90s of the last century but gained on momentum with the launch of the International Symposium for Music Information Retrieval (ISMIR) in October of 2000 \cite{Downie2009}, which was renamed in 2008 to the conference of the International Society for Music Information Retrieval, while retaining the acronym. This change went hand in hand with the founding of the homonymous research society. One of the stepping stones to MIR research was the study by Beth Logan on applying Mel Frequency Cepstral Coefficients (MFCC), an audio feature which emerged from the speech recognition domain, to the digital music domain \cite{logan2000mel}. The next milestone was the \textit{ISMIR 2004 Audio Description Contest (ADC)} \cite{Cano2006} - an international evaluation campaign where researchers could compete in ten different tasks such as audio fingerprinting, music genre classification or music structure analysis. Based on the experiences made at the ADC the\textit{ Music Information Retrieval Evaluation Exchange (MIREX)} was initiated in 2005 \cite{downie20052005}, which has then become the annual evaluation campaign of ISMIR \cite{downie2014ten}.


\subsection{What is Music Information?}

The Oxford Dictionary defines music as \textit{'the art of combining vocal or instrumental sounds (or both) to produce beauty of form, harmony, and expression of emotion'} \cite{oxford_music} and thus refers mainly to the acoustic nature of it. The field of MIR considers a much broader perspective on music information. The following section will thus concentrate on the different modalities and types of music information.

\paragraph{Recorded Audio} - This might be the most palpable form of music. Referring to the cited definition of the Oxford Dictionary, music is a combination of vocal or instrumental sounds. In a more general form this is referred to as a performance. Thus, when we speak of music we generally refer to recorded performances whose copies are distributed by resellers or played on radio stations. The history of recorded music started in 1896 when Thomas Alva Edison established the National Phonograph Company to sell music pressed on wax cylinders for the phonograph. In 1901 the Victor Talking Machine Company introduced the phonographs whose 78-rpm flat disc defined the standard for the future development of music distribution. This further lead to the development of the Long Play (LP) album which was replaced by the cassette in the early 80s of the last century. With the introduction of the Compact Disc (CD) a major shift to digital music production and consumption was performed. Digital audio is the main source for content based audio analysis which is one of the major sub-fields of MIR. We will introduce this in more detail in the following chapters.

\paragraph{Symbolic Music} - Symbolic music is a representation of music in a logical structure based on symbolic elements representing audio events and the relationships among those events. The most commonly known form of symbolic music is the music score which is also referred to as ‘Sheet Music’. Music scores are available as printed sheets or in various digital representations such as MIDI \cite{rothstein1995midi}, MusicXML \cite{good2001musicxml}, ABC \cite{walshaw2014statistical} or GUIDO \cite{hoos2001guido}. The most powerful among them is MusicXML which was released first in 2004 and describes apart from the composition also the layout of the score, lyrics and textual annotations and further features. The main difference between symbolic music and recorded or perceived music is, that scores only describe how a piece of music is intended to be performed. In other words: It is a guideline that is used by a performer to play a certain piece of music. For MIR, symbolic music has some advantageous properties compared to recorded music. The most obvious is the exact knowledge about notes and their durations. Such information needs to be estimated from the audio spectrum of a recorded music track which is a complex and not yet matured task. 

\paragraph{Text} - Text based music related information is presented in form of music lyrics, music metadata such as title, album and artist names, textual information provided on artist web pages, blog posts, music and album reviews or artist biographies, as well as from social tagging platforms (e.g. last.fm\footnote{\url{http://www.last.fm}}, MusicBrainz\footnote{\url{http://musicbrainz.org}}, etc.) and Wikipedia\footnote{\url{http://www.wikipedia.org}}. To harness such information, classical information retrieval is applied to the text extracted from different sources which was collected from the Internet by querying search engines \cite{knees2008document}, monitoring MP3 blogs \cite{celma2006search} or crawling music web-sites \cite{whitman2004automatic}. Most systems proposed in literature use \textit{vector space} document representations such as the Term Frequency-Inverse Document Frequency (TF-IDF) representation \cite{knees2007music,whitman2002combining}.

\paragraph{Collaborative Information} - 
Music as a socio-cultural phenomenon also generates information through interaction. This includes data such as listening and buying behaviors or interactions between users and groups on social media. 
Such information is often used in music recommendation systems. Collaborative filtering techniques are applied to estimate relationships between tracks and listeners \cite{Celma2010, schedl2015music, schedl2015music}. Such techniques are best known from popular recommendation systems such as Amazon\footnote{\url{http://www.amazon.com}} or Netflix\footnote{\url{http://www.netflix.com}}. They depend on a large community of active users to track their habits of interacting with items such as music and to infer similarities between these items or users and to estimate user-items preferences. Thus, the combination of songs that users have listened to or rated provide information about these user's preferences and or 'tastes' and how these songs relate to each other.
Data gathered from social media and microblogs \cite{hauger2013million} can also be used to reveal latent social and collective perspectives on music \cite{iren2016using} by analyzing geographic, temporal, and other contextual information related to listening histories. 

\paragraph{Visual Media} - Visual based music related information is provided by album cover arts, music videos and music advertising. The visual representation has become a vital part of music. Stylistic elements are used as prototypical visual descriptions of genre specific music properties. The rise of music videos in the early 80s provided further momentum to this development. Music marketing uses these visual concepts to promote a new artist. Genre specific visual cues are provided to express its music style or genre. There also exists a strong relationship with fashion.

\subsection{Music Information Retrieval Approaches}

Before giving a brief introduction to a selection of actively researched MIR tasks, a general overview of common methods will be provided. Generally, MIR approaches can be divided into two categories - those relying on \textit{music content} and those on \textit{contextual data}.

\subsubsection{Content Based Approaches}

Content based approaches are based on the idea that the semantically relevant information is provided in the audio itself. Thus the aim is to identify and extract this information in a meaningful representation. The biggest problems of such approaches stem from the digital nature of the recorded audio signal as well as from the standard music distribution format. Although instruments in contemporary music productions are recorded separately, they are finally mixed together into one (mono) or two (stereo) channels. As a consequence in the resulting acoustic signal the originating audio sources cannot be distinguished anymore or have to be separated with great efforts \cite{pedersen2007survey}.

Audio signals as perceived by our ears have a continuous form. Analog storage media were able to preserve this continuous nature of sound (e.g. vinyl records, music cassettes, etc.). Digital logic circuits on the other hand rely on electronic oscillators that sequentially trigger the unit to process a specific task on a discrete set of data units (e.g. loading data, multiplying registers, etc.). Thus, an audio signal has to be fed in small chunks to the processing unit. The process of reducing a continuous signal to a discrete signal is referred to as \textit{sampling}. The continuous electrical signal generated in the microphone is converted into a sequence of discrete numbers that are evenly spaced in time as defined by the \textit{sampling rate}. This process of turning continuous values into discrete values is called quantization. The resulting digital signal represents the changes in air pressure over time. For digitizing audio especially music in CD quality, typically a sampling rate of 44.100 Herz at a bit depth of 16 is used.

\textit{Feature extraction} is a crucial part of content-based approaches. The goal is to transform and reduce the information provided by the digital audio signal into a set of semantically descriptive numbers. A typical CD quality mainstream radio track has an average length of three minutes. This means, that song is digitally described in Pulse-code Modulation (PCM) by 15.9 million numbers (3 [minutes] x 60 [seconds] x 2 [stereo channels] x 44100 [sampling rate]). Using CD-quality 16bit encoding this information requires 30.3MB of memory. Besides music specific reasons, the computational obstacles concerned with processing of huge music collections make raw audio processing a suboptimal solution. Feature design and implementation tries to overcome technological obstacles of digital audio and to extract essential music properties that can be used to analyze, compare or classify music. Common music features are descriptors for timbre \cite{logan2000mel}, rhythm \cite{lidy10_ethnic} or general spectral properties \cite{tzanetakis2000marsyas,lartillot2007matlab}. Because the audio signal is only a description for changes in air pressure over time, spectral analysis such as Fourier transformations is applied. Audio and music features are then calculated from the resulting frequency distributions by subsequently applying further transformations and statistical modeling \cite{casey2008content}.

Recently, feature development has been philosophically challenged. From a pure musicologist point of view the validity of task-oriented approaches of most solutions reported in literature is in question \cite{wiggins2009semantic,sturm2014state}. It is argued that the designed features capture sound properties that are not related to music itself, but to artifacts of the production process and that the evaluation methods applied are inappropriate \cite{urbano2013evaluation,sturm2014kiki}. From a machine learning point of view it seems counterproductive to neglect valuable information due to the fact that it is not directly related to music - especially if it increases the performance of the system in solving a task.

A major obstacle of content-based approaches is its dependency on the access to the actual music file. Due to copyright and intellectual property restrictions music cannot be distributed without permission \cite{mckay2006large}. This restricts reproducibility of research results, because privately assembled test collections cannot be shared \cite{downie2004scientific}. Only a few datasets are available for each of the corresponding MIR tasks, but their sizes, ranging from a few hundred to a few thousand tracks, are often not representative in consideration of contemporary on-line music services which host catalogs of several millions of tracks.

\subsubsection{Context Based Approaches}

Contrary to \textit{content-based} approaches \textit{context-based} ones \cite{knees2013survey} do not rely on the music content itself, but on contextual data which is not included in the audio signal. The most common type of information is meta-data embedded in the music file such as artist name, title, album name, length of the track, etc. The advantage of context-based approaches is their independence from the access to the music files. Further, the distribution of music meta-data does not constitute a copyright infringement \cite{mckay2006large}. The down-side of this approach is, that it relies on the existence of proper meta-data and that its absence entails the in-existence of the corresponding track. In recent years several on-line music-platforms such as Last.fm\footnote{\url{http://last.fm}}, MusicBRainz\footnote{\url{http://www.musicbrainz.org}},
Echonest\footnote{\url{http://the.echonest.com}}, Spotify\footnote{\url{http://www.spotify.com}} or Europeana\footnote{\url{http://www.europeana.eu}} in the cultural heritage domain, are providing access to their meta-data catalogs. These are valuable sources of semantically relevant music information.

\textit{Text-based approaches} retrieve contextual music information from web pages, user tags, or song lyrics and apply methods from traditional Information Retrieval (IR) and Natural Language Processing (NLP).
\textit{Collaborative tags} are user generated annotations for on-line content such as images, videos, audio-tracks, articles, etc. Participants of on-line or social media platforms are usually free to define the labels themselves. Such tags range from brief semantic content description such as the music genre or used instrumentation to lengthly expressions of appreciation and dislike.
\textit{Song Lyrics} are part of music composition and thus comprise semantically relevant information about the song's topic, genre, mood, etc.
\textit{Co-occurrence based approaches} are based on the assumption that objects appearing within the same context are similar. Tracks contained in the same playlist or artists mentioned within the same article supposedly share similar properties. Micro-blogging services such as Twitter\footnote{\url{http://twitter.com}} or peer-to-peer (p2p) networks provide further rich sources for mining music-related data.


\subsection{MIR Tools and Datsets}

One of the biggest obstacles of MIR research was and is the impeded access to data. Copyrighted music must not be shared without license clearing which usually implies high monetary costs. Consequently, only a few datasets emerged over the past decades. Still, some of them are not properly licensed and are shared between research groups on the quiet. One of the first datasets available to the community and relevant for this thesis is the George Tzanetakis Dataset (\textit{GTZAN}) \cite{tzanetakis2000marsyas} which was introduced by his PhD thesis along with the famous audio feature extraction framework MARSYAS in 2000. This was one of the first datasets for Music Genre Recognition (MGR) tasks and has since then been used to evaluate countless approaches to this task \cite{sturm2012analysis}. Further MGR datasets were introduced with the ADC in 2004 \cite{Cano2006}, the \textit{ISMIR Genre}  dataset\footnote{\url{http://ismir2004.ismir.net/genre_contest/}}  and the \textit{ISMIR Rhythm} dataset\footnote{\url{http://mtg.upf.edu/ismir2004/contest/tempoContest/node5.html}} (also known as the \textit{Ballroom Dataset}). Together with the \textit{Latin Music Dataset (LMD)} \cite{Jr2008} released in 2008, these four collections have long been the main benchmark datasets in MIR to test new audio features or classification systems on. The biggest flaw of these datasets was their number of tracks, which ranged from 1000 to 3227 tracks. Because music catalogs of on-line music distributors were already on the scale of millions of tracks, it became problematic to generalize results presented on datasets of a few thousand tracks. Thus, in 2011 the \textit{Million Song Dataset (MSD)} \cite{bertin2011million} was released. Initially, the dataset was only a large collection of meta-data and pre-extracted proprietary audio features which was further lacking of appropriate ground-truth assignments to approach the various MIR tasks. The MSD but contains identifiers which can be used to access different on-line-music repositories from which some of them provide access to audio samples of the corresponding tracks. An effort to use these audio samples and further on-line resources to extend the MSD by state-of-the-art music features and ground-truth assignments is presented in Chapter \ref{ch4.2:intro}. These additions were well received and an improvement of the genre-assignments was introduced in 2015 \cite{schreiber2015improving}. Most recently a new dataset with 77.643 creative commons full audio tracks, downloaded from the \textit{Free Music Archive}\footnote{\url{http://freemusicarchive.org/}} was released \cite{benzi2016fma}.

Besides the introduction of benchmark datasets which can be used to develop new approaches and compare their performance with previously reported results, the provision of tool-sets, frameworks and libraries for audio feature extraction and machine learning algorithms provided further momentum to the development of the MIR research field. One of the first toolsets made available was \textit{MARSYAS}\footnote{\url{http://marsyas.info/}} \cite{tzanetakis2000marsyas}, which implemented among others the MFCC audio features \cite{logan2000mel}. The provision of code and data \cite{tzanetakis2000marsyas} has the advantage that reported results are fully reproducible and thus also reliably comparable to other approaches. Further tool-sets which implemented a wider range of then well established music features was the \textit{MIR-Toolbox}\footnote{\url{https://sourceforge.net/projects/mirtoolbox/}} \cite{lartillot2007matlab} for Matlab\footnote{\url{https://www.mathworks.com/products/matlab.html}} and \textit{jmir}\footnote{\url{http://jmir.sourceforge.net/}} \cite{mckay2010} for the Java. Recently, the Python\footnote{\url{https://www.python.org/}} programming language enjoys great popularity, possibly due to numerous scientific computing libraries and the interactive IPython/Jupyter Notebook\footnote{\url{http://jupyter.org/}} environment, which facilitates structured and well documented experimentation. Recently a couple of MIR libraries were announced which provide implementations of a wide range of state-of-the-art and bleeding edge audio features and machine learning tools. These libraries are \textit{LibRosa}\footnote{\url{https://github.com/librosa/librosa}} \cite{mcfee2015librosa}, 
\textit{Essentia}\footnote{\url{https://github.com/MTG/essentia/}} \cite{bogdanov2013essentia} and 
\textit{MadMom}\footnote{\url{https://github.com/CPJKU/madmom}} \cite{bock2016madmom}. Together with my colleague Thomas Lidy, we also translated the Rhythm Patterns feature extractor \textit{rp-extract} \cite{lidy_ismir07} from its Matlab implementation to Python and made it publicly available\footnote{\url{https://github.com/tuwien-musicir/rp_extract}}.

\subsection{Music Information Retrieval Tasks}
\label{ch2:mir_tasks}

The manifold nature of music is also reflected in the variety of research tasks that emerged in the domain of Music Information Retrieval. The following sections give a few prominent examples including those addressed by the experiments and evaluations which will be elaborated in Chapters \ref{ch6:intro} to \ref{ch8:intro}.

\subsubsection{Music Genre Recognition (MGR)}

\begin{figure}[t]
	\centering
	\includegraphics[width=1.0\linewidth]{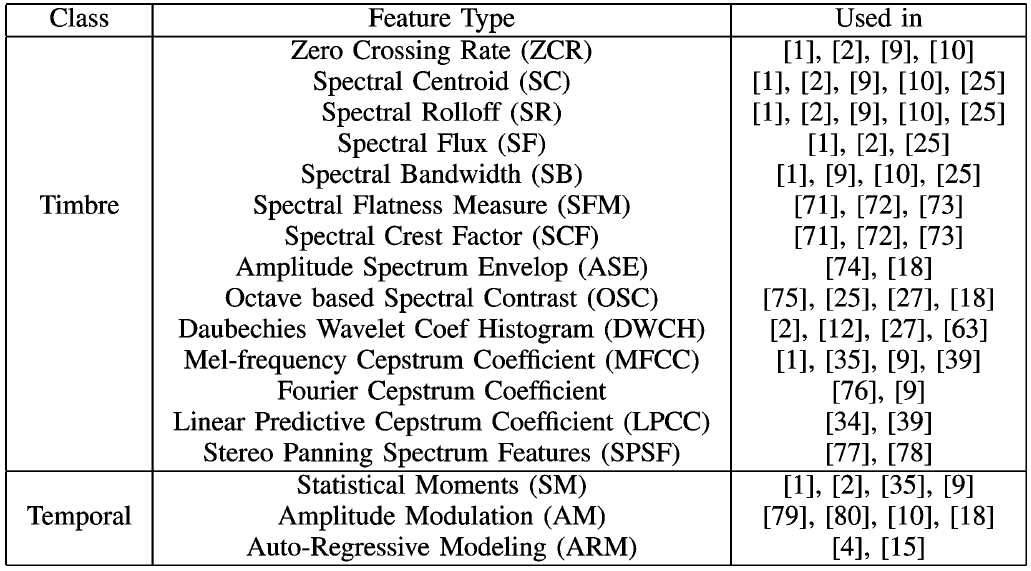}
	\caption{Overview of common low-level audio features used in music classification (taken from \cite{fu2011survey}. Citations in the table refer to the references of the original paper.)}
	\label{fig:fu2011table1}
\end{figure}

Music Genre Recognition is a well researched MIR \cite{downie2003music} task. As for many other content-based MIR tasks the algorithmic design consists of the two parts. In a first step audio-content descriptors are extracted from the audio signal. A comprehensive overview of music features is provided by \cite{lartillot2007matlab,tzanetakis2002musical,lidy2005evaluation}. In a second step these features are used to train machine learning based models, using popular supervised classifiers including k-nearest neighbors (k-NN), Gaussian mixture models (GMM) or Support Vector Machines (SVM). Comprehensive surveys on music and genre classification are provided by \cite{scaringella2006automatic,fu2011survey}. 

Initial approaches to MGR used a variety of low-level audio features \cite{tzanetakis2000marsyas,tzanetakis2002musical,allamanche2004music}. 
An extensive evaluation of these features, which consisted mostly of spectral shape descriptions and MFCCs, was presented in \cite{pohle2005evaluation}. A first survey of MGR approaches was published in 2006 \cite{scaringella2006automatic}. The most commonly used audio features could be categorized into the three groups of \textit{Timbre}, \textit{Melody/Harmony} and \textit{Rhythm} descriptors. The most commonly used classifiers were \textit{K-Nearest Neighbor (KNN)} \cite{tzanetakis2000marsyas}, \textit{Gaussian Mixture Models (GMM)} \cite{burred2003hierarchical}, \textit{Hidden Markov Model (HMM)} \cite{scaringella2005modeling} and \textit{Support Vector Machines (SVM)} \cite{lidy2005evaluation}. Besides providing an overview of state-of-the-art audio features and classification approaches, this survey also summarizes the general discussions about MGR. 

These discussions are still ongoing, because MGR is seen controversially. The one side, mostly coming from signal processing or machine learning background argues, that the aim is to create descriptors which are able to capture the most relevant semantic information of a given content in respect to a given task. Recognizing the genre of an unknown song is thus a prime example of a machine learning approach. The controversy about this is, that MGR is often used to evaluate the performance of audio features with respect to their capability to describe music \cite{sturm2013classification}. Genres are therefor used because they are synonyms which group several music properties such as instrumentation, rhythm, moods, political and religious positions, even historical epochs. This is a convenient but controversial simplification of the problem space. Instead of evaluating if the features are able to capture the music properties, they are evaluated if they are able to group songs which share the same properties. In many cases this is a sufficiently appropriate approach. Yet, from a scientific point of view, there are issues such as at which level genre labels do apply - at artist, album or track level? Further it is now generally agreed upon that there is no general agreement on genre taxonomies and genre assignments \cite{pachet2000taxonomy,aucouturier2003representing}. Further, many genre labels are ill-defined, locally skewed through cross-cultural differences in genre perception \cite{shevy2008music} and only a minimal agreement can be reached among human annotators \cite{mckay2006musical}. Based on the personal listening preferences the choice of individual genre labels can differ as well as the pool of known genre labels. A study on human genre classification showed an inter-participant agreement rate of only 76\% \cite{lippens2004comparison}. Another issue with the simplified assumption that genres group music characteristics is that most genres have their specific production processes which include specific recording and post-processing steps. It has been shown that MGR approaches are prone to capture artifacts of these processes to optimize the genre boundaries instead of learning on actual intrinsic music properties \cite{sturm2014simple}. On the relatively small-sized available datasets it was soon argued that a glass ceiling has been reached \cite{pachet2004improving} and concerns were raised if further pursuing MGR research is still reasonable \cite{mckay2006musical}.
In this thesis genre classification is used to demonstrate the performance of the developed visual and audio-visual features. The aim is not to insist on recognizing the correct genre but rather to use the same methodology frequently used in literature to provide comparable results, by classifying music videos by artificially assigned labels which refer to common acoustic properties (see Section \ref{ch4:dataset_creation}). Through isolated experiments on classes that have clearly defined boundaries based on isolated stylistic elements the rational of this study is to identify and capture extramusical concepts related to music genres \cite{shevy2008music,sturm2013classification}.

\begin{figure}[t]
	\centering
	\includegraphics[width=1.0\linewidth]{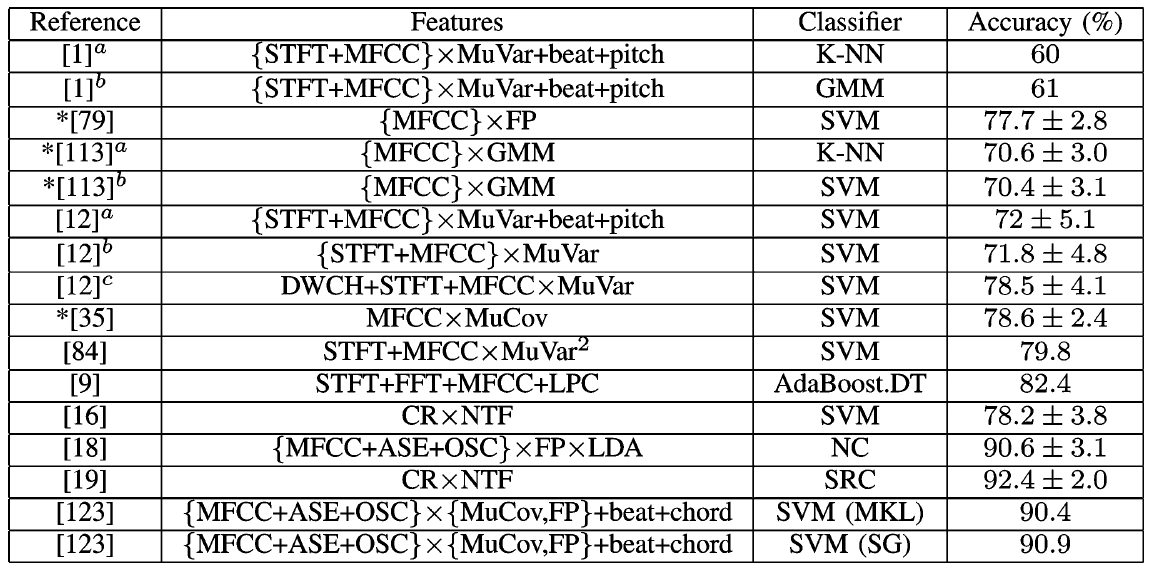}
	\caption{Performance comparison of genre classification algorithms evaluated on the GTZAN dataset (taken from \cite{fu2011survey}. Citations in the table refer to the references of the original paper.)}
	\label{fig:fu2011table2}
\end{figure}

A survey from 2011 \cite{fu2011survey} provides a comprehensive overview of the state-of-the-art in music classification. Their elaborate summary of common low-level audio features is depicted in Figure \ref{fig:fu2011table1}. This table is cited to give a comprehensive overview. The citation numbers within the table refer to the references of the cited paper. For further information please refer to the original publication. Figure \ref{fig:fu2011table2} further provides a good overview of state-of-the-art genre classification systems including comparative performance values (measured in classification accuracy) on the GTZAN dataset. Among the most relevant and top-performing music features for MGR are the \textit{Block-Level Features} \cite{seyerlehner2009block,seyerlehner2010using} and the Rhythm-Patterns feature family \cite{lidy2005evaluation,lidy2007improving}. Both are able to capture music related properties such as timbre and rhythm effectively, which is also demonstrated in the preceding experiments (see Chapter \ref{ch5.2:intro}).

Most recently the attention of the MIR community was brought to \textit{Deep Neural Networks (DNN)} due to their sensational successes in the visual computing domain. Despite their outperforming results, which will be discussed in detail in Section \ref{ch2:soa_cbir}, approaches of the MIR domain are still attempting to find the appropriate architecture to learn music descriptions which outperform conventional hand-crafted music features. This is also the conclusion of a recent comparative study \cite{Costa201728} which provides an evaluation of Convolutional Neural Networks (CNN) for music classification. They suggest to use ensembles of conventional and neural network based approaches. Most DNN based systems use segments of raw or Mel-Log-transformed spectrograms as input for the network \cite{pons2016experimenting}. The Mel-transform as pre-processing step is often applied to rescale a higher dimensional output of a Short-Term Fourier Transform (STFT) to a lower number of frequency bands \cite{shannon2003comparative}. Different architectures have been proposed to modeling temporal \cite{pons2017designing} or timbral features \cite{pons2017timbre} using CNNs.

\begin{table}[t]
	\centering
	\caption{Comparing single handcrafted audio feature classification accuracies (taken from Chapter \ref{ch5.2:intro}) with CNN based approaches (taken from \cite{schindler2016comparing}). Results show top accuracies using different machine learning algorithms for the given feature on the GTZAN \cite{tzanetakis2000marsyas}, ISMIR Genre \cite{Cano2006} and Latin Music Datasets \cite{Jr2008}.}
	\label{tab:comparing_handcrafted_dnn}
	\begin{tabular}{lrrrr}
		\hline
		\rowcolor{gray!30}
		Dataset     & MFCC & RP & TSSD & CNN   \\ 
		\hline
		& & & &  \\ [-1.5ex]
		GTZAN       & 67.8 & 64.9                                            & 66.2 & 82.20 \\ 
		ISMIR Genre  & 62.1 & 75.1                                            & 80.9 & 87.17 \\ 
		Latin Music Dataset      & 60.4 & 86.3                                            & 87.3 & 96.03 \\ 
		& & & &  \\ [-1.5ex]
		\hline
	\end{tabular}
\end{table}

Based on our own experiments with DNNs \cite{schindler2016comparing} where we compared shallow versus deep neural network architectures for automatic music genre classification, or our task-leading 2016 MIREX \cite{lidy2016parallel} or DCASE \cite{lidy2016cqt} contributions, we come to a similar conclusion. DNNs are already outperforming most of the distinct handcrafted audio features, but not approaches using combinations of different audio content descriptors. Table \ref{tab:comparing_handcrafted_dnn} summarizes the top-performing classification results of state-of-the-art hand-crafted feature-sets with prediction accuracies of CNN-based systems. These results were taken from the evaluation presented in Chapter \ref{ch5.2:intro} and the empirical evaluation of CNN-based music genre classification presented in \cite{schindler2016comparing}. Both studies were conducted on the same dataset. 

\subsection{Music Artist Recognition (MAR)}

Early approaches to Music Artist Recognition (MAR) are based on frame-level features such as Mel-Frequency Cepstral Coefficients (MFCCs) and chroma features \cite{ellis2007classifying} in combination with Support Vector Machines (SVM) \cite{kim2002singer,mandel2005song} or ensemble classifiers \cite{bergstra2006aggregate} for classifying music by artist names. In \cite{kim2006towards} a quantitative analysis of the album effect - which refers to effects of post-production filters to create a consistent sound quality across a record -  on artist identification was provided. A Hybrid Singer Identifier (HSI) is proposed by \cite{shen2009novel}. Multiple low-level features are extracted from vocal and non-vocal segments of an audio track and mixture models are used to statistically learn artist characteristics for classification. Further approaches report more robust singer identification through identifying and extracting the singers voice after the track has been segmented into instrumental and vocal sections \cite{mesaros2007singer,tsai2006automatic}. A multi-modal approach using audio and text based features extracted from lyrics was reported in \cite{aryafar2014multimodal}. Recent approaches to artist identification make use of I-vector based factor analysis techniques \cite{eghbal2015vectors}. This approach and its adaption on neural networks \cite{eghbalcosine} show promising improvements on the \textit{artist20} dataset \cite{ellis2007classifying}. DNNs were also reported to perform well using Deep Belief Networks (DBN) \cite{hamel2010learning} or Convolutional Neural Networks (CNN) \cite{dieleman2011audio} to automatically learn feature representations from spectrograms.
Harnessing the visual information of music videos for MAR has not been reported in literature, yet. This approach will be adressed in Chapter \ref{ch6:intro}.


\subsubsection{Music Emotion Recognition}
\label{task_emotion_recognition}

\begin{figure}[t]
	\centering
	\includegraphics[width=0.80\textwidth]{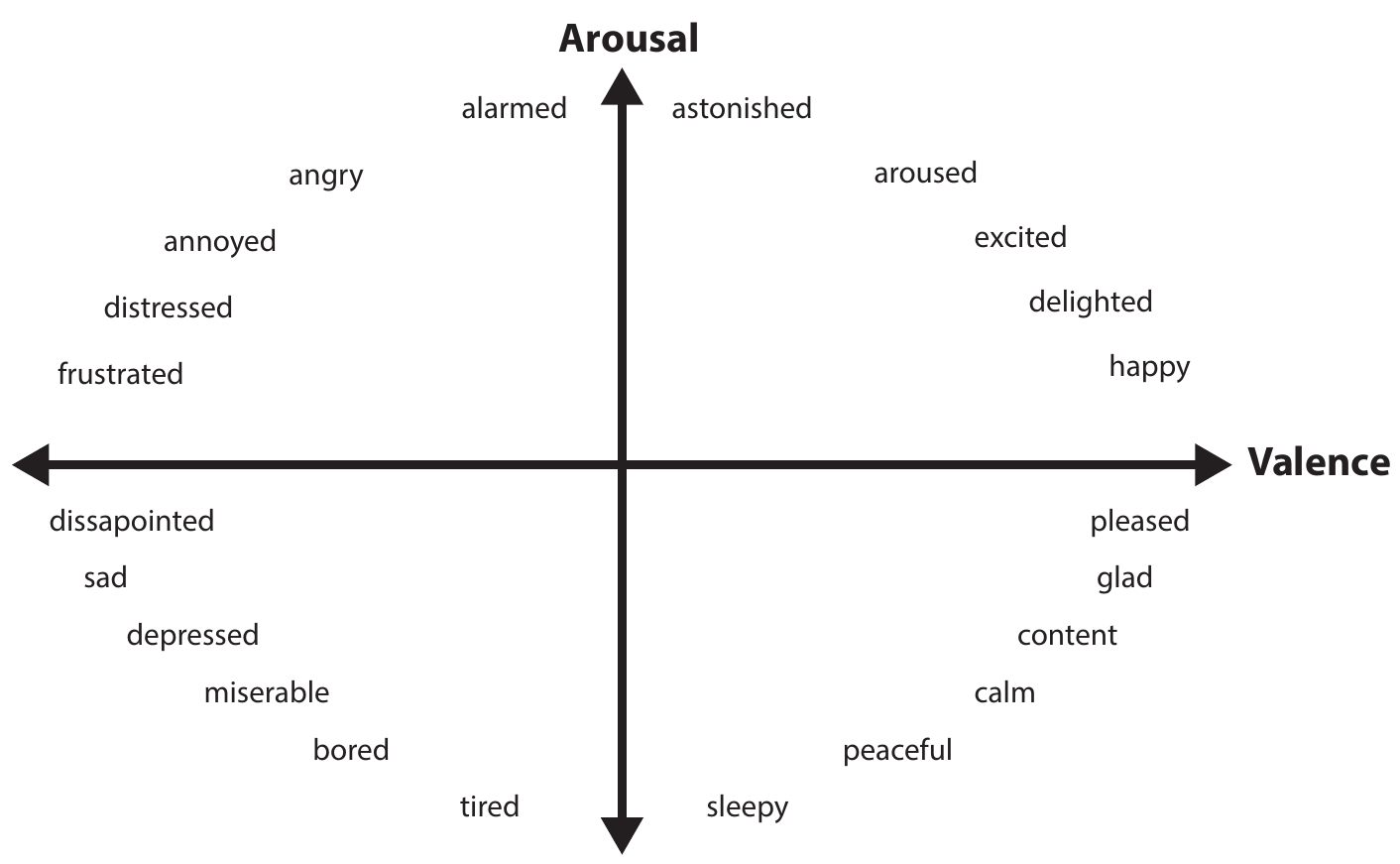}
	\caption{The Valence-Arousal space used in music emotion recognition maps affect words as a function of the degree of valence and arousal \cite{russell1980circumspect}.} 
	\label{fig:ch1:valence_arousal_space}
\end{figure}

There is an affirmed consent going back to Plato about the ability of music to evoke emotional responses in listeners \cite{meyer1956emotion}. A wide range of literature discusses the psychology \cite{deutsch2013} as well as the emotions of music \cite{juslin2011handbook}.
Music emotion recognition attempts to estimate the kind and extent of emotions triggered by a music track.
Emotions are either represented as categorical descriptions (e.g. happy, sad, angry, etc.) or as a psychometric dimensional space. A frequently used representation is based on the Valence-Arousal (V-A) space \cite{russell1980circumspect} (see Figure \ref{fig:ch1:valence_arousal_space}) where affect words are mapped as a function of the degree of valence and arousal. Various further experiments provided word mappings for this model \cite{bradley1999affective} which can be used to estimate the emotions of song's lyrics or tags provided by social platforms \cite{hu2010improving}.
Similar to the previously explained tasks, a common approach is to use regression models based on extracted music descriptors to estimate the corresponding values for valence and arousal. The results are either used directly for similarity estimations or are discretized into categorical labels \cite{kim2010music,yang2011music}.

\subsubsection{Music Similarity Retrieval}

Music Similarity estimation or retrieval is one of the initial research tasks of the MIR domain \cite{logan2001music} and it is still actively researched. The objective is to estimate the notion of similarity between two given tracks. Thus, applying this operation onto an entire collection it should be able to search similar tracks for a given seed song.
The main obstacle researchers focusing on this task face is the subjective nature of human judgments concerning music similarity \cite{berenzweig2004large}. Estimations about song similarities or their categorical belonging to a genre or mood vary in relation to individual tastes and preferences. Further, the concept of music similarity has multiple dimensions including distinct properties such as genre, melody, rhythm, tempo, instrumentation, lyric content, mood, etc. These ambiguities complicate the development of adequate algorithms as well as their valid evaluation \cite{logan2003toward,urbano2010crowdsourcing}. 
A central part of such approaches is the definition of a measure for similarity \cite{berenzweig2004large} which is further affected by the approach taken to extract the relevant information. 

\textit{Content based} approaches largely focus on the notion of \textit{acoustic similarity}. Music features are extracted from the audio content. The resulting music descriptors are high-dimensional numeric vectors and the accumulation of all feature vectors of a collection populates a vector-space. The general principle of content based similarity estimations is based on the assumption that numerical differences are an expression of perceptual dissimilarity. Different metrics such as the Manhattan (L1) or the Euclidean Distance (L2) or non-metric similarity functions such as the Kullback-Leibler divergence are used to estimate the numerical similarity of the feature vectors \cite{berenzweig2004large}.

\textit{Context based} approaches harvest music related information from various sources \cite{knees2013survey}. \textit{Text-based} approaches rely on mined web-pages, music-tags or lyrics \cite{mayer2008rhyme} and apply methods from traditional information retrieval or natural language processing to extract text-based features such as TF-IDF \cite{zobel1998exploring}.
\textit{Co-occurrence based} approaches examine playlists or microblogs \cite{schedl2014harvesting} and estimate the similarity of tracks or artists by the notion how often they co-occur within a dedicated context or by the conditional probability that one track or artist is found within the same context as the other \cite{schedl2005web}.


\subsubsection{Music Recommendation and Discovery}

The recent shift of the music industry from traditional to digital distribution through online music services such as Spotify, Deezer or Amazon has made automatic music recommendation and discovery an increasingly interesting and relevant problem \cite{song2012survey}.
Its objective is to let listeners discover new music according their tastes and preferences. An ideal recommender system should automatically detect current as well as adapt to changing preferences and suggest songs or generate playlists accordingly.

Most music recommendation systems rely on Collaborative Filtering (CF) \cite{Celma2010,schedl2015music}. One of the biggest problems of these approaches is known as the \textit{cold start problem} \cite{schein2002methods}. If no usage data for a given item is available, the system fails to provide results. In the music domain this refers to new or unpopular songs, albums or artists. 
On the opposite, songs with already assigned usage data will be included in recommendations more often and thus popular songs become more popular while unknown ones remain further so. This is also referred to as the \textit{long tail problem} \cite{celma2009music} where a small percentage of the collection is recommended with high frequency, whereas the majority of it - the \textit{long tail} - is seldom discovered.

Generally, a recommender system consists of the three components \cite{song2012survey}. 
\textit{User modeling}, as one of the key elements, attempts to capture the variations in user-profiles such as gender, age, geographic region, life style or interests. All of these parameters might affect the user's music preferences. Additionally to user-profile models, user models depend also on user listening experience models.
\textit{Item modeling} takes advantage of all music information available and accessible in MIR research including music context as well as content.
\textit{Meta-data retrieval} based systems rely on curated textual information provided by the creators or content owners, such as artist, title, album name, lyrics, etc. The obstacle of this approach is the cumbersome task of creating and maintaining the required meta-data. Further, recommendations are only based on information annotated in the meta-data, which often does not reflect intrinsic music properties such as rhythm or mood nor does it include descriptions about the users.
\textit{Collaborative filtering} approaches can be generalized into \textit{memory based} systems which predict items based on all previously collected ratings and \textit{model based} systems which use machine learning to train user preference models for the item predictions.
\textit{Content based approaches} rely on the audio content and thus require access to the audio file. A wide range of already discussed techniques is used such as estimating similarity of songs or predicting social tags \cite{eck2008automatic} through single or multilabel classification. 
A final distinction can be made between \textit{emotion based}, \textit{context based} and \textit{hybrid} models. The first attempts to map emotions expressed by a song onto the valence-arousal space (see Section \ref{task_emotion_recognition}) and to recommend songs based on minimal distance to a query song.
\textit{Hybrid} models try to combine two or more	 approaches to overcome limitations and improve precision.

\subsubsection{Score following / Music Alignment}
Score following relates to the process of aligning the notes of a score to its interpretation during a specific performance. This might be during a live performance, or to an audio recording. Research on automatic score following through computers distinguishes between different scenarios \textit{Symbolic to MIDI} and \textit{Symbolic to Audio}.
Their main differences are based on the digital representation of music. Symbolic music is usually represented in a machine processable form, such as MIDI or MusicXML. Recorded music is represented as a time-series of sampled audio. More generally speaking, sampled audio is an exhaustive sequence of numbers representing the measured auditive energy at a certain time. Information provided in this form is not suitable for automatic processing at first hand and has to be reduced and transformed into an appropriate representation. This process is the focus of the research domain music information retrieval (MIR). While, in the symbolic representation of music, it is clear which note is being played at a certain time, in digital audio this information has to be deduced from the spectral properties of the sampled audio through digital signal processing.

\paragraph{Symbolic to MIDI alignment:} This scenario is based on the prerequisite that the performed music is available in the same symbolic format of the score. This is the case for set ups where a musical instrument is connected to the computer through a Musical Instrument Digital Interface (MIDI). Music events such as hitting a key on an electronic piano, are immediately communicated to the connected computer and are available in an interpretable from. Having both sources - scores and performance - in a comparable format makes it more convenient for further processing (e.g. score following, automatically assisted training, etc.).

\begin{figure}[t]
	\centering
	\includegraphics[width=1.0\linewidth]{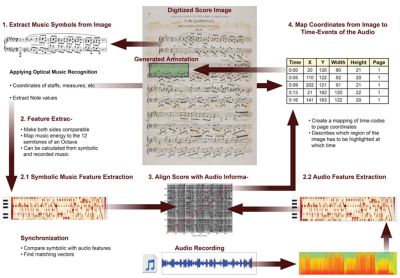}
	\caption{Visualization of a audio-to-score alignment system developed during the Europeana-Sounds project (www.europeanasounds.eu). The system consists of two main parts. 1. applying Optical Music Recognition (OMR) to extract symbolic music information from scanned score images. 2. extract Chroma features from symbolic and recorded music. 3. apply Dynamic Time-Warping (DTW) to align the extracted features. 4. map synchronized time information to coordinates and page-numbers of the scanned images.}
	\label{fig:scorefollowing}
\end{figure}

\paragraph{Symbolic to Audio alignment:} While symbolic music unambiguously describes which note is played at which time of the track, this does not apply to recorded music. The main challenge with sampled audio is that it is a mix of frequencies, originating usually from a multitude of individual instruments and voices, which sources currently cannot fully be separated again, after they have been fixed in an audio mix. Musical notes refer to audio frequencies (e.g. concert pitch = 440Hz). Thus, it seems obvious, that sampled audio can be transcribed into symbolic music by assigning note values to audio frequencies. In a simplified approach this works for monophonic tunes played by a single instrument. Having multiple instruments playing polyphonic tunes (like chords and harmonies) creates overlapping frequencies, partial frequencies caused by the instrument’s timbre and other influences in the overall audio mix, which cause complex distributions of the sound energy over the frequency spectrum of the recording. Thus, it is not computationally distinguishable anymore which notes have been played by the distinct instruments. To align symbolic music to sampled audio, both types have to be transformed into a representation that can be compared directly. Common approaches to audio-to-score alignment systems is to convert both music modalities into a comparable representation. A common choice are Chroma features \cite{ellis2007classifying} because they can easily be extracted from both modalities. The time-synchronization is then performed on this feature-space. Figure \ref{fig:scorefollowing} depicts an audio-to-score alignment system that has been implemented for the Europeana Sounds\footnote{\url{https://www.europeanasounds.eu/}} project.


\subsection{Multi-modal Approaches to MIR}

Multi-modal systems approach MIR tasks by utilizing music information of different modalities such as song lyrics \cite{mayer2008rhyme,hu2010improving}, web pages \cite{schedl2006assigning} and social media/tagging \cite{lamere2008social}. Visual related music information extraction has been reported utilizing album art images for MGR \cite{mayer2011}, artist identification \cite{Libeks2011} and similarity retrieval \cite{brochu2003sound}. Image features extracted from promotional photos of artists are used to estimate the musical genre of the image \cite{libeks2010exploring}. A multi-modal approach to mood classification using audio and image features is reported in \cite{dunker2008content}. Perceptual relationships between art in the auditory and visual domains were analyzed in \cite{mattek2011cross}.

The difficulties of developing high-level audio features have triggered the search for different sources to pull information from - introducing a multi-modal view of music. Approaches combining audio with text using information extracted from lyrics have been reported for automatic genre \cite{mayer2008rhyme} and mood \cite{laurier2008multimodal} recognition. Further attempts to improve music genre recognition by adding information from different modalities included combining audio content with symbolic features extracted from midi \cite{lidy2007improving} as well as social media data \cite{celma2006foafing}. In the video retrieval domain using music to discriminate video genres has recently gained a lot of attention \cite{brezeale2008automatic}. Different acoustic styles are used to estimate the video genre, as certain types of music are chosen to create specific emotions and tensions in the viewer \cite{moncrieff2003horror}. The video retrieval domain generally observes music videos as a video genre of its own and further sub-categorization in different classes (e.g. musical genre, type, etc.) has been of no interest, yet. Using the visual layer to make estimations about audio is a nearly unexplored field. Multi-modal approaches have only been reported for automatic music video summarization \cite{shao2006automatic} and indexing \cite{gillet2007correlation}. The MIRES road-map for music information research \cite{MiresRoadmap2013} identifies music videos as a potential rich source of additional information. As an open research it requires more attention of the audio/visual communities as well as appropriate data-sets.

\section{Content Based Image Retrieval}
\label{ch2:soa_cbir}

Content Based Image Retrieval (CBIR) is a research domain with a very long tradition. On that account a series of survey papers has already been provided \cite{tamura1984image,doermann1998indexing,veltkamp2001content,liu2007survey,singhai2010survey,rehman2012content,dharani2013survey}. Generally, CBIR systems were categorized into those using low-level images features such as \textit{color} \cite{manjunath2001color,plataniotis2000color,schettini2001survey}, \textit{image segmentation} \cite{ma1997edge,shi2000normalized,feng2001curve} or \textit{texture} \cite{sethi2001mining,tamura1978textural,liu1996periodicity} including Local Binary Patterns (LBP) \cite{ojala2002multiresolution}. CBIR experienced a huge boost of performance \cite{mikolajczyk2005performance} with the introduction of interest point based approaches such as the original Scale Invariant Feature Transforms (SIFT) \cite{lowe2004distinctive} which is acknowledged to be one of the most influential publications in CBIR or one of their successors such as the Speeded-Up Robust Features (SURF) \cite{bay2008speeded}. 

Recently, CBIR has experienced remarkable progress in the fields of image recognition by adopting methods from the area of deep learning using convolutional neural networks (CNNs). A full review of deep learning and convolutional neural networks is provided by \cite{chatfield2014return}. Neural networks and CNNs are not new technologies, but with early successes such as LeNet \cite{lecun1989optimal}, it is only recently that they have shown competitive results for tasks such as in the ILSVRC2012 image classification Challenge \cite{krizhevsky2012imagenet}. With this remarkable reduction in a previously stalling error-rate there has been an explosion of interest in CNNs. Many new architectures and approaches were presented such as \textit{GoogLeNet} \cite{szegedy2015going}, \textit{Deep Residual Networks (ResNets)} \cite{he2016deep} or the \textit{Inception Architecture} \cite{szegedy2015going}. Neural networks have also been applied to metrics learning \cite{jain2012metric} with applications in image similarity estimation and visual search. \textit{Siamese network} architectures are trained on pairs \cite{bell2015learning} or triplets \cite{hoffer2015deep} of images. By using a distance metric as a loss function the network train on general image similarity instead of discrete concepts. 
The similarity of images is then directly estimated by resulting model  or by the provided learned embeddings which can be used for visual search. Also multimodal models using visual and textual input have been reported recently. These models are trained for image captioning and visual scene understanding \cite{mao2014explain}.

\subsection{Face Recognition}

Good summaries of state-of-the-art approaches and challenges in face recognition are provided by \cite{li2011handbook,phillips2005overview,zhao2003face}. Face detection, tracking and recognition is also used in multi-modal video retrieval \cite{lew2006content,snoek2005multimodal}. Faces are either used to count persons or to identify actors. Most common methods used to recognize faces are Eigenfaces \cite{turk1991eigenfaces} and Fisherfaces \cite{belhumeur1997eigenfaces}. In \cite{dimitrova2000video} face tracking and text trajectories are used with Hidden Markov Models (HMM) for video classification. A face recognition approach based on real-world video data is reported in \cite{stallkamp2007video}.

\section{Content Based Video Retrieval (CBVR)}
\label{ch2:soa_cbvir}

\begin{figure}
	\centering
	\includegraphics[width=1.0\linewidth]{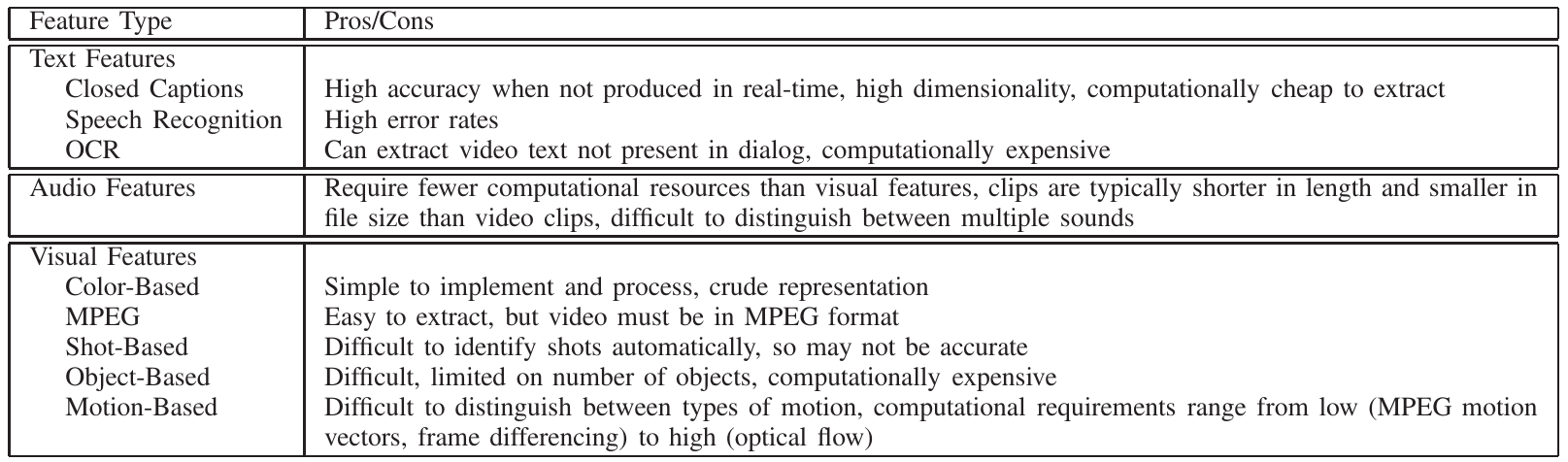}
	\caption{Comparison of pros and cons of different classification approaches and the correspondingly used features (cited from \cite{brezeale2008automatic})}
	\label{fig:brezeale2008automatictable1}
\end{figure}

A comprehensive survey on Content Based Video Retrieval (CBVR) is provided by \cite{hu2011survey}. Generally, CBIR approaches can be categorized into \textit{video structure analysis}, including the tasks \textit{Shot Boundary Detection} \cite{smeaton2010video}, \textit{Key Frame Extraction} and \textit{Scene Segmentation} \cite{koprinska2001temporal},\textit{semantic video search} \cite{hu2011survey} and \textit{classification} \cite{brezeale2008automatic}. Especially, \cite{brezeale2008automatic} provides a very comprehensive overview of relevant concepts of video classification methods. Video classification follows the same paradigm of almost all multimedia retrieval approaches. First descriptive features are extracted from the content, then a model is trained usually using elaborated machine learning algorithms. Generally, three major approaches can be identified in literature: \textit{Audio-based} \cite{roach2001classification}, \textit{Text-based} \cite{brezeale2006using}, \textit{Visual-based} \cite{hauptmann2002video} and ensemble-methods utilizing multiple modalities \cite{wang2003hybrid} (see Table \ref{fig:brezeale2008automatictable1} which is cited from \cite{brezeale2008automatic}). Visual-based systems can further be grouped into \textit{Color-based} \cite{hauptmann2002video}, \textit{Shot-based} \cite{truong2000automatic}, \textit{Object-based} \cite{wang2003hybrid}, \textit{MPEG-based} \cite{jasinschi2001automatic}, and \textit{Motion-based} \cite{fischer2004automatic} approaches. The approaches presented in this thesis fall into the categories \textit{color-} and \textit{object-based} systems. In Chapter \ref{ch7:intro} a wide range of color- and texture-features are evaluated towards their performance in classifying music videos by their music genre. Object-based approaches are presented in Chapter \ref{ch6:intro} where faces are extracted from the music video-fames to recognize the performing artist and in Chapter \ref{ch8:intro} high-level concepts are extracted from the video-frames to predict the music genre.

\section{Music Videos Analysis}
\label{ch2:soa_mvir}

Multi media analysis of music videos is part of reported automatic \textbf{music video creation systems}. An approach based on audio and video segmentation was presented in \cite{foote2002creating}. Segments of the source video are selected based on a calculated suitability score and are combined to create a new music video. 
The two studies presented in \cite{hua2004automatic} and \cite{yoon2009automated} which build on the work presented in \cite{foote2002creating} automatically select and join music video-like videos from personal home videos. 
In \cite{cai2007automated} the authors describe an approach to extract salient words or phrases from the lyrics and uses them to search for corresponding images on the web. Also contrast features similar to those used in this publication were used.

In \cite{tsukuda2015exploratoryvideosearch} a \textbf{music video retrieval system} is presented based on coordinate terms and diversification, using artist name queries and coordinate terms derived from the Million Song Dataset \cite{bertin2011million}. An approach to automatically determine regions of interest in music videos is reported in \cite{kim2007automatic}. An approach to automatic music video summarization is presented in \cite{shao2006automatic}. An audio-visual approach to segmentations of music videos was proposed in \cite{gillet2007correlation}, including an evaluation of audio-visual correlations with an intended application in audio retrieval from video. Approaches to affective content analysis of music videos are provided by \cite{zhang2010affective} and \cite{yazdani2011affective}. An approach using convolutional neural networks (CNNs), in order to learn mid-level representations from combinations of low-level MFCCs and RGB color-values in order to build higher level audio-visual representations, is presented in \cite{acar2014understanding}. The approach is evaluated on the DEAP dataset \cite{koelstra2012deap}, a music video dataset with valence and arousal valued ground truth data. \cite{Xiangjian2015} describes an audio-visual approach for a recommender system that uses a video as input query. To index music via the video domain mood values calculated from the audio content were matched to mood information derived from the color information. The visual features used by the authors \cite{valdez1994effects} were also evaluated in this publication. Unfortunately we were not able to draw the same conclusions about the effectiveness of these features for music classification. Recently, a similar approach as presented in Chapter \ref{ch8:intro} was reported \cite{hong2017deep}, also using a combination of Convoluational Neural Networks (CNN) and music descriptors to find similar music or videos.

\section{Transfer Learning}

\textit{Transfer Learning} refers to concept of transferring knowledge from a related task that has already been learned within a given context to a problem of a different context or even a different research domain \cite{pan2010survey}. A well known example is the transfer of the well evaluated automatic speech recognition method based on MFCC to the MIR domain \cite{logan2000mel}. A comprehensive summary of further examples is provided by \cite{weninger2012music}. In this study we transfer knowledge from the visual computing domain to MIR by applying trained visual concept detection to music classification tasks.

\section{Summary}

This section provided an overview of the state-of-the-art in topics relevant for this theses. It first provided a broad overview of the research domain Music Information retrieval. Due to the focus of this thesis to augment MIR based approaches by harnessing the information provided by the visual layer of music videos, such an elaborated introduction provides the relevant basis for the discussion of the potential application fields of an audio-visual approach to MIR.

The consecutive sections elaborated which multi-modal, especially audio-visual, approaches have been reported in the specific domains such as image processing, visual computing or MIR as well as their applications to MIR domain.
This literature review indicated that there is a lack of basic research concerning audio-visual approaches to MIR problems. Identified studies reported improvements including visual information in classification or retrieval experiments, but often failed to analyze the contribution of the distinct visual features on the result.



\chapter{Audio-Visual Relationships in Music}
\label{ch:visualmusic}
\label{ch3:intro}

\epigraph{
	``Albums should be as bold and dashing as we can make them; they should stand out in dealers' windows screaming for attention, yet always reflecting the spirit of the music inside. Color should be violent and strong. Copy should be pared to a minimum, and each album should reflect the quality of the Columbia name''} 
{--- Pat Dolan, CBS, 1940}


Music is an acoustic phenomenon. To experience music it is unavoidable to listen to it. In times of cellular networks with permanently connected devices, where music is streamed over the Internet and can be consumed at any place any time, this condition does not seem to be a problem. But, listening to music was not always that convenient. Looking back two decades, the major music distribution media was the Compact Disc (CD) which replaced the cassette in the early 90s which in turn succeeded vinyl LPs in the 80s. Music was not sold on-line where it is possible to conveniently browse through catalogs and listen to provided samples. One needed to go to record stores browse through shelves of records, pick some interesting albums, queue up in front of a record player provided by the store and listen to the record. Thus, discovering new music was a time-consuming task. Music genres are a convenient aid for describing music tastes or for navigating trough on-line and off-line music stores. They are synonyms for music styles and aggregate several music attributes which are distinct for the certain genre. Large signs in record stores with music genres printed in large letters show the way to the desired music. The record shelves themselves are structured according diversified sub-genre categories which provide a more fine-grained discrimination of music styles. But, despite all these subsidiary synonymic music descriptions, finally the customer still has to make a decision between a large set of records - and, although a bit further, is still facing the same problem - the music is not directly tangible. No matter if it is available on Vinyl, CD, cassette, DVD or even as printed scores - it is not easily possible to get an impression of its content. The only information provided at hand is of visual kind and the most prevalent of it is the cover.

\begin{figure}[t]
	
	\centering
	
	\subfloat[Music sheet cover 'The Brigand's Ritornella' from Auber's opera Masaniello, showing the image of the singer Madam Vestris who was a hugely popular star of burlesque at the Paris Olympic Theatre, 1835.]{\includegraphics[width=0.435\textwidth]{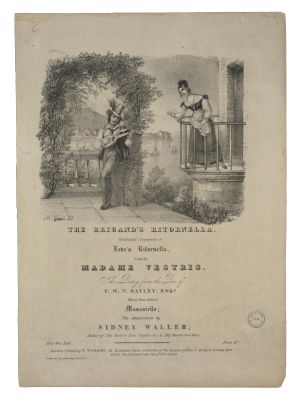}}
	\qquad
	\subfloat[Music sheet cover 'The National Football Song' designed by H. G. Banks depicts popular football players, about 1880.]{\includegraphics[width=0.44\textwidth]{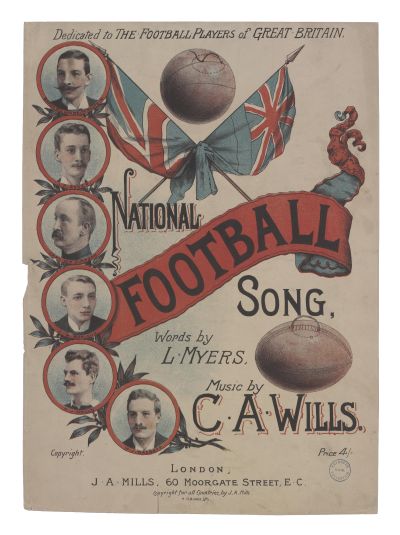}}
	\caption{Early examples of sheet music and song book illustrations.}%
	\label{ch3:early_sheet_music}
\end{figure}

\noindent
Using decorative images and paintings to advertise music is not a modern concept. In 1820 artists became famous for their cover designs of songbooks and music scores \cite{axford2004song} (see Figure \ref{ch3:early_sheet_music}). Further, the development of lithography had a strong impact on cover art design for music sheets \cite{piola2014}.

This chapter provides a structured overview of how visual accentuation is used in music marketing and how this has developed over time. As a first example album cover arts will be explored by giving a brief introduction to their history and their common themes. The role of visual accentuation will then be discussed in reference to music marketing strategies. Finally, the broad visual aspects provided by music videos are elaborated.

\section{When Packaging Music Became asn Art-Form}
\label{ch3:albumart_history}

This section briefly summarizes Jones and Sorger's article \cite{jones1999steve}. For a detailed insight in the history of album cover design please refer to this excellent article.\\

\begin{figure}[t]
	\centering
	\subfloat[Edison Gold Moulded record made of relatively hard black wax, 1904. \textsuperscript{\textcopyright} Wikipedia]{\includegraphics[width=0.47\textwidth]{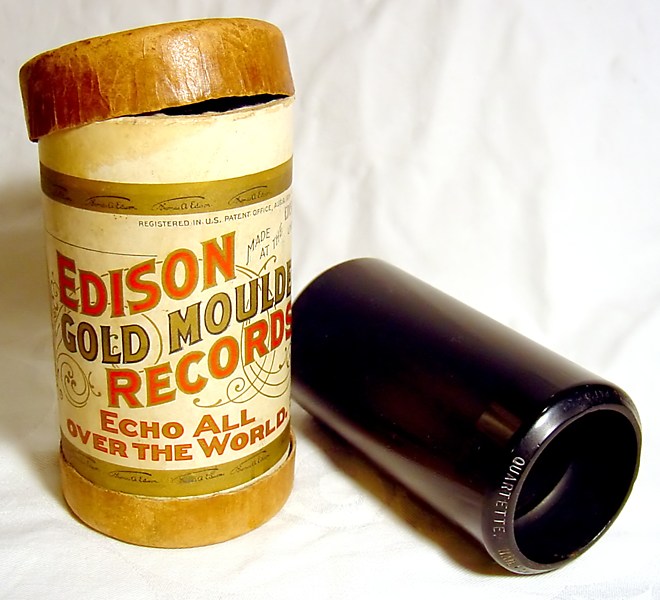}}
	\qquad
	\subfloat[78rpm record with sleeve by Victor Talking Machine Company, 1928. \textsuperscript{\textcopyright} Jeff Crompton ]{\includegraphics[width=0.435\textwidth]{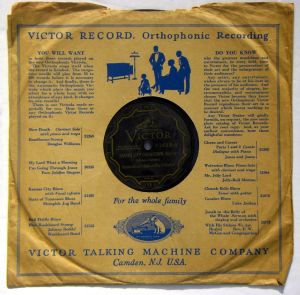}}
	
	\caption{Examples of early record covers. Their initial purpose was to protect the media. Less attention was set on design.}
	\label{fig:ch3:early_covers}
\end{figure}

The history of album cover design begins with the establishment of phonograph companies which started to sell prerecorded musical selections. In 1896 Thomas Edison established the \textit{National Phonograph Company} to sell music pressed on wax cylinders for the phonograph. Due to the fragility of the cylinders cardboard boxes lined with felt were used for packaging (see Figure \ref{fig:ch3:early_covers} a). Their standardized design focused mainly on highlighting the company name.
In 1901 the \textit{Victor Talking Machine Company} introduced the 78-rpm flat disc phonographs. Although, early discs were shipped sleeveless and thus, without protection, sleeves have become an industry-wide standard by 1910. The more common design type consisted of a blank sleeve with a cut out hole in the center to reveal the label of the record (see Figure \ref{fig:ch3:early_covers} b).
The first durable sleeves made of cardboards were produced by record stores which initially only had the store's name printed on it but later incorporated photographs and portraits of the performing artist or the composer.
The denotation of the record as an album originates from that epoch. Recordings of classical music such as concerts or operas exceeded the length of a single 78-rpm shellac disc. Usually three to five discs were required for longer orchestral performances. The design of the card-boxes used to distribute these recordings reminded of photo albums which soon became a frequently used synonym.

\begin{figure}[t]
	\centering
	\subfloat[Cover of \textit{The Velvet Underground \& Nico} designed by Andy Warhol, 1967]{\includegraphics[width=0.44\textwidth]{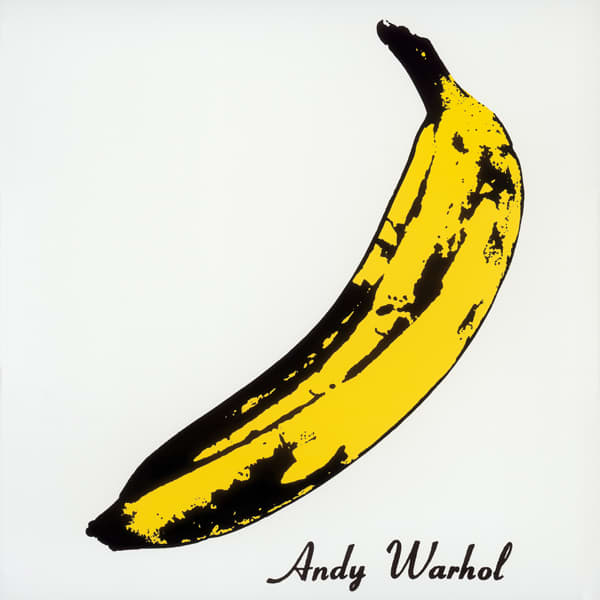}}
	\qquad
	\subfloat[Cover of \textit{The Rolling Stones - Sticky Fingers. The cover featured a functional zipper which was embedded in the cardboard.}, 1971]{\includegraphics[width=0.435\textwidth]{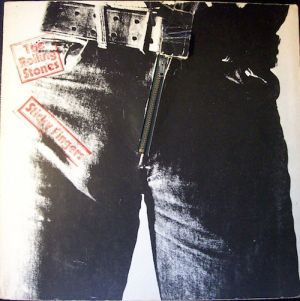}}
	
	\caption{Examples of album cover designs of the late 1960s.}
	\vspace{-2.5mm}
	\label{fig:ch3:1960s_covers}
\end{figure}

After the \textit{Great Depression} album cover design developed and intensified from abstract wallpaper-like designs in the 1930s to a refined marketing tool of major music companies. Designing cover arts became an attractive field for artists. 
Especially, as record stores changed to self-service selection in the 1940s, providing customers with racks where they could flip through the records on their own, the album cover became a direct marketing tool.
In 1948 Columbia introduced the \textit{Long-Playing record} (LP) which quickly became the new standard. This more fragile LP required a new packaging which now hid the entire record and thus provided more space for designers. Rock-and-roll marketing of the 1950s was related to movies. Records were covered with large portraits of the performing artists. Only Jazz album covers were artistically more creative.
While LPs in rock and pop music were mostly used as a collection of hit-singles, bands in the 1960s started to sell complete albums with new compositions. Further, they began to get involved in album cover design themselves, using well-known designers such as Richard Hamilton or Andy Warhol (see Figure \ref{fig:ch3:1960s_covers} a). The Beatle's \textit{Sgt. Pepper's Lonely Hearts Club Band} is often cited as one of the most familiar and influential records of this time. It was the first to contain an inner sleeve, printed lyrics and a gate-fold to contain cards with cut-outs.
By taking over the complete design of their album, the Beatles have set a new measure of album cover design which now more and more acted as an \textit{audio-visual experience} (see Figure \ref{fig:ch3:1960s_covers} b).
The visual design continuously gained importance and became a part of the marketing strategy. Artists developed individual styles that were repetitively used on their releases such as stylized letters or band-logos.

The appearance of Punk Rock at the end of 1975 introduced a wide range of small independent labels. Punk Rock cover design reflected the revolting nature of the genre which was in clear opposition to established aesthetic norms.  Similar to the music, whose simple structure opposed the overly complex compositions of Hard and Glam Rock bands, Punk album covers were simple. \textit{New Wave} which is the succeeding movement of Punk created a wide range of new sounds and styles which again put more emphasis on album cover design. 

In 1983 the \textit{Compact Disc (CD)} was introduced and thus the digital age of music. The new packaging format, the \textit{jewel box}, only offered space for a small square booklet. Besides the smaller front to advertise the record, there was also no space to add additional properties such as posters or cards. This minimalistic visual experience was often criticized.
With the advent of online music, the media completely lost its haptic property. Listening to music was reduced to its intrinsic nature and the holistic experience of flipping through album cover pages or CD booklets was withdrawn. 

\begin{figure}[t]
	
	\centering
	\subfloat[\textbf{Sex}, \textit{The Cars - Candy O, 1979}]{\includegraphics[width=0.29\textwidth]{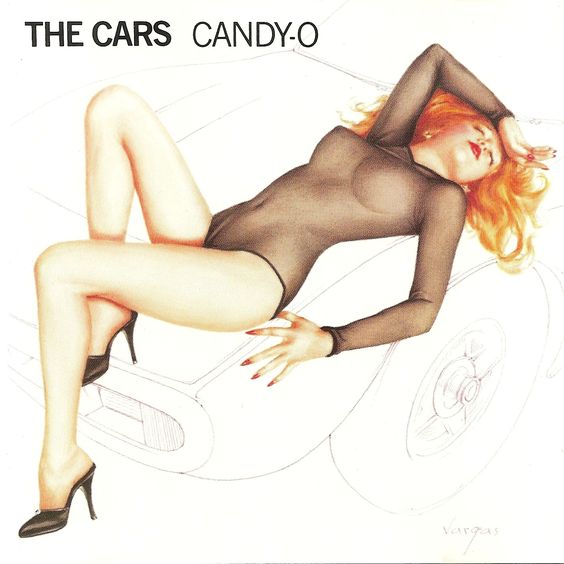}}
	\qquad
	\subfloat[\textbf{Art}, \textit{Led Zeppelin - III, 1970}]{\includegraphics[width=0.29\textwidth]{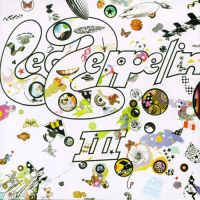}}
	\qquad
	\subfloat[\textbf{Identity}, \textit{Chicago - X, 1976}]{\includegraphics[width=0.29\textwidth]{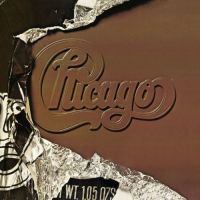}}
	\qquad
	\subfloat[\textbf{Drugs}, \textit{Cream - Disraeli Gears, 1967}]{\includegraphics[width=0.29\textwidth]{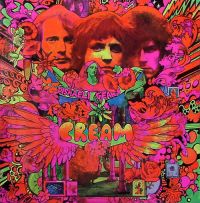}}
	\qquad
	\subfloat[\textbf{Artist}, \textit{The Ramones - Ramones, 1976}]{\includegraphics[width=0.29\textwidth]{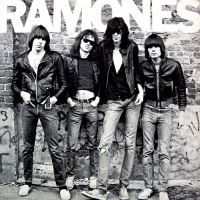}}
	\qquad
	\subfloat[\textbf{Politics}, \textit{U2 - War, 1983}]{\includegraphics[width=0.29\textwidth]{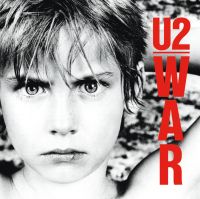}}
	\qquad
	\subfloat[\textbf{Death}, \textit{Slayer - Seasons in the Abyss, 1990}]{\includegraphics[width=0.29\textwidth]{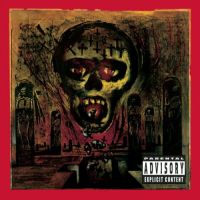}}
	\qquad
	\subfloat[\textbf{Face Close-ups}, \textit{Bob Marley \& The Wailers - Legend, 1984}]{\includegraphics[width=0.29\textwidth]{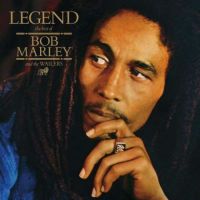}}
	\qquad
	\subfloat[\textbf{Text Only}, \textit{John Mayer - Continuum, 2006}]{\includegraphics[width=0.29\textwidth]{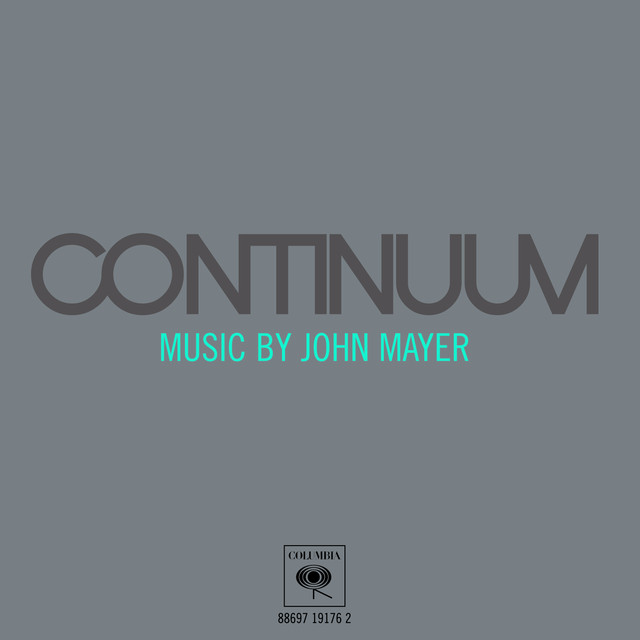}}

	\caption{Overview of common types, styles and subjects shown on album cover-arts.}
	\label{fig:ch3:album_cover_subjects}
\end{figure}

\noindent
Figure \ref{fig:ch3:album_cover_subjects} gives an overview of common album-cover themes. 

\clearpage

\section{Music Videos}
\label{ch3:mv_overview}

Initially called \textit{promotional videos} music video production had its break through in the early 80s. Music videos were one of the major music inventions of that decade, although they were initially criticized to provoke a diminishing of the interpretative liberty of the individual music listener by imposing the narrative visual impression upon him or her \cite{frith2005sound}. In that period the structure of the prevalent music business changed completely by shifting the focus from selling whole records to hit singles \cite{frith2005sound}. The entire marketing campaign of the related record was built around that specific song and with the new intensification of celebrity around pop music, music videos played an important role in the development of the \textit{pinup culture}.
With their increasing popularity music videos got more complex and soon became an art form themselves. Contemporary video production typically uses a wide range of film making techniques and roles, including screen-play, directors, producers, director of photography, etc. Further, as cited in \cite{caston2014fine} ``academics are finally starting to recognize that music videos are a \textit{persistent cultural form} that has outlived one of their initial commercial functions as a promotional tool''.
This section provides a brief over\-view of music video types, concepts and properties, exceeding those relevant for the experiments performed in this thesis to provide a basis for potential future work.

\subsection{History of Music Videos}

The hybrid nature of music videos puts them in an overlapping space between being musical and visual. The two major positions on whether the one or the other part is privileged are represented by E. Ann Kaplan's book \textit{Rocking around the Clock: Music Television} \cite{kaplan1987rocking} and Andrew Goodwin's \textit{Dancing in the Distraction Factory} \cite{goodwin1992dancing}. Kaplan argues from a film theoretical standpoint, favoring the visual dimension while Goodwin argues that these visual aspects are only secondary to the music itself.

\begin{figure}[t]
	
	\centering
	
	\subfloat[The \textit{Movie Machine} or \textit{Panoram Soundie} by Mills Novelty Company, 1940.]{\includegraphics[width=0.335\textwidth]{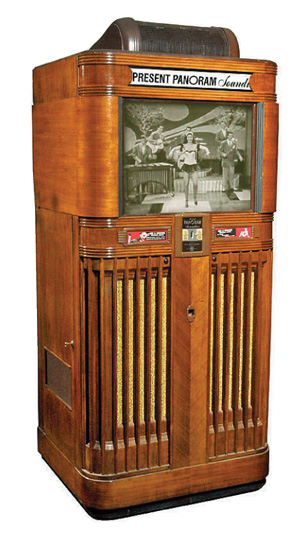}}
	\qquad
	\subfloat[Scopitone machine, 1960.]{\includegraphics[width=0.31\textwidth]{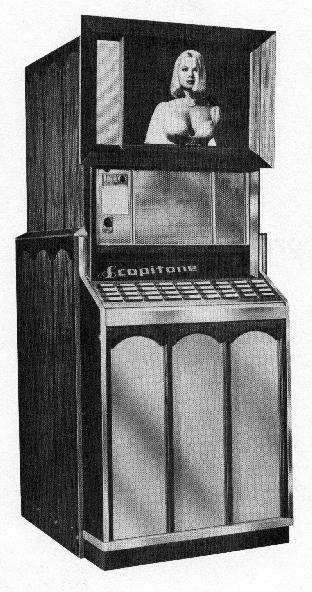}}
	\caption{Early cinematic jukeboxes. These coin-operated machines showed short music films.}%
	\label{ch3:early_music_videos}
\end{figure}

Although the history of music videos is often brought into context with the launch of MTV, the relationship of music and moving images started much earlier \cite{dickinson2007medium}. In 1940 the Panoram \textit{Movie Machine} was introduced by the Mills Novelty Company (see Figure \ref{ch3:early_music_videos}a). Those were coin-operated cinematic jukeboxes containing a screen on which three minute musical films could be watched. Those so called \textit{Soundies} \cite{herzog2004musical} featured musical styles ranging from big-band swing, American country, romantic ballads to Latin and Jazz. Like music videos they were promotional films produced and edited to pre-recorded music. Their visual style included vaude-ville-style and burlesque but also world-war I related and comedian acts, but did not differ much from \textit{musical shorts} which were popular in cinemas in this period. In the 1960s a similar sound film projection system, the \textit{Scopitone} \cite{Duval1965}, was introduced and marketed (see Figure \ref{ch3:early_music_videos}b). The improvement was, that the \textit{music shorts} were not played sequentially but could be chosen from among 36 different 16-millimeter Technicolor films of B-list pop stars performing their latest hits. Scopitone films were deliberately designed to appeal to a target audience of men in bars which was where these machines could primarily be found. Soundies and Scopitones set themselves apart from conventional film by being entirely music based whereas most of the literature on film music looks at music's role as a supplement to the visual presentation \cite{herzog2004musical}. Further were their paths of distribution and consumption more similar to recorded music.

In the 1960s various music bands began to create promotional short films for their songs. In 1964 \textit{The Moody Blues} promoted their single ``Go Now'' using a short clip where the band was portrayed while singing the song and playing instruments. This clip already featured many elements that are common in contemporary music videos and described in Chapter \ref{ch3:mv_types}. \textit{The Beatles} created feature films for their albums such as ``A Hard Day's Night'' (1964), ``Help!'' (1965) and ``Yellow Submarine'' (1968) where they played the main roles and also performed to their own songs. Further the Beatles creates several short music films for their singles such as ``Strawberry Fields Forever'' and ``Penny Lane'' (1967) which also featured many concepts of contemporary music video productions.
In the 1970s many music artists started to create promotional videos which were shown in music shows such as BBC production \textit{Top of the Pops} \cite{keazor2010music}. The most popular example of these is the video for \textit{Queen}'s ``Bohemian Rhapsody'' (1975).

Finally, music television \textit{MTV} was launched in August 1981 with the video ``Video Killed the Radio Star'' by \textit{The Buggles}. Still, music videos were low in number and the quality of the produced clips was low due to lack of experience of film makers with the medium of music and the yet not elaborated processes. Style and design features as explained in Chapter \ref{ch3:mv_types} and \ref{ch3:mv_styles} in more detail, slowly developed through the explosive increase of interest and productions. Music videos became a forerunner in visual aesthetics and influenced other media genres such as musicals, advertisement and feature films. While in the 1990s their popularity reached a zenith, due to the budget spent as well as the artistic and technological complexity, music videos emerged into a high-art form, from which many elements found its way into cinematic productions such as the \textit{Frozen Moment} of Michel Gondry's video for \textit{Like a Rolling Stone} by \textit{The Rolling Stones} in 1995 which was used in the Hollywood movie \textit{The Matrix} in 1999 \cite{keazor2010music}.

The downturn the music industry experienced in the late 1990s led to huge cuts in the budgets for music videos which is one of many reason that led to a still ongoing crisis of this medium. Further, MTV continuously changed its program from music videos to teenager shows and finally lost its significance to the internet where music videos can be accessed on demand.

\subsection{Types of Music Videos}
\label{ch3:mv_types}

Music videos are short films intended to promote a song and its performing artist. The accentuation of the music video as an artistic work added a further semantic layer to a	composition. Above the traditional audio and lyrical layers the visual part can be used to either illustrate, amplify or contradict the meaning of the underlying song. There are typically three ways to approach this goal \cite{frith1993sound}:

\begin{description}[leftmargin=!,itemsep=1pt,labelwidth=\widthof{\bfseries Amplification a},labelindent=\descleftmargin,rightmargin=\descrightmargin] 
	
	\item [Illustration]
	The visual layer illustrates and further explains the meaning of the song by either using a narrative plot or visual clues that are related to lyrics or harmonics.
	
	\item [Amplification] The meaning of the song or certain relevant messages of its lyrics are emphasized and reinforced through constant repetition and dominating visual accentuation.
	
	\item [Contradiction] The underlying meaning of lyrics and melodics is ignored. The visual layer either counterpoints the meaning of the song or becomes completely abstract.
	
\end{description}

\begin{figure}[t]
	
	\centering
	
	\subfloat[\textbf{Performance Videos}, \textit{Van Halen - Jump, 1970.} The band is filmed while performing in synch to the music.]{\includegraphics[width=0.29\textwidth]{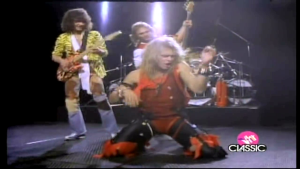}}
	\hfill
	\subfloat[\textbf{Concept Videos}, \textit{Red Hot Chili Peppers - Californication, 1999}. Video is realized as a fictional 3D video game.]{\includegraphics[width=0.22\textwidth]{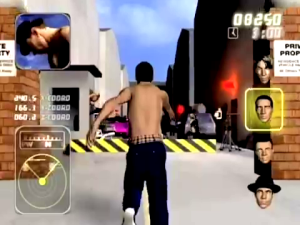}}
	\hfill
	\subfloat[\textbf{Narrative Videos}, \textit{Beyonce - If I Were A Boy, 2008.} The video plays with gender stereotypes by switching the roles. ]{\includegraphics[width=0.29\textwidth]{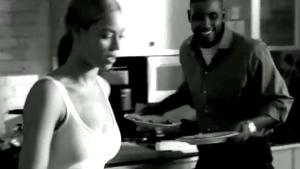}}
	\hfill
	\subfloat[\textbf{Dance Videos}, \textit{Michael Jackson - Bad, 1987.} Choreagraphed dance in reference to the musical `West Side Story'. ]{\includegraphics[width=0.29\textwidth]{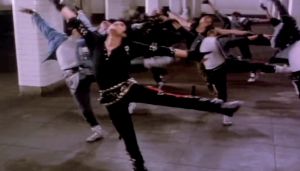}}
	\hfill
	\subfloat[\textbf{Animated Videos}, \textit{Gorillaz - Clint Eastwood, 2001.} Animated band members are performing with dancing zombie gorillas. ]{\includegraphics[width=0.29\textwidth]{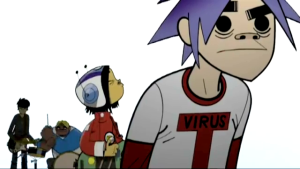}}
	\hfill
	\subfloat[\textbf{Lyrics Videos}, \textit{Neil Young - My Pledge, 2016.} Lyrics appear on array of postcards. ]{\includegraphics[width=0.29\textwidth]{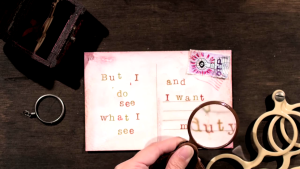}}
	
	\caption{Example frames of different music video types.}
	\label{fig:ch3:music_video_types}
\end{figure}

\noindent
There are \textbf{three main types of music videos} \cite{caston2014fine}:

\begin{description}[leftmargin=!,itemsep=1pt,labelwidth=\widthof{\bfseries 12345},labelindent=0cm,rightmargin=0.1cm] 
	
	\item [Performance videos]
	The artist is presented performing in sync with the song - typically in an environment corresponding to the song or genre (e.g. with or without friends or instruments, live or staged performance, studio recording, etc.). Performance videos were predominantly used in early pre-MTV music shows such as BBC's \textit{Top of the Pops} (see Figure \ref{fig:ch3:music_video_types} a).
	
	\item [Concept videos]
	The meaning of the lyrics is not narrated but illustrated through visual concepts and metaphors. The plot of the video is mostly obscure and surreal, trying to attract and entertain the audience by constantly keeping their attention on the screen. Concept videos could include performance, but make use of unusual settings or film techniques such as trick photographic and novel editing techniques to suspend end invert the laws of nature for novel entertainment value \cite{gunning1989primitive} (see Figure \ref{fig:ch3:music_video_types} b).
	
	\item [Narrative videos]
	The visual part commonly illustrates or portraits the story told by the lyrics, but may also contradict the underlying song by telling a completely different story. Usually the narrator is the performing artist who often acts as the protagonist of the story (see Figure \ref{fig:ch3:music_video_types} c).
	
	\item [Dance videos]
	In dance videos the artist is performing alone or with additional background dancers a rehearsed dance choreography (see Figure \ref{fig:ch3:music_video_types} d). Famous examples are the influencing music videos by \textit{Michael Jackson} which feature complex synchronized choreographies.
	
	\item [Animated Music Videos]
	Animated music videos make use of all kinds of styles and technologies such as drawing, stop motion photography, claymation, computer graphics, etc. (see Figure \ref{fig:ch3:music_video_types} e) Animated videos provide more options to express emotions reflected by the corresponding song, but can be generally categorized by the same types as non-animated music videos.
	
	\item [Lyrics Videos]
	Lyric videos display the lyrics in sync with the singer, often in artistic ways (see Figure \ref{fig:ch3:music_video_types} f). A famous example is \textit{Bob Dylan's} video to his song \textit{Subterranean Homesick Blues} (1965), which was also one of the first ``promotional'' music clips. In the video Dylan stands in front of the camera flipping through white cards with selected words and phrases of the lyrics. 
	
\end{description}

\subsection{Stylistic Properties of Music Videos}
\label{ch3:mv_styles}

A wide range of cinematographic styles and effects can be spotted in music videos. While film making has developed standardized rules how to aesthetically apply them, music videos do not follow these rules but give these elements different functions and meanings \cite{vernallis2004experiencing}. The following list summarizes the most important stylistic features of music videos:

\begin{description}[leftmargin=!,itemsep=1pt,labelwidth=\widthof{\bfseries 12345},labelindent=0cm,rightmargin=0.1cm] 
	
	\item [Camera Work and Movement]Abiding the continuing progression of the musical track movement is one of the most common properties of music videos. Most of the shots of a video make use of one or more of the following techniques: \textit{tracking shot} - the camera moves towards or away from the subject, \textit{pan} - the camera rotates to the left or right to reveal more of the environment, \textit{tilt} - similar to panning but camera rotates horizontally, \textit{zoom in} or \textit{out} - magnify or reduce subject by changing lens focal length. Tempo of fore- and background movement is also used to express moods and emotions of the track.
	
	\item [Lighting and Color]Lighting and color can influence our interpretation of an individual scene. Color can be used to attract attention or express moods (e.g. warm colors are often used to create romantic or pleasurable ambiance). Lighting in visual production has developed many techniques to create all kinds of effects and moods (e.g. directional light, back and spot lighting, shadows, etc.).
	
	\item [Visual and stylistic coherence]A common aimed at property of music videos. Shots of a video provide a coherent impression that should correspond to the underlying track. Usually accomplished in post-production through global filter (e.g. color, blur, etc.)
	
	\item [Close Ups]Usually a full-face shot composed from below the shoulder line. There is often a pressing demand of music labels to promote and iconize the contracted artist through a proliferated presence on screen. Close ups are further applied to underscore a musical hook or the peak of a phrase or, more recently, to advertise brand names or products added to the video to earn additional revenue.
	
	\item [Editing and Shots]Contrary to film making, music video editing does not intend to remain unnoticed but is perceived as an art-form. Traditional rules to guide the viewer in time and space are ignored due to its short form and the demand to showcase the star. Shots are joined by favoring compositional elements (e.g. color, shape, etc.) over content. \textit{Jump Cuts} - intentionally disjunctive edits - are frequently employed using drastic shifts in color, scale or content. Low angle shots reproduce the impression of looking up to an artist performing on stage. High-angle shots harmonize with key moments of a song. Mixing those in a series disorients the viewer who seeks additional guidance in the music. Shot length and boundaries commonly correspond to tempo and rhythm of the track.
	
\end{description}

\pagebreak

\subsection{Significance of Music Videos}

Music videos were initially criticized to provoke a diminishing of the interpretative liberty of the individual music listener by imposing the narrative visual impression upon him or her \cite{frith2005sound}. Yet, over the past four decades music videos shaped our cultural life. Movie editors speak of MTV-style editing when referring to short shot sequences. Artists use huge video walls at life performances displaying complex beat synchronized visuals that turn a whole concert into a music video like experience.\\


\noindent
\textit{\smaller Parts of this introduction to Music Videos as presented in this section was first published and presented in \textit{Alexander Schindler. ``A picture is worth a thousand songs: Exploring visual aspects of music''} at the 1st International Digital Libraries for Musicology workshop (DLfM 2014), London, UK, September 12 2014 \cite{Schindler2014dlfm}. }

\newpage

\section{Summary}

This chapter provided an in-depth overview of how visual media is utilized to promote new artists and advertise new music. 

\begin{itemize}
	
	\item A broad overview from early utilization of illustrations on music sheets to the history of album cover art is provided in Section \ref{ch3:albumart_history}. It is outlined how visual media has been intentionally used in the pre-online-music-streaming era to advertise new and unknown music acts in magazines, which were the prevalent distribution channels.
	
	\item The experimental focus of this thesis lies on Music Videos. Section \ref{ch3:mv_overview} provides an extensive introduction to this medium, including its historical development, different music video types and stylistic properties used in music video production. 
	
\end{itemize}

\noindent
The information compiled in this Chapter is relevant to understand the intentions and production processes of music videos. On the one hand, this is relevant to develop appropriate approaches to harness the information of music videos by deriving or extracting corresponding representations. On the other hand this knowledge is relevant for the evaluation of these approaches.\\

\noindent
Finally, this chapter closes with a discussion on the relevance of music videos for the Music Information Retrieval research domain.



\chapter{Data}
\label{ch4:intro}

\epigraph{
	``It is a capital mistake to theorize before one has data. Insensibly one begins to twist facts to suit theories, instead of theories to suit facts.''} 
{--- Sir Arthur Conan Doyle, ``The Adventures of Sherlock Holmes '', 1892}


Music Information Retrieval (MIR) research has historically struggled with issues of publicly available benchmark datasets that would allow for evaluation and comparison of methods and algorithms on the same data base. Most of these issues stem from the commercial interest in music by record labels, and therefore imposed rigid copyright issues, that prevent researchers from sharing their music collections with others. Subsequently, only a limited number of data sets has risen to a pseudo benchmark level, i.e. where most of the researchers in the field have access to the same collection.

Another reason identified as a major challenge for providing access to research data in general is the lack of esteem and valuation of these kind of activities. While preparing, maintaining and providing access to massive data collections requires significant investments in terms of system maintenance and data (pre-)processing, it is considered administrative rather than research work (in spite of several even research-affine challenges emerging during such activities), and thus does not gain acceptance in classical research-oriented publication venues. Such lack of career rewards is one of the many factors, next to legal limitations and lack of expertise, limiting sharing of research data \cite{jubb08_share}. Several initiatives have been started in the research infrastructures area to mitigate this problem and foster collaborative research. These areas span across virtually all topical areas, from Astronomy, via meteorology, chemistry to humanities\footnote{\url{http://ec.europa.eu/research/infrastructures/}}.

\noindent
This chapter introduces the datasets that have been created or enhanced as part of this thesis in order to investigate on the introduced research questions, to facilitate experimentation and to foster further research.

\pagebreak

\section{The Million Song Dataset}
\label{ch4.2:intro}

The Million Song Dataset (MSD) is a music dataset presented in 2011 \cite{Bertin-Mahieux2011}. It provides a huge collection of meta-data for one million contemporary popular songs. This was the first dataset which facilitated large-scale experimentation for a wide range of Music Information Retrieval (MIR) related tasks. Meta-data information include essential attributes such as track titles, artist and album names, but also references to third party metadata-repositories such as MusicBrainz or 7Digital. Additionally, a number of descriptive feature-sets extracted from the original audio files using a proprietary feature extractor from the former start-up \textit{The Echo Nest}\footnote{\url{http://the.echonest.com}} are provided. In the meanwhile several additions have been provided to complement the initial MSD data such as the \textit{Second Hand Dataset} \cite{bertin2011large} for cover-song identification, the \textit{musiXmatch dataset}\footnote{\url{http://labrosa.ee.columbia.edu/millionsong/musixmatch}} which links a huge collection of song lyrics to the MSD, and of course the additional contemporary features extracted from samples downloaded from 7Digital (see Chapter \ref{ch4.2:intro}).

Unfortunately, there are no easily obtainable audio files available for this dataset, and therefore, researchers are practically restricted to benchmarking on algorithms that work on top of the existing features, such as recommendation of classification, but cannot easily develop new or test existing feature sets on this dataset. The availability of just one feature set also does not allow an evaluation across multiple feature sets. As previous studies showed, however, there is no single best feature set, but their performance depends very much on the dataset and the task. This section therefore aims to alleviate these restrictions by providing a range of features extracted for the Million Song Dataset, such as MFCCs and a set of low-level features extracted with the jAudio feature extraction software as well as the Mar\-syas framework \cite{Tzanetakis2000}, and the Rhythm Patterns and derived feature sets~\cite{LID05_ismir}.

A second shortcoming of the MSD is that it initially did not contain mappings to ground truth labels such as genres or moods. Thus, experimental evaluations such as musical genre classification, a popular task in MIR research, are not possible. This section therefore further introduces ground truth assignments created from human annotated metadata, obtained from the All Music Guide\footnote{\url{http://allmusic.com}}. Specifically, different assignments on two levels of detail, with 13 top-level-genres and 25 sub-genres are proposed which are provided as a number of partitions for training and test-splits, with different filters applied, allowing several evaluation settings. Both the feature sets and the partitions are available for download from our website\footnote{\url{http://www.ifs.tuwien.ac.at/mir/msd/}}. The features are provided in WEKA Attribute-Relation File Format (ARFF)~\cite{Witten99weka:practical}, with one attribute being the unique identifier of the song in the MSD. Further, a set of scripts are provided to join features with genre assignments so they can be used in classification experiments.

The remainder of this section is structured as follows. Section~\ref{ch4.2:audioSamples} introduces the dataset and the properties of the audio samples, while Section~\ref{ch4.2:features} describes the sets of features extracted from them. Section~\ref{ch4.2:genre:Allmusic} details on the genre assignment obtained, and in Section~\ref{ch4.2:partitions}, the benchmark partitions and how they aim to facilitate ex\-cha\-nge between researchers are described.

\subsection{The original MSD}
\label{ch4.2:audioSamples}

\begin{figure}[t]
	\centering
	\includegraphics[width=0.8\columnwidth]{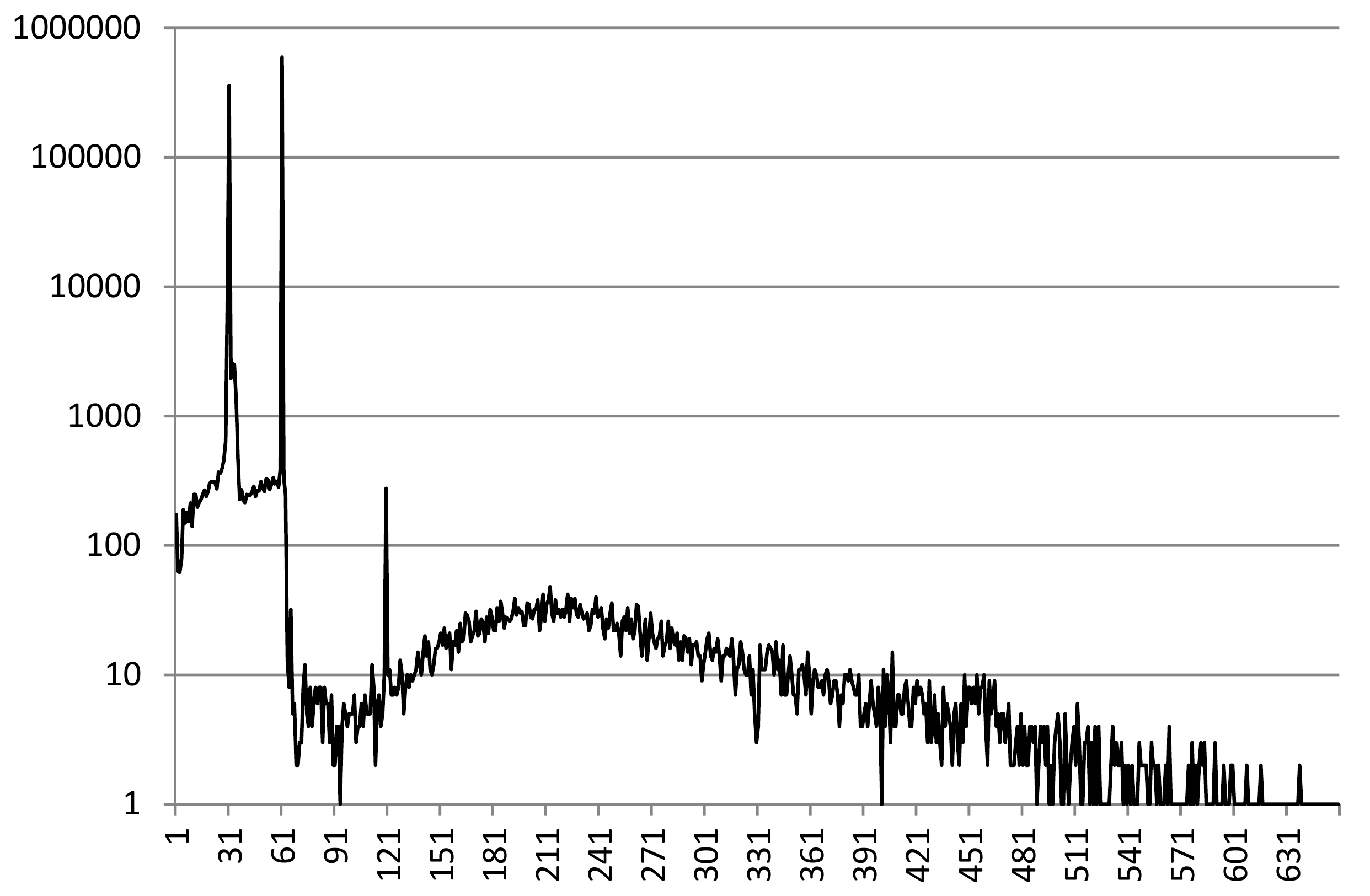}
	\caption{Distribution of sample length. The x-axis represents the length of the audio samples. The logarithmic-scaled y-axis represents the number of corresponding files.}
	\label{fig:ch4:Sample_Length_Distribution}
\end{figure}

The biggest asset of the MSD is its size. Until its introduction all research datasets ranged in the size of a few thousand tracks. Comparing these to catalogs of of millions of tracks of contemporary music streaming services or collections of cultural heritage institutions such as Europeana\footnote{\url{http://www.europeana.eu}} \cite{schindler2016europeana} the question comes up if approaches evaluated on such small datasets can generalize on these large collections as well.

The MSD was initially distributed with high- and low-level music features which were extracted using a proprietary feature extractor developed by the company \textit{The Echonest}. These features include tempo, loudness, timings of fade-in and fade-out, and MFCC-like features for a number of segments (for a full description of the provided features, please refer to \cite{Bertin-Mahieux2011}). Moreover, a range of other meta-data has been published recently, such as song lyrics (for a subset of the collection), or tags associated to the songs from Last.fm\footnote{\url{http://www.last.fm}}. Its meta-data further contains a unique identifier to the content provider 7digital\footnote{\url{http://www.7digital.com}}, which could be used to download a sample of the original audio file. 

\subsubsection{Audio}

A major part of the effort described in this section is based on the acquisition of all of these samples. For some songs no sample could be downloaded, as the identifier was unknown to 7digital. This may be due to data curation efforts of 7Digital or result from changed or expired licensing contracts. Although, the data collection was performed less than one year after the publication of the MSD, it was only possible to obtain a total of 994,960 audio samples, i.e. a coverage of 99.5\% of the dataset (the list of missing audio samples is available on the website). This points to an important issue related to the use of external on-line resources for scientific experimentation. Especially when the provider is not genuinely interested in the actual research performed, there is little motivation to maintain the data accessible in unmodified manners, and it is thus susceptible to changes and removal. Thus, maintaining a copy of a fixed set of data is essential in benchmarking to allow the evaluation of newly developed feature sets, and for acoustic evaluation of the results. In total, the audio files amount to approximately 625 gigabyte of data.

The audio samples do not adhere to a common encoding quality scheme, i.e. they differ in length and quality provided. Figure \ref{fig:ch4:Sample_Length_Distribution} shows a plot of the sample lengths; please note that the scale is logarithmic. It can be observed that there are two peaks at sample lengths of 30 and 60 seconds with 366,130 and 596,630 samples, respectively, for a total of 96,76\% of all the samples. These shorter snippets normally contain a section in the middle of the song. Many other well-known collections in the MIR domain also contain only 30 second snippets, and feature extraction algorithms normally deal with this. Table~\ref{tab:SampleRatesOfAudioSamples} gives an overview on the audio quality of the samples. The majority, more than three quarters, of the audio snippets have a sample rate of 22khz, the rest has a sample rate of 44khz (with the exception of 81 songs, of which the majority have 24 and 16khz). Regarding the bitrate, approximately two-thirds of the songs are encoded with 128kbit, the majority of the rest with 64kbit; only about half a percent of the songs come with higher (192 or 320kbps) or variable bitrates (anywhere between 32 and 275kpbs). Almost all samples are provided in some form of stereo encoding (stereo, joint stereo or dual channels) -- only 0.6\% of them have only one channel. These characteristics, specifically the sample rate, may have significant impact on the performance of the data analysis algorithms. This had to be considered for stratification purposes when designing the benchmark splits.

\begin{table}[t]
	\caption{Audio properties of 7Digital Samples}
	\centering
	\begin{tabular}{l r r | l r r}
		
		\hline
		\rowcolor{gray!30}
		\textbf{Sample Length} & & & \textbf{Samplerate} & & \\ 
		\hline
		
		& & & & & \\ [-0.9ex]
		
		30 Seconds & 366,130 & 36.80\% & 22.050 Hz & 768,710 & 77.26\% \\
		60 Seconds & 596,630 & 59.97\% & 44.100 Hz & 226,169 & 22.73\% \\
		other      &  32,200 &  3.24\% & other  & 81      & 0.01\% \\
		
		& & & & & \\ [-0.9ex]
		
		\hline
		\rowcolor{gray!30}
		\textbf{Bitrate} & & & \textbf{Channels} & & \\
		\hline
		
		& & & & & \\ [-0.9ex]
		
		128 kBits        & 646,120 & 64.94\% & Mono         & 6,342   & 0.64\%  \\
		64 kBits         & 343,344 & 34.51\% & Stereo       & 150,779 & 15.15\% \\
		Variable Bitrate < 64 kBits   & 650     & 0.07\%  & Joint stereo & 837,702 & 84.20\% \\
		Variable Bitrate 64-128 kBits & 935     & 0.09\%  & Dual channel & 134     & 0.01\%  \\
		Variable Bitrate > 128 kBits  & 3,909   & 0.39\%  & & & \\
		
		& & & & & \\ [-0.9ex]
		
		\hline
		
	\end{tabular}
	\label{tab:SampleRatesOfAudioSamples}
\end{table}

\subsection{Feature Sets} 
\label{ch4.2:features}

The MSD was originally introduced as a collection of pre-extracted low- to high-level features-sets and the corresponding metadata. This section first describes the original feature-sets provided by the MSD and then the contributions added along with this thesis.

\subsubsection{Echonest Features (Original MSD Feature-Sets)}
\label{ch4.2:echonestfeatures}

The Echonest Analyzer \cite{analyzer2011} is a music audio analysis tool available as a free Web service which is accessible over the Echonest API\footnote{http://developer.echonest.com - API offline since May 2016 (see Infobox)}. In a first step of its analysis the Echonest Analyzer uses audio fingerprinting to locate tracks in the Echonest's music metadata repository. Music metadata returned by the Analyzer includes artist information (name, user applied tags including weights and term frequencies, a list of similar artists), album information (name, year) and song information (title). Additionally a set of identifiers is provided that can be used to access complimentary metadata repositories (e.g. musicbrainz\footnote{http://musicbrainz.org}, playme\footnote{http://www.playme.com},7digital).

\begin{figure}[t]
	\centering
	\includegraphics[width=1.00\textwidth]{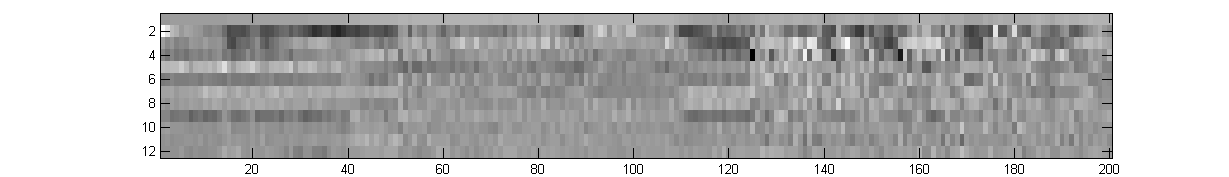}
	\caption{First 200 timbre vectors of 'With a little help from my friends' by 'Joe Cocker'}
	\label{fig:ch5.2:timbre}
\end{figure}

Further information provided by the Analyzer is based on audio signal analysis. Two major sets of audio features are provided describing timbre and pitch class information of the corresponding music track. Unlike conventional MIR feature extraction frameworks, the Analyzer does not return a single feature vector per track and feature. The Analyzer implements an onset detector which is used to localize music events called \textit{Segments}. These \textit{Segments} are described as sound entities that are relatively uniform in timbre and harmony and are the basis for further feature extraction. For each \textit{Segment} the following features are derived from musical audio signals:

\begin{description}[leftmargin=!,itemsep=1pt,labelwidth=\widthof{\bfseries Segments Loudness Max Tim},labelindent=\descleftmargin,rightmargin=0.8cm] 
	
	\item[Segments Timbre] 
	are casually described as MFCC-like features. A 12 dimensional vector with unbounded values centered around 0 representing a high level abstraction of the spectral surface (see Figure \ref{fig:ch5.2:timbre}).
	
	\item[Segments Pitches] 
	are casually described as Chroma-like features. A normalized 12 dimensional vector ranging from 0 to 1 corresponding to the 12 pitch classes C, C\#, to B.
	
	\item[Segments Loudness Max] 
	represents the peak loudness va\-lue within each segment.
	
\end{description}


\begin{description}[leftmargin=!,itemsep=1pt,labelwidth=\widthof{\bfseries Segments Loudness Max Time},labelindent=\descleftmargin,rightmargin=0.8cm] 
	
	\item[Segments Loudness Max Time] 
	describes the offset with\-in the segment of the point of maximum loudness.
	
	\item[Segments Start] 
	provide start time information of each segment/onset.
	
\end{description}

\noindent
Onset detection is further used to locate perceived musical events within a \textit{Segment} called \textit{Tatums}. \textit{Beats} are described as multiple of \textit{Tatums} and each first \textit{Beat} of a measure is marked as a \textit{Bar}.  Contrary to \textit{Segments}, that are usually shorter than a second, the Analyzer also detects \textit{Sections} which define larger blocks within a track (e.g. chorus, verse, etc.). From these low-level features some mid- and high-level audio descriptors are derived (e.g. tempo, key, time signature, etc.). Additionally, a confidence value between 0 and 1 is provided indicating the reliability of the extracted or derived values - except for a confidence value of '-1' which indicates that this value was not properly calculated and should be discarded. Based on the audio segmentation and additional audio descriptors the following features provide locational informations about music events within the analyzed track:

\medskip
\begin{description}[leftmargin=!,itemsep=1pt,labelwidth=\widthof{\bfseries Bars/Beats/Tatums start ed},labelindent=\descleftmargin,rightmargin=\descrightmargin] 
	
	\item[Bars/Beats/Tatums start] the onsets for each of the detected audio segments
	
	\item[Sections start] the onsets of each section. 
	
	\item[Fadein stop] the estimated end of the fade-in 
	
	\item[Fadeout start] the estimated start of the fade-out
	
\end{description}


\medskip
\noindent
Additionally a set of high-level features derived from previously described audio descriptors is returned by the Analyzer:

\medskip
\begin{description}[leftmargin=!,itemsep=1pt,labelwidth=\widthof{\bfseries Bars/Beats/Tatums start ed},labelindent=\descleftmargin,rightmargin=\descrightmargin] 
	
	\item[Key] the key of the track (C,C\#,...,B)
	
	\item[Mode] the mode of the track (major/minor)
	
	\item[Tempo] measured in beats per minute
	
	\item[Time Signature] three or four quater stroke
	
	\item[Danceability] a value between 0 and 1 measuring of how danceable this song is 
	
	\item[Energy] a value between 0 and 1 measuring the perceived energy of a song
	
	\item[Song Hotttnesss] a numerical description of how hot a song is (from 0 to 1)
	
\end{description}


\begin{table*}
	\caption{Overview on features extracted from the MSD samples. \textit{Dim.} denotes the dimensionality, \textit{Deriv.} de\-ri\-vat\-ives computed from the base features }
	
	\centering
	\begin{tabular}{l|p{8.5cm}|l|l|l}
		\hline
		\rowcolor{gray!30}
		\textbf{\#}	& \textbf{Feature Set} 													& \textbf{Extractor} 	& \textbf{Dim} & \textbf{Deriv.}\\
		\hline
		1 			& MFCCs~\cite{Rabiner1993Fundamentals}							& MARSAYS		& 52	& \\
		2 			& Chroma~\cite{Goto06}											& MARSAYS		& 48	& \\
		3 			& Timbral~\cite{Tzanetakis2000}									& MARSAYS		& 124	& \\
		\hline
		4 			& MFCCs~\cite{Rabiner1993Fundamentals}							& jAudio		& 26	& 156 \\
		5 			& Low-level spectral features~\cite{mckay2010}
		\tiny{(Spectral Centroid, Spectral Rolloff Point, Spectral Flux, Compactness, and Spectral Variability, Root Mean Square, Zero Crossings, and Fraction of Low Energy Windows)}
		& jAudio		& 16	& 96 \\
		6 			& Method of Moments~\cite{mckay2010}								& jAudio		& 10	& 60 \\
		7 			& Area Method of Moments~\cite{mckay2010}							& jAudio		& 20	& 120 \\
		8 			& Linear Predictive Coding~\cite{mckay2010}						& jAudio		& 20	& 120 \\
		\hline
		9 			& Rhythm Patterns~\cite{LID05_ismir}								& rp\_extract	& 1440	& \\
		10			& Statistical Spectrum Descriptors~\cite{LID05_ismir}				& rp\_extract	& 168	& \\
		11			& Rhythm Histograms~\cite{LID05_ismir}							& rp\_extract	& 60	& \\
		12			& Modulation Frequency Variance Descriptor~\cite{lidy10_ethnic}	& rp\_extract	& 420	& \\
		13			& Temporal Statistical Spectrum Descriptors~\cite{lidy10_ethnic}	& rp\_extract	& 1176	& \\
		14			& Temporal Rhythm Histograms~\cite{lidy10_ethnic}					& rp\_extract	& 420	& \\
		\hline
	\end{tabular}
	\label{tab:features}
\end{table*}


A wide range of audio features are extracted from the downloaded samples, namely features provided by the jAudio feature extraction software (which is a part of the jMIR package~\cite{mckay2010}), the MARSYAS feature extractor~\cite{Tzanetakis2000}, and the Rhythm Patterns family of feature sets~\cite{LID05_ismir}. An overview on these features is given in Table~\ref{tab:features}.

\subsubsection{jAudio}
The \textit{jAudio} software provides a range of 28 features associated with both the frequency and time domains. It includes several intermediate-level musical features, mainly related to rhythm, as well as lower-level signal processing-oriented features. It also provides an implementation of MFCC features~\cite{Rabiner1993Fundamentals}, using 13 coefficients. Further features are for example the Spectral Flux, a measure of the amount of spectral change in a signal from frame to frame, or Root Mean Square (RMS), a measure of the average energy of a signal calculated over an analysis window. In addition, a set of statistics is computed, including mean, standard deviations and derivatives. Specifically extracted were the low-level features Spectral Centroid, Spectral Rolloff Point, Spectral Flux, Compactness, and Spectral Variability, Root Mean Square, Zero Crossings, and Fraction of Low Energy Windows (16 dimensions), and derivatives thereof (96 dimensions). The \textbf{Method of Moments} extractor captures the first five statistical moments of the magnitude spectrum, and then computes average and standard deviations over all segments (10 dimensions), and derivatives thereof (60 dimensions). The \textbf{Area Method of Moments} extractor additionally analyses the time series of the spectrum (20 dimensions, and derivatives thereof (120 dimensions)). \textbf{Linear Predictive Coding} estimates formants (spectral bands corresponding to resonant frequencies in the vocal tract), filters them out, and estimates the intensity and frequency of the residual ``buzz'' that is assumed to be the original excitation signal (20 dimensions, and derivatives thereof (120 dimensions)). jAudio computes in general mean and standard deviations over the sequence of frames, and provides for most measures also derivatives, i.e. additional statistical moments over the basic measures. For the extraction, jAudio as bundled in the jMIR 2.4 release\footnote{available from \url{http://jmir.sourceforge.net/}} was utilized.

\subsubsection{MARSYAS}
\label{ch4.1:marsyas}
A very popular audio feature extraction system is MARSAYS, one of the first comprehensive software packages to be available to MIR researchers. A very popular set from this audio extractor is the so-called ``timbral'' set, which is com\-posed of 13 MFCC co\-ef\-fi\-cients, and the twelve chroma features and the average and minimum chroma value, and the four low-level features zero crossings, and rolloff, flux and centroid of the spectrum. For these 31 values, four statistical moments are computed, resulting in a 124 dimensional vector. Chroma features \cite{Goto06} aim to represent the harmonic content (e.g, keys, chords) of an audio by computing the spectral energy present at frequencies that correspond to each of the 12 notes in a standard chromatic scale (e.g., black and white keys within one octave on a piano). For the extraction, MARSYAS version 0.4.5\footnote{available from \url{http://sourceforge.net/projects/marsyas/}} was utilized.

\subsubsection{rp\_extract}
\label{ch4.1:rp_extract}
The Rhythm Patterns and related features sets are extracted from a spectral representation, partitioned into segments of 6 sec. Features are extracted segment-wise, and then aggregated for a piece of music computing the median (Rhythm Patterns, Rhythm Histograms) or mean (Statistical Spectrum Descriptors, Modulation Frequency Variance Descriptor) from features of multiple segments. For details on the computation - the feature extraction will only be described very briefly - please refer to \cite{LID05_ismir}. The feature extraction for a \textbf{Rhythm Pattern} is composed of two stages. First, the specific loudness sensation on 24 critical frequency bands is computed through a Short Time FFT, grouping the resulting frequency bands to the Bark scale, and successive transformation into the Decibel, Phon and Sone scales. This results in a psycho-acoustically modified Sonogram representation that reflects human loudness sensation. In the second step, a discrete Fourier transform is applied to this Sonogram, resulting in a spectrum of loudness amplitude modulation per modulation frequency for each critical band. After additional weighting and smoothing steps, a Rhythm Pattern exhibits magnitude of modulation for 60 modulation frequencies on the 24 critical bands~\cite{LID05_ismir}. A \textbf{Rhythm Histogram} (RH) aggregates the modulation amplitude values of the critical bands computed in a Rhythm Pattern and is a descriptor for general rhythmic characteristics in a piece of audio \cite{LID05_ismir}. The first part of the algorithm for computation of a \textbf{Statistical Spectrum Descriptor} (SSD), the computation of specific loudness sensation, is equal to the Rhythm Pattern algorithm. Subsequently at set of statistical values\footnote{minimum, maximum, mean, median, variance, skewness, and kurtosis\label{footnote:statistics}} are calculated for each individual critical band. SSDs describe fluctuations on the critical bands and capture additional timbral information very well~\cite{LID05_ismir}. The \textbf{Modulation Frequency Variance Descriptor} (MVD)~\cite{lidy10_ethnic} measures variations over the critical frequency bands for a specific modulation frequency (derived from a Rhythm Pattern). It is computed by taking statistics for one modulation frequency over the critical bands, for each of the 60 modulation frequencies. For the \textbf{Temporal Rhythm Histograms} (TRH)~\cite{lidy10_ethnic}, statistical measures are computed over the individual Rhythm Histograms extracted from the segments in the audio. Thus, the TRHs capture change and variation of rhythmic aspects in time. Similarly, the \textbf{Temporal Statistical Spectrum Descriptor} (TSSD)~\cite{lidy10_ethnic} capture the same statistics over the individual SSD segments, and thus describe timbral variations and changes over time in the spectrum on the individual critical frequency bands. For the extraction, the Matlab-based implementation, version 0.6411\footnote{available from \url{http://www.ifs.tuwien.ac.at/mir/downloads.html}} was utilized.

\medskip
It was intentional to provide two different ver\-sions of the MFCCs features, as this will allow for interesting insights in how these implementations differ on various MIR tasks.

\subsubsection{Publication of Feature Sets}

All features described above were made available for download, encoded in the WEKA Attribute-Relation File Format (ARFF)~\cite{Witten99weka:practical}. The features are available under the Creative Commons Attribution-NonCommercial-ShareAlike 2.0 Generic License\footnote{\url{http://creativecommons.org/licenses/by-nc-sa/2.0/}}.

To allow high flexibility when using them, one ARFF file for each type of features is provided. These can then be combined in any particular way when performing experimental evaluations. A set of scripts is provided as well on the website for this. In total, the feature files amount to approximately 40 gigabyte of uncompressed text files. The files contain the numeric values for each feature, and additionally the unique identifier assigned in the MSD. This way, it is possible to generate various feature files with different ground truth assignments. Scripts to accomplish this are also provided. The proposed assignment into genres for genre classification tasks is described in Section~\ref{ch4.2:genre:Allmusic}. 

\subsection{Allmusic Ground-Truth Assignments}
\label{ch4.2:genre:Allmusic}

The All Music Guide (AMG) \cite{datta2002} was initiated by an archivist in 1991 and emerged 1995 from its book form into a data\-base which can be accessed through the popular commercial Web page allmusic.com. The Web page offers a wide range of music information, including album reviews, artist biographies, discographies as well as classifications of albums according to genres, styles, moods and themes. This information is provided and curated by music experts. 

The genre information is provided as a single tag for each album. The genre taxonomy is very coarse. The two main categories Pop and Rock are combined into a single genre 'Pop/Rock'. Additionally to genre labels, style tags are provided allowing for a more specific categorization of the annotated albums. Labels are not applied exclusively. Multiple style tags are applied for each album, but unfortunately no weighting scheme can be identified and in many cases only one tag is provided. Style tags also tend to be even more generic than genre labels. Especially non-American music is frequently tagged with labels describing the originating country or region as well as the language of the lyrics. Instrumentation, situational descriptions (e.g. \textit{Christmas}, \textit{Halloween}, \textit{Holiday}, etc.) as well as confessional or gender attributes (e.g. \textit{Christian}, \textit{Jewish}, \textit{Female}, etc.) are also provided. Unfortunately these attributes are not used as isolated synonyms, but are concatenated with conventional style information (e.g. \textit{Japanese Rock}, \textit{Christian Punk}, \textit{Classic Female Blues}, etc.). 

Allmusic.com assembles styles to meta-styles which can be interpreted as a hierarchy of sub genres used to diversify the major genre labels. Meta-styles are not distinctive and are used overlapping in many meta-styles (e.g. \textit{Indie Electronic} is contained in the meta-styles \textit{Indie Rock}, \textit{Indie Pop} and \textit{Alternative/Indie Rock}). 

\subsubsection{Genre Data Collection}

Data was collected automatically from Allmusic.com using a web-scraping script based on direct string matching to query for artist-release combinations. From the resulting Album Web page genre and style tags were collected. It was possible to retrieve 21 distinct genre labels for 62,257 albums which initially provided genre tags for 433,714 tracks. Style tags were extracted attributing only 42,970 albums resulting in 307,790 labeled tracks. An average of 3.25 tags out of a total of 905 styles were applied to each album, but 5,742 releases were only tagged with a single style label. The most popular genre with 32,696 tagged albums, was \textit{Pop/Rock} - this is 10\% more than the sum of all other genres. Considering tracks the difference rises to 30\%. Further, the granularity of Rock is very coarse, including Heavy Metal, Punk, etc. A similar predominating position of this genre was also reported by \cite{Bergstra06predictinggenre}. The most popular style tag is \textit{Alternative/Indie Rock} applied to 12,739 albums, which is more than twice as much as the second most popular style \textit{Alternative Pop/Rock}. About 120 tags describe the country of the performing artist or the language of the interpretation - the most common among them is \textit{Italian Music} which has been applied to 610 albums.

\subsubsection{Allmusic Genre Dataset}

The Allmusic Genre Dataset is provided as an unoptimized expert annotated ground truth dataset for music genre classification. Two partitions of this set are provided. The \textit{MSD Allmusic Genre Dataset (MAGD)} assembles all collected genres including generic and small classes. The second partition - \textit{MSD Allmusic Top Genre Data\-set (Top-MAGD)} - consists of 13 genres - the 10 major genres of Allmusic.com (Pop/Rock, Jazz, R\&B, Rap, Country, Blues, Electronic, Latin, Reggae, International) including the three additional genres Vocal, Folk, New Age (see Table \ref{tab:AllmusicGenrePartition}). Generic genres as well as classes with less than 1\% of the number of tracks of the biggest class Pop/Rock are removed. Due to the low number of tracks, the Classical genre is also removed from the Top Genre dataset.


\begin{table}[t]
	\centering
	\caption{MSD Allmusic Genre Dataset (MAGD) - upper part represents the MSD Allmusic Top Genre Dataset (Top-MAGD)}
	\begin{tabular}{l r }
		
		\hline
		\rowcolor{gray!30}
		\textbf{Genre Name} & \textbf{Number of Songs} \\
		\hline
		
		Pop/Rock      & 238,786 \\
		Electronic    & 41,075 \\
		Rap           & 20,939 \\
		Jazz          & 17,836 \\
		Latin         & 17,590 \\
		R\&B          & 14,335 \\
		International & 14,242 \\
		Country       & 11,772 \\
		Reggae        & 6,946 \\
		Blues         & 6,836 \\
		Vocal         & 6,195 \\
		Folk          & 5,865 \\
		New Age       & 4,010 \\
		
		\hline
		
		Religious     &	8814 \\
		Comedy/Spoken &	2067 \\
		Stage         & 1614 \\
		Easy Listening&	1545 \\
		Avant-Garde   &	1014 \\
		Classical     &	556 \\
		Childrens     &	477 \\
		Holiday       &	200 \\
		
		\hline
		Total         & 422,714 \\
		
	\end{tabular}
	\label{tab:AllmusicGenrePartition}
\end{table}


\subsubsection{Allmusic Style Dataset}

The Allmusic Style Dataset attempts to more distinctively separate the collected data into different sub-genres, alleviating dominating classes. For the compilation of the dataset genre labels are omitted and solely style tags are used. In a first step metastyle description as presented on the Allmusic.com Web site are used to map multiple style tags to a single label name - in this case the metastyle name is used. This simple aggregation approach generates a total of 210 labels many of them highly generic or hierarchical specializing (e.g. \textit{Electric Blues} and \textit{Electric Chicago Blues}. The \textit{MSD Allmuisc Metastyle Dataset - Multiclass (MAMD)} is derived from these 210 resulting metaclasses. Each track is matched to one or more metaclasses according to its style tags. In a second step confessional, situational and language specific labels are removed from the initial set of 905 style tags. Regional tags are discarded if they do not refer to a specific traditional cultural music style (e.g. \textit{African Folk}). Popular music attributed with regional information is discarded due to extensive genre overlaps (e.g. \textit{Italian Pop} ranges from Hip-Hop to Hard-Rock). Finally, these genres are successfully merged into general descriptive classes until the dataset is finalized into the \textit{MSD Allmusic Style Dataset (MASD)} presented in Table \pageref{tab:AllmusicStylePartition}. For completeness also the \textit{MSD Allmuisc Style Dataset - Multiclass (Multi-MASD)} is provided. This set contains the pure track-style mapping as collected from Allmusic.com.


\begin{table}
	\centering
	\caption{The MSD Allmusic Style Dataset (MASD)}
	\begin{tabular}{l r }
		
		\hline
		\rowcolor{gray!30}
		\textbf{Genre Name} & \textbf{Number of Songs} \\
		\hline
		
		Big Band             & 3,115 \\
		Blues Contemporary   & 6,874 \\
		Country Traditional  & 11,164 \\
		Dance                & 15,114 \\
		Electronica          & 10,987 \\
		Experimental         & 12,139 \\
		Folk International   & 9,849 \\
		Gospel               & 6,974 \\
		Grunge Emo           & 6,256 \\
		Hip Hop Rap          & 16,100 \\
		Jazz Classic         & 10,024 \\
		Metal Alternative    & 14,009 \\
		Metal Death          & 9,851 \\
		Metal Heavy          & 10,784 \\
		Pop Contemporary     & 13,624 \\
		Pop Indie            & 18,138 \\
		Pop Latin            & 7,699 \\
		Punk                 & 9,610 \\
		Reggae               & 5,232 \\
		RnB Soul             & 6,238 \\
		Rock Alternative     & 12,717 \\
		Rock College         & 16,575 \\
		Rock Contemporary    & 16,530 \\
		Rock Hard            & 13,276 \\
		Rock Neo Psychedelia & 11,057 \\
		
		\hline
		
		Total & 273,936 \\
		
	\end{tabular}
	\label{tab:AllmusicStylePartition}
\end{table}


\subsubsection{Derivates of the Allmusic Dataset}

The genre annotations of the Allmusic Genre Datset, specifically the Top-MAGD genre assignments, are further processed by Schreiber \cite{schreiber2015improving}. The Top-MAGD annotations are combined with additional crowd sourced genre labels retrieved from Last.fm\footnote{\url{http://last.fm}} and Beatunes\footnote{\url{http://www.beatunes.com/}}. Latent Semantic Analysis (LSA) is applied to infer genre taxonomies which are then matched with genre assignments provided for the MSD (including Top-MAGD). Matching is approached by using majority voting and truth by consensus. Assignments are provided for both approaches. They are also partitioned corresponding to the stratified 90\%, 80\%, 66\% and 50\% splits, presented in this chapter.


\subsection{Benchmark Partitions} 
\label{ch4.2:partitions}

Influenced by the tremendous success in the text classification domain, specifically with the landmark Reuters-21578 corpus, a number of benchmark partitions are provided that researchers can use in their future studies, in order to facilitate repeatability of experiments with the MSD beyond x-fold cross validation. The following categories of splits are provided:

\begin{itemize}
	\item Splits with all the ground truth assignments into genre and style classes, described in Section~\ref{ch4.2:genre:Allmusic}.
	\item Splits with just the majority classes from these two ground truth assignments.
	\item Splits considering the sample rate of the files, i.e. only the 22khz samples, only the 44khz samples, and a set with all audio files.
\end{itemize}

\noindent
In particular, the following size partitions were provided:

\begin{itemize}
	\item ``Traditional'' splits into training and test sets, with 90\%, 80\%, 66\% and 50\% size of the training set, applying stratification of the sampling to ensure having the same percentage of training data per class, which is important for minority classes.
	\item A split with a fixed number of training samples, eq\-ually sized for each class, with 2,000 and 1,000 samples per class for the genre and style data sets, respectively. This excludes minority classes with less than the required number of samples.
\end{itemize}

\noindent
Finally, stratification on other criteria than just the ground truth class was applied, namely:

\begin{itemize}
	\item Splits into training and test sets with an artist filter, i.e. avoiding to have the same artist in both the training and test set; both stratified and non-stratified sets are provided
	\item As above, but with an album filter, i.e. no songs from the same album appear in both training and test set, to account for more immediate production effects
	\item As above, but with a time filter, i.e. for each genre using the earlier songs in the training set, and the later releases in the test set.
\end{itemize}

\noindent
\textit{\smaller This dataset and its corresponding evaluation presented in this chapter was published and presented in \textit{Alexander Schindler, Rudolf Mayer and Andreas Rauber. ``Facilitating comprehensive benchmarking experiments on the million song dataset''} at the 13th International Society for Music Information Retrieval Conference (ISMIR 2012), pages 469-474, Porto, Portugal, October 8-12 2012 \cite{schindler2012}. }


\clearpage

\section{The Music Video Dataset}
\label{ch4.1:intro}

To facilitate comparable results and reproducible research on music related visual analysis of music videos the \textit{Music Video Dataset (MVD)} is introduced sequentially in \cite{SCHINDLER_2013CMMR,schindler2015} and \cite{Schindler_2016tist} (see Chapters \ref{ch6:intro} to \ref{ch8:intro}). The MVD follows the Cranfield paradigm \cite{cleverdon1967cranfield} and provides test collections of multimedia documents and corresponding ground truth assignments within the context of well-defined tasks. The main focus of the dataset is set on the development and evaluation of visual or audio-visual features that can be used to augment or substitute audio-only based approaches. The MVD consists of four major subsets that can be combined into two bigger task related collections. The strong emphasis on classification experiments is motivated by their facilitation of rapid content descriptor development. The data-sets are carefully selected to be specialized tools in this process. Despite their different focuses, all sub-sets are non-overlapping and thus can be mutually combined. This section provides a detailed overview of the distinct dataset properties and class/genre descriptions as well as its creation. \\

\noindent
The four major sub-sets of the MVD are:

\begin{description}[leftmargin=!,itemsep=3pt,labelwidth=\widthof{\bfseries Bollywood},labelindent=\descleftmargin,rightmargin=\descrightmargin]
	
	\item[MVD-VIS:] The \textit{M}usic \textit{V}ideo \textit{D}ata\-set for \textit{VIS}ual content analysis and classification (see Chapter \ref{ch4:mvdvis}) is intended for classifying music videos by their visual properties only. 
	
	\item[MVD-MM:] The \textit{M}usic \textit{V}ideo \textit{D}ataset for \textit{M}ulti\textit{M}odal content analysis and classification (see Chapter \ref{ch4:mvdmm}) is intended for multi-modal classification and retrieval tasks. 
	
	\item[MVD-Themes:] The MVD-Themes data-set is a collection based on music themes which span across multiple genres such as ``Christmas Music'' or ``Portest Songs''.
	
	\item[MVD-Artists:] The MVD-Artists data-set is a collection of music videos by 20 popular western music artists.
	
\end{description}

\noindent
Further these sub-sets can be combined into the following two larger sets:

\begin{description}[leftmargin=!,itemsep=3pt,labelwidth=\widthof{\bfseries Bollywood},labelindent=\descleftmargin,rightmargin=\descrightmargin]

	\item[MVD-MIX:] The MVD-MIX data-set is a combination of the data-sets MVD-VIS and MVD-MM (see Chapter \ref{ch4:mvdmix}) which can be used to evaluate the performance and stability of a classification approach according to a higher number of classes.
	
	\item[MVD-Complete:] The MVD-Complete data-set is the combination of all the sub-sets of the Music Video Dataset.
	
\end{description}

\pagebreak

\subsection{Dataset Creation}
\label{ch4:dataset_creation}

The dataset creation is preceded by the selection of genres. To align the dataset to contemporary music repositories a pre-selection is based on the Recording Industry Association of America's (RIAA) report on consumer expenditures for sound recordings \cite{united2009statistical} which separates profiles into the following genres: Rock, Pop, Rap/Hip Hop, R\&B/Urban, Country, Religious, Classical, Jazz, Soundtracks, Oldies, New Age, Children's and Other. For the \textit{MVD-VIS} dataset (see Section \ref{ch4:mvdvis}) eight orthogonal classes with minimum overlap are defined. This aim is accomplished by restricting the search on clearly defined sub-genres. For the \textit{MVD-MM} dataset (see Section \ref{ch4:mvdmm}) eight top-level genres with high inter-genre overlaps are selected. Additional avoidance of overlaps between the genres of these two subsets allow for a combination into the bigger \textit{MVD-MIX} (see Section \ref{ch4:mvdmix}) dataset. Each class consists of 100 videos which are primarily selected by their acoustic properties. The class labels are manually assigned and only refer to commonly known music genres. They do not infer to be accurate in musicological terms and are not result of a common agreement. This decision is based on the introducing definition of Music Video Information Retrieval as \textit{a cross-domain approach to MIR problems}. After listening to the tracks the video properties are inspected. A set of criteria has been strictly applied to the selection process (see Figure \ref{list:ch4.1:quality_criteria}).

These criteria and the variance constraints of the subset's genres makes the selection process complex and exhaustive. More than 6000 videos are examined by listening and watching to them. Especially older music video productions are filtered out due to insufficient sound or video quality. Videos are downloaded from Youtube in MPEG-4 format.

\begin{figure}[t]
	\begin{framed}

		\begin{itemize}[noitemsep, topsep=5pt, leftmargin=10pt]
			\itemsep0.4em
			\renewcommand\labelitemi{--}
			
			\item \textbf{Quality filter:} 
			
			\begin{itemize}[noitemsep]
				\itemsep0em
				\renewcommand\labelitemi{--}
				\item A minimum of 90 kBits/s audio encoding
				\item A video resolution ranging from QVGA (320x240) to VGA (640x480) 
			\end{itemize}
			
			\item \textbf{Content filter:}
			
			\begin{itemize}[noitemsep]
				\itemsep0em
				\renewcommand\labelitemi{--}
				\item Only official music videos
				\item No or minimal watermarking
				\item No lyric-videos (Videos showing only lyrics)
				\item No non-representational (not showing artists)
				\item No live performance, abstract or animated videos
				\item No videos with intro/outro longer than 30 seconds
			\end{itemize}

			\item \textbf{Stratification:}
			\begin{itemize}[noitemsep]
				\itemsep0em
				\renewcommand\labelitemi{--}
				\item Only two tracks by the same artist in \textit{MVD-VIS}, \textit{MVD-MM} and \textit{MVD-MIX} (exceptions: Bollywood, Opera)
				\item Artists of the \textit{MVD-Artists} dataset do not feature other artists of this set.
				\item Tracks of the \textit{MVD-Themes} dataset are not contained in the \textit{MVD-VIS} or \textit{MVD-MM} set.
			\end{itemize}
			
		\end{itemize}
		
		The stratification rule for the Bollywood and Opera genre of the \textit{MVD-VIS} dataset was substituted by:
		
		\begin{itemize}[noitemsep, topsep=5pt]
			\itemsep0em
			\renewcommand\labelitemi{--}
			
			\item Bollywood: only two tracks from the same movie
			\item Opera: only two tracks of the same opera/performance
			
		\end{itemize}
		
		For the \textbf{MVD-ARTISTS} dataset the following additional criteria were considered:
		
		\begin{itemize}[noitemsep, topsep=5pt]
			\itemsep0em
			\renewcommand\labelitemi{--}
			
			\item has to be an official music video produced by the artist
			\item the lead singer has to appear in the video
			
		\end{itemize}
		
		\caption{List of quality criteria and filter rules considered during the selection and accumulation of the Music Video Dataset.}
		\label{list:ch4.1:quality_criteria}
		
	\end{framed}
\end{figure}

\begin{table}[t]
	
	\begin{tabularx}{\textwidth}{*{12}{X}}
		

		\hline
		\rowcolor{gray!30}
		\multicolumn{6}{c}{\textbf{MVD-VIS}} & \multicolumn{6}{|c}{\textbf{MVD-MM}} \\
		\hline
		
		\multicolumn{6}{l}{} & \multicolumn{6}{|l}{\textbf{}} \\ [-1.5ex]
		
		\multicolumn{3}{l}{\textbf{Genre}} & \multicolumn{2}{r}{\textbf{Videos}} & \multicolumn{1}{c}{\textbf{Artists}} & \multicolumn{3}{|l}{\textbf{Genre}} & \multicolumn{2}{r}{\textbf{Videos}} & \multicolumn{1}{l}{\textbf{Artists}} \\
		
		\multicolumn{6}{l}{} & \multicolumn{6}{|l}{\textbf{}} \\ [-1.5ex]
		\multicolumn{3}{l}{Bollywood} & \multicolumn{2}{r}{ 100 } & \multicolumn{1}{c}{32} & \multicolumn{3}{|l}{80s    }  & \multicolumn{2}{r}{ 100 } & \multicolumn{1}{c}{72} \\ 
		\multicolumn{3}{l}{Country  } & \multicolumn{2}{r}{ 100 } & \multicolumn{1}{c}{70} & \multicolumn{3}{|l}{Dubstep  }  & \multicolumn{2}{r}{ 100 } & \multicolumn{1}{c}{78} \\ 
		\multicolumn{3}{l}{Dance    } & \multicolumn{2}{r}{ 100 } & \multicolumn{1}{c}{84} & \multicolumn{3}{|l}{Folk     }  & \multicolumn{2}{r}{ 100 } & \multicolumn{1}{c}{66} \\ 
		\multicolumn{3}{l}{Latin    } & \multicolumn{2}{r}{ 100 } & \multicolumn{1}{c}{72} & \multicolumn{3}{|l}{Hard Rock}  & \multicolumn{2}{r}{ 100 } & \multicolumn{1}{c}{69} \\ 
		\multicolumn{3}{l}{Metal    } & \multicolumn{2}{r}{ 100 } & \multicolumn{1}{c}{76} & \multicolumn{3}{|l}{Indie    }  & \multicolumn{2}{r}{ 100 } & \multicolumn{1}{c}{64} \\ 
		\multicolumn{3}{l}{Opera    } & \multicolumn{2}{r}{ 100 } & \multicolumn{1}{c}{NA} & \multicolumn{3}{|l}{Pop Rock }  & \multicolumn{2}{r}{ 100 } & \multicolumn{1}{c}{65} \\ 
		\multicolumn{3}{l}{Rap      } & \multicolumn{2}{r}{ 100 } & \multicolumn{1}{c}{81} & \multicolumn{3}{|l}{Reggaeton}  & \multicolumn{2}{r}{ 100 } & \multicolumn{1}{c}{69} \\ 
		\multicolumn{3}{l}{Reggae   } & \multicolumn{2}{r}{ 100 } & \multicolumn{1}{c}{75} & \multicolumn{3}{|l}{RnB      }  & \multicolumn{2}{r}{ 100 } & \multicolumn{1}{c}{67} \\ 
		\multicolumn{6}{l}{} & \multicolumn{6}{|l}{}\\[-1.5ex]
		
		\hline
		\rowcolor{gray!30}
		\multicolumn{6}{c}{\textbf{MVD-MIX}} & \multicolumn{6}{|c}{\textbf{MVD-Themes}} \\
		\hline
		
		\multicolumn{6}{l}{} & \multicolumn{6}{|l}{}\\[-1.5ex]
		\multicolumn{3}{l}{                } & \multicolumn{2}{r}{    } & \multicolumn{1}{c}{    } & \multicolumn{3}{|l}{Christmas    } & \multicolumn{2}{r}{56} & \multicolumn{1}{c}{42}  \\ 
		\multicolumn{3}{l}{MVD-VIS + MVD-MM} & \multicolumn{2}{r}{1600} & \multicolumn{1}{c}{1040} & \multicolumn{3}{|l}{K-Pop        } & \multicolumn{2}{r}{50} & \multicolumn{1}{c}{39}  \\ 
		\multicolumn{3}{l}{16 Genres}        & \multicolumn{2}{r}{    } & \multicolumn{1}{c}{    } & \multicolumn{3}{|l}{Broken Heart } & \multicolumn{2}{r}{56} & \multicolumn{1}{c}{48}  \\ 
		\multicolumn{3}{l}{         }        & \multicolumn{2}{r}{    } & \multicolumn{1}{c}{    } & \multicolumn{3}{|l}{Protest Songs} & \multicolumn{2}{r}{50} & \multicolumn{1}{c}{42}  \\ 
		\multicolumn{6}{l}{} & \multicolumn{6}{|l}{}\\[-1.5ex]

		\hline
		\rowcolor{gray!30}
		\multicolumn{12}{c}{\textbf{MVD-Artists (v2.0)}} \\
		\hline

		\multicolumn{6}{l}{} & \multicolumn{6}{l}{}\\[-1.5ex]
		\multicolumn{2}{l}{\textbf{Artist Name}} & \multicolumn{2}{r}{\textbf{Videos}} & \multicolumn{2}{l}{\textbf{Artist Name}} &  \multicolumn{2}{r}{\textbf{Videos}} & \multicolumn{2}{l}{\textbf{Artist Name}} &  \multicolumn{2}{c}{\textbf{Videos}} \\ 
		
		\multicolumn{12}{c}{} \\[-1.5ex]
		\multicolumn{3}{p{0.24\textwidth}}{Aerosmith         }  & 23 & \multicolumn{3}{l}{Jennifer Lopez    }  & 23 & \multicolumn{2}{l}{Nickelback     } & & 18  \\
		\multicolumn{3}{p{0.24\textwidth}}{Avril Lavigne     }  & 20 & \multicolumn{3}{l}{Justin Timberlake }  & 12 & \multicolumn{2}{l}{P!nk           } & & 23  \\
		\multicolumn{3}{p{0.24\textwidth}}{Beyonce           }  & 26 & \multicolumn{3}{l}{Katy Perry        }  & 12 & \multicolumn{2}{l}{Rihanna        } & & 25  \\
		\multicolumn{3}{p{0.24\textwidth}}{Bon Jovi          }  & 27 & \multicolumn{3}{l}{Madonna           }  & 30 & \multicolumn{2}{l}{Shakira        } & & 24  \\
		\multicolumn{3}{p{0.24\textwidth}}{Britney Spears    }  & 25 & \multicolumn{3}{l}{Maroon 5          }  & 14 & \multicolumn{2}{l}{Taylor Swift   } & & 20  \\
		\multicolumn{3}{p{0.24\textwidth}}{Christina Aguilera}  & 15 & \multicolumn{3}{l}{Matchbox Twenty   }  & 13 & \multicolumn{2}{l}{Train          } & & 11  \\
		\multicolumn{3}{p{0.24\textwidth}}{Foo Fighters      }  & 23 & \multicolumn{3}{l}{Nelly Furtado     }  & 16 & \multicolumn{2}{l}{               } & &     \\
		\multicolumn{12}{c}{} \\[-3.5ex]                  \\

		\hline
		\rowcolor{gray!30}
		\multicolumn{12}{c}{\textbf{MVD-Complete}} \\
		\hline
		
		\multicolumn{12}{c}{} \\[-1.5ex]
		\multicolumn{9}{l}{MVD-VIS + MVD-MM + MVD-THEMES + MVD-ARTISTS}  & \multicolumn{2}{r}{ } & \multicolumn{1}{c}{2212}  \\
		\multicolumn{12}{c}{} \\[-1.5ex]
		
		\hline                      
		
	\end{tabularx}
	
	\caption{The Music Video Dataset - Detailed Overview of structure including class description, number of artists, average Beats per Minutes and standard deviation per genre.}
	\label{tab:MVDMMSOverview}
\end{table}

\subsection{MVD-VIS}
\label{ch4:mvdvis}

The \textit{M}usic \textit{V}ideo \textit{D}ataset for \textit{VIS}ual content analysis (MVD-VIS) is intended for feature development and optimization. To facilitate this, 800 tracks of eight clearly defined and well differentiated sub-genres were aggregated (see Table \ref{tab:MVDMMSOverview}). Their tracks were selected concerning minimal with-in class variance in acoustic characteristics, thus sharing high similarity in instrumentation, timbre, tempo, rhythm and mood. Audio classification results provided in Table \ref{tab:Results} reflect that state-of-the-art audio-content based approaches can accurately discriminate such well differentiated classes. Based on the premise that the tracks of a class sound highly similar, these results should serve as a baseline for the task of identifying patterns of similarity within the visual layer as well as developing means to extract this information. Music genre classification based on conventional audio features provides accuracy results above-average (see Table \ref{tab:ch7:audiobenchmarks}) compared to current benchmarks of the Music Information Retrieval domain as presented in Chapter \ref{ch5.2:results} and by \cite{Fu2011}.

\subsubsection{Class Descriptions}

\medskip
\begin{description}[leftmargin=!,itemsep=3pt,labelwidth=\widthof{\bfseries Bollywood},labelindent=\descleftmargin,rightmargin=\descrightmargin]

	\item [Bollywood]
	This class represents a collection of Hindi songs that have either featured in Bollywood films or are music videos on their own. The music is based on Indian ragas but includes western harmonies and melodies. The traditional rhythm is sometimes merged with contemporary electronic dance music. The instrumentation is a mixture of the predominating Indian Sitar, Tabla, Bansuri, Shehnai and percussions, as well as western instruments including synthesizers and drum computers.
	
	\item [Country]
	A collection of contemporary north American popular music with its roots in American folk music, with focus on the sub-genres Honky-Tonk, Hillbilly and Country Rock. Dominating instruments are pedal steel guitar, fiddle and piano. Tracks have an strong emphasis on mid-tempo upbeat rhythms with a gently-swinging shuffle.
	
	\item [Dance]
	This class focuses on House music, a sub-genre of electronic dance-music. With an emphasis on Deep-, Electro- and Progressive-House its sound is characterized by repetitive 4/4 base drum beats, off-beat hi-hat cymbals and ambient synthesizers.
	
	\item [Latin]
	Bachata is the traditional folk music of the Dominican Republic. It was strongly influenced by Merengue and features a high pitched \textit{requinto} guitar playing main theme and interludes as well as a characteristic rhythm using Guira and Bongos.
	
	\item [Metal]
	This class is dominated by aggressive sub-genres of Heavy Metal music (e.g. Metalcore, Thrash Metal, etc.) characterized by heavily distorted guitars and aggressive drum plays and singing.
	
	\item [Opera]
	Single tracks or short excerpts of opera performances recorded by official TV stations. The video is edited using the recording of several cameras shot from different angles. Characteristics of these recordings are similar to music videos.
	
	\item [Rap]
	This class focuses on Rap music of the 90ies also known as Gangsta Rap which is a sub-genre of Hip-Hop music. The music is typically described as noisy, dominated by drum machine rhythms that are accompanied by simple repeated synthesizer melodies and bass lines.
	
	\item [Reggae]
	The Reggae collection refers to traditional Jamaican Reggae style with its typical staccato chords played on the offbeats including upbeat tempo Ska and slower Rocksteady songs. Many recordings originate from TV broadcasts and have lower video quality due to lighting.
	
\end{description}

\subsection{MVD-MM}
\label{ch4:mvdmm}

The structure of the \textit{M}usic \textit{V}ideo \textit{D}ataset for \textit{M}ulti\textit{M}odal content analysis (MVD-MM) is aligned to the \textit{MVD-VIS} but its classes are less well differentiated. The heterogeneous distributions of inter and intra class variance refer to problems of imprecision and subjectivity of music genre definitions \cite{mckay2006musical} which are observed in current music classification datasets \cite{sturm2013classification}. This var\-iance was intentionally introduced to facilitate comparability with results reported on these datasets \cite{SCHINDLER_2012amr}. The task is to eval\-uate and improve the performance of visual features in such environments and to analyze if state-of-the-art audio-only based approaches can be improved through audio-visual combinations.

\subsubsection{Class Descriptions}

\medskip
\begin{description}[leftmargin=!,itemsep=3pt,labelwidth=\widthof{\bfseries Hard-Rock a},labelindent=\descleftmargin,rightmargin=\descrightmargin]
	
	\item [80ies]
	New Wave music characterized by punk influences, increased synthesizer usage and electronic production.
	
	\item [Dubstep]
	Sub-genre of electronic dance music with characteristic reverberating sub-bass modulated at different speeds - also referred to as wobble bass - combined with syncopated drum and percussion patterns.
	
	\item [Folk]
	Folk-influenced romantic or melancholic music ranging from Indie-Folk to slow-tempo Indie-Pop/Rock.
	
	\item [Reggaeton]
	Latin American genre with a characteristic percussion rhythm referred to as \textit{Dem Bow}.
	
	\item [RnB]
	Contemporary Rhythm and Blues - a progression of classic R\&B with strong electronic influences including Hip Hop elements and drum-machine rhythms.
	
	\item [Indie]
	A broad class spreads from Indie Pop/Rock/Folk to Alternative Rock. Sound can be described as sensitive, melancholic, low-fidelity with experimental influences.
	
	\item [Hard-Rock]
	A mix of loud aggressive guitar rock and the pop-oriented Glam Rock/Metal genre.
	
	\item [Pop-Rock]
	A broad mixture of contemporary dance and mainstream rock music.
	
\end{description}

\subsection{MVD-MIX}
\label{ch4:mvdmix}

The \textit{MVD-MIX} dataset is a combination of the datasets MVD-VIS and MVD-MM. The distinct genres of the subsets have been selected to facilitate a union of the two sets providing a non-overlapping bigger set. While MVD-VIS is intended for feature development and optimization and MVD-MM is for evaluation, the MVD-MIX set is for evaluating the performance of the features concerning their stability towards a higher number of classes.

\subsection{MVD-Themes}
\label{ch4:mvdthemes}

The MVD-Themes set is a collection of thematically tagged classes that span across musical genres. The task is aligned to the MusiClef multi-modal music tagging task \cite{orio2012musiclef}. The strong contextual and non-audio connotations of the themes should be captured by information extracted from the visual layer.  To address cross-lingual and cross-cultural challenges \cite{lee2005challenges} of multi-modal approaches analysing song lyrics \cite{mayer2008rhyme,hu2010improving} (most of these approaches were evaluated only for the English language) the MVD-THEMES set includes performances in various languages. The following Themes are provided: 

\medskip
\begin{description}[leftmargin=!,itemsep=3pt,labelwidth=\widthof{\bfseries Protest Songs a},labelindent=\descleftmargin,rightmargin=\descrightmargin]
	
	\item [Christmas]
	Tracks that can be related to Christmas. \textit{Genres} covered: Alternative-, Indie- and Hard-Rock, 60s-, 80s- and 90s-Pop, Dance, Rock 'n Roll, RnB, Soul, Big Band, Country, A capella. \textit{Langauges:} English, Thai.
	
	\item [K-Pop]
	Korean-Pop is strongly influenced by western music \cite{lee2013k} and characterized through visual content (synchronized dance formations, colorful out\-fits). \textit{Genres} covered: Pop, Dance, RnB, Rap. \textit{Langauges:} Korean, English (chorus).
	
	\item [Broken Heart]
	Songs about sorrowfully loosing someone beloved either through death or end of a relationship. \textit{Genres} covered: Rock, Hard Rock, Metal, Pop, RnB, Country, Folk, 80s. \textit{Languages:} English, Taiwanese.
	
	\item [Protest Songs]
	Songs protesting against war, racism, police power and social injustice. \textit{Genres} covered: Pop, Folk, Rap, Reggae, Rock, Punk Rock, Metal, Punjabi, Indie, 80s. \textit{Languages}: English, French, German, Egyptian Arabic, Hindi.
	
\end{description}

\subsection{MVD-ARTISTS}
\label{ch4:mvdartists}

This dataset for audio-visual based artist identification using music videos is a set of 20 popular western music artists listed in Table \ref{tab:MVDMMSOverview}. This set was initially introduced with only 14 artists (version 1.0, see Table \ref{tab:MVDArtists1.0}) for the evaluations presented in Chapter \ref{ch6:intro}. This initial set was further expanded to contain now 20 artists and 124 additional music videos (version 2.0, see Table \ref{tab:MVDMMSOverview}). Popular artists were chosen to meet the requirement of collecting enough music videos for each musician which predominately belong to the two genres Pop and Rock.

\subsection{MVD-COMPLETE}
\label{ch4:mvdcomplete}

The \textit{MVD-Complete} dataset is a combination of the MVD-MIX and the MVD-Artists datasets providing 2212 music videos for similarity search and recommendation. Since the dedicated tasks of the sub-sets are not overlapping, their classes are not semantically related and thus no class-labels are provided.

\begin{table}[t]
	
	\begin{tabularx}{\textwidth}{*{12}{X}}
		
		\hline
		\rowcolor{gray!30}
		\multicolumn{4}{c}{\textbf{MVD-Artists (v1.0)}} \\
		\hline

		\multicolumn{4}{l}{}\\[-1.5ex]
		\textbf{Artist Name} & \textbf{Videos} & \textbf{Artist Name} &  \textbf{Videos}  \\ 
		
		\multicolumn{4}{c}{} \\[-1.5ex]
		Aerosmith           & 23 & Jennifer Lopez   & 21  \\
		Avril Lavigne       & 20 & Madonna          & 25  \\
		Beyonce             & 19 & Maroon 5         & 10  \\
		Bon Jovi            & 26 & Nickelback       & 18  \\
		Britney Spears      & 24 & Rihanna          & 21  \\
		Christina Aguilera  & 14 & Shakira          & 20  \\
		Foo Fighters        & 19 & Taylor Swift     & 19    \\
		\multicolumn{4}{c}{} \\[-3.5ex]                  \\                                             
		
		\hline                      
		
	\end{tabularx}
	
	\caption{MVD-Artists (v1.0) - Music Video Dataset subset for 14 Artists as used in Chapter \ref{ch6:intro}.}
	\label{tab:MVDArtists1.0}
\end{table}

\newpage

\section{Summary}

This chapter introduced the main datasets which are used in this thesis to develop and test visual features and which are major contributions of this dissertation.

\begin{itemize}
	
	\item The \textbf{Million Song Dataset (MSD)} is one of the biggest datasets made available to the MIR research domain. This chapter introduced additional acoustic and music content descriptors for the MSD to overcome some initial shortcomings and to facilitate comprehensive large scale experiments for various content based MIR tasks. Additionally, further ground-truth assignments for music genres and music styles were introduced. To foster exchange between different researchers, for a number of tasks standardized splits between training and test data are provided.
	
	\item The \textbf{Music Video Dataset (MVD)} is the first dataset intended entirely for developing audio-visual approaches to MIR. It is composed of four combine-able subsets. These subsets of the MVD are assembled to facilitate the development and evaluation of visual and audio-visual content descriptors.
	
\end{itemize}

\noindent
Baseline results for these datasets will be provided in the next Chapter.

\chapter{Audio-only Analysis of the MSD}
\label{ch5:intro}

The Music Video Information Retrieval (MVIR) approach presented in this thesis represents a novel way to approach the Music Information Retrieval (MIR) problem space. As discussed in Chapter \ref{ch2:intro} music videos have not yet received appropriate attention and research towards MIR related video analysis is lacking necessary resources such as datasets, ground truth assignments or benchmark results to compare new results with. Further, the research questions presented in Chapter \ref{ch1:research_questoins} open a new perspective to the MIR problem space with no preceding research available to properly base the necessary experimentation on.

This thesis is a step towards providing these resources. The previous chapter already introduced a set of datasets that can be used for developing and evaluating new approaches to MIR tasks using the additional information provided by the visual layer of music videos. One requirement was to obtain audio files and ground truth assignments for the Million Song Dataset in order to facilitate large scale experiments. A second was to create a representative collection of music videos to facilitate audio-visual analysis, feature development and experiments. Due to the novelty of these created datsets, they are lacking comparable results. Further, for some datasets, proprietary feature sets are publicly made available which also require proper pre-analysis to reliably use them in experiments and to compare their performance with results reported in literature. This chapter explicates initial experiments that are carried out in order to asses these results, to evaluate proper pre-processing steps for different feature sets, as well as to provide baseline results for succeeding experiments.

This chapter presents two separate evaluations which support the audio-visual analysis of music videos. Section \ref{ch5.1:intro} provides audio-only results for the MSD which serve as baselines for the audio-visual approaches. Section \ref{ch5.2:intro} empirically evaluates the pre-extracted feater-set, which is provided by the MSD, and provides comparative results on standard music genre classification datsaets. These results contribute especially to the audio-visual artist identification approach presented in Chapter \ref{ch6:intro}.

\pagebreak

\section{Million Song Dataset Experiments}
\label{ch5.1:intro}

The experiments on the Million Song Dataset (MSD) aimed at providing baseline results for the audio-visual experiments. In a first step, baseline results for the ground-truth assignments presented in \ref{ch4.2:intro} are provided as a general overview. This sections only details the experimental setup and the results. For full details on the dataset and the evaluated ground-truth assignments please refer to Chapter \ref{ch4.2:intro}. More precisely, the results of a musical genre classification experiment on the MSD Allmusic Guide Style Dataset (MASD) with a frequently-used 2/3 training and 1/3 test set split will be discussed.

\subsection{Initial Music Genre Classification Experiments}
\label{ch5.1:MSD}

Table~\ref{tab:resultsStyle} shows classification accuracies obtained with five different classifiers using the WEKA Machine Learning Toolkit~\cite{Witten99weka:practical}, version 3.6.6. Specifically, Na\"ive Bayes, Support Vector Machines, k-nearest Neighbours, a J48 Decision Tree and Random Forests (specific configuration detailed in the subsequent paragraphs) were applied. The number in parentheses after the feature set name corresponds to the number given in Table~\ref{tab:features}. Bold print indicates the best, italics the second best result per feature set (column-wise). 

The following feature sets are used in the experiments to evaluate the performance of the feature:

\begin{description}[leftmargin=!,itemsep=1pt,labelwidth=\widthof{\bfseries 12345},labelindent=\descleftmargin,rightmargin=0cm] 
	
	\item[jMir:] The jMir feature set \cite{mckay2010} as described in detail and extracted as described in Chapter \ref{ch4.1:marsyas} is included due to its popularity in literature. For the following features mean and standard deviation values were calculated: The set of \textbf{Spectral Features (spfe)} containing Spectral Centroid, Spectral Flux and Spectral Rolloff. \textbf{Mel-Frequency Cepstral Coefficients (mfcc)}. 
	
	\item[Psychoacoustic Features:] Psychoacoustics feature sets as described in detail in Chapter \ref{ch4.1:rp_extract}. From this \textit{rp\_extract} feature family, the following feature sets are extracted: \textbf{Rhythm Histograms (RH)} aggregating the modulation values of the critical bands computed in a Rhythm Pattern (RP). \textbf{Statistical Spectrum Descriptors (SSD)} capture both timbral and rhythmic information. The features were extracted using the Matlab implemetation of rp\_extract\footnote{http://www.ifs.tuwien.ac.at/mir/downloads.html} - version 0.6411.
	
\end{description}

\noindent
The classifiers used in this evaluation represent a selection of machine learning algorithms frequently used in MIR research.

\begin{description}[leftmargin=!,itemsep=1pt,labelwidth=\widthof{\bfseries 12345},labelindent=\descleftmargin,rightmargin=0cm]
	
	\item [K-Nearest Neighbors (KNN):] the nonparametric classifier has been applied to various music classification experiments and has been added to this evaluation due to its popularity. It was tested with Eucledian (L2) distance with one nearest neighbor (k = 1).
	
	\item [Support Vector Machines (SVM):] have shown remarkable performance in supervised music classification tasks. SVMs were tested with Linear PolyKernels.
	
	\item [J48] The C4.5 decision tree is not as widely used as KNN or SVM, but it has the advantage of being relatively fast to train, which might be a concern processing one million tracks. J48 was tested with a pruning confidence factor of 0.25 and a minimum of two instances per leaf.
	
	\item [Random Forest (RF)] The ensemble classification algorithm is inherently slower than J48, but is superior in precision. It was tested with unlimited depth of the trees, ten generated trees and the number of attributes to be used in random selection set to 0.
	
	\item [NaiveBayes (NB)] The probabilistic classifier is efficient and robust to noisy data. It is included in the evaluation due to comparability with experiments reported in literature \cite{mayer2008rhyme}.
	
\end{description}

\begin{table}[b]
	\caption{Classification results on MSD Allmusic Guide Style dataset (MASD), 66\% training set split}
	\label{tab:resultsStyle}
	\footnotesize
	{\centering \begin{tabular}{lrrrrr}
			\\
			\hline
			\rowcolor{gray!30}
			Dataset 					& NB				& SVM				& k-NN				& DT				& RF \\
			\hline
			MFCC (4)					& \textbf{15.04} 	& 20.61 			& \textit{24.13}	& 14.21 			& 18.90 \\
			Spectral (5)				& \textit{14.03}	& 17.91 			& 13.84 			& 12.81 			& 17.21 \\
			Spectral Derivates (5)		& 11.69 			& \textit{21.98}	& 16.14 			& \textit{14.09}	& \textit{19.03} \\
			MethodOfMoments (6)			& 13.26 			& 16.42 			& 12.77				& 11.57 			& 14.80 \\
			LPC (8)						& 13.41 			& 17.92 			& 15.94 			& 11.97 			& 16.19 \\
			SSD (10)					& 13.76 		& \textbf{27.41} 	& \textbf{27.07}	& \textbf{15.06}	& \textbf{20.06} \\
			RH (11)						& 12.38 		& 17.23 			& 12.46 			& 10.30 			& 13.41 \\
			\hline
		\end{tabular} \footnotesize \par}
\end{table}

For this classification task, we have 25 categories, for which the biggest ``Pop Indie'' accounts for 6.60\% of the songs, which is thus the lowest baseline for our classifiers. It can be noted from the results that the jMIR MFFC features provide the best results on the Na\"ive Bayes classifier, followed by the jMIR low-level spectral features. However, all results on this classifier are just roughly twice as good as the baseline and low in absolute terms. Better results have been achieved with Support Vector Machines and k-NN classifiers, on both the Statistical Spectrum Descriptors achieve more than 27\% accuracy. Also on the other two classifiers, Random Forests and Decision Trees, the SSD feature set is the best, followed by either the derivatives of the jMIR spectral features, or the jMIR MFFC implementation.\\


\noindent
\textit{\smaller This evaluation and its experimental results were published along with the introduction of the Million Song Dataset feature and ground-truth enhancements (see Chapter \ref{ch4.2:intro}) in \textit{Alexander Schindler, Rudolf Mayer and Andreas Rauber. ``Facilitating comprehensive benchmarking experiments on the million song dataset''} at the 13th International Society for Music Information Retrieval Conference (ISMIR 2012), pages 469-474, Porto, Portugal, October 8-12 2012 \cite{schindler2012}. }


\subsection{Neural Network based Classifiers}

Music classification is commonly accomplished in two major steps. First, semantically meaningful audio content descriptors are extracted from the sampled audio signal. Second, a machine learning algorithm is applied, which attempts to discriminate between the classes by finding separating boundaries in the multidimensional feature-spaces. Especially the first step requires extensive knowledge and skills in various specific research areas such as audio signal processing, acoustics and/or music theory. Recently many approaches to MIR problems have been inspired by the remarkable success of Deep Neural Networks (DNN) in the domains of computer vision \cite{krizhevsky2012imagenet}, where deep learning based approaches have already become the \textit{de facto standard}. The major advantage of DNNs are their \textit{feature learning} capability, which alleviates the domain knowledge and time intensive task of crafting audio features by hand. Predictions are also made directly on the modeled input representations, which is commonly raw input data such as images, text or audio spectrograms. Recent accomplishments in applying Convolutional Neural Networks (CNN) to audio classification tasks have shown promising results by outperforming conventional approaches in different evaluation campaigns such as the Detection and Classification of Acoustic Scenes and Events (DCASE) \cite{Lidy2016} and the Music Information Retrieval Evaluation EXchange (MIREX) \cite{Lidy_Schindler_MIREX2016}.

An often mentioned paradigm concerning neural networks is that deeper networks are better in modeling non-linear relationships of given tasks \cite{szegedy2015going}. So far preceding MIR experiments and approaches reported in literature have not explicitly demonstrated the advantage of deep over shallow network architectures in a magnitude similar to results reported from the computer vision domain. This may be related to the absence of similarly large datasets as they are available in the visual related research areas. In this chapter shallow and deep neural network architectures are presented and evaluated on the Million Song Dataset.

\subsubsection{Method}
\label{ch5.1:method}

\begin{figure}[t]
	\centering
	\includegraphics[width=0.9\columnwidth]{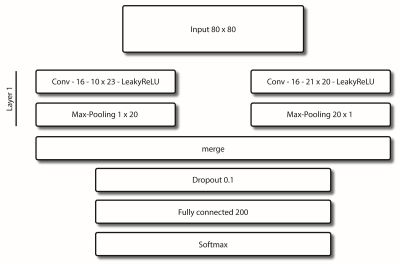}
	\caption{Shallow CNN architecture}
	\label{fig:CNNarchitecture_original} 
\end{figure}

The parallel architectures of the neural networks used in the evaluation are based on the idea of using a time and a frequency pipeline described in \cite{pons_cbmi2016}, which was successfully applied in two evaluation campaigns \cite{Lidy2016,Lidy_Schindler_MIREX2016}. The system is based on a parallel CNN architecture where a log-amplitude scaled Melspectrogram is fed into separate CNN layers which are optimized for processing and recognizing music relations in the frequency domain and to capture temporal relations (see Figure \ref{fig:CNNarchitecture_original}). 

\begin{figure}[t]
	\centering
	\includegraphics[width=0.7\textwidth]{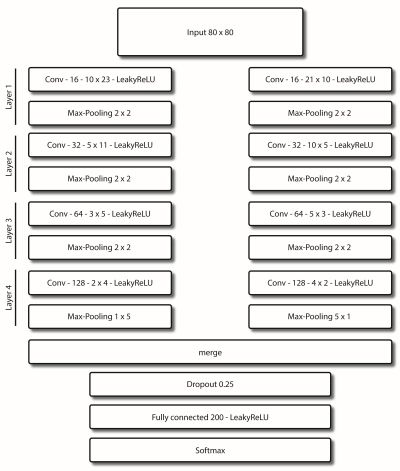}
	\caption{Deep CNN architecture}
	\label{fig:CNNarchitecture_deep} 
\end{figure}

\begin{description}[leftmargin=!,itemsep=1pt,labelwidth=\widthof{\bfseries 12345},labelindent=0cm,rightmargin=0.1cm] 
	
	\item [The Shallow Architecture:]
	In our adaption of the CNN architecture described in \cite{pons_cbmi2016} we use two similar pipelines of CNN Layers with 16 filter kernels each followed by a Max Pooling layer (see Figure \ref{fig:CNNarchitecture_original}). The left pipeline aims at capturing frequency relations using filter kernel sizes of 10\x 23 and Max Pooling sizes of 1\x 20. The resulting 16 vertical rectangular shaped feature map responses of shape 80\x 4 are intended to capture spectral characteristics of a segment and to reduce the temporal complexity to 4 discrete intervals. The right pipeline uses a filter of size 21\x 20 and Max Pooling sizes of 20\x 1. This results in  horizontal rectangular shaped feature maps of shape 4\x 80. This captures temporal changes in intensity levels of four discrete spectral intervals. The 16 feature maps of each pipeline are flattened to a shape of 1\x 5120 and merged by concatenation into the shape of 1\x 10240, which serves as input to a 200 units fully connected layer with a dropout of 10\%.
	
	\pagebreak

	\item [The Deep Architecture:]
	This architecture follows the same principles of the shallow approach. It uses a parallel arrangement of rectangular shaped filters and Max-Pooling windows to capture frequency and temporal relationships at once. But, instead of using the information of the large feature map responses, this architecture applies additional CNN and pooling layer pairs (see Figure \ref{fig:CNNarchitecture_deep}). Thus, more units can be applied to train on the subsequent smaller input feature maps. The first level of the parallel layers are similar to the original approach. They use filter kernel sizes of 10\x 23 and 21\x 10 to capture frequency and temporal relationships. To retain these characteristics the sizes of the convolutional filter kernels as well as the feature maps are sub-sequentially divided in halves by the second and third layers. The filter and Max Pooling sizes of the fourth layer are set to have the same rectangular shapes with one part being rotated by 90\degree. As in the shallow architecture the same sizes of the final feature maps of the parallel model paths balances their influences on the following fully connected layer with 200 units with a 25\% dropout rate.
	
\end{description}

\subsubsection{Training and Predicting Results}
\label{ch5.1:training}

In each epoch during training the network multiple examples sampled from the segment-wise log-transformed Mel-spectrogram analysis of all files in the training set are presented to both pipelines of the neural network architecture. Each of those parallel pipelines uses the same 80 $\times$ 80 log-transformed Mel-spectrogram segments as input. These segments have been calculated from a fast Fourier transformed spectrogram using a window size of 1024 samples and an overlap of 50\% from 0.93 seconds of audio transformed subsequently into Mel scale and Log scale. For each song of the dataset 15 segments have been randomly chosen.

All trainable layers used the \textit{Leaky ReLU} activation function \cite{Maas2013}, which is an extension to the ReLU (Rectifier Linear Unit) that does not completely cut off activation for negative values, but allows for negative values close to zero to pass through. It is defined by adding a coefficient $\alpha$ in $f(x) = \alpha{x}$, for $x < 0$, while keeping $f(x) = x$, for $x \geq 0$ as for the ReLU. In our architectures, we apply Leaky ReLU activation with $\alpha=0.3$. $L_1$ weight regularization with a penalty of 0.0001 was applied to all trainable parameters. All networks were trained towards \textit{categorical-crossentropy} objective using the stochastic \textit{Adam} optimization \cite{Diederik2014} with $beta_{1}=0.9$, $beta_2=0.999$, $epsilon=1e-08$ and a learning rate of 0.00005. 

\medskip\noindent
The system was implemented in Python and using \textit{librosa} \cite{mcfee2015librosa} for audio processing and Mel-log-transforms and \textit{Theano}-based library \textit{Keras} for Deep Learning.

\subsubsection{Evaluation}
\label{ch5.1:eval}

The presented system analyzes and predicts multiple audio segments per input file, there are several ways to perform the final prediction of an input instance:

\begin{description}[leftmargin=!,itemsep=1pt,labelwidth=\widthof{\bfseries 12345},labelindent=0cm,rightmargin=0.1cm] 
	
	\item[Raw Probability:] The raw accuracy of predicting the segments as separated instances ignoring their file dependencies. 
	
	\item[Maximum Probability:] The output probabilities of the Softmax layer for the corresponding number of classes of the datasets are summed up for all segments belonging to the same input file. The predicted class is determined by the  maximum probability among the classes from the summed probabilities.
	
	\item[Majority Vote:] Here, the predictions are made for each segment processed from the audio file as input instance to the network. The class of an audio segment is determined by the maximum probability as output by the Softmax layer for this segment instance. Then, a majority vote is taken on all predicted classes from all segments of the same input file. Majority vote determines the class that occurs most often in the predictions (in a tie, it will decide for the first one to appear in the list of classes). 
	
\end{description}

\noindent
Stratified 4-fold cross validation was used. Multi-level stratification was applied paying special attention to the multiple segments used per file. It was ensured that the files were distributed according to their genre distributions and that no segments of a training file were provided in the corresponding test split.\\

\noindent
The performance of the shallow and deep architecture is evaluated according their accuracy in classifying the correct genre labels of the MSD. The architectures are further evaluated according their performance after a different number of training epochs. The networks are trained and evaluated after 100 epochs without early stopping. Preceding experiments showed that test accuracy could improve despite rising validation loss though on smaller sets no significant improvement was recognizable after 200 epochs. 

\subsubsection{Data}
\label{ch5.1:datasets}

The evaluation is based on the  \textbf{Million Song Dataset (MSD)} \cite{bertin2011million} using the CD2C genre assignments \cite{schreiber2015improving} (see also Chapter \ref{ch4.2:genre:Allmusic}) which are an adaptation of the MSD genre label assignments presented in Chapter \ref{ch4.2:intro}. For the experiments a sub-set of approximately 49,900 tracks is sub-sampled.

\subsubsection{Results}
\label{ch5.1:results}

The presented results for the Million Song Dataset are part of a larger evaluation on different datasets. This evaluation is presented in more detail in \cite{schindler2016a}.

\medskip\noindent
The results of the experiments provided in Table \ref{tab:cnnmsdresults} were tested for significant difference using a Wilcoxon signed-rank test. None of the presented results showed a significant difference for $p < 0.05$. Thus, we tested at the next higher level $p < 0.1$. The following observations on the MSD was made: A not significant advantage of deep over shallow models was observed. Experiments using data augmentation and longer training were omitted due to the already large variance provided by the MSD which multiplies the preceding datasets by factors from 15 to 50.

\begin{table}[t]
	\center
	\footnotesize
	
	\setlength{\tabcolsep}{2pt}
	\begin{tabularx}{0.58\textwidth}{p{0.35cm}|p{1.5cm}p{1.75cm}p{1.75cm}p{1.8cm}|r}
		
		\hline
		\rowcolor{gray!30}
		\textbf{D} & \textbf{Model} & \textbf{raw}  & \textbf{max}  & \textbf{maj}  & \textbf{ep} \\ 
		\hline
		& & & & & \\[-0.9ex]
		
		& & & & & \\[-0.9ex]
		\multirow{2}{*}{\rotatebox{90}{MSD}}   
		& shallow        & 58.20 (0.49)          & 63.89 (0.81)          & 63.11 (0.74)          & 100  \\ %
		& deep           & \textbf{60.60} (0.28) & \textbf{67.16} (0.64) & \textbf{66.41} (0.52) & 100  \\ %
		& & & & & \\
		\hline
		
	\end{tabularx}
	\caption{Experimental results for the evaluation datasets (D) at different number of training epochs (ep): Mean accuracies and standard deviations of the 4-fold cross-evaluation runs calculated using raw prediction scores (raw) and the file based maximum probability (max) and majority vote approach (maj).}
	\label{tab:cnnmsdresults}
\end{table}


\subsubsection{Conclusions}
\label{sec:conclusions}

The observations of the evaluations presented in \cite{schindler2016a} showed that for smaller datasets shallow models seem to be more appropriate since deeper models showed no significant improvement. Deeper models performed slightly better in the presence of larger datasets, but a clear conclusion that deeper models are generally better could not be drawn. Data augmentation using time stretching and pitch shifting significantly improved the performance of deep models on the small datasets. For shallow models on the contrary it showed a negative effect on the small datasets. Thus, deeper models should be considered when applying data augmentation. Comparing the presented results with previously reported evaluations on the same datasets \cite{SCHINDLER_2012amr} shows, that the CNN based approaches already outperform handcrafted music features such as the Rhythm Patterns (RP) family \cite{lidy10_ethnic} or the in the referred study presented Temporal Echonest Features \cite{SCHINDLER_2012amr}.

\medskip
The experimental results presented in this chapter provided initial results for large scale experiments on the Million Song Dataset. Especially the results of Table \ref{tab:resultsStyle} were represented the first results on the new introduced genre assignments presented in Chapter \ref{ch4.2:intro}. High classification accuracies are reported using feature-set combinations which cover multiple music characteristics such as rhythm, timbre and harmony with Support Vector Machine (SVM) classifiers \cite{fu2011survey}. Due to the large number of instances and the high dimensionality of music content descriptors, it was not feasible to compute such state of the art feature-set/SVM combinations in \cite{schindler2012}.\\

\noindent
\textit{\smaller This evaluation of different Deep Neural Network (DNN) architectures for music classification was published and presented in \textit{Alexander Schindler, Thomas Lidy and Andreas Rauber. ``Comparing shallow versus deep neural network architectures for automatic music genre classification''} at the 9th Forum Media Technology (FMT2016), St. Poelten, Austria, November 23 - November 24 2016 \cite{schindler2016a}. }


\clearpage

\pagebreak 

\section{The Echonest Featuresets}
\label{ch5.2:intro}

Due to copyright restriction, the source audio files of the Million Song Datset (MSD) cannot be distributed. Instead a collection of one million metadata entries and features extracted from the source audio files is provided (an introduction and overview of these features was provided in Chapter \ref{ch4.2:echonestfeatures}). Due to the restrictions concerning the access to the audio content of the MSD (see Chapter \ref{ch4.2:intro}), this set of features represents the only general accessible basis to utilize this dataset. Although the two main audio feature sets are described as similar to \textit{Mel-Frequency Cepstral Coefficients (MFCC)} \cite{logan2000mel} and \textit{Chroma} Features (also referred to as \textit{Pitch class profiles}), the absence of accurate documentation of the extraction algorithms makes such a statement unreliable. Especially, no experiments are reported that verify that the Echo Nest features perform equivalent or at least similar to MFCC and Chroma features from conventional state-of-the-art MIR tools as Marsyas \cite{Tzanetakis2000} or Jmir \cite{McKay2009}. Further, several audio descriptors (e.g. MFCCs, Chroma, loudness information, etc.) are not provided as a single descriptive feature vector. Using an onset detection algorithm, the Echonest's feature extractor returns a vector sequence of variable length where each vector is aligned to a music event. To apply these features to standard machine learning algorithms a preprocessing step is required. The sequences need to be transformed into fixed length representations using a proper aggregation method. Approaches proposed so far include simply calculating the average over all vectors of a song \cite{Dieleman2011}, as well as using the average and covariance of the timbre vectors for each song \cite{bertin2011million}. An explicit evaluation of which method provides best results has not been reported, yet.

This section provides a performance evaluation of the Echonest audio descriptors. Different feature set combinations as well as different vector sequence aggregation methods are compared and recommendations towards optimal combinations are presented. The evaluations are based on four well known and researched MIR genre classification datasets in order to make the results comparable to results reported in literature.

\subsection{Evaluation Method}
\label{ch5.2:eval}

At the time of the introduction of the Million Song Dataset its audio content descriptors were relatively new to the community. Although the Echonest was a spin-off of the \textit{MIT Media Lab\footnote{\url{https://www.media.mit.edu/}}}  little information could be found about these features. In the provided user manual of their proprietary closed source feature extractor, the two main feature sets were only described as \textit{Mel-Spectrum Cepstral Coefiicients (MFCC)} and \textit{Pitch class / Chroma} like.

This section gives a description of the evaluation environment used in the experiments described in Section \ref{ch5.2:experiments}. First, the Echonest features are compared against the two conventional feature sets provided by the Marsyas and Rhythm Patterns feature extractors. The evaluation is performed on four well known datasets that have been widely used and evaluated in music genre classification tasks. The performance of the different features is measured and compared by classification accuracy that has been assessed applying five commonly used classifiers. In a second experiment, different aggregation methods are evaluated according their accuracy in discriminating music genres (see Section \ref{ch5.2:TEN}).
The following feature sets are used in the experiments to evaluate the performance of features provided by the Echnoest Analyzer.

\begin{description}[leftmargin=!,itemsep=1pt,labelwidth=\widthof{\bfseries 12345},labelindent=\descleftmargin,rightmargin=0cm] 
	
	\item[Marsyas Features:] The Marsyas feature set \cite{Tzanetakis2000} (described in detail in Chapter \ref{ch4.1:marsyas}) is included due to its popularity in literature. Marsyas features are extracted using a version of the Marsyas framework\footnote{http://marsyas.info} that has been compiled for the Microsoft Windows operating system. Using the default settings of \texttt{bextract} the complete audio file is analyzed using a window size of 512 samples without overlap, offset, audio normalization, stereo information or downsampling. For the following features mean and standard deviation values were calculated: \textbf{Chroma Features (chro)} corresponding to the 12 pitch classes C, C\#, to B. The set of \textbf{Spectral Features (spfe)} containing Spectral Centroid, Spectral Flux and Spectral Rolloff. \textbf{Timbral Features (timb)} containing Time ZeroCrossings, Spectral Flux and Spectral Rolloff, and Mel-Frequency Cepstral Coefficients (MFCC). \textbf{Mel-Frequency Cepstral Coefficients (mfcc)}. 
	
	\item[Psychoacoustic Features:]
	Psychoacoustics feature sets (described in detail in Chapter \ref{ch4.1:rp_extract}) deal with the relationship of physical sounds and the human brain's interpretation of them. From this \textit{rp\_extract} feature family, the following feature sets are extracted: \textbf{Rhythm Patterns (RP)} representing fluctuations per modulation frequency on 24 frequency bands according to human perception. \textbf{Rhythm Histograms (RH)} aggregating the modulation values of the critical bands computed in a Rhythm Pattern. \textbf{Statistical Spectrum Descriptors (SSD)} capture both timbral and rhythmic information. \textbf{Temporal Statistical Spectrum Descriptor (TSSD)} and dimension. \textbf{Temporal Rhythm Histogram (TRH)} describe variations over time by including a temporal. The features were extracted using the Matlab implemetation of rp\_extract\footnote{http://www.ifs.tuwien.ac.at/mir/downloads.html} - version 0.6411.

	\item[Echonest features] (described in detail in Chapter \ref{ch4.2:echonestfeatures}) of all four datasets were extracted using the Echonest's open source Python library Pyechonest\footnote{https://github.com/echonest/pyechonest/}. This library provides methods for accessing the Echonest API. Python code provided by the MSD Web page\footnote{\url{https://github.com/tb2332/MSongsDB/tree/master/PythonSrc}} was used to store the retrieved results in the same HDF5\footnote{http://www.hdfgroup.org/HDF5/} format which is also used by the MSD.
	
\end{description}

\noindent
For the evaluation four data sets that have been extensively used in music genre classification over the past decade have been used.

\begin{description}[leftmargin=!,itemsep=1pt,labelwidth=\widthof{\bfseries 12345},labelindent=\descleftmargin,rightmargin=0cm]
	
	\item[GTZAN] This data set was compiled by George Tzane\-takis \cite{Tzanetakis2002e} in 2000-2001 and consists of 1000 audio tracks equ\-ally dist\-ributed over the 10 music genres: blues, classical, country, disco, hiphop, pop, jazz, metal, reggae, and rock.
	
	\item[ISMIR Genre] This data set has been assembled for training and development in the ISMIR 2004 Genre Classification contest \cite{Cano2006}. It contains 1458 full length audio recordings from Magna\-tune.com distributed across the 6 genre classes: Classical, Electronic, JazzBlues, MetalPunk, RockPop, World.
	
	\item[ISMIR Rhythm] The ISMIR Rhythm data set, also known as the \textit{Ballroom} data set, was used in the ISMIR 2004 Rhythm classification contest \cite{Cano2006}. It contains 698 excerpts of typical ball\-room and Latin American dance music, covering the genres Slow Waltz, Viennese Waltz, Tango, Quick Step, Rumba, Cha Cha Cha, Samba, and Jive.
	
	\item[Latin Music Database (LMD)] \cite{Jr2008} contains 3227 songs, categorized into the 10 Latin music genres Ax\'e, Bachata, Bolero, Forr\'o, Ga\'ucha, Merengue, Pagode, Salsa, Sertaneja and Tango. The data was labeled by human experts with experience in teaching Latin American dances.
	
\end{description}

\begin{table}
	\centering
	\begin{tabular}{lrrr}
		
		\hline
		Dataset      & Genres & Files & Ref \\ 
		\hline
		GTZAN        & 10     & 1000  & \cite{Tzanetakis2002e} \\ 
		ISMIR Genre  & 6      & 1458  & \cite{Cano2006} \\ 
		ISMIR Rhythm & 8      & 698   & \cite{Cano2006} \\ 
		LMD          & 10     & 3227  & \cite{Jr2008} \\
		\hline
		
	\end{tabular}
	
	\caption{Datasets used in evaluations}
	\label{tab:usedDataset}
\end{table}

\noindent
The classifiers used in this evaluation represent a selection of machine learning algorithms frequently used in MIR research. They are the same set of classifiers as used in the experiments described in Section \ref{ch5.1:MSD}. The classifiers were evaluated using their implementations in the Weka machine learning framework \cite{hall2009} version 3.7 with 10 runs of 10-fold cross-validation.

\begin{description}[leftmargin=!,itemsep=1pt,labelwidth=\widthof{\bfseries 12345},labelindent=\descleftmargin,rightmargin=0cm]
	
	\item [K-Nearest Neighbors (KNN):] Because the results of this classifier rely mostly on the choice of an adequate distance function it was tested with Eucledian (L2) and Manhatton (L1) distance with one one nearest neighbor (k = 1).
	
	\item [Support Vector Machines (SVM):] Linear PolyKernel and RBFKernel (RBF) are used in this evaluation, both with standard parameters: penalty parameter set to 1, RBF Gamma set to 0.01 and c=1.0.
	
	\item [J48] J48 was tested with a confidence factor used for pruning from 0.25 and a minimum of two instances per leaf.
	
	\item [Random Forest (RF)] It was tested with unlimited depth of the trees, ten generated trees and the number of attributes to be used in random selection set to 0.
	
	\item [NaiveBayes (NB)] The probabilistic classifier is efficient and robust to noisy data and has several advantages due to its simple structure.
	
\end{description}

\pagebreak

\subsection{Experiments and Results}
\label{ch5.2:experiments}

\subsubsection{Comparing Echonest features with conventional implementations}

The features \textit{Segments Timbre} and \textit{Segments Pitches} provided by the Echonest's Analyzer are described as MFCC and Chroma 'like'. Unfortunately no further explanation is given to substantiate this statement. The documentation \cite{analyzer2011} gives a brief overview of the characteristics described by these feature sets, but an extensive description of the algorithms used in the implementation is missing.
Compared to conventional implementations of MFCC and Chroma features the most obvious difference is the vector length of \textit{Segments Timbre} - which is supposed to be a MFCC like feature. Most of the available MFCC implementations in the domain of MIR are using 13 cepstral coefficients as described in \cite{logan2000mel} whereas the Echonest Analyzer only outputs vectors with dimensionality 12. Although the number of coefficients is not strictly defined and the use of 12 or 13 dimensions seems to be more due to historical reasons, this makes a direct comparison using audio calibration/benchmark testsets impossible.
To test the assumption, that the Echonest features are similar to conventional implementations of MFCC and Chroma features, the audio descriptors are evaluated on four different datasets using a set of common classifiers as described in the evaluation description (see Sect. \ref{ch5.2:eval}). Echonest Segments Timbre were extracted as described in Section \ref{ch5.2:eval}. The beat aligned vector sequence was aggregated by calculating mean and standard deviation for each dimension. The MFCC implementation of the Marsyas framework was used as reference. Mean and standard deviations of the MFCC features were extracted using \texttt{bextract}.

Table \ref{tab:compENMAR} shows the genre classification accuracies for MFCC and Chroma features from the Echonest Analyzer and Marsyas. Significance testing with a significance level $\alpha = 0.05$ is used to compare the two different features. Significant differences are highlighted in bold letters. According to these results the assumption that \textit{Segments Timbre} are similar to MFCC does not hold. There are significant differences on most of the cases and except for the GTZAN dataset the Echonest features outperform the Marsyas MFCC implementation. Even more drastic are the differences between \textit{Segments Pitches} and Marsyas Chroma features except for the ISMIR Rhythm dataset. Similar to \textit{Segments Timbre} \textit{Segments Pitches} perform better except for the GTZAN dataset.


\begin{table*}[t]
	\caption{Comparing MFCC and Chroma implementations of the Echonest Analyzer (EN) and Marsyas (MAR) by their classification accuracy on the GTZAN, ISMIR Genre (ISMIR-G), ISMIR Rhythm (ISMIR-R) and Latin Music Dataset (LMD) datasets. (EN1) corresponds to mean and variance aggregations of \textit{Segments Timbre} and (EN2) to mean and variance aggragations of \textit{Segments Pitches} (see Page \pageref{ch5.2;TEN_desc}). Significant differences ($\alpha = 0.05$) between EN and MAR are highlighted in bold letters.}
	\footnotesize
	{\centering \begin{tabular}{l|rr|rr|rr|rr}
			
			\multicolumn{9}{c}{Segments Timbre / MFCC}  \\ 
			\hline
			\rowcolor{gray!30}
			& \multicolumn{2}{|c}{\textbf{GTZAN}} & \multicolumn{2}{|c}{\textbf{ISMIR-G}} & \multicolumn{2}{|c}{\textbf{ISMIR-R}} & \multicolumn{2}{|c}{\textbf{LMD}} \\
			\hline
			Dataset       & EN1    & MAR  & EN1   & MAR   & EN1    & MAR  & EN1  & MAR \\
			\hline                      
			SVM Poly      & 61.1 & \textbf{69.0} & \textbf{75.1} & 62.1 & \textbf{63.1} & 57.1 & \textbf{78.4} & 60.4 \\
			SVM RBF       & 35.1  & 39.3 & \textbf{46.8} & 44.1 & 30.3  & 31.0                 & \textbf{41.2} & 38.0 \\
			KNN K1 L2     & 58.1 & \textbf{63.4} & \textbf{77.0} & 64.2 & \textbf{49.2} & 43.3 & \textbf{78.7} & 58.4 \\
			KNN K3 L2     & 57.6  & 61.4 & \textbf{77.0} & 63.0 & 51.8  & 46.8                 & \textbf{79.4} & 56.9 \\
			KNN K1 L1     & 56.6 & \textbf{63.0} & \textbf{77.9} & 63.0 & \textbf{49.4} & 44.0 & \textbf{79.1} & 57.9 \\
			KNN K3 L1     & 56.4 & \textbf{62.3} & \textbf{76.6} & 61.5 & 50.0  & 47.1         & \textbf{79.9} & 57.4 \\
			J48           & 44.7 & \textbf{49.7} & \textbf{69.4} & 52.9 & 40.4  & 37.4         & \textbf{62.5} & 44.4 \\
			Rand-Forest   & 54.7 & \textbf{59.1} & \textbf{75.8} & 60.8 & 50.8  & 45.3         & \textbf{74.7} & 54.0 \\
			NaiveBayes    & 50.5 & \textbf{55.9} & \textbf{63.2} & 49.6 & \textbf{53.3} & 38.3 & \textbf{68.4} & 46.7 \\
			\hline  
			
			\multicolumn{9}{c}{}  \\                                              
			\multicolumn{9}{c}{}  \\     
			
			\multicolumn{9}{c}{Segments Pitches / Chroma}  \\   
			\hline
			\rowcolor{gray!30}
			& \multicolumn{2}{|c}{\textbf{GTZAN}} & \multicolumn{2}{|c}{\textbf{ISMIR-G}} & \multicolumn{2}{|c}{\textbf{ISMIR-R}} & \multicolumn{2}{|c}{\textbf{LMD}} \\
			\hline
			Dataset       & EN2    & MAR  & EN2   & MAR  & EN2    & MAR  & EN2   & MAR \\
			\hline                               
			SVM Poly      & 37.0 & \textbf{41.2} & \textbf{64.3} & 50.3 & 38.7  & 38.6 & \textbf{54.1} & 39.4 \\
			SVM RBF       & \textbf{26.1} & 22.0 & \textbf{50.2} & 46.7 & 22.6  & 24.9 & \textbf{32.6} & 26.1 \\
			KNN K1 L2     & 38.0 & \textbf{42.7} & \textbf{62.1} & 46.0 & 31.8  & 28.9 & \textbf{57.1} & 37.3 \\
			KNN K3 L2     & 35.0 & \textbf{41.1} & \textbf{63.0} & 50.6 & 28.9  & 28.9 & \textbf{54.7} & 36.0 \\
			KNN K1 L1     & 38.2 & \textbf{44.1} & \textbf{62.8} & 45.4 & 32.5  & 27.4 & \textbf{56.4} & 37.3 \\
			KNN K3 L1     & 36.5 & \textbf{42.5} & \textbf{62.7} & 51.3 & 32.5  & 28.4 & \textbf{53.8} & 36.7 \\
			J48           & 27.8 & \textbf{40.1} & \textbf{53.7} & 43.8 & 29.0  & 26.9 & \textbf{41.5} & 33.6 \\
			Rand-Forest   & 37.0 & \textbf{48.5} & \textbf{62.1} & 50.5 & 35.2  & 30.6 & \textbf{53.3} & 39.1 \\
			NaiveBayes    & \textbf{34.1} & 28.6 & \textbf{59.7} & 46.8 & \textbf{39.7} & 22.7 & \textbf{47.0} & 26.9 \\
			
			\hline
		\end{tabular} \footnotesize \par}
	\label{tab:compENMAR}
\end{table*}


\subsubsection{Feature selection and proper aggregation of beat algined vector sequences}
\label{ch5.2:TEN}

The second part of the experiments conducted in this study deals with the huge amount of information provided by the by the MSD respectively the Echonest Analyzer. No initial evaluations were provided giving reliable benchmarks on how to achieve best performing results on these features sets.

\paragraph{Scope of selected Features:}
Due to number of features provided by the MSD only a subset of them was selected for the experiments. A comprehensive comparison of all possible feature combinations is beyond the scope of this experiment. The focus was set on the beat aligned vector sequences \textit{Segments Timbre}, \textit{Segments Pitches}, \textit{Segments Loudness Max} and \textit{Segments Loudness Max Time}. Further \textit{Segments Start} was used to calculate the length of a segment by subtracting the onsets of two consecutive vectors.

\paragraph{Aggregation of Echonest vector sequences:}
A further focus has been set on the feature sets that are provided as beat aligned vector sequences. Such sequences represent time series of feature data that can be harnessed in various MIR scenarios (e.g. audio segmentation, chord analysis). Many classification tasks in turn require a fixed-length single vector representation of the feature data. Consequently, the corresponding Echonest features need to be preprocessed. A straight forward approach would be to simply calculate an average of all vectors resulting in a single vector, but this implies discarding valuable information. Lidy et. al. \cite{lidy10_ethnic, Lidy2010} demonstrated how to effectively harness temporal information of sequentially retrieved feature data by calculating statistical measures. The temporal variants of Rhythm Patterns (RP), Rhythm Histograms (RH) and Statistical Spectrum Descriptor (SSD) describe variations over time reflecting rhythmical, instrumental, etc. changes of the audio spectrum and have previously shown excellent performance on conventional MIR classification benchmark sets as well as non-western music datasets. For this evaluation the vector sequences provided by the Echonest Analyzer were aggregated by calculating the statistical measures mean, median, variance, skewness, kurtosis, min and max.

\paragraph{Temporal Echonest Features}
\label{ch5.2;TEN_desc}
Temporal Echonest Features (TEN) follow the approach of temporal features by Lidy et. al. \cite{lidy10_ethnic}, where statistical moments are calculated from Rhythm Pattern features. To compute Temporal Rhyhtm Patterns (TRP) a track is segmented into sequences of 6 seconds and features are extracted for each consecutive time frame. This approach can be compared to the vector sequences retrieved by the Echonest Analyzer, except for the varying time frames caused by the onset detection based segmentation. To capture temporal variations of the underlying feature space, statistical moments (mean, median, variance, min, max, value range, skewness, kurtosis) are calculated from each feature of the feature-sets.
Different combinations of Echonest features and statistical measures were analyzed. The combinations were evaluated by their effectiveness in classification experiments measured in accuracy. The experiments conclude with a recommendation of a featureset-combination that achieves maximum performance on most of the testsets and classifiers used in the evaluation. 

\begin{description}[leftmargin=!,itemsep=1pt,labelwidth=\widthof{\bfseries Segmen},labelindent=\descleftmargin,rightmargin=0.8cm]
	
	\item [EN0] This represents the trivial approach of simply calculating the average of all \textit{Segments Timbre} descriptors (12 dimensions).
	
	\item [EN1] This combination is similar to EN0 including variance information of the beat aligned \textit{Segments Timbre} vectors already capturing timbral variances of the track (24 dimensions).
	
	\item [EN2] Mean and variance of \textit{Segments Pitches} are calculated (24 dimensions).
	
	\item [EN3] According to the year prediction benchmark task presented in \cite{bertin2011million} mean and the non-redundant values of the covariance matrix are calculated (90 dimensions).
	
	\item [EN4] All statistical moments (mean, median, variance, min, max, value range, skewness, kurtosis) for \textit{Segments Timbre} are calculated (96 dimensions)
	
	\item [EN5] All statistical moments of \textit{Segments Pitches} and \textit{Segments Timbre} are calculated (192 dimensions).
	
	\item [Temporal Echonest Features (TEN)] All statistical moments of \textit{Segments Pitches}, \textit{Segments Timbre}, \textit{Segments Loudness Max}, \textit{Segments Loudness Max Time} and lengths of segments calculated from \textit{Segments Start} are calculated (216 dimension). 
	
\end{description}

\subsubsection{Results}
\label{ch5.2:results}

Table \ref{tab:reults2} shows the results of the evaluations for each dataset. Echonest features are located to the right side of the tables. Only EN0 and EN3-TEN are displayed, because EN1 and EN2 are already presented in Table \ref{tab:compENMAR}. Bold letters mark best results of the Echonest features. If a classifier shows no bold entries, EN1 or EN2 provide best results for it. Conventional feature sets on the left side of the tables provide an extensive overview of how the Echonest features perform in general. Table \ref{tab:bestEN} provides a summary of all Echonest feature combinations EN0 - EN5 and Temporal Echonest Features (TEN).

\paragraph{Good results with simple but short feature sets}
The trivial approach of simply averaging all segments (EN0) provides expectedly the lowest precision results of the evaluated combinations. As depicted in Table \ref{tab:reults2}, the values range between MFCC and Timbre features. On the other hand, taking the low dimensionality of the feature space into account, this approach represents a good choice for implementations focusing on runtime behavior and performance. Especially the non-parametric K-Nearest-Neighbors classifier provides good results. Adding additional variance information (EN1) provides enhanced classification results on \textit{Segments Timbre} features. Specifically Support Vector Machines gain from the extra information provided. As pointed out in Table \ref{tab:bestEN}, this combination already provides top or second best results for K-Nearest Neighbors and Decision Tree classifiers. Again, addressing performance issues, the combinations EN0 and EN1 with only 12 or 26 dimensions may be a good compromise between computational efficiency and precision of classification results.

Chroma features are reported to show inferior music classification performance compared to MFCC \cite{ellis2007classifying}. This behavior was reproduced. Marsyas Chroma features as well as Echonest \textit{Segments Pitches} (EN2) provide the lowest results for their frameworks.

\paragraph{Better results with complex feature sets}
Providing the classifier with more information expectedly results in better performance. Adding more statistical measures to simple feature sets (EN4) provides no significant performance gain but increases the length of the vector by a factor of 4. Also combining \textit{Segments Timbre} with \textit{Segments Pitches} and calculating the statistical moments (EN5) only provides slightly better results. The 192 dimensions of this combination may alleviate this result when performance issues are taken into consideration. Only the initially as benchmark proposed approach by \cite{bertin2011million} (EN3) provides inferior results.

\paragraph{Recommendation: Temporal Echonest Features}
Including additional information of loudness distribution and the varying lengths of segments in the feature set (TEN), enhances performance for all classifiers and provides the best results of the experiments (see Table \ref{tab:bestEN}). For many testset-classifier combinations the Temporal Echonest Features provide best performing results for all feature sets. Compared to similar performing features like TSSD - which have a dimension of 1176 - TENs outperform concerning precision and computational efficiency. Table \ref{tab:bestEN} summarizes the best performing Echonest feature combinations.


\begin{table*}[t]
	\caption{Comparing Echonest, Marsyas and Rhythm Pattern features by their classification accuracy. Best performing Echonest feature combinations are highlighted in bold letters.}
	\scriptsize
	{\centering \begin{tabular}{l|rrrr|rrrrr|rrrrr}
			
			\hline
			\rowcolor{gray!30}
			\multicolumn{15}{c}{\textbf{ISMIR Genre Dataset}}\\
			\hline
			& & & & & & & & & & & & & & \\
			Classifiers        & chro & spfe & timb & mfcc & rp   & rh   & trh  & ssd  & tssd & EN0  & EN3  & EN4  & EN5  & TEN  \\
			\hline                                                                                    
			
			& & & & & & & & & & & & & & \\
			SVM Poly       & 50.3 & 54.9 & 67.7 & 62.1 & 75.1 & 64.0 & 66.5 & 78.8 & 80.9 & 67.0 & 67.2 & 78.5 & 80.4 & \textbf{81.1} \\
			SVM RBF        & 46.6 & 44.2 & 50.0 & 44.1 & 69.0 & 55.5 & 64.5 & 64.1 & 72.0 & 44.3 & 49.1 & 64.9 & 69.4 & \textbf{70.9} \\
			KNN K1 L2      & 46.0 & 56.3 & 65.8 & 64.2 & 72.9 & 60.7 & 63.3 & 77.8 & 76.6 & 76.8 & 64.0 & 75.5 & 75.9 & \textbf{77.8} \\
			KNN K1 L1      & 45.4 & 56.5 & 65.9 & 63.0 & 71.5 & 60.8 & 63.3 & 78.5 & 77.6 & 77.1 & 60.8 & 77.6 & 78.3 & \textbf{81.3} \\
			J48            & 43.8 & 53.3 & 56.5 & 52.9 & 61.9 & 56.9 & 56.7 & 69.6 & 68.3 & 68.5 & 64.5 & 67.4 & 66.5 & 68.0 \\
			Rand-Forest    & 51.5 & 60.4 & 62.3 & 60.8 & 69.8 & 65.2 & 65.4 & 75.7 & 74.6 & 74.3 & 65.9 & 74.7 & 73.2 & 74.4 \\
			NaiveBayes     & 46.8 & 53.2 & 52.3 & 49.6 & 63.5 & 56.7 & 60.2 & 61.0 & 40.2 & 66.1 & 45.5 & \textbf{63.8} & 56.0 & 63.3 \\
			& & & & & & & & & & & & & & \\
			
			\hline                                                                                    
			
			\rowcolor{gray!30}
			\multicolumn{15}{c}{\textbf{Latin Music Database}}\\                                          
			\hline                                                                                        
			& & & & & & & & & & & & & & \\
			SVM Poly       & 39.4 & 38.2 & 68.6 & 60.4 & 86.3 & 59.9 & 62.8 & 86.2 & 87.3 & 70.5 & 69.6 & 82.9 & 87.1 & \textbf{89.0} \\
			SVM RBF        & 26.1 & 19.1 & 51.0 & 38.0 & 79.9 & 36.6 & 53.2 & 71.6 & 83.3 & 29.2 & 40.9 & 69.4 & 76.6 & \textbf{79.3} \\
			KNN K1  L2     & 37.3 & 42.5 & 62.7 & 58.4 & 74.3 & 58.7 & 49.5 & 83.1 & 78.4 & 73.5 & 52.2 & 77.3 & 79.0 & \textbf{80.9} \\
			KNN K1  L1     & 37.3 & 43.2 & 61.5 & 57.9 & 73.8 & 59.0 & 53.1 & 83.8 & 81.7 & 72.6 & 49.8 & 79.8 & 81.6 & \textbf{83.0} \\
			J48            & 33.6 & 38.4 & 48.8 & 44.3 & 57.1 & 43.3 & 43.8 & 64.7 & 64.4 & 58.7 & 53.9 & 60.5 & 61.7 & \textbf{64.8} \\
			Rand-Forest    & 39.4 & 46.4 & 58.1 & 53.6 & 58.8 & 50.3 & 47.5 & 76.3 & 73.0 & 69.9 & 54.9 & 74.1 & 73.5 & \textbf{75.9} \\
			NaiveBayes     & 26.9 & 35.7 & 43.5 & 46.7 & 66.0 & 47.0 & 49.9 & 64.1 & 67.8 & 66.5 & 40.4 & 70.8 & 71.1 & \textbf{73.3} \\
			& & & & & & & & & & & & & & \\
			
			\hline
			\rowcolor{gray!30}
			\multicolumn{15}{c}{\textbf{GTZAN}}\\
			\hline
			
			& & & & & & & & & & & & & & \\        
			SVM Poly       & 41.1 & 43.1 & 75.2 & 67.8 & 64.9 & 45.5 & 38.9 & 73.2 & 66.2 & 56.4 & 53.6 & 63.9 & 65.2 & \textbf{66.9}  \\
			SVM RBF        & 22.0 & 27.1 & 52.1 & 37.7 & 56.7 & 31.4 & 39.9 & 53.1 & 63.3 & 36.7 & 22.3 & 46.6 & 56.3 & \textbf{56.5}  \\
			KNN K1 L2      & 41.9 & 42.1 & 67.8 & 61.8 & 51.5 & 40.2 & 32.7 & 63.7 & 53.4 & 56.3 & 39.9 & 56.8 & 56.1 & \textbf{58.2}  \\
			KNN K1 L1      & 43.6 & 43.0 & 68.2 & 61.7 & 53.4 & 39.8 & 35.8 & 64.1 & 60.6 & 55.1 & 36.4 & 56.9 & 56.3 & \textbf{58.7}  \\
			J48            & 38.6 & 39.2 & 53.6 & 48.9 & 38.3 & 32.6 & 31.6 & 52.0 & 50.6 & \textbf{45.0} & 39.1 & 44.3 & 43.6 & 44.1  \\
			Rand-Forest    & 48.0 & 47.2 & 64.2 & 57.9 & 45.9 & 39.6 & 38.0 & 63.4 & 59.3 & 54.7 & 41.1 & 54.0 & 53.2 & \textbf{55.0}  \\
			NaiveBayes     & 28.1 & 40.0 & 52.2 & 54.9 & 46.3 & 36.2 & 35.6 & 52.4 & 53.0 & 53.1 & 29.5 & \textbf{53.6} & 52.5 & 53.3  \\
			& & & & & & & & & & & & & & \\
			
			\hline                                                                                    \rowcolor{gray!30}
			\multicolumn{15}{c}{\textbf{ISMIR Rhythm}}\\                                                  
			\hline                                                                                        
			& & & & & & & & & & & & & & \\
			SVM Poly       & 38.1 & 41.4 & 60.7 & 54.5 & 88.0 & 82.6 & 73.7 & 58.6 & 56.0 & 55.1 & 51.7 & 62.7 & 63.7 & \textbf{67.3} \\
			SVM RBF        & 25.1 & 27.9 & 36.4 & 29.7 & 79.6 & 36.6 & 63.2 & 42.1 & 55.3 & 24.7 & 26.6 & 37.1 & 46.5 & \textbf{53.1} \\
			KNN K1 L2      & 28.3 & 34.8 & 43.9 & 37.3 & 73.7 & 77.7 & 51.5 & 45.5 & 39.8 & 43.5 & 34.6 & 44.5 & 43.0 & \textbf{45.7} \\
			KNN K1 L1      & 26.8 & 35.8 & 44.3 & 38.9 & 71.4 & 73.9 & 60.3 & 43.4 & 42.1 & 44.0 & 32.9 & 46.9 & 44.7 & \textbf{49.2} \\
			J48            & 26.9 & 33.7 & 37.6 & 37.1 & 64.3 & 67.6 & 65.9 & 37.6 & 35.8 & 38.5 & 34.0 & 38.5 & 40.5 & \textbf{48.0} \\
			Rand-Forest    & 31.0 & 38.1 & 44.4 & 43.8 & 64.9 & 71.6 & 68.2 & 46.6 & 44.1 & 47.5 & 37.1 & 47.9 & 48.8 & \textbf{53.5} \\
			NaiveBayes     & 23.3 & 37.0 & 37.7 & 36.5 & 75.9 & 69.0 & 69.3 & 44.4 & 46.8 & 52.8 & 25.1 & 52.8 & 49.9 & \textbf{55.1} \\
			& & & & & & & & & & & & & & \\
			
			\hline
		\end{tabular}  \par}
	\label{tab:reults2}
\end{table*}


\clearpage


\begin{table}[t]
	\caption{Overview of which Echonest feature combination performs best for a certain classifier on the datasets (a) GTZAN, (b) ISMIR Genre, (c) ISMIR Rhythm and (d) LMD}
	{\centering \begin{tabular*}{0.9\textwidth}{p{2.855cm}|p{1cm}p{1cm}p{1cm}p{1cm}p{1cm}p{1cm}p{1cm}}
			
			\hline
			\rowcolor{gray!30}
			Dataset     & EN0 & EN1   & EN2 & EN3 & EN4 & EN5 & TEN  \\
			\hline                          
			
			& & & & & & & \\
			SVM Poly    &     &       &     &     &     &     & a,b,c,d \\
			SVM RBF     &     &       &     &     &     &     & a,b,c,d \\
			KNN K1 L2   &     & c     &     &     &     &     & a,b,d   \\
			KNN K3 L2   &     & a,b,c &     &     &     &     & d       \\
			KNN K1 L1   &     & c     &     &     &     &     & a,b,d   \\
			KNN K3 L1   &     & c     &     &     &     &     & a,b,d   \\
			J48         & a   & b     &     &     &     &     & c,d     \\
			Rand-Forest &     & a,b   &     &     &     &     & c,d     \\
			NaiveBayes  & b   &       &     &     & a   &     & c,d     \\
			& & & & & & & \\
			
			\hline
			
		\end{tabular*} \par}
	\label{tab:bestEN}
\end{table}

\subsection{Conclusion and Future Work}
\label{ch5.2:conclusion}

This section presented a comparison of Echonest features - as provided by the Million Song Dataset - with feature sets from conventionally available feature extractors. Due to the absence of audio samples, researchers at first hand solely rely on these Echonest features as well as the feature sets described in Chapter \ref{ch4.2:intro} which were published a year later. Thus, the aim was to provide empirically determined reference values for further experiments based on the Million Song Dataset. Six different combinations of Echonest features were used and their statistical moments were calculated to capture their temporal domain. Experiments showed that Temporal Echonest Features (TEN) - a combination of MFCC and Chroma features combined with loudness information as well as the distribution of segment lengths - complimented by all calculated statistical moments - outperforms almost all datasets and classifiers - even conventional feature sets, with a prediction rate of up to 89\%. Although higher percentages have been reported on these datasets based on other feature sets or hybrid combinations of different feature sets, these additional audio descriptions were or are still not available on the MSD. Additionally it was observed, that top results can already be obtained calculating average and variance of \textit{Segments Timbre} features. This short representation could be considered for retrieval systems with constrained computing resources.


Generally, it can be stated that the features provided with the Million Song Dataset perform comparable to conventional MIR feature-sets and due to the official lack of audio files, they represent quite a good choice for experimentation. Unofficially, many research groups used the 7-Digital identifier to download audio samples, which offers a wider range of options and facilitate the extraction of more complex features. On the other hand, after The Echonest has been aquired by Spotify, their music data APIs have been incorporated into the Spotify platform. Although the option to upload local tracks to API to have them analyzed has been removed, the features described in this section are now provided by the Spotify API. This facilitates rapid prototyping and show-casing on a very large corpus. The presented feature aggregation methods apply to the Spotify API too.

\subsection{Distribution of Data}

All feature sets described in Section \ref{ch5.2:experiments}, including Temporal Echonest Features (TEN) and the different aggregated feature combinations EN0 - EN5, are provided for download on the Million Song Dataset Benchmarking platform presented in Chapter \ref{ch4.2:intro} and in \cite{schindler2012}:

\begin{center}
	\url{http://www.ifs.tuwien.ac.at/mir/msd/}
\end{center}

The aggregated Echonest features are provided as single files containing all vectors for the tracks of the MSD and are stored in the WEKA Attribute-Relation File Format (ARFF) \cite{hall2009}. Additionally different benchmark partitions based on different genre label assignments are provided for instant use and comparability.\\

\noindent
\textit{\smaller These experiments and their corresponding evaluations presented in this chapter were presented in \textit{Alexander Schindler and Andreas Rauber. `` Capturing the temporal domain in echonest features for improved classification effectiveness''} at the 10th International Workshop on Adaptive Multimedia Retrieval (AMR2012), Copenhagen, Denmark, October 24-25, 2012. The article was published in \textit{Adaptive Multimedia Retrieval: Semantics, Context, and Adaptation}, Lecture Notes in Computer Science, Volume 8382, pp 214-227, October 29, 2014 \cite{SCHINDLER_2012amr}. }


\newpage

\section{Summary}

This chapter presented results for the datasets introduced in Chapter \ref{ch4:intro}. The experiments presented in this thesis attempt to evaluate if information extracted from music related visual media such as music videos can be harnessed to improve Music Information Retrieval (MIR) tasks as well as to approach new tasks. To facilitate these evaluations this chapter provides audio-analytical results which serve as baselines.

\begin{itemize}
	
	\item Section \ref{ch4.2:intro} provides baseline results for large scale classification experiments on the Million Song Dataset (MSD). First results are calculated using standard feature-sets with common classifiers reported in literature. Recent approaches based on Deep Neural Networks (DNN) are then applied to assess the maximum performance gain in the acoustic domain.
	
	\item The featureset provided for the MSD is the most complete and most accessible set available. Unfortunately, these features have been extracted using proprietary closed source and insufficiently documented feature extractors. Section \ref{ch5.2:intro} provides comparative results for these featuresets. The provided features are described to be comparable to state-of-the-art music content descriptors. In empirical experiments the performance of the provided features was compared against open source implementation of the mentioned features. Results verified this and variances were discussed. Further, methods were evaluated and recommended on how to aggregate the provided featuressets for improved classification performance.
	
\end{itemize}

\noindent
The results of these preceding experiments serve as baseline for the evaluations of the experiments of the consecutive chapters.



\chapter[Visually improved Artist Identification]{Visually improved\\ Artist Identification}
\label{ch6:intro}

\epigraph{
	``The secret to modeling is not being perfect. What one needs is a face that people can identify in a second. You have to be given what’s needed by nature, and what’s needed is to bring something new.''} 
{--- Karl Lagerfeld, Fashion Designer}


In this chapter the problem of music artist identification - the task of identifying the performing musician of a given track - is addressed to demonstrate the opportunities of a Music Video Information Retrieval (MVIR) approach. Artist recognition is an important task for music indexing, browsing and content based retrieval originating from the Music Information Retrieval (MIR) domain. Typically it is subdivided into the tasks \textit{artist identification}, \textit{singer recognition}, and \textit{composer recognition}. Recent achievements in music classification and annotation including artist recognition are summarized in \cite{fu2011survey}. A typical content based approach to this problem is to extract audio features from the corresponding tracks, train a machine learning based classifier and predict the artist name for the given track. This approach is similar to music genre classification, but whereas respectable results are reported from genre prediction, artist identification is still failing to achieve comparable levels of performance. The problem is that audio features used in many of the evaluations reported in literature are statistical descriptions of the audio signal correlating mainly to sound properties such as brightness, timbre or frequency/amplitude modulations over a period of time. All these features describe sound characteristics that are rather related to genre properties. Although an artist is mainly dedicated to a specific genre, its distinct songs are not. Tracks of a record may vary in tempo, instrumentation and rhythm. Further, stylistic orientations of the artists may change over time. The most intrinsic problem is that audio features are low level description of the audio content. Thus, two artists with similar sounding repertoire of songs will get confused, because the discriminating unique qualities of the singers voices are lost during the data reduction phase. The solution provided in \cite{kim2002singer} attempts to extract vocal segments of a song to identify the singer. The vocal segmentation is accomplished through harmonic segmentation in the vocal frequency regions. The identification is based on features from speech coding.

\textbf{Video Information Retrieval (VIR)} on the other side, pursues the same goals in the video domain as MIR does in the music domain. A lot of effort is put into categorizing videos into different genres. A good summary of video classification is provided by \cite{brezeale2008automatic}. Typically, these approaches draw from more than one modality - the most common among them are text-based, audio-based and visual-based. Different properties of videos in conjunction with cinematic principles (e.g., light, motion, transitions from one scene to the other) are explored to estimate the genre of the video. Fast motion and short shot sequences are a good indicator for music videos. Although it is easy to distinguish music videos from other video genres, approaches directly addressing the problem of categorizing videos by their musical genres have not been reported yet. Different acoustic characteristics such as localised sound energy patterns associated with cinematic events are nevertheless used to estimate the video genre \cite{moncrieff2003horror}. 
The multimodal VIR approaches indicate that such a combined approach could also be suitable for the MIR domain. To approach the addressed task of artist identification, the most applicable visual computing approach is \textbf{Face recognition}, which is the task of identifying or verifying a person from still or video images. These tasks has received increased attention from academic and industrial communities over the past three decades due to potential applications in security systems, law enforcement and surveillance, and many other domains. 

In this chapter a system is proposed which uses portrait images of artists to train a face recognition classifier which is applied to frames of music videos. From the video's corresponding audio track, music features are extracted and a classifier is trained on artist name labels. These two classifiers are combined to an ensemble to produce the final prediction on the performing artist. The audio classifier represents a standard MIR approach to the task of artist identification and its results further serve as baseline to estimate if the visually augmented ensemble classifier shows an improved performance over the audio-only approach.

\medskip
The remainder of this chapter is organized as follows. In the next section a brief overview of the state-of-the-art in the different domains and modalities is given. In Section \ref{ch6:architecture} the layout of the classification approach as well its evaluation is described. Section \ref{ch6:dataset} discusses the different datasets used in the evaluation. In Section \ref{ch6:audio} and \ref{ch6:video} the separate classification approaches based on the two modalities audio and video are explained. In Section \ref{ch6:ensemble} the ensemble classification method that combines the previous two classifiers is outlined and the final results of the evaluation are provided which are further discussed in Section \ref{ch6:discussion}. Conclusions with suggestions for future work are provided in Section \ref{ch6:conclusions}.

\section{Related Work}
\label{relatedwork}

Early approaches to artist identification are based on the Mel-Frequency Cepstral Coefficients (MFCCs) feature set in combination with Support Vector Machines (SVM) for classification \cite{kim2002singer,mandel2005song}. In \cite{kim2006towards} a quantitative analysis of the album effect - effects of post-production filters to create a consistent sound quality across a record -  on artist identification was provided. A Hybrid Singer Identifier (HSI) is proposed by \cite{shen2009novel}. Multiple low-level features are extracted from vocal and non-vocal segments of an audio track and mixture models are used to statistically learn artist characteristics for classification. Further approaches report more robust singer identification through identifying and extracting the singers voice after the track has been segmented into instrumental and vocal sections \cite{mesaros2007singer,tsai2006automatic}.

Good summaries of state-of-the-art approaches and challenges in face recognition are provided by \cite{li2011handbook,phillips2005overview,zhao2003face}. Face detection, tracking and recognition is also used in multi-modal video retrieval \cite{lew2006content,snoek2005multimodal}. Faces are either used to count persons or to identify actors. Most common methods used to recognize faces are Eigenfaces \cite{turk1991eigenfaces} and Fisherfaces \cite{belhumeur1997eigenfaces}. In \cite{dimitrova2000video} face tracking and text trajectories are used with Hidden Markov Models (HMM) for video classification. A face recognition approach based on real-world video data is reported in \cite{stallkamp2007video}.
Despite the reported achievements and promising results  of systems in relatively controlled environments, most face recognition approaches are still limited by variations in different image or face properties (e.g., pose, illumination, mimic, occlusions, age of the person, etc.) - properties that are extensively used as artistic and stylistic features of music videos. The predominating approaches to face recognition are Principal Component Analysis (PCA) (e.g., Eigenfaces \cite{turk1991eigenfaces}) and Linear Discriminant Analysis (LDA) (e.g., Fisherfaces \cite{belhumeur1997eigenfaces}). A good summary of video-based face recognition is provided by \cite{wang2009video}.

\section{Classification Architecture}
\label{ch6:architecture}

Due to the expenses of producing music videos the number of productions per artist or album is marginally low compared to the number of songs recorded. Videos are typically produced for single releases of records to promote the track. As a consequence only a few videos can be collected and especially for relative young artists not enough entries might be found to reliably train a classifier. A common evaluation method in classification experiments is to assess classification accuracies using k-fold cross-validation with k usually set to 10. This requires at least 10 videos per artist, which for many artists are not available.

\begin{figure}
	\centering
	\includegraphics[width=1.00\textwidth]{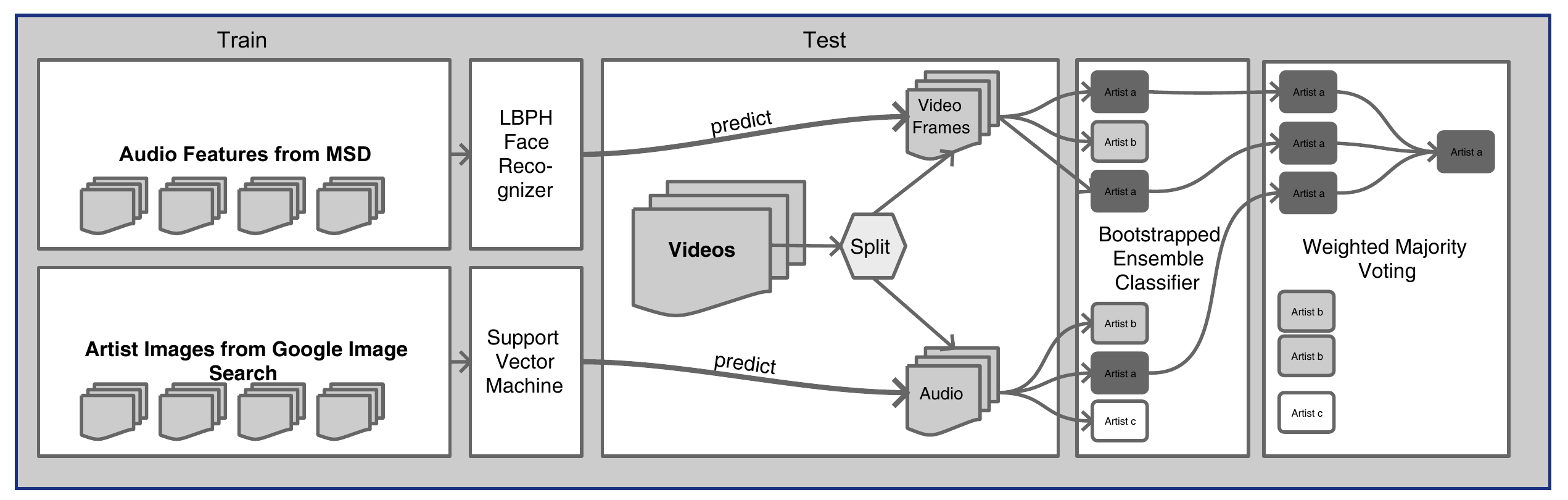}
	\caption{The classification architecture used for the evaluation. 1) two classifiers are trained on features provided by the Million Song Dataset (MSD) and artist images downloaded from Google Images. 2) the same features are extracted from the music videos. 3) These features are applied to the trained classifiers to predict the artist. The predictions of the separate modalities are combined for the final classification results.}
	\label{fig:Evaluation_Chart}
\end{figure}

The method presented in this chapter is a three-tiered classification approach based on two separate training data-sets to take advantage of multiple sources to predict the performing artist of a video. Figure \ref{fig:Evaluation_Chart} depicts the architecture of the classification system. The two modalities of the systems are trained independently on their data-sets and combined by an ensemble classifier to make a final prediction.
The audio classifier is trained on all available songs of an artist, that have not been released as music video. This takes advantage of the broad spectrum of the artist's work and provides a richer set of information. The face recognition system is trained on artist images downloaded from Google Image Search. The separated audio data and the image data represent the training data for the presented multi-modal classification system. Both classifiers are cross-validated on their data-sets to assess their confidences.

An ensemble classification approach based on bootstrapped aggregation is used. Instead of using the complete training-set, the classifiers for each modality are trained only on sub-samples. The trained audio and visual classifiers are applied to the music video test data-set. Using the bootstrapping approach this classification step is repeated $n$ times resulting in $2n$ predictions for each music video. These predictions are aggregated through a weighted majority vote, using the previously evaluated confidence values of the classifiers as weights.

\pagebreak

\section{Dataset}
\label{ch6:dataset}

The evaluation data-set used for the experiments is an initial version of the Music Video Dataset (MVD) subset \textit{MVD-Artists} presented in Chapter \ref{ch4:mvdartists}. It contains videos of 14 popular western music artists listed in Table \ref{tab:ArtistsAndTrainingData}. Popular artists were chosen to meet the requirement of collecting enough music videos for each musician. To demonstrate typical problems of content based artist identification the selected artists belong predominately to the two non-overlapping genres Pop and Rock.

\subsection{Training Data}

As described in the classification architecture in Section \ref{ch6:architecture} training and test data do not originate from the same data-set. The training data is drawn from two different sources - an audio and an image data-set.

\paragraph{Artist Tracks} - For the audio modality, the artist tracks provided by the Million Song Dataset (MSD) \cite{bertin2011million} have been used. For each artist all tracks available in the MSD have been selected excluding those that are present in the music video test-set. Table \ref{tab:ArtistsAndTrainingData} lists the number of tracks for each artist. A total of 645 tracks was used with an average number of 46 tracks per artist - ranging from a minimum of 19 to a maximum of 83 tracks.

\paragraph{Artist Images} - For each artist, portrait images have been downloaded. If the performing artist was a band, only images of the lead singer were used. Bulk download from Google Image Search was used to retrieve images for each artist. In a second step the face detection algorithm described in Section \ref{ch6:face_recognition} was applied to each image to filter out photos that do not contain detectable faces or where the resolution of the detected face was below 120x120 pixels. The resulting subset was manually analyzed to remove duplicates and images where the portrait person does not look frontal into the camera. It was also verified that the remaining images are not, in fact, screen-shots from the music videos used for the evaluation. Further images with low resolutions, occlusions, exaggerated smiles or arbitrary illuminations were removed. Such deviations from pass-photo like portrait images will degrade the performance of the recognition system by introducing too much variance. Further problems concerning face recognition in music video will be addressed in Section \ref{ch6:obstacles}. The resulting set of training images contains approximately 50-150 portraits per artist (see Table \ref{tab:ArtistsAndTrainingData}).


\begin{table*}[t]
	\caption{Artists and training data}
	\centering
	\begin{tabular}{lccc}
		\hline
		\multicolumn{1}{c}{\textbf{Artist Name}} & \textbf{MSD Tracks} & \multicolumn{1}{c}{\textbf{Images}} & \multicolumn{1}{c}{\textbf{Music Videos}} \\ 
		\hline
		Aerosmith          & 83 & 104 & 23 \\ 
		Avril Lavigne      & 29 & 105 & 20 \\ 
		Beyonce            & 32 & 117 & 19 \\ 
		Bon Jovi           & 59 & 54  & 26 \\ 
		Britney Spears     & 57 & 160 & 24 \\ 
		Christina Aguilera & 46 & 123 & 14 \\ 
		Foo Fighters       & 64 & 55  & 19 \\ 
		Jennifer Lopez     & 45 & 92  & 21 \\ 
		Madonna            & 62 & 47  & 25 \\ 
		Maroon 5           & 20 & 78  & 10 \\ 
		Nickelback         & 57 & 47  & 16 \\ 
		Rihanna            & 24 & 122 & 21 \\ 
		Shakira            & 48 & 123 & 20 \\ 
		Taylor Swift       & 19 & 117 & 19 \\ 
		\hline
		& 645 & 1344 & 277 \\
	\end{tabular}

	\label{tab:ArtistsAndTrainingData}
\end{table*}


\subsection{Test Data}

The test-data consists of music videos that have been downloaded from Youtube\footnote{\url{http://www.yoututbe.com}}. As described in Chapter \ref{ch4.1:intro} a set of quality criteria and rules was considered for selecting the videos. Additionally to the default MVD criteria, the following two rules were added for the MVD-ARTISTS dataset:

\begin{itemize}[noitemsep, topsep=5pt]
	\itemsep0em
	\renewcommand\labelitemi{--}
	\item has to be an official music video produced by the artist
	\item the lead singer has to appear in the video
\end{itemize}

\pagebreak

\paragraph{Audio Data} was retrieved directly from the video files by separating the audio stream using FFMPEG\footnote{\url{http://www.ffmpeg.org/}}. The audio was converted to mp3 format with a sample-rate of 44100 Hz and a bitrate of 128 kBit/s. 

\paragraph{Visual Data} from the videos was retrieved frame by frame using the Open Computer Vision Library (OpenCV)\footnote{\url{http://opencv.org}} \cite{opencv_library} that was also used for the further video processing. 

\section{Audio Content Analysis}
\label{ch6:audio}

The audio content analysis task is based on audio features provided by the Million Song Dataset (MSD) \cite{bertin2011million} (see Chapter \ref{ch4.2:intro}). The MSD provides a rich set of low level features (e.g., timbre, chroma) and mid level features (e.g., beats per minute, music key, audio segmentation) as explained in detail in Chapter \ref{ch4.2:echonestfeatures}. For each artist of the evaluation test-set all tracks available in the MSD that do not overlap with the test-set are used. The number of tracks used for each artist is summarized in Table \ref{tab:ArtistsAndTrainingData}.

\subsection{Audio Features}

Content based artist classification is based on the audio features provided by the Million Song Dataset (MSD). These MSD features were extracted and provided by the Echonest\footnote{\url{http://echonest.com/}} using the Echonest API\footnote{\url{http://developer.echonest.com/}} to extract the audio features from the files which were stored equivalent to the MSD format. This evaluation makes use of the \textit{Temporal Echonest Features (TEN)} as introduced in Chapter \ref{ch5.2:TEN}). These audio descriptors summarize an empirically selected set of MSD features by calculating all statistical moments of the features \textit{Segments Pitches}, \textit{Segments Timbre}, \textit{Segments Loudness Max}, \textit{Segments Loudness Max Time} and lengths of segments calculated from \textit{Segments Start}. The resulting feature vector has 224 dimensions. As shown in Chapter \ref{ch5.2:intro})  these features perform comparable to state-of-the-art music feature sets in genre classification tasks on conventional data-sets.

\subsection{Audio Classification Results}

Audio classification was conducted using the Python machine learning library Scikit Learn\footnote{\url{http://scikit-learn.org}}. Training and test data was separately normalized to have zero mean and unit variance. A Support Vector Machine (SVM) with a linear kernel and a penalty parameter of $C=0.1$ was trained on the data from the MSD and used to classify the audio test-data set of the music videos. The classification in Table \ref{tab:separateresults} results show, that these audio features are able to discriminate the performing artists with a precision of 37\% and a recall of 36\%. Such a value was to be expected do to the high variance in musical style of some of the artists. This can be seen in the high variance of the distinct values for all artists in Table \ref{tab:separateresults} and is also illustrated by the corresponding confusion-matrix of the classification result in Figure \ref{fig:audio_confusion}.

\section{Visual Content Analysis}
\label{ch6:video}

The visual content analysis part of this evaluation is focused on face recognition. A frame-by-frame analysis of still images, ignoring their spatio-temporal relationships, is applied. In a first step faces from the training-set - the portrait images collected from the Web - were detected and extracted to train a face recognizer. In a second step faces in video frames were detected and the trained recognizer was applied to predict the performing artist.

\subsection{Face Detection}
\label{ch5:facedetection}

Detecting faces in the video data is the first step of the recognition task. Frame based face detection using boosted cascades of Haar-like features as proposed by Viola and Jones \cite{viola2001rapid} and Lienhart \cite{lienhart2002extended} is used. Their method uses a set of simple features based on pixel value differences between neighboring adjacent rectangles. These features are rapidly calculated from an intermediate representation - the \textit{integral image} - that already pre-computes neighborhood statistics for each pixel of the original image. The classifier for the face detection task is constructed by selecting a small subset of important features using AdaBoost \cite{freund1997decision}. Finally more complex classifiers are combined in a cascade structure. This approach for object detection minimizes computation time while achieving high prediction accuracy.

In an automatic post-processing step detected faces are further analyzed to eliminate false positives. Three additional cascaded detectors are used to locate eye-pairs, noses and mouths within the region of the detected face. If all sub-components are recognized, the detected region is verified as a face and used for predicting the performing artist.

\subsection{Obstacles in Face Detection / Recognition}
\label{ch6:obstacles}

\begin{figure}[t]
	\centering
	\includegraphics[width=1.00\textwidth]{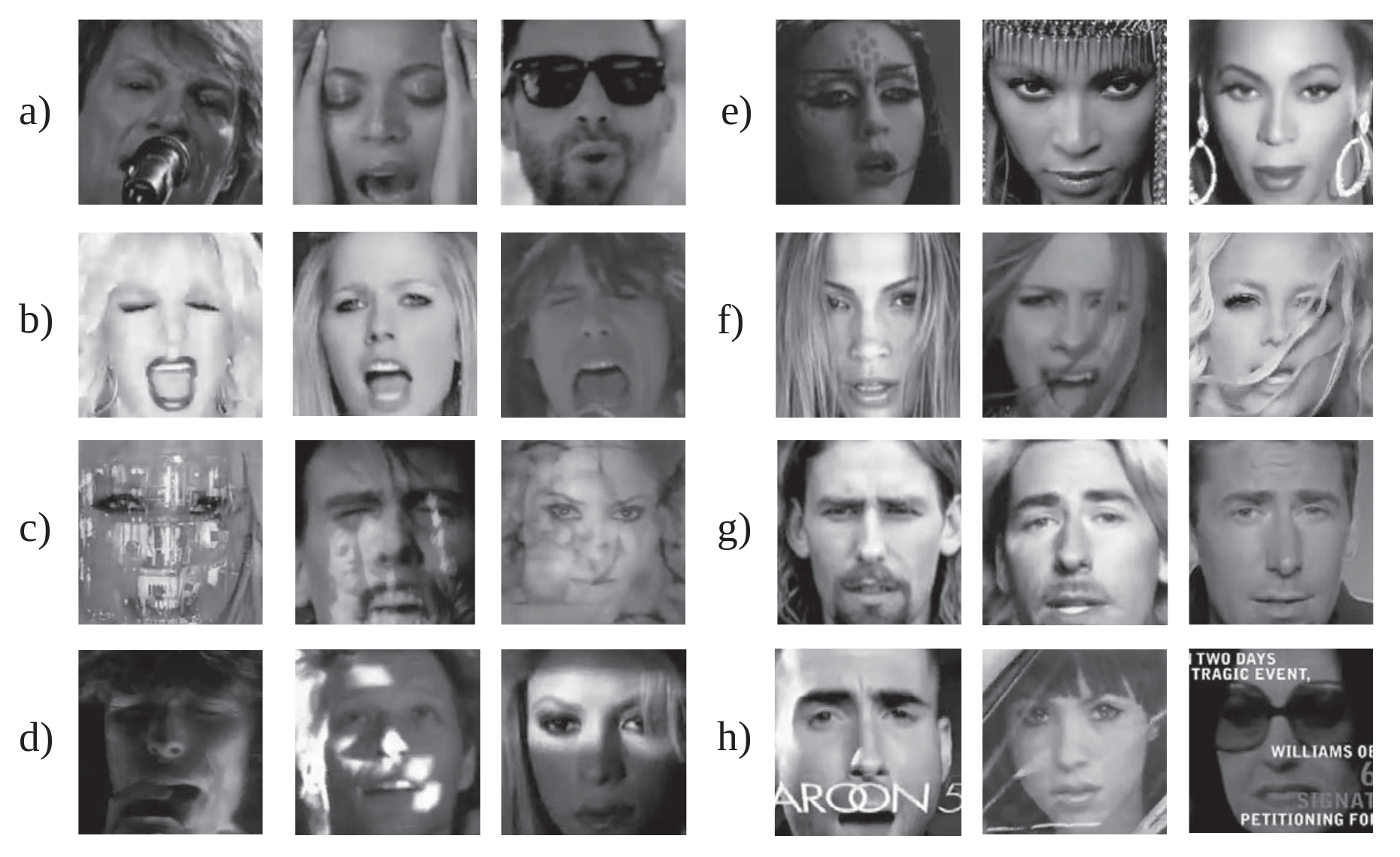}
	\caption{Examples of problematic faces - a) occlusions b) distortions c) video transitions d) varying illuminations e) make up and ornaments f) hair g) beards h) stylistic elements}
	\label{fig:Bad_Faces}
\end{figure}

Despite the remarkable progress in face recognition over the last decades \cite{li2011handbook,phillips2005overview,wang2009video,zhao2003face} most of the reported work has been evaluated in laboratory environments. The most influencing factors for the accuracy of face recognition systems are illumination, occlusions and distortions - all properties that are very common in music videos. See Figure \ref{fig:Bad_Faces}. The following list describes the most common face detection obstacles in music videos:

\pagebreak

\begin{description}[leftmargin=!,itemsep=1pt,labelwidth=\widthof{\bfseries Blending Effects a},labelindent=\descleftmargin,rightmargin=0.8cm]
	
	\item[Occlusions] of the face are one of the biggest problems in face detection and recognition and unfortunately very common in music videos (e.g., microphone, hands touching the face, sunglasses, caps, hats, etc.) (see Figure \ref{fig:Bad_Faces}a). Further excessive usage of Makeup and jewelry could also lead to failed detections and predictions (see Figure \ref{fig:Bad_Faces}e).
	
	\item[Distortions] of the face due to singing, screaming, expressive mimic or fast movements (see Figure \ref{fig:Bad_Faces}b).
	
	\item[Pose] Face recognition systems work optimal when subjects look frontal into the camera, but in video or photography frontal camera shots are not considered to flatter the photographed person. Further, poses and specific camera angles are used for acting purposes to express emotions such as grief, sorrow or thinking.
	
	\item[Illumination] changes are a stylistic tool in many music videos. A common effect is to use stage lighting to create the impression of live performance. This results in fast illumination changes even within a short sequence of video frames (see Figure \ref{fig:Bad_Faces}d).
	
	\item[Facial Hair] in the form of locks of hair hanging into the face is a similar problem to occlusions (see Figure \ref{fig:Bad_Faces}f). Another, more severe problem are beards of male artists. Those may change over time or disappear completely. Because they are not excluded during the face extraction process, beards influence the training-set or prediction. Figure \ref{fig:Bad_Faces}g shows the same artist with different beard styles and without beard.
	
\end{description}

\medskip\noindent
Special video related problems:

\begin{figure}[b]
	\centering
	\includegraphics[width=1.00\textwidth]{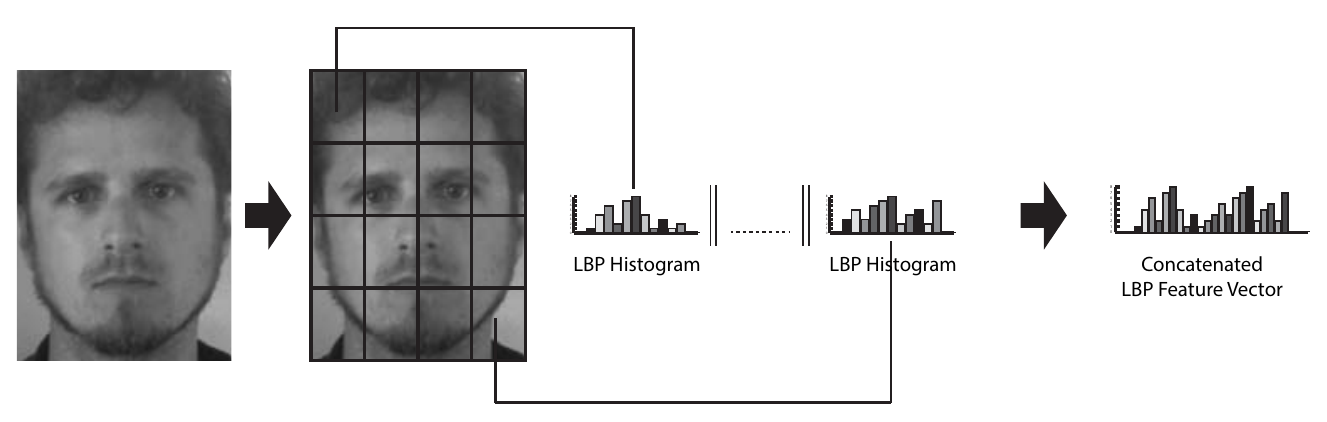}
	\caption{Face recognition with Local Binary Patterns}
	\label{fig:LBP_overview}
\end{figure}

\medskip\medskip
\begin{description}[leftmargin=!,itemsep=1pt,labelwidth=\widthof{\bfseries Blending Effects a},labelindent=\descleftmargin,rightmargin=0.8cm]
	
	\item[Blending Effects] between scenes or cuts. Smooth transitions with image-cross fading effects can overlay the content of consecutive frames onto the face (see Figure \ref{fig:Bad_Faces}c). In such cases the face detector recognizes valid properties of a face, but the overlaid content distorts the face similar to make-up or illumination changes.
	
	\item[Overlays] of visual blocks (e.g. text, images, logos) have similar effects as occlusions (see Figure \ref{fig:Bad_Faces}h).
	
\end{description}

\noindent
Further problems arise through aging of the artist. Music videos are typically produced in combination with new records which are released in a time-span of one to three years on average \cite{Mortimer20123}. Artists that have begun to produce videos in the early stages of the music video trend and are still actively doing so have aged more than thirty years now. The effects of aging are reflected in the face and even when surgery is used to overcome them, the effects on the face recognizer are the same - the person might get misclassified.

\subsection{Face Recognition}
\label{ch6:face_recognition}

The face recognition approach used in this evaluation is based on \textit{Local Binary Patterns (LBP)} as proposed by Ahonen et al. \cite{ahonen2004face}. This approach was chosen due to its robustness against different facial expressions, illumination changes and aging of the subjects. LBP is a simple but very efficient gray-scale invariant texture descriptor that combines properties of structural and statistical texture analysis. It labels each pixel of an image by thresholding their 3x3-neighborhood with its center value and considers the result as an 8 bit binary number. The texture of an image is described by a histogram representation of the frequency of the 256 different labels. For efficient face recognition the image is divided into regions to retain also spatial information. As depicted in Figure \ref{fig:LBP_overview} the resulting histograms of the different image regions are normalized to have uniform vector sum and concatenated to form the final face descriptor. Recognition based on these descriptors is performed using a nearest neighbor classifier in the corresponding feature space with Chi square as a dissimilarity measure.

\begin{table*}[t]
	\caption{Classification results of the separate modalities}
	\centering
	\begin{tabular}{l|ccc|ccc}
		\hline
		\textbf{Artist Name} & \multicolumn{3}{c|}{\textbf{Audio}} & \multicolumn{3}{c}{\textbf{Video}} \\
		
		\hline
		& \textbf{Precision} & \textbf{Recall} & \textbf{F1-Score} & \textbf{Precision} & \textbf{Recall} & \textbf{F1-Score} \\ 
		\hline
		
		Aerosmith & 0.33 &  0.52 & 0.39 & 0.14 &  0.33 & 0.20 \\
		Avril Lavigne & 0.50 &  0.45 & 0.47 & 0.62 &  0.25 & 0.36 \\
		Beyonce & 0.33 &  0.26 & 0.29 & 0.28 &  0.42 & 0.33 \\
		Bon Jovi & 0.28 &  0.36 & 0.32 & 0.20 &  0.04 & 0.07 \\
		Britney Spears & 0.32 &  0.33 & 0.33 & 0.16 &  0.17 & 0.16 \\
		Christina Aguilera & 0.48 &  0.71 & 0.57 & 0.18 &  0.43 & 0.26 \\
		Foo Fighters & 0.41 &  0.47 & 0.44 & 0.00 &  0.00 & 0.00 \\
		Jennifer Lopez & 0.22 &  0.24 & 0.22 & 0.33 &  0.14 & 0.20 \\
		Madonna & 0.27 &  0.28 & 0.24 & 0.50 &  0.12 & 0.19 \\
		Maroon 5 & 0.20 &  0.10 & 0.13 & 0.12 &  0.80 & 0.20 \\
		Nickelback & 0.55 &  0.38 & 0.44 & 1.00 &  0.18 & 0.30 \\
		Rihanna & 0.29 &  0.19 & 0.23 & 0.40 &  0.10 & 0.15 \\
		Shakira & 0.44 &  0.40 & 0.41 & 0.25 &  0.21 & 0.23 \\
		Taylor Swift & 0.60 &  0.32 & 0.41 & 0.50 &  0.06 & 0.10 \\
		
		\hline
		\textbf{avg} & 0.37 &  0.36 & 0.35 & 0.34 &  0.21 & 0.20\\		
		\hline
	\end{tabular}
	\label{tab:separateresults}
\end{table*}

\subsection{Video Classification Results}

\begin{figure}
	\centering
	\subfloat[Audio Classifier]{\includegraphics[width=0.49\textwidth]{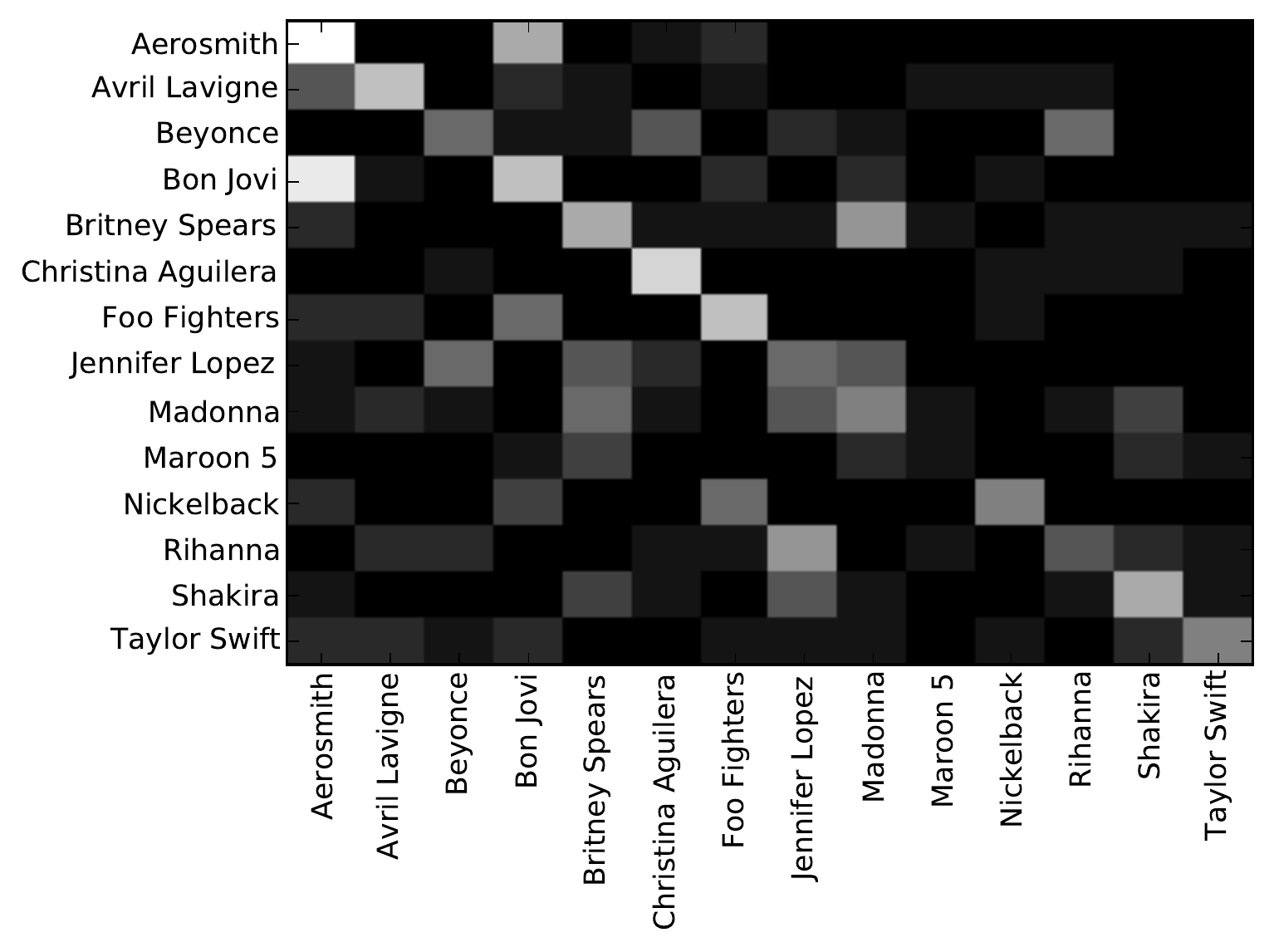}}\hfill
	\subfloat[Video Classifier]{\includegraphics[width=0.49\textwidth]{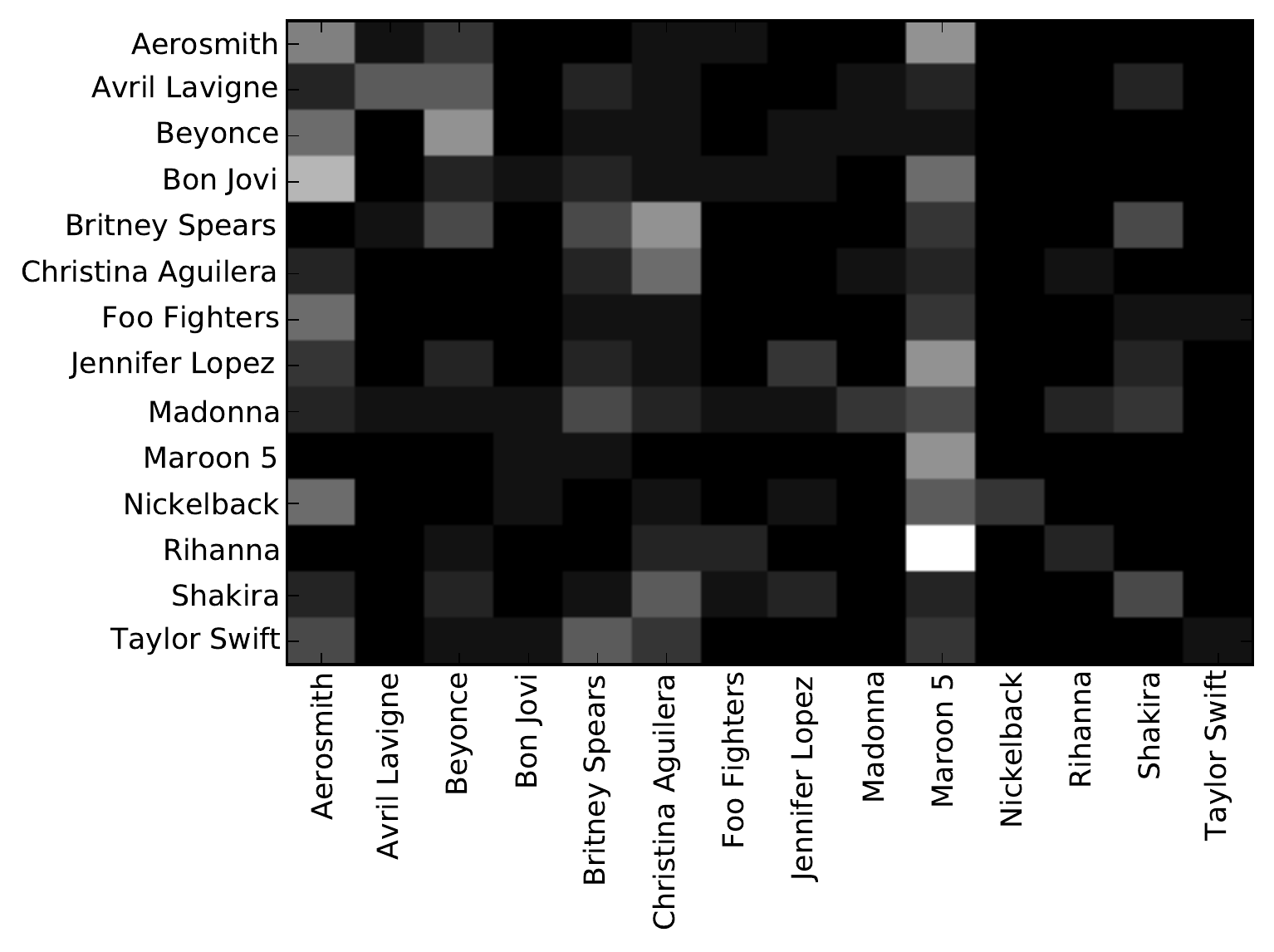}}
	\caption{Confusion matrices of the classification results.}
	\label{fig:audio_confusion}
\end{figure}

The face recognition based visual classifier was implemented using the Python programming language bindings of the \textit{Open Computer Vision (OpenCV)} library \cite{opencv_library}. This library provides implementations for the cascaded classifier based on Haar-like features which is used for face detection (see Section \ref{ch6:face_recognition}). In a preceding step the images were converted to gray-scale and their color histograms were normalized to obtain better results from the face detector. The detected and verified faces were extracted. Contrast Limited Adaptive histogram equalization (CLAHE) \cite{pizer1987adaptive} was applied to the face images to further enhance contrasts and normalize the images for the identification step. The OpenCV implementation of the LBP face recognition approach described in Section \ref{ch6:face_recognition} was used to predict the corresponding artist name. The recognizer was initiated with the radius of 1 and 8 neighbors used for building the \textit{Circular Local Binary Pattern}. A grid of 8x8 cells was applied to the image resulting in a LBP descriptor consisting of 64 concatenated histograms. Each extracted and post-processed face got an artist name label assigned by the face recognizer. For each label the average prediction confidence is calculated. To punish supposed isolated mis-classifications and to favor frequent assignments the average confidence is divided by the natural logarithm of the number of how often this label has been assigned. The distinct values of the artists are listed in Table \ref{tab:separateresults}. The corresponding confusion-matrix of the classification result is depicted in Figure \ref{fig:audio_confusion}. The results show, that the visual data from the videos can be predicted with a precision of 34\% and a recall of 21\% where \textit{Precision} describes the confidence a video classified as artist $a$ to be truly from $a$ whereas \textit{Recall} describes how reliably all videos of artist $a$ are recognized to be from $a$. 

\section{Cross-Modal Results}
\label{ch6:ensemble}

The previous sections \ref{ch6:audio} and \ref{ch6:video} describe the functionality, implementation and performance of the classifiers for the separate audio and video modalities. In this section a combined classification approach is presented that is capable to combine the distinct modalities and provide enhanced classification accuracy.

\subsection{Ensemble Classification Method}

The presented ensemble classifier is based on the \textit{Bootstrap Aggregation (Bagging)} approach as introduced by Breiman \cite{breiman1996bagging}. Bagging generates multiple versions of a predictor by making bootstrap replicates of the training set through random sub-sampling. In our approach subset selection on $Train_{Audio}$ and $Train_{Video}$ was applied to generate $i=10$ classifiers for each modality. Each Support Vector Machine classifier $C_{Audio_{i}}$ and $C_{Video_{i}}$ was trained on a selected sub-set $Train_{Audio_{i}}$ and the remainder of the training set $Train_{Audio} - Train_{Audio_{i}}$ was used to estimate its confidence $Conf_{Audio_{i}}$.

\pagebreak

The resulting 20 predictions were aggregating through weighted majority voting. Each classifier $C_{Audio_{i}}$ and $C_{Video_{i}}$ predicts a music video $mv$ of the test-set. Each prediction is now assigned a weight that is defined through the confidence of the used classifier $Conf_{Audio_{i}}$ or $Conf_{Video_{i}}$. 

\begin{align}
	weight_{mv_{i}} = Conf_{Audio_{i}}
\end{align}

For each music video $mv$ the sum of the weights of all labels is calculated. The label with the highest sum wins the vote and is the result of the ensemble classifier for the music video $mv$.

\section{Results and Discussion}
\label{ch6:discussion}
\label{ch6:results}

The bootstrap aggregation ensemble classification approach as described in the previous section has been implemented using the Python Scientific Machine Learning Kit (SciKit Learn). For each modality bootstrapped sub-sampling with 10 iterations and 10\% test-set size was applied to the according training-set. The results are summarized in Table \ref{tab:EnsembleResults} and show improvement in precision using the multi-modal ensemble classification approach.

\begin{table*}[t]
	\caption{Results of the Ensemble Classification}
	\centering
	\begin{tabular}{lcccl}
		\hline
		\textbf{Artist Name} & \textbf{Precision} & \textbf{Recall} & \textbf{f1-score}  \\ 
		\hline
		Aerosmith & 0.36 & 0.57 & 0.44 \\
		Avril Lavigne & 0.64 & 0.45 & 0.53 \\
		Beyonce & 0.55 & 0.32 & 0.40 \\
		Bon Jovi & 0.24 & 0.27 & 0.25 \\
		Britney Spears & 0.34 & 0.42 & 0.38 \\
		Christina Aguilera & 0.33 & 0.50 & 0.40 \\
		Foo Fighters & 0.62 & 0.53 & 0.57 \\
		Jennifer Lopez & 0.27 & 0.19 & 0.22 \\
		Madonna & 0.30 & 0.24 & 0.27 \\
		Maroon 5 & 0.35 & 0.70 & 0.47 \\
		Nickelback & 0.58 & 0.44 & 0.50 \\
		Rihanna & 0.75 & 0.14 & 0.24 \\
		Shakira & 0.28 & 0.65 & 0.39 \\
		Taylor Swift & 1.00 & 0.16 & 0.27 \\
		
		\hline                             
		avg  & 0.47 & 0.38 & 0.37 \\
		\hline
	\end{tabular}

	\label{tab:EnsembleResults}
\end{table*}

The presented approach demonstrates how to improve common approaches to artist identification through information extracted from music videos. The scenario set for this experiment was a typical audio content based approach using the audio feature set presented in Chapter \ref{ch5.2:intro} and its performance values as baseline. According to this the precision of the audio based classifier could be increased by 27\% while recall values were only slightly improved by 5\%. Thus, the ensemble approach did not increase the number of correctly identified tracks, but did enhance the reliability.

As described in Section \ref{ch6:architecture} this evaluation was intentionally based on two simple approaches. The audio classifier uses song-level features describing temporal statistics of timbral and chromatic music properties. Using audio segmentation to separate voiced from un-voiced sections \cite{mesaros2007singer,tsai2006automatic} may enhance the performance of the audio classifier. The visual classification approach was based on frame-by-frame face recognition and prediction was made by a majority vote. This approach might be improved through considering spatio-temporal relationships. By applying shot-detection music videos can be segmented and faces tracked within one shot could be verified and summarized more reliably. A further limiting factor of this evaluation was the low resolution of the music videos which has been chosen as a compromise to collect enough videos. Face recognition systems highly depend on the information provided in the images. The minimum resolution of 120x120 pixel is sub-optimal and a verification test-set using high-definition videos might provide better results.

Still, the presented results showed that the performance of audio based artist identification can be improved through information extracted from music videos.

\section{Summary}
\label{ch6:conclusions}

This chapter introduced a cross-modal approach to music artist identification. Audio content and visual based classifiers were combined using an ensemble classifier. The audio classifier used Temporal Echonest Features which were introduced in Chapter \ref{ch4.2:echonestfeatures} and evaluated in Chapter \ref{ch5.2:intro} to predict artist labels. Its precision of 37\% and recall of 36\% was used as benchmark for the further experiments. The visual content classifier used face recognition based on a Local Binary Patterns (LBP) predictor. The two modalities were combined through bootstrap aggregation. For each modality 10 classifiers were created and trained on sub-samples of their according training-sets. The final prediction for a music video was calculated on the basis of weighted majority voting of the resulting 20 predictions. The proposed cross-modal approach showed that the initial audio content based baseline could be increased by 27\% through information extracted from the visual part of music videos.

The presented approach relies on a predefined dataset of artist images - thus, still requiring manual interaction. Future work will include automatic identification of lead singers to train the face recognition algorithm directly on faces extracted from the music videos. Such an approach would provide the possibility to use k-fold cross-validation on a single music video dataset.\\


\noindent
\textit{\smaller This approach and its corresponding evaluation was published and presented at the 10th International Symposium on Computer Music Multidisciplinary Research (CMMR2013) in Marseille, France on October 14-18 2013 \cite{SCHINDLER_2013CMMR}.}




\chapter{Shot Detection for Music videos}
\label{ch9:intro}


\epigraph{
	``[Cinema] combines so many other art forms, as do theater and opera, but the essence of cinema is editing. It’s the combination of what can be extraordinary images, images of people during emotional moments, or just images in a general sense, but put together in a kind of alchemy. A number of images put together a certain way become something quite above and beyond what any of them are individually''} 
{--- Francis Ford Coppola}


One of the most obvious features of music videos, which discriminates them from other visual media, are their specific characteristics and styles of editing. Scenes in music videos are generally short, ranging in length from only a few seconds to even milliseconds. In movies such short-scene sequences are often referred to as ``MTV-style editing''. The intentions behind this style of editing and how it diverges from traditional movie editing have been discussed in Chapter \ref{ch3:intro}. 
The shot-length of music videos is a discriminative feature to distinguish them from other video categories such as movies, news, cartoons or sports \cite{iyengar1997models}. Music Videos are more comparable to commercials and movie trailers which have similar short edited sequences and shot transitions.
A shot in a video is defined as an unbroken sequence of images captured by a recording operation of a camera \cite{hampapur1994digital}. Shots are joined during editing to compose the final video. 
The simplest method to accomplish this are \textit{sharp cuts} where two shots are simply concatenated. Gradual visual effects such as fades and dissolves are common ways to smooth the transition from one shot to the other. 

During the assembly of the Music Video Dataset (MVD) it was observed that there are characteristic styles in editing music videos for certain music genres. Also, style and complexity of shot transition changed over time with new technologies becoming available. Thus, it was intended to use state-of-the-art shot-detection approaches \cite{cotsaces2006video,yuan2007formal,smeaton2010video} to extract features such as \textit{Shots per Minute}, \textit{Average Shot Length}, \textit{Variance in Shot Length} as well as further statistics.
Several successful approaches reported to literature \cite{cotsaces2006video,yuan2007formal,smeaton2010video} 
were implemented and applied to the MVD. During this implementation and evaluation cycle it was observed, that these approaches only apply to a limted set of shot transition styles which are commonly used in movies, sport and news broadcast - such as commonly used in shot-detection evaluation datasets. Music videos utilize shot transitions in a far more artistic way and use them for example to create tension or express emotions such as distress, horror or melancholy. The biggest challenge in developing an appropriate shot-detection system for music videos is this vast number of transition styles which is further complemented by unconventional camera work such as rapid zooming or panning. A major problem in this regard is the definition of a shot-boundary or transition itself. For example in one of the videos of the MVD-VIS category \textit{Dance} the editor used to skip three or four video frames of a recorded scene to create a rhythmic visual effect. The woman walking from left to right has an unnatural ``jump'' in her movement while the original scene remains the same. While some of the applied shot-detection systems identified this as shot-boundary, it is still unclear if this can be defined as a such.

\begin{figure}
	\centering
	\includegraphics[width=1.0\linewidth]{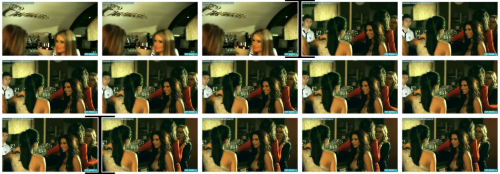}
	\caption{Example of \textbf{Skip Frames}. Replaying these frames with the 25 frames per second of the original video reveals, that this scene was recorded in a single camera movement, starting from the left corner of the bar and panning to the right until the focus is on two women. The depicted frames show that a large segment was skipped after frame number 3 and a small segment after frame number 11.}
	\label{fig:skip_frames}
\end{figure}

\section{Transition-Types in Music Videos}
\label{transition_types}

This section lists transition types found in music videos. The following types are most common:

\begin{itemize}[noitemsep, topsep=5pt, leftmargin=0pt]
	\itemsep0.7em \renewcommand\labelitemi{}
	
	\item \textbf{Sharp Cuts:}
	Two successive scenes are just concatenated. This is the most common transition in music videos.
	
	\item \textbf{Gradual Transitions:}
	One scene gradually dissolves into the other. This is a common cinematic effect and also frequently used in music videos.
	
	\item \textbf{Fade-In / Fade-Out:}
	These are gradual transition usually applied to the beginning (Fade-In) or end (Fade-Out) of a video. Scenes dissolve from or to a single-chromatic frame such as a blank black screen.
	
\end{itemize}

\pagebreak

These effects are common types of scene transitions which are also applied in movies or television broadcasts. As mentioned in the introduction, shot boundary detection for these types has been extensively studied. More problematic are the following types of transitions. These are artistic variations of common types or new transitions which still need to be defined.

\begin{figure}
	\centering
	\includegraphics[width=1.0\linewidth]{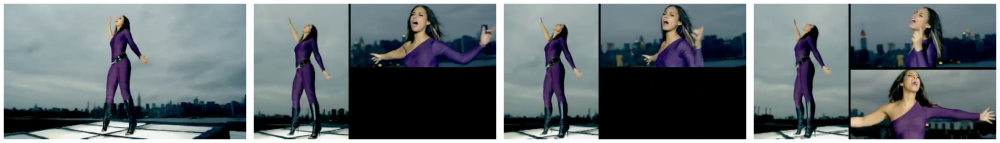}
	\caption{Example of \textbf{Split-Screen} scenes. The video frame is split into various, mostly geometric, regions which display different recorded scenes.}
	\label{fig:shot_splitscreen}
\end{figure}

\begin{itemize}[noitemsep, topsep=5pt, leftmargin=0pt]
	\itemsep0.7em \renewcommand\labelitemi{}
	
	\item \textbf{Skip-Frames:}
	A few video frames of a scene are skipped to create rhythmic effects or to increase the pace of the visual flow. Figure \ref{fig:skip_frames} depicts a 0.6 seconds long scene with skipped frames. The scene was originally recorded as a single camera movement. During editing a large and a small segment was removed. While the large edit can be recognized as a cut in the scene, the smaller skipped edit is hardly recognizable in the depicted sequence of images of the figure. On the other hand, when watching the video with its defined frame-rate, the entire scene is perceived as coherent, although both edits are clearly recognizable.
	Variations include sequences of skip-frames rhythmically aligned to a music sequence or beat.
	
	\item \textbf{Jumping back and forth:}
	This effect is similar to skip-frames. The scene does not change but the order of the recorded video frames is altered. This results in perceived back- and forward jumps in the temporal progression. This effect is used intuitively and no generalizable pattern could be recognized. Observed examples include repetitively jumping back to visualize musical crescendos which dissolved in a new scene in synch with the music or jumping back and forth to express confusion.
	
	\item \textbf{Distortion Overlays:} 
	Distortion overlays are visual effects applied to video frames and appear in various forms: \textit{heavy blurring} or \textit{distortion}, \textit{diffusion}, \textit{rippling,} \textit{white noise} and many more. Simulation of analog TV screen errors such as \textit{Vertical Roll} or \textit{horizontal or vertical synchronization failures} and \textit{vertical deflection problems} (see Figure \ref{fig:shot_distortions}) are a good example to show the complexity of shot detection in music videos. Vertical roll is caused by a loss of vertical synchronization and results in a split screen, where the upper part of the image is shifted with the lower part.
	
	\begin{figure}
		\centering
		\includegraphics[width=1.0\linewidth]{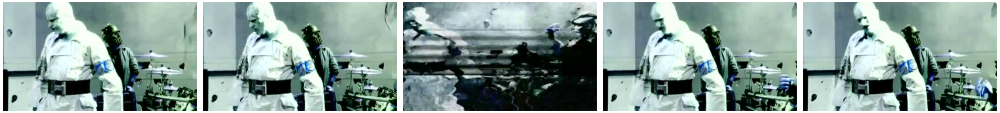}
		\caption{}
		\label{fig:shot_distortions}
	\end{figure}
	
	\item \textbf{Fast zoom in/out:}
	This effect is partially a transition to a new shot. Within the same scene, the camera quickly zooms in to or out of a certain subject or object. Zoom levels can vary from such as zooming in to the lead singers head and zooming out again the show the entire band, to only minor zoom levels. Again, this effect can be applied several times to the same scene.
	
	\item \textbf{Split-Screen:}
	The video frame is subdivided into multiple, usually rectangular, regions which display different scenes (see Figure \ref{fig:shot_splitscreen}). These scenes can change independently and frequently within such a split-screen video scene. Also the number of split-segments can change within such a scene.
	
	\item \textbf{Abrupt tempo changes:} 
	The tempo of the recorded scene is altered (e.g. from normal to slow motion or high-speed). Tempo changes can last several seconds or change back quickly (e.g. speeding up for a short time or fast forward like skip-frames). Slowing down is sometimes used to emphasize on a musical transition (e.g. transition from bridge to chorus).
	
	\item \textbf{Spotlights:}
	Spotlights or stage lighting are common equipment used in music videos, especially in performance focused videos. The artists are performing on a stage or in a club like environment where bright spots and lightings are used. These spots often shine directly into the camera which results in several highly or completely illuminated video frames. Some videos put the artists in front of large spots pointed directly at the camera and artistically play with the shadow thrown. Such abrupt changes in illumination were often misclassified as shot boundaries in the experiments.
	
	\item \textbf{Frame-Swapping / Flickering:} 
	Transition between two scenes where several frames of one video are shown, then several of the other, then again some of the first, and so on. This creates a flickering sensation. This effect is often used with build-ups in electronic dance music, which are crescendoing parts used to create tension and excitement often preceding main themes or choruses.
	
	\begin{figure}
		\centering
		\includegraphics[width=1.0\linewidth]{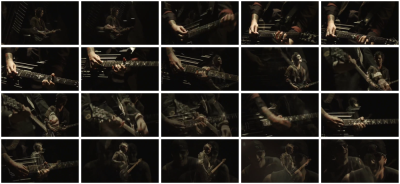}
		\caption{Example of \textit{Blending} video frames. Within 0.7 seconds (20 frames at 29 frames per seconds) 7 different scenes are blended.}
		\label{fig:shotexample5}
	\end{figure}
	
	\item \textbf{Blending / Fading / Dissolving:}
	Blending, fading and dissolving are effects where one scene gradually transitions to another. This well researched problem \cite{truong2000new,yuan2007formal,hu2011survey,mohanta2012model} is partly deemed to be solved for movies, news and sport videos \cite{smeaton2010video}. Again, music videos cross borders by exploiting the use of dissolves in an exaggerated artistic manner. Figure \ref{fig:shotexample5} shows 20 consecutive video frames of a music video. Within these 20 frames - which correspond to 0.7 seconds at 29 frames per seconds - 7 different scenes are blended in and out. It is not clear when one scene starts and the other ends.
	
	\pagebreak
	
	\item \textbf{Abrupt focal changes:}
	The focus shifts abruptly between fore- and background within a scene. Thus, one area becomes blurred and the other clears. Focus shifts could appear several times within the same scene (e.g. to shift between main and background vocals).
	
	\item \textbf{Overlays:}
	Overlays are a popular artistic tool that is applied in manifold ways such as fire and flames blended over Heavy Metal music videos or parts of lyrics as text overlays.
	
	\item \textbf{Camera Tilts:} 
	Camera tilts are fast abrupt pans by the camera. In some videos this effect is applied between words or lines of the lyrics. The camera tilts away from the singer and back when he or she continues to sing.
	
	\item \textbf{Dancing in front of Green-Box:}
	The artist or a group of people is dancing or acting in front of a green-box. The scenery projected onto the green surface changes rapidly in music videos. While to the viewer it is clear that this is a connected and coherent dancing scene, the visual change in the background may confuse a shot-detection system. In such and similar cases it could be considered to abstract from the original shot concept and to derive additional categories such as \textit{Background Shot-changes}.
	
	\item \textbf{Freezing on a frame:}
	Slowing down to a complete halt and freezing a frame for a certain amount of time. Sometimes followed by a sped up section. A variant of this effect is to freeze in a photography. In this case, the video frame is surrounded by a photo frame such as the borders of a Polaroid photo and the still image is shown for several video frames until the video progresses naturally again.
	
	\item \textbf{Dropping to black:}
	Illumination is dropped to zero over two or three frames. This effect is sometimes used in \textit{Dance} or \textit{Dubstep} music videos to simulate drop beats or to emphasize \textit{drops}. These are sudden changes in the rhythm or bass line and are usually preceded by a build-up section. In music videos such drop frames do not usually indicate a scene change, although it does occur.
	
\end{itemize}

\section{Examples}

\begin{figure}
	\centering
	\includegraphics[width=1.0\linewidth]{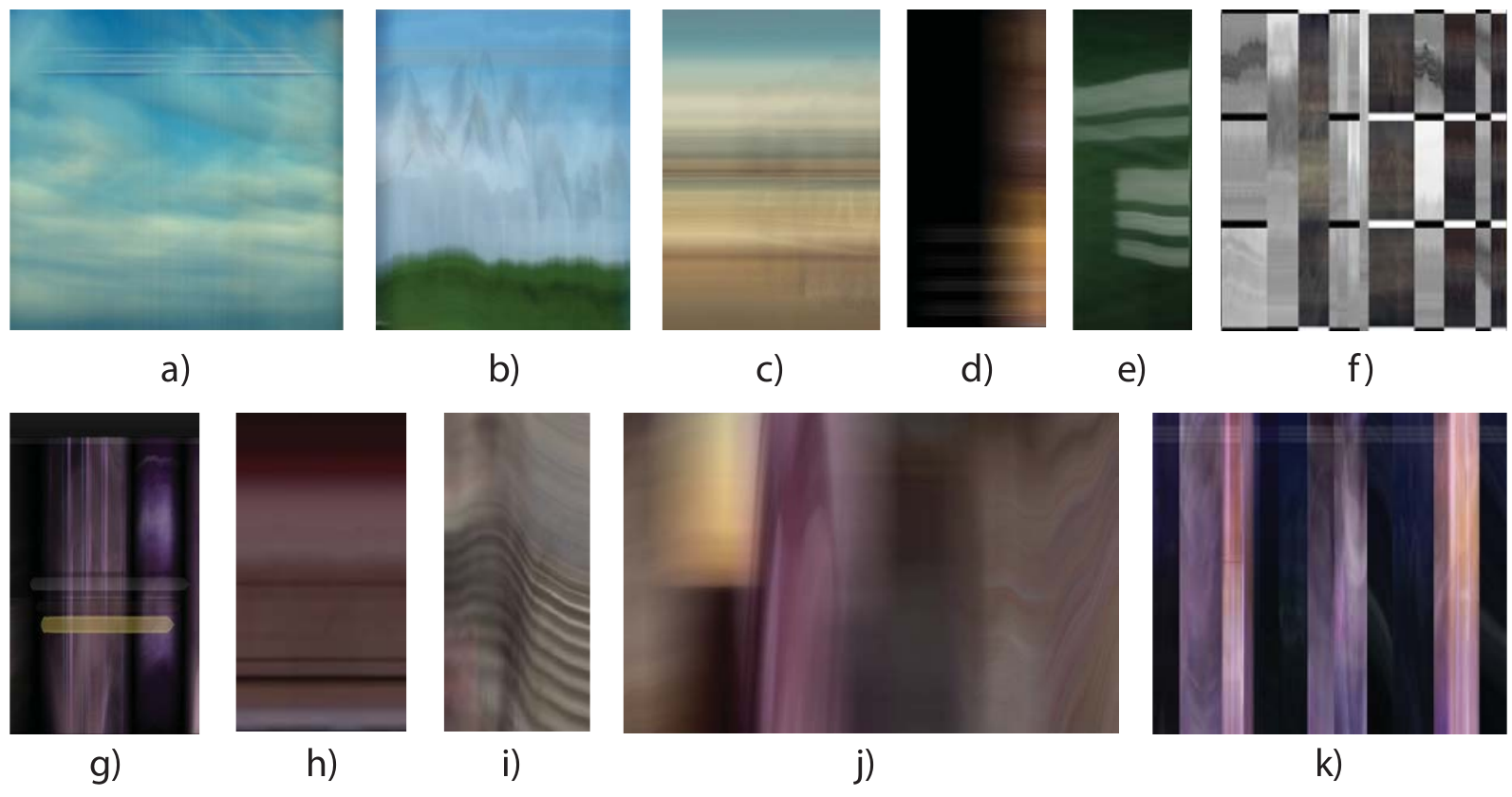}
	\caption{Mean-color bar examples - each column of an example visualization corresponds to the mean color vector of a video frame projected against the vertical axis in RGB color space. a) video sequence of the sky with slowly moving clouds. b) video sequence showing trees and sky. c) beach, sea and sky. d) fade-in effect. e) zooming in on object. f) split screen video sequences. g) object or text overlays. h) camera fixed on object or scene. i) moving camera focus. j) gradual and dissolving transitions. k) sharp cuts.} 
	\label{fig:meanbar_examples}
\end{figure}

This section picks some examples of music video to visualize some of the introduced problems. Therfore a mean-color bar is generated to summarize music video content and to visualize its progression over time. This is achieved by projecting the mean pixel values against the vertical axis of each video frame and each color dimension. This results in a vector representation of the mean color of a video frame in the RGB color space. The mean-color bar is generated by concatenating the mean color vectors of all consecutive video frames. These bars are a convenient tool and provide a rough overview of the video content. There exist a few easy to recognize patterns which can be directly related to the displayed content. Figure \ref{fig:meanbar_examples} shows examples of mean-color bar generated from music videos of the MVD.

\subsection{Example 1 - Split Screens and Frame-Swapping / Flickering}

\begin{figure}
	\centering
	\includegraphics[width=1.0\linewidth]{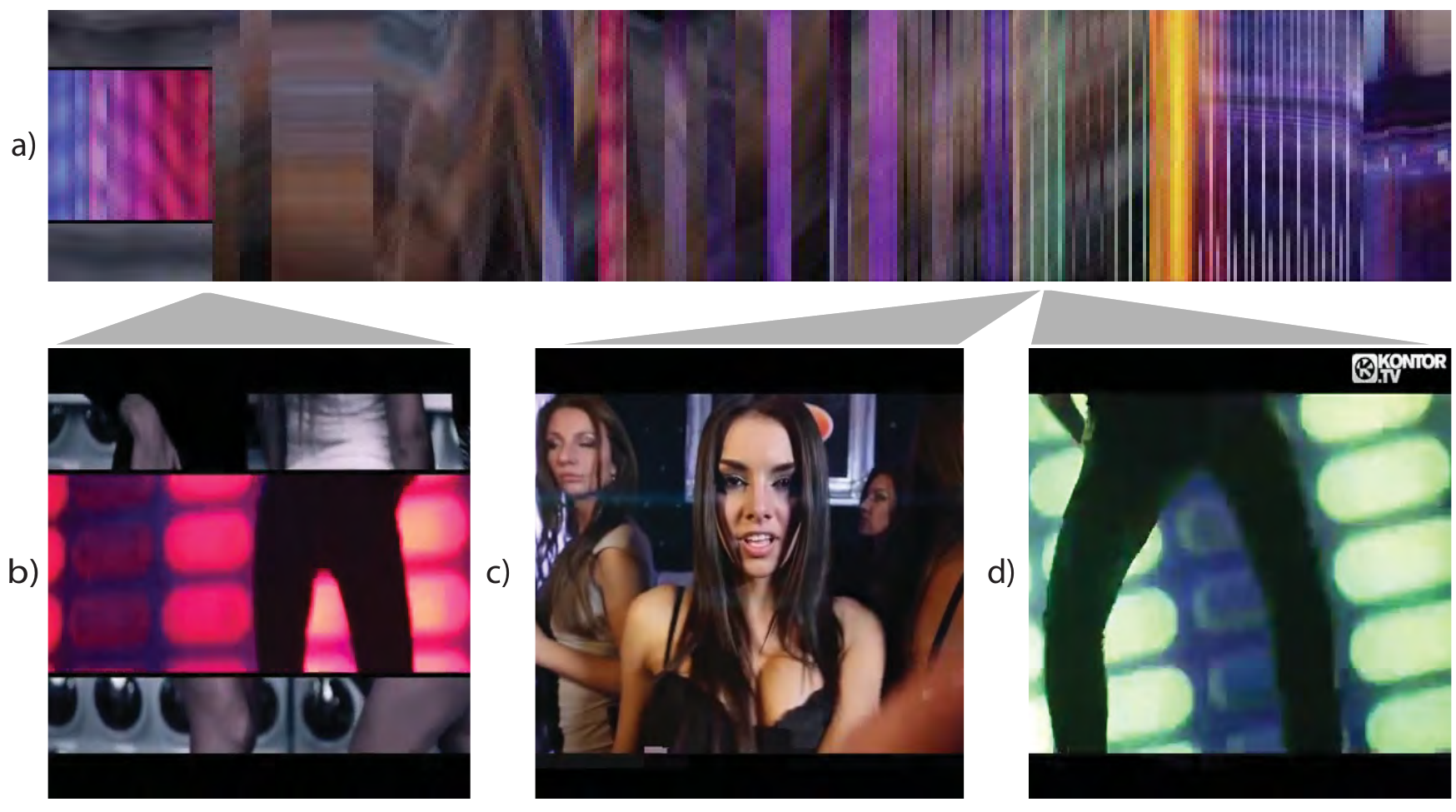}
	\caption{Example 1: a) Mean-color-bar to visualize music video activity over time. b) vertical split-screen section (first segment in a). b) and c) in the greenish segment of a) the video swaps quickly between scene b) (darker columns) and scene c) (brighter columns).}
	\label{fig:example1}
\end{figure}

Example 1 is track 92 from the \textit{Dance} category taken from the \textit{MVD-VIS} dataset. It is a standard electronic dance music (EDM) track. The video is situated in a dance club. The main plot of the video is to show women dancing to the music. The discussed part of the track is visualized in Figure \ref{fig:example1} a). The leading segment is a split-screen sequence. Figure \ref{fig:example1} b) shows an example frame of this sequence. The screen is split horizontally and each sequence shows a different scene. This sequence is followed by a segment with interchanging slow and normal motion recording of a dancing woman which seems to serve as a preparation for the build-up which starts at about the center of the sequence. The acoustic part is a typical EDM build-up with dropped bass, amplified midst and a crescendoing progression towards the \textit{drop}. The visual part mimics this progression by synchronously swapping between several scenes. An example is given in the greenish segment of Figure \ref{fig:example1} a). In this segment the scenes containing Figure \ref{fig:example1} b) and c) are swapped six times over 12 consecutive frames. Based on the videos frame-rate of 25 frames per second (fps) this corresponds to 0.48 seconds or 0.08 seconds per scene. The hazy ascending regions between the purple and the yellow segment depict, that there is a coherent background scene, over which various different scenes are swapped. This scene is the dancing woman shown in Figure \ref{fig:example1} b). The build-up ends with the drop in the yellow segment - which are yellow flares that are synthetically layed over the captured video frames. After the drop the scene changes to show a crowded dance-floor with numerous people dancing. The same patterns of sequential highly illuminated video frames seem continues, but these illuminations originate from the synchronous disco lights in the club.

This example was chosen because it addresses three problems defined in Section \ref{transition_types} - \textit{Split-screen} sections, \textit{frame-swapping} and \textit{spotlights}.

\subsection{Example 2 - Multiple mini-cut scenes with static camera view}

\begin{figure}
	\centering
	\includegraphics[width=1.0\linewidth]{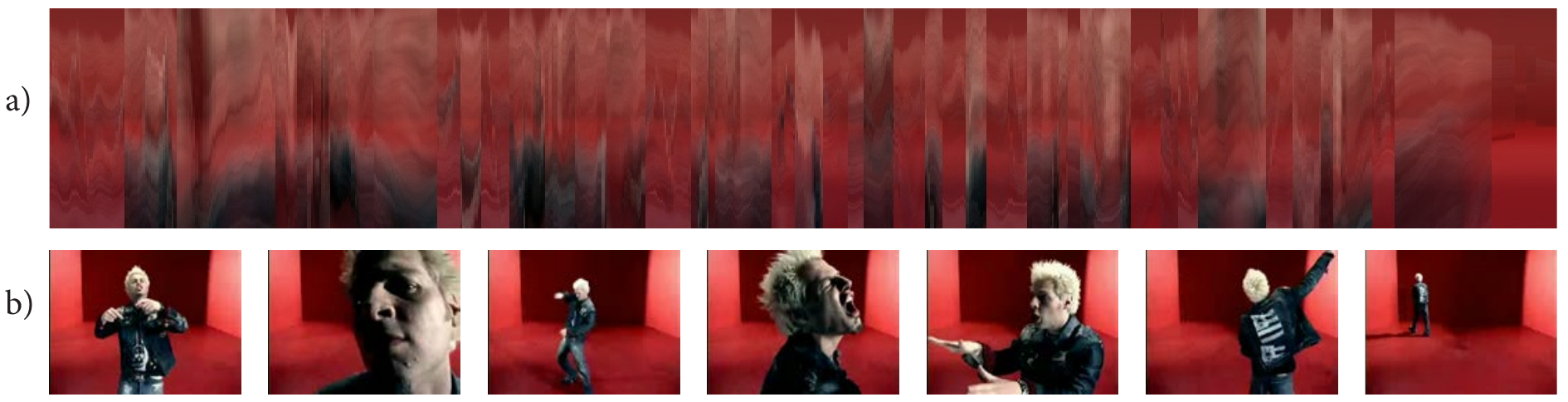}
	\caption{Example 2: Cut scene of multiple independent takes with static camera. a) Mean-color-bar to visualize music video activity over time. Shot edges are recognizable for large positional changes. b) Example frames of the consecutive shots which are not longer than a few video frames. Singer abruptly changes position and orientation with every sharp cut.}
	\label{fig:example2}
\end{figure}

Example 2 is track 64 from the \textit{Metal} category taken from the \textit{MVD-VIS} dataset. The video features the performing band and the scenery is reduced to a red painted room with graffiti on the wall. In the discussed sequence, which is the bridge section of the song, the singer is in another room with clean red painted walls. The sequence is cut together of multiple independently shot takes in front of a static camera. The order of how these shots are cut together should express the distress the protagonist of the lyrics is currently in. Figure \ref{fig:example2} b) shows video frames of such a sequence. The singer changes abruptly position and posture. A shot lasts between 10 to 20 video frames which are 0.3 - 0.6 Seconds at 30fps or approximately 3 shots per second. 

\pagebreak

This example was chosen because it illustrates the controversy towards the current definition of shot boundaries. In Figure \ref{fig:example2} a) large positional changes of the singer are clearly recognizable as sharp cuts. On the other hand, the static camera position creates the impression of a coherent scene. Taking into consideration that this room is only shown for the bridge of the song and changes back with the transition to the chorus, this coherent impression increases further. Thus, this is a good example of how cinematic techniques are used in an artistic way to express emotion and rhythm.

\subsection{Example 3 - One-shot Music Videos}

\begin{figure}
	\centering
	\includegraphics[width=1.0\linewidth]{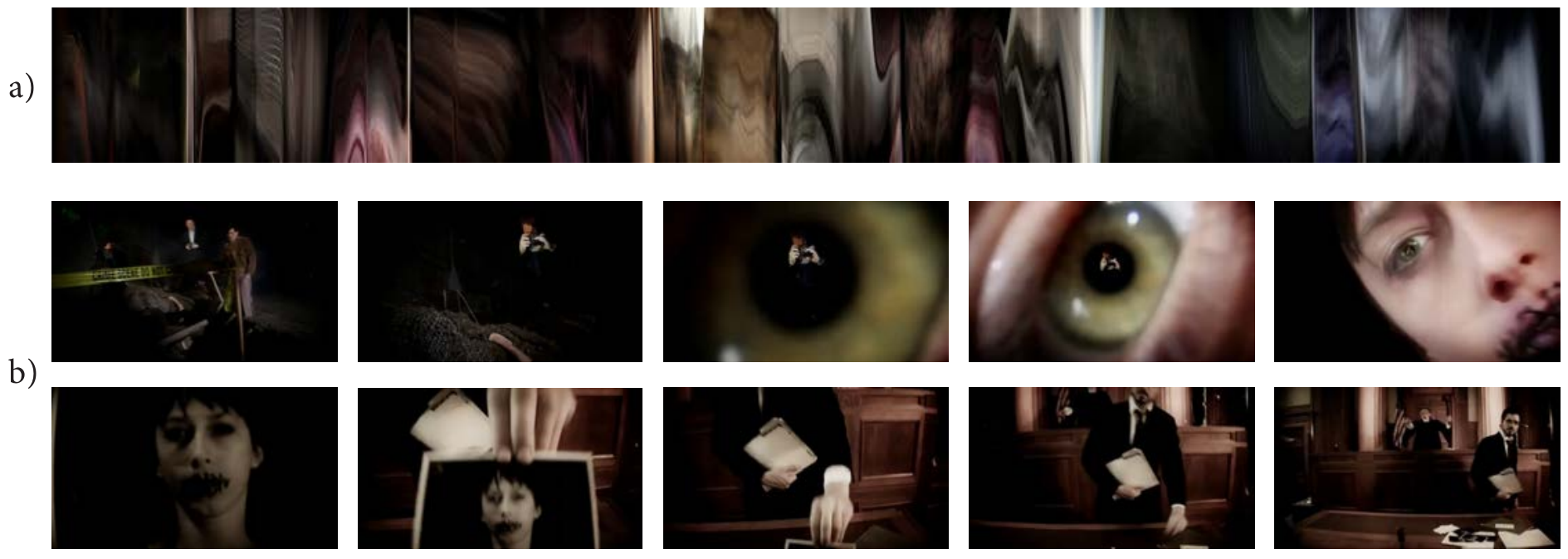}
	\caption{Example 3: One-shot Music Video. a) Mean-color-bar depicting that there are no sharp cuts in the video. b) example video frames of the starting sequence of the music video. These frames demonstrate how zooming out is used to transition between scenes.}
	\label{fig:example3}
\end{figure}

One-shot or one-take music videos consist literally of one long take. Nevertheless, to present different scenes to different parts of the song, it requires good preparation and many helping hands. In a usual setup different szenes are prepared along a trail on which the camera progresses. Artists and stage props move along with the camera and walk in and out of view.
Example 3 is track 17 from the \textit{Folk} category taken from the \textit{MVD-VIS} dataset. The video opens with the investigation of a crime scene where a woman has been murdered and continues to tell the story of the curt trial, public media coverage and perception, and ends by lynching the accused. The video uses various visual effects to simulate a one-shot music video. 
Figure \ref{fig:example3} b) depicts such an effect by showing an example sequence of video frames. This sequence corresponds to the mentioned opening sequence. The camera follows a photographer as he approaches the crime scene. Then the camera zooms out of this scene. While the photographer vanishes in the distance, an iris appears as frame around the image. The iris evolves to an eye and further to the face of the victim. This face turns into a frames photograph taken by the photographer of the previous scene. The camera still keeps zooming out, so it can be recognized that the photograph is held by an attorney in a court room.

This short sequence features three different scenes. Transitions are created by harnessing dark sceneries which hide sharp transitions. The new scene emerges out of the shadow. The other effect zooms in onto small objects to hide the background scene. When zooming out again, the object is part of a different scene. These two effects are frequently used in this video.
Figure \ref{fig:example3} a) depicts that there are no sharp cuts. Further, it is hard to find the transitions at all using this visualization.

\subsection{Example 4 - Artistic Effects}

This example discusses four artistic effects applied to music videos.

\begin{figure}
	\centering
	\includegraphics[width=1.0\linewidth]{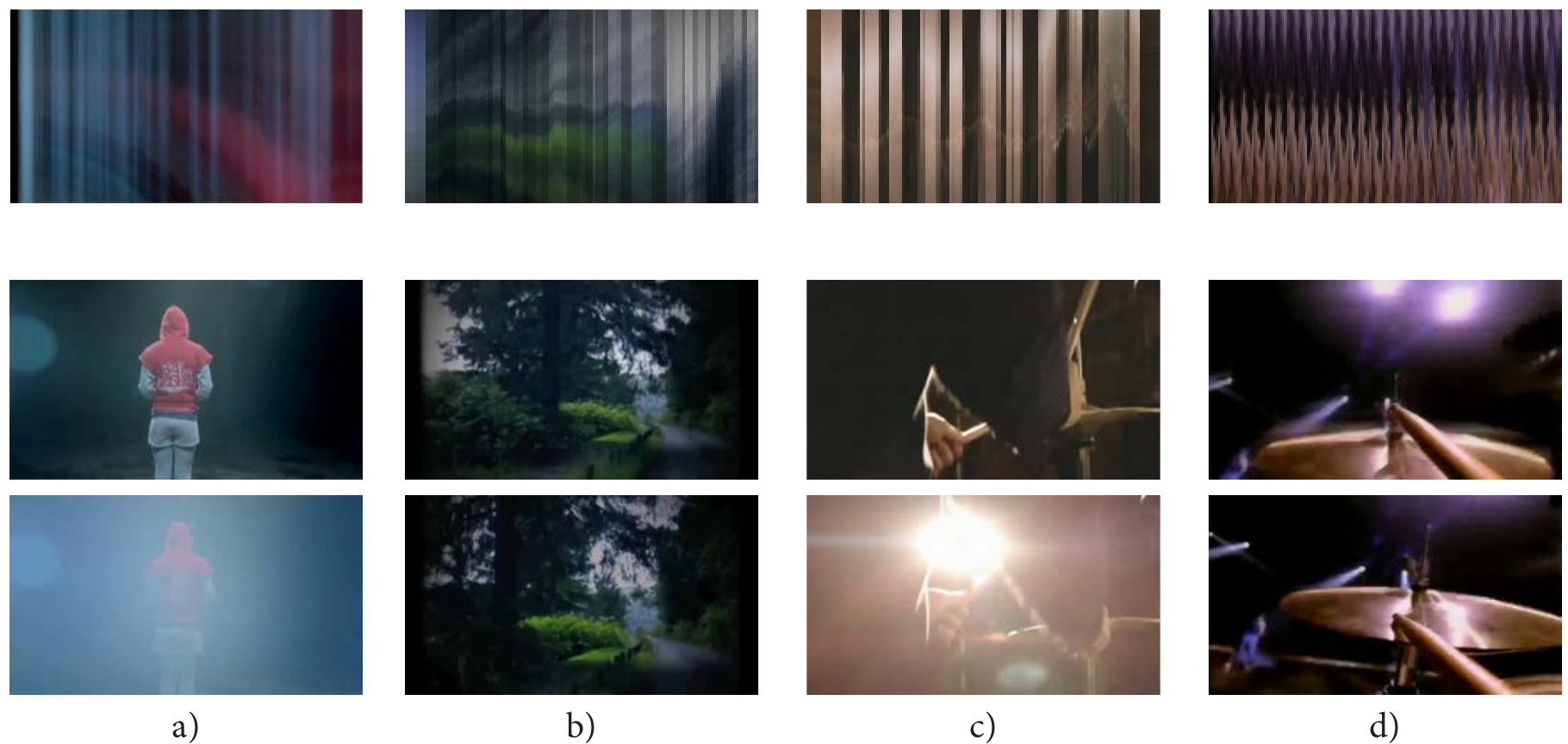}
	\caption{Example 4: Four examples of visual effects applied to music video frames. a) flashlights illuminating the entire video frame. b) silent-film effect of degrading celluloid film. c) Spotlight pointed at camera, randomly hidden by drummer. d) Camera mounted on drum-stick while playing the Hi-Hat cymbals.}
	\label{fig:example4}
\end{figure}

\begin{itemize}[noitemsep, topsep=5pt, leftmargin=0pt]
	\itemsep0.7em \renewcommand\labelitemi{}
	
	\item Figure \ref{fig:example4} a) - \textit{Folk} 06:
	The example is taken from the opening of the video. The camera circles around the person while the screen is randomly illuminated by bright white flashes. These flashes are easy recognizable in the mean-color bar. 
	
	\item Figure \ref{fig:example4} b) - \textit{Indie} 48:
	This video uses an effect that simulates the degradation of old celluloid films, as they are known from old Silence films. This results in random flickering through alternating illumination and saturation values of successive video frames (as can be seen in the mean-color bar).
	
	\item Figure \ref{fig:example4} c) - \textit{Indie} 95:
	In the chosen scene, a drummer plays on his drums. The camera is aimed directly at a glaring headlight, which is alternately covered by the drummer's arm when playing. The pattern visualized in the mean-color bar is very similar to sharp cut sequences.
	
	\item Figure \ref{fig:example4} d) - \textit{Hard Rock} 1:
	In this video sequence the camera is mounted on a drum-stick while the drummer plays the Hi-Hat cymbals. The abrupt changes create a rhythmic visual pattern depicted in the mean-color bar.	
	
\end{itemize}

The intention behind these four examples is to illustrate the influence of different effects. Naive color based approaches to shot boundary might be prone to wrong detections. While Figure \ref{fig:example4} a) and b) can be solved with minor modifications, Figure \ref{fig:example4} c) requires dedicated approaches to distinguish this effect from real transitions.

\section{Discussion}

The authors of summaries on shot boundary detection \cite{cotsaces2006video,yuan2007formal,smeaton2010video} list the most common shot transitions as to be \textit{sharp Cut}, \textit{Dissolve}, \textit{Fade in/out} and \textit{Wipe}. Further transition types are labeled as \textit{Other transition types} and are stated to be difficult to detect, but are rather rare. The experience from assembling the Music Video Dataset (MVD) and the experiments performed to detect shot boundaries showed, that \textit{Other transition types} are more commonly used in music video production including effects applied during editing or recording, which are yet not clearly defined in the context of the shot boundary detection task. Among the identified problems and challenges \cite{yuan2007formal} are:

\begin{itemize}[noitemsep, topsep=5pt, leftmargin=0pt]
	\itemsep0.7em \renewcommand\labelitemi{}
	
	\item \textbf{Detection of Gradual Transitions:}
	A comprehensive overview on the difficulties of detecting dissolving transitions is provided in \cite{lienhart2001reliable}. \textit{Threshold-Based approaches} detects transition by comparing similarities between frames in feature space. To detect \textit{Fade In/out} monochrome frame detection based on mean and standard deviation of pixel intensity values \cite{lienhart2001reliable,zheng2006novel} is used. Thresholds are commonly set globally \cite{cernekova2006information} which is generally estimated empirically or adaptively \cite{xia2007shot} using a sliding window function - or a combination of both \cite{qusenot2003clips}. Combinations of Edge Detectors and Motion Features are used to train a Support Vector Machine (SVM) \cite{zhao2007bupt} which is applied in a sliding window to detect gradual changes.
	Most of these approaches are not invariant towards the artistic effects described in the previous section. Especially global threshold based approaches will provide inaccurate predictions on the various kinds of \textit{Overlays} applied to music videos. Another problem is, that many blended music video sequences do not dissolve in a new shot, but the faded in sequence is faded out again.
	Combinations with motion and audio features are reported including thresholding with Hidden Markov Models (HMM) \cite{boreczky1998hidden}. In music videos audio features are not reliable because the transitions are not always aligned nor correlated with changes in song structures such as chord changes or progressions from verse to chorus.
	
	\item \textbf{Disturbances of Abrupt Illumination Change:}
	Most features used in shot boundary detection are not invariant to abrubt changes in illumination. Especially \textit{color based features} such as color histograms or color correlograms \cite{amir2003ibm} are based on luminance information of different color channels. Abrubt changes such as spotlights or overlays cause discontinuities in inter-frame distances which are often classified as shot boundaries. \textit{Texture based features} such as Tamura features, wavelet transform-based texture features or Gabor wavelet filters \cite{hauptmann2004confounded} are more robust against changes in illumination but are vulnerable to abrubt changes in textures such as motion blurring caused by fast camera panning and tilts.
	
	\item \textbf{Disturbances of Large Object/Camera Movement:}
	As mentioned in the previous section, fast camera movements or large moving objects in front of the camera affect the feature response of most features used in shot boundary detection, resulting in erroneous predictions of shot boundaries. Especially, fast camera movements are very frequent in music videos. Movement in any way is used to create tension or to bring a person into scene by circling around them.
	
\end{itemize}

Generally it can be summarized, that most approaches presented in literature harness or rely on a wide range of rules and definitions. These may either be based on physical conditions such as spatio-temporal relationships between consecutive frames, or on rules developed by the art of film-making. For example the use of audio features \cite{boreczky1998hidden} is based on the observation that dissolving transitions are more often used with scene changes than with transitions within the same scene. This includes a change of the sound scene, which is harnessed to augment the detector.
As extensively elaborated in Chapter \ref{ch3:intro} music videos deliberately do not stick to these rules. Of course, many of the challenges listed in Section \ref{transition_types} can already be solved, but not with a general approach. Most of them are exceptions to commonly known problems and require distinct detection approaches. For example, concerning the problem of rapid sequences of dissolving scenes as depicted in Figure \ref{fig:shotexample5}, one solution could be to interpret this sequence as a scene by itself. Again, a custom model or an exception handling to existing models has to be implemented for this. Further, some listed points still require a broader discussion on whether they should be considered a transition and if so, how it should be labeled. 

This chapter discussed observations made during an effort to develop a shot boundary detector for music videos. The number of shots per seconds as well as the type of transition is considered to be a significant feature for discriminating music videos by genre and mood. This development stalled when it was obvious, that state-of-the-art shot boundary detection approaches are not generally applicable to music videos - for the reasons elaborated in this Chapter. These issues are not insoluble. However, many of these effects require dedicated solutions or detectors to process them. Some issues require a broader discussion to define their category such as whether or not they are shot transitions. More problematic is, that this is only an abstract of examples and that music video creators regularly develop new creative effects.

\section{Summary}

This Chapter discussed open issues of shot boundary detection for music videos. The number of shots per seconds as well as the type of transition is considered to be a significant feature for discriminating music videos by genre or mood. 
Various transition types observed in the Music Video Dataset (MVD) are listed. They are further discussed concerning in which way those pose a problem for state-of-the-art shot boundary detection approaches. 
These issues are not insoluble. However, many of these effects require dedicated solutions or detectors to process them. Some issues require a broader discussion to define their category such as whether or not they are shot transitions. More problematic is, that this is only an abstract of examples and that music video creators regularly develop new creative effects. 
It is yet conceivable that approaches based on Recurrent Convolutional Neural Networks \cite{baraldi2017hierarchical} are able to learn the different visual effects and transition types. To pursue such experiments, ground truth labels are required for the Music Video Data-set or another data-set. To facilitate the creation of such annotations we have crated an interactive tool, which is provided as open-source software\footnote{https://blinded.for.review}.\\

For \textit{future work} it would be required to come to mutual agreement on the labeling the various artistic effects applied in music videos and whether or not they are considered to be types of transitions. Based on these definitions, it would then be of interest to evaluate if on the one hand approaches can be found to detect these transitions, and on the other hand, the frequency of their application is correlated with music characteristics such as genre, style or mood.\\

\noindent
\textit{\smaller This summary of open issues in shot boundary detection for music videos as presented in this chapter was published in \textit{Alexander Schindler and Andreas Rauber. ``On the unsolved problem of Shot Boundary Detection for Music Videos''} in the Proceedings of the 25th International Conference on MultiMedia Modeling, January 8th 2019, Thessaloniki, Grece \cite{schindler2019}. }




\chapter[The Color of Music: Evaluating Color and Affective Image Descriptors]{The Color of Music\\Evaluating Color and Affective\\Image Descriptors}
\label{ch7:intro}

\epigraph{
	``'Begin at the beginning', the King said, very gravely, 'and go on till you come to the end: then stop.' ''} 
{---  Lewis Carroll, Alice's Adventures in Wonderland, 1865}


Over the past decades music videos distinctively influenced our pop-culture and became a significant part of it. Since their global inception in the early 1980s music videos emerged from a promotional support medium into an art form of their own. 
The potential advantages of harnessing music-related visual information are extensively discussed in Chapter \ref{ch1:mvir_opportunities}.
A first example of this potential has been demonstrated in Chapter \ref{ch6:intro} on music video based artist identification. A precision improvement of 27\% over conventional audio features shows that the additional information benefits this audio-related task.
In order to use the visual domain for music retrieval tasks, it has to be linked to the acoustic domain. Since substantial research on audio-visual correlations in music videos is yet scarce or not available, the approach presented in this chapter is based on the simplified assumption that both layers intend to express the same music characteristics. Those can be emotions such as happy or sad, but also themes of the song such as party or hanging out at the beach.
The underlying hypothesis here is that there exists a correlation between the chosen visual aesthetics and the characteristics of the corresponding song.

So far, no basic research has been conducted to investigate this hypothesis.
An audio-visual approach to music video segmentation \cite{gillet2007correlation} was reported, which addresses some of the issues discussed in Chapter \ref{ch9:intro}.
Approaches to affective content analysis of music videos are provided by \cite{yazdani2011affective} and \cite{zhang2010affective}. Both use visual information to augment the acoustic information in order to improve the results of an audio-only approach in the dedicated tasks. The specificity of these tasks does not allow to draw general conclusions to other MIR related tasks. 

Thus, in this chapter a bottom-up evaluation of affective visual features is presented. More specifically the color information is analyzed, if it is sufficient to discriminate music by their genres. This type of evaluation is chosen, because music genre is an abstraction of music characteristics and implicetly describes information about instrumentation, charachteristics of sound and production, style and typical rhythms, etc. The evaluation focuses on low-level color descriptors and attempts to provide fundamental insights on correlation between the audio and the visual representation of music in music videos. Using color in content-based image retrieval has been extensively studied \cite{manjunath2001color,plataniotis2000color,schettini2001survey} and is yet described as problematic since it is highly influenced by lighting conditions during image acquisition. In music videos different illumination settings and colors are usually desired artistic effects (see Chapter \ref{ch3:mv_styles}).


This approach and its corresponding evaluation presented in this chapter was published and presented in \textit{Alexander Schindler and Andreas Rauber. ``An audio-visual approach to music genre classification through affective color features''} at the 37th European Conference on Information Retrieval (ECIR'15), Vienna, Austria, March 29 - April 02 2015 \cite{SCHINDLER_2015ecir}.
In the following sections seven feature sets are introduced that derive from psychological experiments, art-theory or try to model human perception. Section \ref{ch7:evaluation} lays out the evaluation. After discussing the results in Section \ref{ch7:results} conclusions and outlooks to future work are provided in Section \ref{ch7:conclusions}.

\section{Acoustic and Visual Features}
\label{ch7:method}

This section introduces a collection of audio and visual feature-sets that are used in the succeeding evaluation. The audio features are extracted from the separated audio channel of the music videos. Visual features are extracted from each frame of a video. To abstract from the varying lengths of the videos the extracted frame-level features are aggregated during post-processing by calculating the statistical measures mean, median, standard deviation, min, max skewness, kurtosis. As a pre-processing step the black bars at the borders of video frames, also called \textit{Letterboxing} or \textit{Pillarboxing}(see Figure \ref{fig:ch7:image_enhancement}a), are removed. Removing this often applied stylistic effect is a normalization step to enhance comparability of videos with and without bars (see Figure \ref{fig:ch7:image_enhancement}b).

\subsection{Audio Features}
\label{ch7:Audio_Features}

\begin{figure*}[t!]
	\centering
	\includegraphics[width=1.0\textwidth]{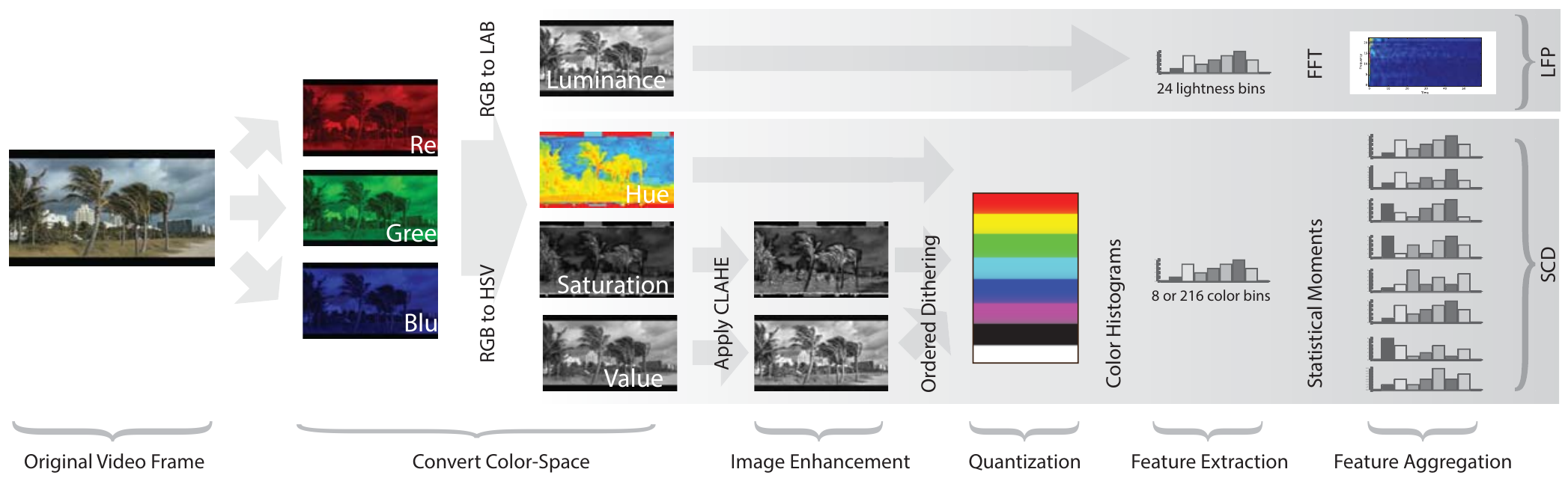}
	\caption{ Visualization of the feature extraction process described in Section \ref{ch7:method}. }
	\label{fig:ch7:feature_extraction}
\end{figure*}

The audio features used for the experiments have been chosen due to comparability with results presented in Chapter \ref{ch5.2:intro}.

\paragraph{Psycho-accoustic Music Descriptors} 
as proposed by \cite{lidy2005evaluation} are based on a psycho-acoustically modified Sonogram representation that reflects human loudness sensation (described in more detail in Chapter \ref{ch4.1:rp_extract}). From the Rhythm Patterns feature family \textit{Statistical Spectrum Descriptors (SSD)}, \textit{Rhythm Patterns (RP)}, \textit{Rhythm Histograms (RH)}, \textit{Temporal Statistical Spectrum Descriptor (TSSD)} and \textit{Temporal Rhythm Histograms (TRH)} are used in the evaluation.
For the extraction, the Matlab-based implementation\footnote{\url{http://www.ifs.tuwien.ac.at/mir/downloads.html}}, version 0.6411 was employed.

\paragraph{Mel Frequency Cepstral Coefficients (MFCC)} are well known audio features derived from speech recognition.
\textbf{Chroma} features project the spectrum onto 12 bins representing the semitones of the musical octave. Both are added to the evaluation as an outline, because their performance on conventional data-sets is well known. The features are extracted using the well known MARSAYS toolset \cite{tzanetakis2000marsyas}. For the feature extraction MARSYAS version\footnote{\url{http://sourceforge.net/projects/marsyas/}}. 0.4.5. was utilized.\\

Music genre classification results are presented to serve as baseline performance values for the evaluation of the color descriptors. Table \ref{tab:ch7:Results} shows mean accuracy values of 10-fold cross-validation experiments for four different classifiers on all genre specific sub-sets of the MVD.

\subsection{Visual Features}
\label{ch7:visual_features}

The selection for this part-evaluation presents consists of well-known low-level image processing features to provide a comprehensible estimation of their expressibility. Further, color based features derived from art-theory and empirical psychological studies, used for affective image retrieval, are applied to study possible effects of intentional color usage in music videos.

\paragraph{Global Color Statistics (GCS)}
are a set of eight features. They calculate \textit{Mean Saturation} and \textit{Mean Bightness} based on the Improved Hue, Luminance and Saturation (IHLS) color space \cite{hanbury20023d} which has the advantages of low saturation values of achromatic pixels and independence of saturation from the brightness function. After converting the frame to the IHLS color space \textit{Mean Saturation} and \textit{Mean Brightness} values are globally calculated for a video frame. Because Hue in IHLS is an angular value circular statistics has to be applied \cite{hanbury2003circular} to assess \textit{angular mean Hue} and \textit{angular deviation of Hue}. For weakly saturated colors the hue channel behaves unpredictably in the presence of color changes induced by image noise. \textit{Saturation weighted mean Hue} and \textit{deviation of Hue} are more robust towards weakly saturated colors. It introduces the relationship between hue and saturation by weighting the unit hue vectors by their corresponding saturation.

\paragraph{Global Emotion values (GEV)}
refer to a Pleasure-Arousal-Dominance model based on investigated emotional reactions presented in \cite{valdez1994effects}. The authors introduce a linear relationship between saturation and brightness as a model for the emotional variables. The values were calculated from the luminance (B) and saturation (S) channel of the previously described independent IHLS color space:

\begin{align}
	Pleasure  &= 0.69  \cdot B + 0.22 \cdot S \\
	Arousal   &= -0.31 \cdot B + 0.60 \cdot S \\
	Dominance &= 0.76  \cdot B + 0.32 \cdot S 
\end{align}

\paragraph{Colorfulness (CF)}
is one of the features used in \cite{datta2006studying} to computationally describe aesthetics in photographies. The proposed method is based on a partitioned RGB palette using Earth Mover's Distance (EMD) \cite{rubner2000earth} to calculate a single valueds dissimilarity of a supplied image to an \textit{ideal} color distribution of a \textit{colorful} image.

\begin{figure*}[t!]
	\centering \includegraphics[width=1.0\textwidth]{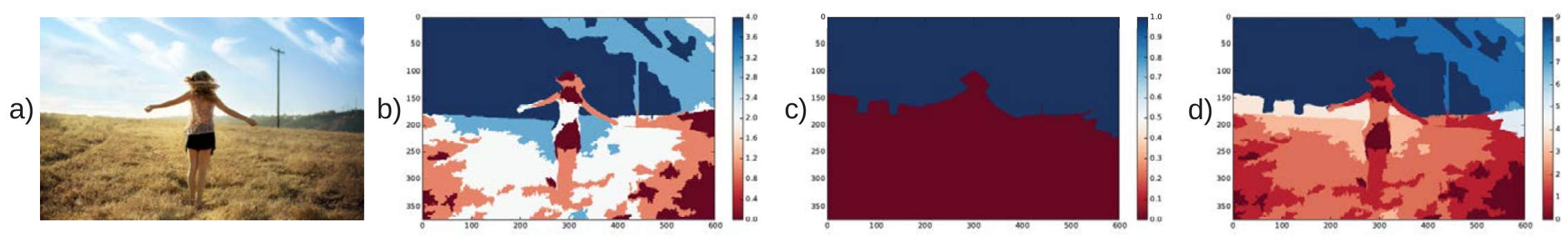}
	\caption{Wang Emotional Factors: a) Original Image. b) Lightness. c) Warm-Cool. d) Lightness-Warm-Cool }
	\label{fig:ch7:wang_emotional_factors}
\end{figure*}

\paragraph{Wang Emotional Factors (WEF)}
Wang et al. \cite{wei2006image} identified three factors based on emotional word correlations that are relevant for emotion semantics based image retrieval and defined three corresponding feature sets. Fuzzy membership functions are used to assign values of the perceptual psychology motivated L*C*H* color space to discrete semantic words. \textit{Feature One} includes lightness description of image segments  ranging from \textit{very dark} to \textit{very bright} (Figure \ref{fig:ch7:wang_emotional_factors}b). These are combined with the classification of hue into \textit{cold} and \textit{warm} colors, resulting in a 10 dimensional histogram. \textit{Feature Two} provides a description of warm or cool regions with respect to different saturations as well as a description of contrast (Figure \ref{fig:ch7:wang_emotional_factors}c). \textit{Feature Three} combines lightness contrast with an sharpness estimation (Figure \ref{fig:ch7:wang_emotional_factors}d). A no-reference perceptual blur measure \cite{crete2007blur} was used to calculate the sharpness. The contrast description of the third factor overlaps with the \textit{Itten contrasts} and is omitted.


\begin{table}[t]
	\small
	\begin{tabularx}{\textwidth}{p{0.3cm}|p{2.0cm} c p{2.0cm}}

		\hline
		\rowcolor{gray!30}
		& \multicolumn{1}{l}{\textbf{Short Name}} & \multicolumn{1}{l}{\textbf{\#}} & \multicolumn{1}{p{8.25cm}}{\textbf{Descriptiom}} \\ 
		\hline
		
		& \multicolumn{1}{p{3.8cm}}{} & \multicolumn{1}{l}{} & \multicolumn{1}{p{8.25cm}}{} \\
		
		\multirow{6}{*}{\rotatebox[origin=c]{90}{\textbf{Audio}}} 
		& \multicolumn{1}{p{3.8cm}}{Statistical Spectrum Descriptors (SSD)} & \multicolumn{1}{l}{168}  & \multicolumn{1}{p{8.25cm}}{Statistical description of a psycho-accoustic transformed audio spectrum} \\
		& \multicolumn{1}{p{3.8cm}}{Rhythm Patterns (RP)}                   & \multicolumn{1}{l}{1024} & \multicolumn{1}{p{8.25cm}}{Description of spectral fluctuations} \\
		& \multicolumn{1}{p{3.8cm}}{Rhythm Histograms (RH)}                 & \multicolumn{1}{l}{60}   & \multicolumn{1}{p{8.25cm}}{Aggregated Rhythm Patterns} \\
		& \multicolumn{1}{p{3.8cm}}{Temporal SSD and RH}                    & \multicolumn{1}{l}{}     & \multicolumn{1}{p{8.25cm}}{Temporal variants of RH (TRH \#420), SSD (TSSD \#1176)} \\
		& \multicolumn{1}{p{3.8cm}}{MFCC}                                   & \multicolumn{1}{l}{12}   & \multicolumn{1}{p{8.25cm}}{Mel Frequency Cepstral Coefficients} \\
		& \multicolumn{1}{p{3.8cm}}{Chroma}                                 & \multicolumn{1}{l}{12}   & \multicolumn{1}{p{8.25cm}}{12 distinct semitones of the musical octave} \\
		
		& \multicolumn{1}{p{3.8cm}}{} & \multicolumn{1}{l}{} & \multicolumn{1}{p{8.25cm}}{} \\
		\hline
		& \multicolumn{1}{p{3.8cm}}{} & \multicolumn{1}{l}{} & \multicolumn{1}{p{8.25cm}}{} \\
		
		\multirow{9}{*}{\rotatebox[origin=c]{90}{\textbf{Visual}}} 
		& \multicolumn{1}{p{3.8cm}}{Global Color Statistics}        & \multicolumn{1}{l}{6}  & \multicolumn{1}{p{8.25cm}}{mean saturation and brightness, mean angular hue, angular deviation, with/without saturation weighting} \\
		& \multicolumn{1}{p{3.8cm}}{Colorfulness}                   & \multicolumn{1}{l}{1}  & \multicolumn{1}{p{8.25cm}}{colorfulness measure based on Earth Movers Distance} \\
		& \multicolumn{1}{p{3.8cm}}{Color Names}                    & \multicolumn{1}{l}{8}  & \multicolumn{1}{p{8.25cm}}{Magenta, Red,Yellow,Green,Cyan Blue, Black, White} \\
		& \multicolumn{1}{p{3.8cm}}{Pleasure, Arousal, Dominance}   & \multicolumn{1}{l}{3}  & \multicolumn{1}{p{8.25cm}}{approx. emotional values based on brightness and saturation} \\
		& \multicolumn{1}{p{3.8cm}}{Itten Contrasts}                & \multicolumn{1}{l}{4}  & \multicolumn{1}{p{8.25cm}}{Contrast of Light and Dark, Contrast of Saturation, Contrast of Hue and Contrast of Warm and Cold} \\
		& \multicolumn{1}{p{3.8cm}}{Wang Emotional Factors}         & \multicolumn{1}{l}{18} & \multicolumn{1}{p{8.25cm}}{Features for the 3 affective factors by Wang et al. \cite{wei2006image}}  \\
		& \multicolumn{1}{p{3.8cm}}{Lightness Fluctuation Patterns} & \multicolumn{1}{l}{80} & \multicolumn{1}{p{8.25cm}}{Rhythmic fluctuations in video lightness} \\
		
		& \multicolumn{1}{p{3.8cm}}{} & \multicolumn{1}{l}{} & \multicolumn{1}{p{8.25cm}}{} \\
		\hline
	\end{tabularx}
	\caption{Overview of all features. The column '\#' indicates the dimensionality of the corresponding feature set.}

\end{table}


\paragraph{Itten's Contrasts (IC)}
are a set of art-theory concepts defined by Johannes Itten \cite{itten1973art} for combining colors to induce emotions. The contrasts are based on an proportional opponent color model with 180 distinct colors that are mixtures from 12 pure colors. The contrast calculation is aligned to the method presented in \cite{machajdik2010affective} which uses Wang's feature extraction \cite{wei2006image} as a predecessor. Instead of a waterfall segmentation a Quick Shift \cite{vedaldi2008quick} approach was used due to better performance at a more reasonable processing time. The following sets of contrasts were calculated: \textit{Contrast of Light and Dark}, \textit{Contrast of Saturation}, \textit{Contrast of Hue} and \textit{Contrast of Warm and Cold}. The corresponding literature \cite{wei2006image} fails to mention the descriptions of aggregation methods used to calculate the distinct values from the fuzzy membership functions. Thus, the average value was used to calculate these values.

\begin{figure*}[t!]
	\centering \includegraphics[width=0.8\textwidth]{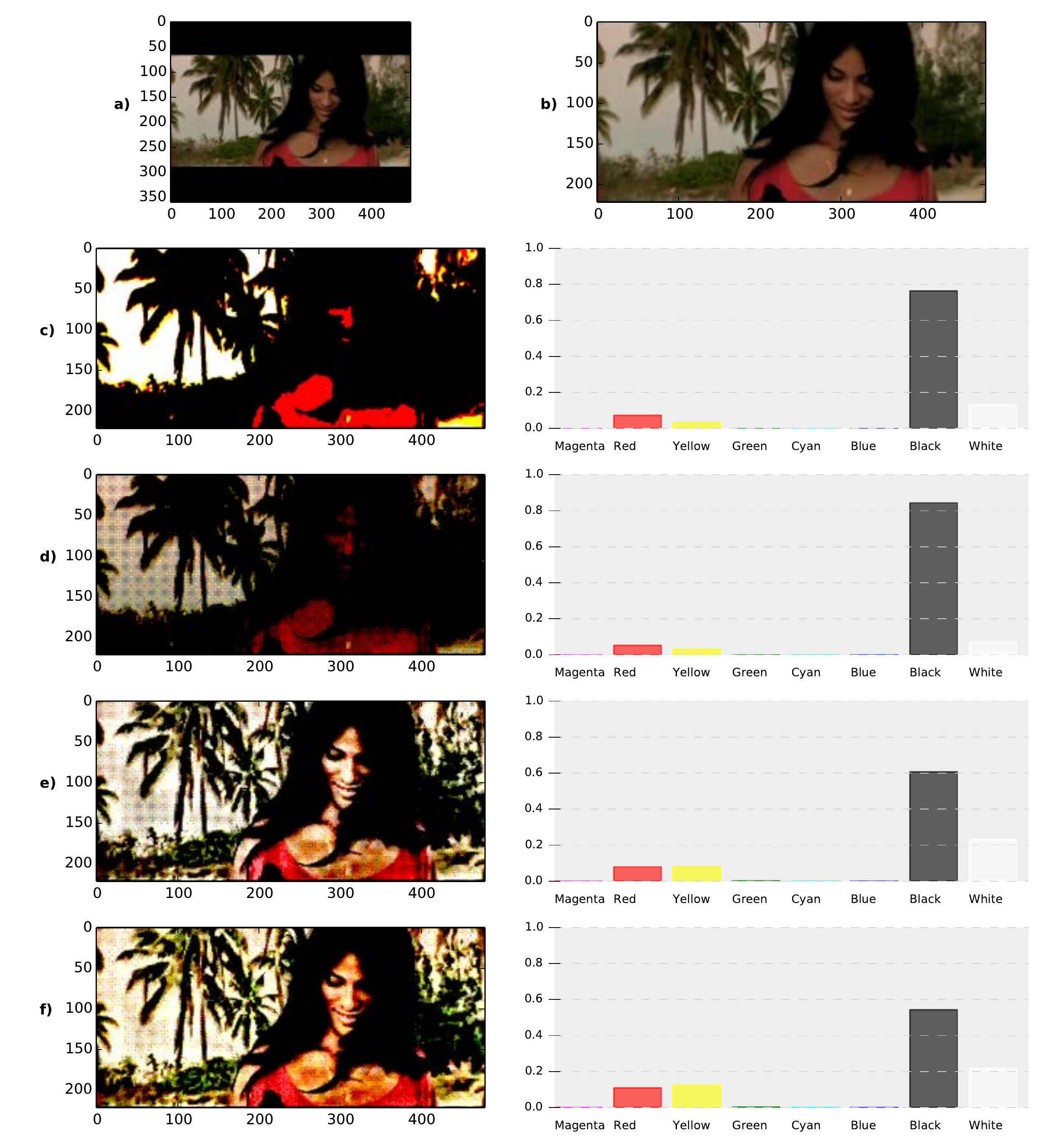}
	\caption{Visualization of the image enahncement steps: a) original image with black horizontal bars (letterboxing), b) pre-processed image without letterboxing, c) simple nearest neighbor match color quantization, d) Ordered Dithering (OD), e) Ordered Dithering with enhanced brightnes, f) Orderd Dithering with enhanced brightness and saturation.}
	\label{fig:ch7:image_enhancement}
\end{figure*}

\paragraph{Color Names (CN)}
\label{ch7:color_names}
describe color distributions of the reduced Web-safe Elementary-color palette consisting of the 8 elementary colors Magenta, Red, Yellow, Green, Cyan, Blue, Black and White. To map a frame of a video to this palette it is converted to Hue Value Saturation (HSV) color-space. \textit{Contrast, brightness and color enhancement} is applied through application of Contrast Limited Adaptive Histogram Equalization (CLAHE) \cite{zuiderveld1994contrast} to the value channel \textit{V} using a region size of 22x22 pixels and a clip limit of 1.0. Saturation and color enhancement was applied similarly to the corresponding channels with slightly adapted values. \textit{Color Quantization} to reduce the number of distinct colors to a desired palette is obtained by applying \textit{error diffusion} or \textit{dithering} which computes the mean square error between the original pixel value and its closest match which is then propagated locally to its surrounding pixels. \textit{Ordered Dithering} was used since it reduces the effect of contouring but stays more consistent with the original colors. A 32x32 Bayer pattern{bayer1976color} matrix was used as threshold map. Figure \ref{fig:ch7:image_enhancement}d) shows a quantized image using ordered dithering compared to a naive nearest-neighbor-match approach in Figure \ref{fig:ch7:image_enhancement}c). \textit{Feature Calculation} as depicted in Figure \ref{fig:ch7:feature_extraction} is concluded by calculating the statistical moments mean, median, variance, min, max, skew and kurtosis of the reduced palette.


\paragraph{Lightness Fluctuation Patterns (LFP)}
are calculated analogous to the music feature Rhythm Patterns (RP) \cite{lidy2005evaluation}. In a first step each frame of a music video is converted to the perceptually uniform LAB color space. This corresponds to the psychoacoustic transformations applied to the audio data as a pre-processing step of RP feature calculation. For each frame a 24 bin histogram of the lightness channel is calculated. Fast Fourier Transform (FFT) is applied to the histogram space of all video frames. This results in a time-invariant representation of the 24 lightness levels capturing reoccurring patterns in the video. Only amplitude modulations in the range from 0 to 10 Hz are used for the final feature set, since rhythm cannot be perceived from higher modulation frequencies. Based on the observation that light effects, motions and shots are usually beat synchronized in music videos, LFPs can be assumed to express rhythmic structures of music videos.

\section{Evaluation}
\label{ch7:evaluation}

The empirical evaluation is based on the Music Video Dataset (MVD) (see Chapter \ref{ch4.1:intro}). Empirical classification experiments and Chi-square feature selection were used to analyze the performance of the visual and audio-visual feature-spaces. The following sub-sets are used in this evaluation. For a full overview of the MVD please refer to Chapter \ref{ch4.1:intro}.

\begin{description}[leftmargin=!,itemsep=3pt,labelwidth=\widthof{\bfseries Bollywood},labelindent=\descleftmargin,rightmargin=\descrightmargin]
	
	\item[MVD-VIS:] The \textit{M}usic \textit{V}ideo \textit{D}ata\-set for \textit{VIS}ual content analysis and classification (see Chapter \ref{ch4:mvdvis}) is intended for classifying music videos by their visual properties only. 
	
	\item[MVD-MM:] The \textit{M}usic \textit{V}ideo \textit{D}ataset for \textit{M}ulti\textit{M}odal content analysis and classification (see Chapter \ref{ch4:mvdmm}) is intended for multi-modal classification and retrieval tasks. 
	
	\item[MVD-MIX:] A combination of the data-sets MVD-VIS and MVD-MM (see Chapter \ref{ch4:mvdmix}).
	
\end{description}

\section{Results and Conclusions}
\label{ch7:results}


%
\begin{table*}[t!]
	
	\scriptsize
	\begin{tabularx}{\textwidth}{p{0.2cm}|X|rrr|rrr|rrr}
		
		\multicolumn{11}{c}{} \\ [-4.0ex]
		\hline
		\rowcolor{gray!30}
		\hline
		& & \multicolumn{3}{c}{ \textbf{MVD-VIS} } & \multicolumn{3}{c|}{ \textbf{MVD-MM} } &  \multicolumn{3}{c}{ \textbf{MVD-MIX} } \\
		\hline
		& & \multicolumn{3}{c|}{} & \multicolumn{3}{c|}{} &  \multicolumn{3}{c}{} \\[-0.9ex]
		
		& & \textbf{SVM} & \textbf{KNN} &  \textbf{NB} & \textbf{SVM} & \textbf{KNN} &  \textbf{NB} & \textbf{SVM} & \textbf{KNN} &  \textbf{NB} \\

		\hline
		
		\multirow{7}{*}{\rotatebox[origin=c]{90}{\parbox[t]{8mm}{\textbf{Audio}}}}
		& & \multicolumn{3}{c|}{} & \multicolumn{3}{c|}{} &  \multicolumn{3}{c}{} \\
		
		& \scriptsize TSSD-RP-TRH & 93.79 & 80.85 & 71.46 &       74.76 & 55.00 & 52.20 &      75.91 & 54.16 & 48.32 \\
		& \scriptsize TSSD        & 86.81 & 72.58 & 62.61 &       69.97 & 53.33 & 53.65 &      66.19 & 47.40 & 44.22 \\
		& \scriptsize RP          & 87.26 & 69.81 & 64.04 &       60.35 & 42.38 & 41.63 &      63.19 & 43.06 & 41.39 \\
		& \scriptsize SSD         & 85.78 & 73.18 & 58.81 &       68.74 & 50.28 & 48.41 &      65.11 & 44.64 & 38.92 \\
		& \scriptsize TRH         & 71.04 & 55.83 & 53.86 &       49.50 & 38.28 & 39.66 &      46.61 & 33.02 & 35.70 \\
		& \scriptsize MFCC        & 62.28 & 48.58 & 46.95 &       42.14 & 29.16 & 34.17 &      37.02 & 26.60 & 27.11 \\
		& \scriptsize Chroma      & 36.34 & 28.09 & 23.03 &       25.26 & 20.11 & 19.41 &      19.64 & 14.68 & 12.08 \\
		
		& & \multicolumn{3}{c|}{} & \multicolumn{3}{c|}{} &  \multicolumn{3}{c}{} \\

		\hline
		
		& & \multicolumn{3}{c|}{} & \multicolumn{3}{c|}{} &  \multicolumn{3}{c}{} \\ [-1.9ex]
		
		\multirow{7}{*}{\rotatebox[origin=c]{90}{\parbox[t]{8mm}{\textbf{Visual}}}}
		& LFP      &         33.21  &         23.59  &         25.45  &         20.38  &         16.74  &         16.46  &         16.93  &         11.71  &         13.36  \\
		& CF       &         34.89  &         25.49  &         31.50  &         21.84  &         17.06  &         20.41  &         18.53  &         11.92  &         16.49  \\
		& IC       &         36.80  &         27.55  &         27.51  &         24.83  & \textbf{19.43} &         19.68  &         21.44  &         13.54  &         12.66  \\
		& GEV      &         39.45  & \textbf{29.84} & \textbf{34.15} &         20.81  &         17.04  &         18.51  &         20.27  &         14.47  & \textbf{17.89} \\
		& GCS      &         40.55  &         29.76  &         33.91  &         24.08  &         17.29  &         18.15  & \textbf{23.72} & \textbf{15.40} &         17.34  \\
		& WAF      &         41.01  &         26.43  &         29.86  &         26.01  &         19.08  & \textbf{21.38} &         22.86  &         13.90  &         16.60  \\
		& CN       & \textbf{43.68} &         29.04  &         32.23  & \textbf{26.74} &         19.13  &         18.77  &         23.48  &         14.76  &         15.99  \\[0.9ex]
		
		& & \multicolumn{3}{c|}{} & \multicolumn{3}{c|}{} &  \multicolumn{3}{c}{} \\ [-1.9ex]
		
		\hline

		& & \multicolumn{3}{c|}{} & \multicolumn{3}{c|}{} &  \multicolumn{3}{c}{} \\[-0.9ex]
		
		& \textbf{Visual \mbox{Features} combined} & 50.13 & 34.04 & 39.38 &     31.69 & 21.16 & 23.38 &     32.22 & 17.89 & 21.16 \\
		
		\hline
		
		\multirow{11}{*}{\rotatebox[origin=c]{90}{\parbox[t]{8mm}{\textbf{\mbox{Audio-Visual}}}}}
		
		& & \multicolumn{3}{c|}{} & \multicolumn{3}{c|}{} &  \multicolumn{3}{c}{} \\
		
		& TSSD-RP-TRH & \textbf{94.86} & \textbf{81.38} & \textbf{71.65} &     \textbf{75.69} & \textbf{55.78} &         51.36  &     \textbf{76.53} & \textbf{55.76} & \textbf{49.08} \\
		& TSSD        & \textbf{88.45} &         71.65  & \textbf{64.75} &     \textbf{70.55} &         52.60  &         52.25  &     \textbf{69.46} &         46.15  & \textbf{45.16} \\
		& RP          & \textbf{89.80} & \textbf{71.99} & \textbf{65.78} &     \textbf{62.79} & \textbf{43.93} &         41.61  &     \textbf{66.59} & \textbf{44.47} & \textbf{41.68} \\
		& SSD         &         85.25  &         62.05  &         57.80  &             65.34  &         42.28  &         44.24  &     \textbf{65.21} &         36.13  & \textbf{38.76} \\
		& TRH         & \textbf{77.84} & \textbf{55.98} & \textbf{59.71} &     \textbf{58.50} &         32.79  & \textbf{41.40} &     \textbf{56.31} &         31.39  &         40.09 \\
		& MFCC        & \textbf{63.71} &         41.53  &         46.28  &     \textbf{42.88} &         24.38  &         27.35  &     \textbf{43.11} &         22.33  & \textbf{25.62} \\
		& Chroma      & \textbf{55.70} & \textbf{39.28} & \textbf{43.13} &     \textbf{35.29} & \textbf{24.16} & \textbf{25.51} &     \textbf{35.43} & \textbf{20.10} & \textbf{24.14} \\
		
		& & \multicolumn{3}{c|}{} & \multicolumn{3}{c|}{} &  \multicolumn{3}{c}{} \\

		\hline
	\end{tabularx}

	\caption{\label{tab:ch7:audiobenchmarks}\scriptsize Classification results for audio, visual and audio-visual features showing accuracies for Support Vector Machines (SVM), K-Nearest Neighbors (KNN) and Naive Bayes (NB) classifiers. Bold-faced values highlight improvements of audio-visual approaches over audio features.} 
	\label{tab:ch7:Results}
	
\end{table*}

\begin{figure}[t]
	
	\centering
	\includegraphics[width=1.0\textwidth, angle=0]{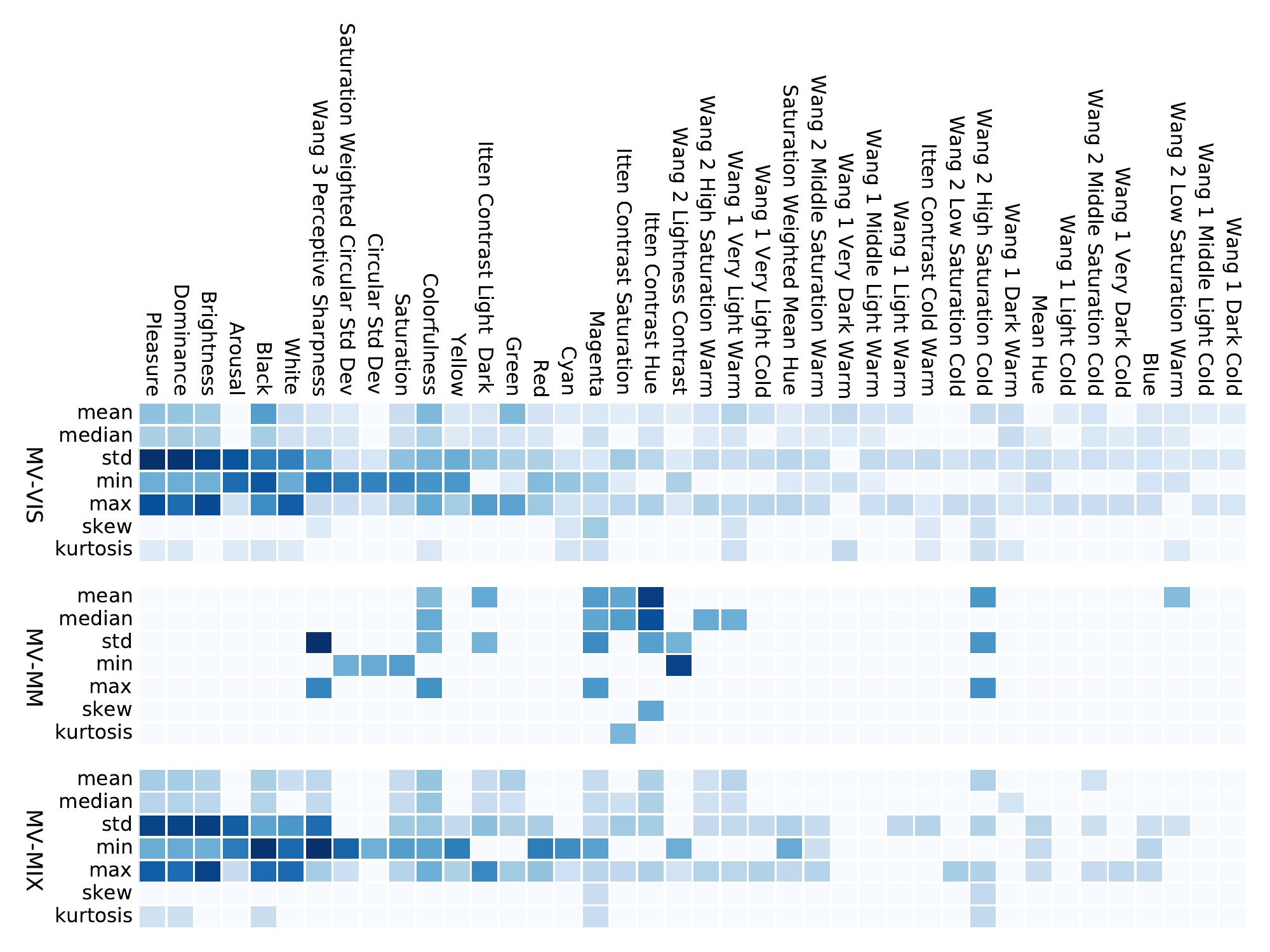}
	\caption{\scriptsize Chi Square Feature Evaluation in descending order from left to right. Dark blue areas correspond with high $\chi^2$ values.}
	\label{fig:ch7:feature_selection}
	
\end{figure}

Table \ref{tab:ch7:Results} summarizes the results of the comparative classification experiments. The top segment of the table provides audio only results which serve as baseline for evaluating the visual and audio-visual approaches. 
Using visual features only an accuracy of 50.13\% could be reached for Support Vector Machines (SVM) for the MVD-VIS set. Accuracies for other sets or classifiers range from 17.89\% to 39.38\%. Because all classes are equal in size these results are above a baseline of 12.5\% or 6.25\% respectively.
Although, the performance of the visual features alone is not representative, the audio-visual results show interesting effects. Generally, there is insignificant or no improvement of the performance over the top performing audio features. The results show that combining the visual features with chroma and rhythm descriptors has a positive effect on the accuracy while it is negative with spectral and timbral features. 
Applying ranked Chi-square attribute selection (see Figure \ref{fig:ch7:feature_selection}) on the visual features shows, that affective features such as \textit{Pleasure}, \textit{Dominance} and \textit{Arousal} as well as the \textit{frequencies of black and white pixels} have the highest values. Further, more information is provided by variance and min/max aggregated values than by mean values. 
The Color Names (CN) feature-set performs best of the presented visual features. Especially, minimum, maximum and standard-deviation aggregations of the features \textit{black} and \textit{white} are identified to have an high impact on the classification results in the combined visual feature-set (see Figure \ref{fig:ch7:feature_selection}).
This indicates that the frequency and intensity of illumination changes is a discriminative feature for music genre classification. The Lightness Fluctuation Patterns (LFP) are intended to capture this information, though according to the experimental results, they perform not accordingly. 


\begin{figure*}[t]
	\centering
	\includegraphics[width=1.00\textwidth]{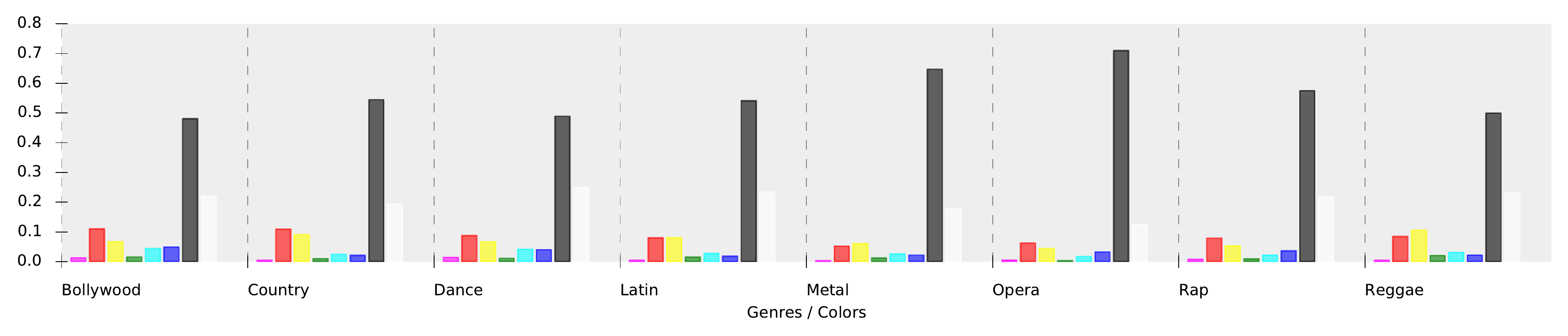}
	\includegraphics[width=1.00\textwidth]{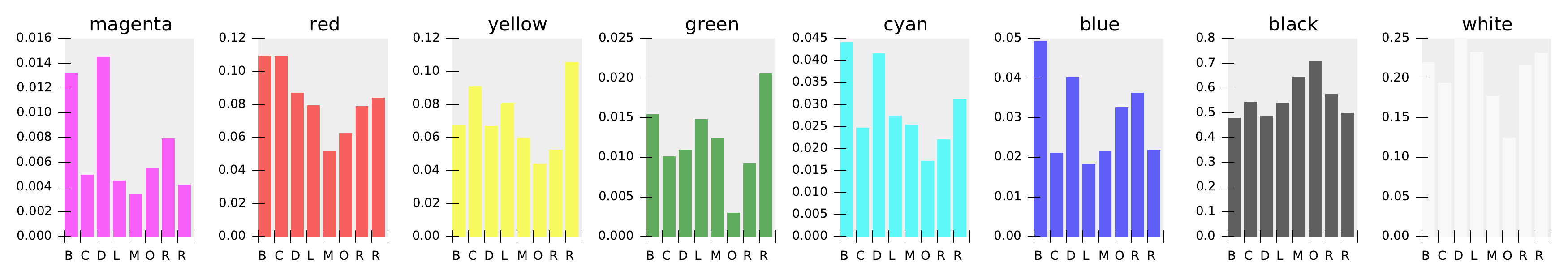}
	\caption{ Overview of the color distributions for the genres of the MV-VIS dataset. The upper chart shows average color histograms per genre. The lower chart outlines the color distribution over genres per color. The characters in the lower chart represent the initial letters of the color names of the upper chart in corresponding order. }
	\label{fig:mvd_vis_color_dist}
\end{figure*}

Figure \ref{fig:mvd_vis_color_dist} provides an overview of the color distributions for the genres of the MV-VIS dataset according the reduced color space which is described by a small set of color names (see Section \ref{ch7:color_names}). The upper chart shows average color histograms per genre. The lower chart outlines the color distribution over genres per color. The characters in the lower chart represent the initial letters of the color names of the upper chart in corresponding order. Although contrast and colors have been enhanced there is still a dominance of low saturated values which are mapped to the color black.

An assumption preceding this evaluation was, that color is a relevant feature to distinguish different music genres. ``Dark Sounds'' are often referred to noisy low frequency tones while ``light sounds'' refer to clear high frequency tones. The assumption was that these linguistic references are also reflected in the choice for colors and illumination in music videos. This assumption could not be confirmed analytically. \textit{Opera} videos have more pixels mapped to the color black than \textit{Heavy Metal} music videos. This rather results from poor lighting conditions in the production process and seems not to be an intended feature of \textit{Opera} videos. \textit{Reggae} music videos show the highest values for the color green. That's because many of the videos in this category of the data set were filmed outdoors. This ``outdoor'' characteristic is one of the most discriminating features of \textit{Reggae} music videos as will also be discussed in the next Chapter.

\section{Summary}
\label{ch7:conclusions}

In this chapter a comparative evaluation of audio-visual music genre classification is presented. The main focus is on the color information of music videos. A broad set of diverging visual feature extraction methods based on psychological or perceptive models has been applied to extract different kinds of low-level semantic information. Further a descriptor to capture rhythmical changes in illumination is introduced. 
The performance of the visual and audio-visual features is compared to audio-only based features in classification experiment to predict the music genre of the music videos.

Although the mean average accuracy for the visual feature based classification results are above the baseline of 12.8\% (\textit{MVD-VIS} and \textit{MVD-MM}) and 6.25\% (\textit{MVD-MIX}), an accuracy of 50.13\% cannot be considerd a well performing approach. The experiments showed, that not enough information can be extracted from color values alone.
This leads to the conclusion, that such low-level visual features are not capable of describing the highly abstract concept of music genre. This conclusion is supported by the small improvemnets in the combined audio-visual experiments.

Future work on music videos will extend the semantic space to include texture, local features and object detection. This will be explained in detail in Chapter \ref{ch8:intro}.



\chapter[High-level Visual Concepts and Visual Stereotypes]{High-level Visual Concepts and \\ Visual Stereotypes}
\label{ch8:intro}

\epigraph{
	``Music videos are a pervasive feature of modern popular culture. They are everywhere; on television, in department stores and shopping malls, sports arenas, dance clubs.''} 
{---  Steve Jones, Cohesive but not coherent: Music videos, narrative and culture \cite{jones1988cohesive}, 1988}


Malcolm McLaren, the manager of the punk rock band ``The Sex Pistols'', stated in 1977 ``Christ, if people bought the records for the music, this thing would have died a death long ago''. In his provoking way he indicated that there are far more dimensions to music purchase behavior than the artistic quality of an act. Punk rock was a result of social-political differences of the late 1970s and dressing in leather jackets, spike bands and Mohawk hairstyle was a visual commitment to the views and rebellious attitudes expressed by the genre. Such visual stereotypes are adopted by and learned over generations from mass media. As a consequence stereotypes such as cowboy- shoes or hats are often visualized when referring to Country music \cite{shevy2008music}. Such visual associations are constantly rehashed to recognizably shape the image of a music act \cite{negus2011producing}. Using these visual associations one could sketch the soundtrack of the past decades without playing a single note. Flowery clothes, twinkling mirror-balls, dreadlocks, shoulder pads, skulls and crosses, Cowboy hats, Baseball caps and lots of jewelry are visual synonyms for music styles ranging from the early sixties to contemporary Hip-Hop. Visual stereotypes play an essential role in our social interaction with unfamiliar others \cite{haake2008visual}. They trigger a categorization process in which we quickly form expectations on a person's likely behavior, attitudes, opinions, personality, manners, etc. and thus shape our personal attitude towards that person. Drama theory and film make profound usage of such concepts \cite{finkelstein2007art}. The vicious antagonist is often recognizable at a glimpse of their first appearance by capitalizing on stereotypes such as ragged clothes, scars and over-exaggerated armament.

\noindent
Accordingly, in his influential book ``Emotion and Meaning in Music'' \cite{meyer1956emotion} Meyer addresses ``extramusical'' connotations, explicitly referring to classes of visual associations which are cognitive schema that are culturally shared by groups of individuals. Half a century later it had been shown that these associations became culturally independent as a consequence of global mass media exposure \cite{shevy2008music}. Artist development departments utilize visual concepts to categorize a new act by an appropriate genre and to create its visual image that is used for promotion in mass media \cite{negus2011producing}. Consequently, as discussed in Chapter \ref{ch3:albumart_history}, easily identifiable genres are a desired goal of music labels. Music videos not only make use of stereotypical visual themes, but also take part in their development and propagation. 

In this chapter a high-level approach to facilitate visual stereotypes based on visual concept detection is introduced. This approach decomposes the semantic content of music video frames into concrete concepts such as Guitars, Cars, or Tools. To provide comprehensive and plausible results the experiments are evaluated on the Music Video Dataset (MVD) (see Chapter \ref{ch4:intro}). Based on this dataset a series of evaluations is provided in Section \ref{ch8:eval_results} on a selected set of audio-content based music descriptors (see Section \ref{ch8:audio_features}), the low-level image processing features introduced in Chapter \ref{ch7:visual_features} (see Section \ref{ch8:visual_features}) and the new introduced high-level semantic descriptors based on visual vocabularies (see Section \ref{ch8:visual_vocabulary}). Finally, conclusions and future work is presented in Section \ref{ch8:conclusions}.

\section{Dataset}
\label{ch8:datasets}

To facilitate comparable results and reproducible research on music related visual analysis of music videos the Music Video Dataset (MVD) introduced in Chapter \ref{ch4:intro} is used. The MVD provides test collections with corresponding ground truth assignments within the context of well defined tasks. Of the four major subsets of the MVD the following three will be used for these evaluations. For a full overview of the MVD please refer to \ref{ch4.1:intro}:


\begin{description}[leftmargin=!,itemsep=3pt,labelwidth=\widthof{\bfseries Bollywood},labelindent=\descleftmargin,rightmargin=\descrightmargin]
	
	\item[MVD-VIS:] The \textit{M}usic \textit{V}ideo \textit{D}ata\-set for \textit{VIS}ual content analysis and classification (see Chapter \ref{ch4:mvdvis}) is intended for classifying music videos by their visual properties only. 
	
	\item[MVD-MM:] The \textit{M}usic \textit{V}ideo \textit{D}ataset for \textit{M}ulti\textit{M}odal content analysis and classification (see Chapter \ref{ch4:mvdmm}) is intended for multi-modal classification and retrieval tasks. 
	
	\item[MVD-MIX:] A combination of the data-sets MVD-VIS and MVD-MM (see Chapter \ref{ch4:mvdmix}).
	
	\item[MVD-THEMES:] a collection of thematically tagged classes that span across musical genres (see Chapter \ref{ch4:mvdthemes}).

\end{description}

\pagebreak

\section{Audio Content Descriptors}
\label{ch8:audio_features}

The audio features used for the experiments have been chosen due to comparability with results presented in Chapter \ref{ch5.2:intro}.

\paragraph{Psycho-accoustic Music Descriptors} 
as proposed by \cite{lidy2005evaluation} are based on a psycho-acoustically modified Sonogram representation that reflects human loudness sensation (described in more detail in Chapter \ref{ch4.1:rp_extract}). From the Rhythm Patterns feature family \textit{Statistical Spectrum Descriptors (SSD)}, \textit{Rhythm Patterns (RP)}, \textit{Rhythm Histograms (RH)}, \textit{Temporal Statistical Spectrum Descriptor (TSSD)} and \textit{Temporal Rhythm Histograms (TRH)} are used in the evaluation.
For the extraction, the Matlab-based implementation\footnote{\url{http://www.ifs.tuwien.ac.at/mir/downloads.html}}, version 0.6411 was employed.

\paragraph{Mel Frequency Cepstral Coefficients (MFCC)} are well known audio features derived from speech recognition.
\textbf{Chroma} features project the spectrum onto 12 bins representing the semitones of the musical octave. Both are added to the evaluation as an outline, because their performance on conventional data-sets is well known. The features are extracted using the well known MARSAYS toolset \cite{tzanetakis2000marsyas}. For the feature extraction MARSYAS version\footnote{\url{http://sourceforge.net/projects/marsyas/}}. 0.4.5. was utilized.\\

Music genre classification results are presented to serve as baseline performance values for the evaluation of the color descriptors. Table \ref{tab:ch7:Results} shows mean accuracy values of 10-fold cross-validation experiments for four different classifiers on all genre specific sub-sets of the MVD.

\section{Visual Object Vocabulary}
\label{ch8:visual_vocabulary}

This section presents an approach to extract high-level visual concepts from music videos. It is based on the assumption that music video directors make extensive use of genre related items, apparels or sceneries to outline the video's reference to that specific music genre. A good example is the cowboy hat as a reference to country music. Recently the attention of the computer vision research community has been attracted by the success of deep convolutional neural networks (CNN) \cite{krizhevsky2012imagenet}. These new approaches made remarkable improvements in visual concept detection, now facilitating comprehensive image understanding. Using these visual computing approaches it is possible to decompose the semantic content of a music video into concrete encapsulated concepts such as guitars, vehicles, landscapes, etc. and to measure their frequency distribution over video frames.


The \textit{ImageNet Large Scale Visual Recognition Challenge (ILSVRC)} \cite{ILSVRC15} is an annual benchmarking campaign to steer the competition in visual concept recognition and retrieval development. The image classification task, as one of three tasks, focuses on the ability of algorithms to identify objects present in the image. Therefore an ImageNet subset of 1000 categories - also called 'synonym sets' or 'synsets' - was defined such that there is no overlap between the synsets. This was guaranteed by referring to the tree structure of the ImageNet dataset and excluding ancestral synsets. These 1000 synset are used as visual object vocabulary to semantically analyze music videos. A full description of the ILSVRC-2012 categories is provided online\footnote{\url{http://image-net.org/challenges/LSVRC/2012/browse-synsets}} and consists of approximately 40\% of animal categories. The remaining synsets are objects of daily life such as clothing, food, means of transportation, electronic devices, landscapes as well as music instruments.

\subsection{Visual Feature extraction}
\label{ch8:visual_features}

New machine learning frameworks provide means for sharing trained and evaluated models enabling the concept of transfer learning \cite{pan2010survey} by transferring the knowledge of one task to other tasks with little or no prior knowledge such as labeled data. The presented approach utilizes the Caffe deep learning framework \cite{jia2014caffe} developed by the Berkeley Vision and Learning Center\footnote{\url{http://bvlc.eecs.berkeley.edu/}}. This framework provides a series of convenient advantages. Besides its fast computational capabilities, its openness and community engagement, one of its remarkable contributions is a sharing platform for trained models called \textit{Model Zoo}. Researchers of all communities are encouraged to upload their models so they can be re-used and applied to different domains. This concept is also referred to as \textit{Transfer Learning} \cite{pan2010survey}, where learning in a new task is improved through the transfer of knowledge from a related task that has already been learned. 
Consequently, by harnessing the trained models of the visual computing domain, semantic descriptions of music videos can be learned identify high level music concepts such as music genre. The outlined approach is based on a pre-trained deep convolutional neural network which won the ILSVRC-2012 image classification task \cite{krizhevsky2012imagenet} achieving a top-5 test error rate of 15.3\%. The model is online available\footnote{\url{https://github.com/BVLC/caffe/wiki/Model-Zoo}} and consists of eight learned layers, five convolutional and three fully-connected.

\begin{figure*}[t!]
	\centering  
	\includegraphics[width=1.0\textwidth]{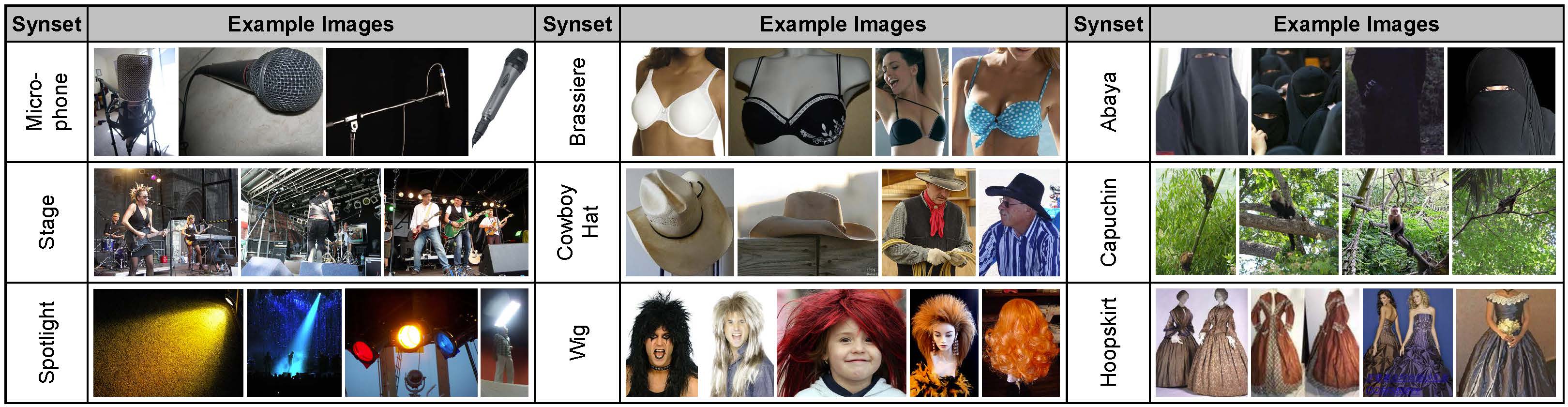}
	\caption{Example ImageNet Synsets which are also included in the ILSVRC2012 competition categories.}
	\label{fig:ch8:synset_examples}	
\end{figure*}

\begin{description}[leftmargin=!,itemsep=3pt,labelwidth=\widthof{\bfseries Places205},labelindent=1pt,rightmargin=0pt]
	
	\item[ImageNet] \cite{imagenet_cvpr09} is an image database organized according to the WordNet \cite{miller1995wordnet} hierarchy in which each node of the hierarchy is depicted by hundreds and thousands of images. The dataset currently consists of more than 10,000,000 labeled images depicting more than 10,000 synsets. The images were collected from the web and labeled by human labelers using Amazon's Mechanical Turk crowd-sourcing tool. Figure \ref{fig:ch8:synset_examples} illustrates example synsets with corresponding images. Each synset is a collection of dozens to thousands of images in different resolutions and variants. As illustrated in Figure \ref{fig:ch8:synset_examples} the synset \textit{Cowboy hat} includes images of hats as well as people wearing hats. 
	
	\item[Places205] \cite{zhou2014learning} is a CNN trained on 205 scene categories of Places Database (used in NIPS'14) with ~2.5 million images. The architecture is based on AlexNet \cite{krizhevsky2012imagenet}.
	
\end{description}

\pagebreak

\noindent


\section{Evaluation}
\label{ch8:eval_results}

The need for multi-modal strategies for MIR has been expressed by \cite{liem2011need,weninger2012music,urbano2013evaluation} and the MIRES roadmap for music information research \cite{MiresRoadmap2013} identified music videos as a potential rich source of additional information. Yet, no relevant work has been reported on evaluating the performance of visual features extracted from music videos towards describing music related properties. In this section we intend to close this gap by describing a series of performance evaluations of the previously introduced audio and visual features performed on the MVD. These evaluations are centered around the assumption that information provided by the visual layer of music videos is music related and can be harnessed to describe music content.

\subsection{General Experimental Setup}

Music genre classification was chosen as a proxy problem to perform the subset related evaluation tasks. Feature and system performance is evaluated in terms of classification accuracy.


\begin{description}[leftmargin=!,itemsep=3pt,labelwidth=\widthof{\bfseries Places205},labelindent=1pt,rightmargin=0pt]
	
	\item[Audio Feature Extraction] was based on audio files from the separated audio channel of the music videos. FFMPEG\footnote{\url{http://www.ffmpeg.org/}} (version 2.1) was used to extract and store the audio streams in mp3-format using a samplerate of 44.100 Hz and a bit rate of 128 kbit/s.
	
	\item[Visual Feature Extraction] was based on frame wise processing of each video. General image pre-processing included the removal of \textit{Letterboxing} or \textit{Pillarboxing} as described in Chapter \ref{ch7:method}.
	To extract the visual features the CNN model is applied to every frame of a video to retrieve the predicted probability values for the visual concepts. The sum of all their values equals 1, resulting in a uniform feature vector. To compare music videos of different length these vectors need to be aggregated into representational vector for the corresponding video. The identification of appropriate aggregations was part of the evaluation and will be discussed in Section \ref{eval_vis_voc}.
	
	\item[Experimental results] of all sections were obtained using the Weka machine learning toolkit \cite{hall2009} (version 3.7.5) based on the following set-up: Stratified 10-fold cross-validation was used to evaluate the mean accuracy of ten repeated runs of the separate feature sets on each of the sub-datasets using the following three classifiers: \textit{Support Vector Machine (SVM):} linear PolyKernel and complexity parameter c=1; \textit{K-Nearest Neighbors (KNN):}  with k=1, using Euclidean Distance (L2); \textit{Naive Bayes (NB):} simple probabilistic classifier based on applying Bayes' theorem
	
\end{description}

\begin{table}
	
	\caption{\small Classification results for audio, visual and audio-visual features showing accuracies for Support Vector Machines (SVM), K-Nearest Neighbors (KNN) and Naive Bayes (NB) classifiers. Bold-faced values highlight improvements of audio-visual approaches over audio features. Bold values in the \textit{Audio} and \textit{Visual Concepts} sections depict top-results for the corresponding classifier. Bold values in the \textit{Audio-Visual} section depict improvement over the corresponding audio-only results. Underlined values are significant on a 0.05 level.}
	
	\scriptsize
	\begin{tabularx}{\textwidth}{c|X|r|rrr|rrr|rrr}
		
		\multicolumn{12}{c}{} \\  [-2.0ex]
		\hline
		\rowcolor{gray!30}
		\hline
		& & & \multicolumn{3}{c}{ \textbf{MVD-VIS} } & \multicolumn{3}{c|}{ \textbf{MVD-MM} } &  \multicolumn{3}{c}{ \textbf{MVD-MIX} } \\
		\hline
		& & & \multicolumn{3}{c|}{} & \multicolumn{3}{c|}{} &  \multicolumn{3}{c}{} \\ [-1.2ex]
		
		& & \textbf{Dim} & \textbf{SVM} & \textbf{KNN} & \textbf{NB} & \textbf{SVM} & \textbf{KNN} &  \textbf{NB} & \textbf{SVM} & \textbf{KNN} & \textbf{NB} \\ 
		
		\hline 
		\rowcolor{gray!30}
		\multicolumn{12}{c}{\textbf{(a) Content Based Audio Features}} \\  
		
		\hline
		
		
		& & & \multicolumn{3}{c|}{} & \multicolumn{3}{c|}{} &  \multicolumn{3}{c}{} \\ [-1.9ex]
		
		$a1$  & Chroma         & 48   &         36.34  &         28.09  &         23.03  &         25.26  &         20.11  &         19.41  &         19.64  &         14.68  &         12.08  \\
		$a2$  & MFCC           & 52   &         62.28  &         48.58  &         46.95  &         42.14  &         29.16  &         34.17  &         37.02  &         26.60  &         27.11  \\
		$a3$  & SSD            & 168  &         85.78  &         73.18  &         58.81  &         68.74  &         50.28  &         48.41  &         65.11  &         44.64  &         38.92  \\
		$a4$  & RP             & 1440 &         87.26  &         69.81  &         64.04  &         60.35  &         42.38  &         41.63  &         63.19  &         43.06  &         41.39  \\
		$a5$  & TRH            & 420  &         71.04  &         55.83  &         53.86  &         49.50  &         38.28  &         39.66  &         46.61  &         33.02  &         35.70  \\
		$a6$  & TSSD           & 1176 &         86.81  &         72.58  &         62.61  &         69.97  &         53.33  & \textbf{53.65} &         66.19  &         47.40  &         44.22  \\
		$a7$  & $a4$+$a6$      & 2616 &         93.08  &         79.47  & \textbf{71.88} &         74.44  &         54.00  &         51.03  &         74.64  &         53.06  & \textbf{48.54} \\
		$a8$  & $a4$+$a3$+$a5$ & 2028 &         92.19  &         75.93  &         67.45  &         71.00  &         50.26  &         44.85  &         72.73  &         49.88  &         43.65  \\
		$a9$  & $a4$+$a3$      & 1608 &         92.55  &         77.74  &         67.36  &         71.64  &         52.44  &         44.40  &         74.38  &         51.60  &         43.52  \\
		$a10$ & $a4$+$a5$+$a6$ & 3036 & \textbf{93.79} & \textbf{80.85} &         71.46  & \textbf{74.76} & \textbf{55.00} &         52.20  & \textbf{75.91} & \textbf{54.16} &         48.32  \\
		
		& & & \multicolumn{3}{c|}{} & \multicolumn{3}{c|}{} &  \multicolumn{3}{c}{} \\ [-1.9ex]
		
		\hline 
		\rowcolor{gray!30}
		\multicolumn{12}{c}{\textbf{(b) Low-level Color and Affect related Image Features}} \\  
		
		\hline
		
		
		& & & \multicolumn{3}{c|}{} & \multicolumn{3}{c|}{} &  \multicolumn{3}{c}{} \\ [-1.9ex]
		
		$v_{co}1$ & LFP     & 60  &         33.21  &         23.59  &         25.45  &         20.38  &         16.74  &         16.46  &         16.93  &         11.71  &         13.36  \\
		$v_{co}2$ & CF      &  7  &         34.89  &         25.49  &         31.50  &         21.84  &         17.06  &         20.41  &         18.53  &         11.92  &         16.49  \\
		$v_{co}3$ & IC      & 28  &         36.80  &         27.55  &         27.51  &         24.83  & \textbf{19.43} &         19.68  &         21.44  &         13.54  &         12.66  \\
		$v_{co}4$ & GEV     & 21  &         39.45  & \textbf{29.84} & \textbf{34.15} &         20.81  &         17.04  &         18.51  &         20.27  &         14.47  & \textbf{17.89} \\
		$v_{co}5$ & GCS     & 42  &         40.55  &         29.76  &         33.91  &         24.08  &         17.29  &         18.15  & \textbf{23.72} & \textbf{15.40} &         17.34  \\
		$v_{co}6$ & WAF     & 126 &         41.01  &         26.43  &         29.86  &         26.01  &         19.08  & \textbf{21.38} &         22.86  &         13.90  &         16.60  \\
		$v_{co}7$ & CN      & 56  & \textbf{43.68} &         29.04  &         32.23  & \textbf{26.74} &         19.13  &         18.77  &         23.48  &         14.76  &         15.99  \\[0.9ex]
		$v_{co}8$ & Combi   & 360 & \textbf{50.13} & \textbf{34.04} & \textbf{39.38} & \textbf{31.69} & \textbf{21.16} & \textbf{23.38} & \textbf{32.22} & \textbf{17.89} & \textbf{21.16} \\

		& & & \multicolumn{3}{c|}{} & \multicolumn{3}{c|}{} &  \multicolumn{3}{c}{} \\ [-1.9ex]

		\hline 
		\rowcolor{gray!30}
		\multicolumn{12}{c}{\textbf{(c) High-level Visual Concepts}} \\  
		
		\hline
		
		
		& & & \multicolumn{3}{c|}{} & \multicolumn{3}{c|}{} &  \multicolumn{3}{c}{} \\ [-1.9ex]
		
		$v_{in}1$ & MEAN                          & 1000 &         66.86  &         42.09  &         53.69  &         51.26  &         31.23  &         37.05  &         46.87  &         23.90  & \textbf{33.07} \\        
		$v_{in}2$ & STD                           & 1000 &         69.78  &         46.76  &         50.08  &         51.95  &         29.99  &         32.88  &         48.29  &         26.83  &         29.63  \\        
		$v_{in}3$ & MAX                           & 1000 &         73.15  &         44.26  &         46.41  &         54.60  &         33.05  &         31.94  &         50.07  &         26.93  &         27.49  \\        
		$v_{in}4$ & $v_{in}3$+$v_{in}2$           & 2000 &         73.61  &         46.53  & \textbf{51.21} &         55.04  &         31.48  &         34.00  &         51.30  &         27.03  &         31.04  \\        
		$v_{in}5$ & $v_{in}3$+$v_{in}1$           & 2000 & \textbf{74.36} &         47.70  &         53.65  &         55.99  & \textbf{33.70} & \textbf{37.83} & \textbf{51.58} & \textbf{28.88} &         33.83  \\[0.9ex] 
		
		$v_{pl}1$ & MEAN                          &  205 &         57.13  &         37.15  &         43.05  &         42.90  &         25.94  &         30.55  &         38.24  &         19.08  &         25.32  \\
		$v_{pl}2$ & MAX                           &  205 &         58.36  &         42.28  &         45.35  &         38.91  &         25.51  &         31.44  &         36.63  &         21.74  &         27.33  \\
		$v_{pl}3$ & STD                           &  205 &         60.74  &         40.70  &         39.39  &         43.95  &         27.58  &         28.99  &         39.33  &         20.90  &         23.03  \\
		$v_{pl}4$ & $v_{pl}1$+$v_{pl}2$           &  510 &         59.46  &         43.11  &         43.85  &         41.25  &         27.08  &         31.58  &         38.26  &         22.28  &         26.72  \\
		$v_{pl}5$ & $v_{pl}1$+$v_{pl}3$           &  510 &         60.49  &         39.74  &         40.99  &         43.40  &         26.45  &         30.33  &         39.72  &         20.50  &         24.88  \\
		
		& & & \multicolumn{3}{c|}{} & \multicolumn{3}{c|}{} &  \multicolumn{3}{c}{} \\ [-1.9ex]
		
		\hline 
		\rowcolor{gray!30}
		\multicolumn{12}{c}{\textbf{(d) Visual Combinations}} \\  
		\hline
		
		$vc1$ & $v_{in}5$+$v_{co}8$            & 2360 &         72.86  &                    45.81   &                    53.75   &         55.59  &         31.84  &         38.08  &         51.94  &         27.48  &                    33.85  \\ 
		$vc2$ & $v_{in}3$+$v_{pl}3$            & 1205 &         72.70  &         \underline{44.38}  &         \underline{47.24}  &         54.11  &         32.54  &         31.86  &         50.51  &         27.46  &         \underline{27.66} \\
		$vc3$ & $v_{in}5$+$v_{pl}3$            & 2205 &         73.80  &                    48.54   &                    52.73   &         55.21  &         33.74  &         36.75  &         52.21  &         28.41  &                    33.03  \\
		$vc4$ & $v_{in}5$+$v_{pl}5$            & 2510 &         73.95  &                    48.35   &                    53.14   &         55.28  &         33.74  &         36.71  &         52.48  &         28.59  &                    33.24  \\
		$vc5$ & $vc4$+$v_{co}8$                & 2870 &         74.25  &                    47.93   &                    54.43   &         56.05  &         32.71  &         37.61  &         54.18  &         28.28  &                    33.79  \\

		\hline 
		\rowcolor{gray!30}
		\multicolumn{12}{c}{\textbf{(e) Audio-Visual Combinations}} \\  
		\hline
		
		
		& & & \multicolumn{3}{c|}{} & \multicolumn{3}{c|}{} &  \multicolumn{3}{c}{} \\ [-1.9ex]
		
		$av1$ & $a10$+$v_{in}5$  & 5036 & \underline{\textbf{96.73}} &         81.13  &         65.00  &            \textbf{81.60}   & \textbf{55.73} &         49.31  &            \textbf{86.73}   & \textbf{59.01} &         47.48  \\ 
		$av2$ &  $a9$+$v_{in}5$  & 3608 &            \textbf{95.63}  &         77.05  &         64.16  &            \textbf{77.83}   &         49.54  &         46.58  &            \textbf{79.44}   &         51.31  &         43.71  \\ 
		$av3$ &  $a9$+$v_{pl}5$  & 2118 &                    94.50   &         79.95  &         68.08  &                    72.96    &         53.29  &         45.99  &                    77.40    &         53.73  &         45.51  \\ 
		$av4$ &  $av2$+$v_{pl}5$ & 4118 &                    95.76   &         75.76  &         61.00  &                    77.55    &         50.31  &         44.59  &                    80.16    &         52.43  &         41.79  \\ 
		$av4$ &  $a6$+$v_{in}5$  & 3176 &            \textbf{94.65}  &         68.61  &         63.64  &            \textbf{78.49}   &         53.01  &         50.41  &            \textbf{82.62}   &         48.94  &         48.53  \\ 
		$av5$ &  $a4$+$v_{in}5$  & 3440 &            \textbf{91.24}  &         68.80  &         63.40  &            \textbf{71.95}   & \textbf{43.78} & \textbf{44.86} &            \textbf{74.14}   & \textbf{45.53} &         42.69  \\ 
		$av6$ &  $a3$+$v_{in}5$  & 2168 &            \textbf{89.85}  &         62.11  &         57.89  &            \textbf{70.13}   &         43.16  &         42.93  &            \textbf{70.30}   &         37.98  &         38.88  \\ [0.4ex] 
		
		
		\hline
		
	\end{tabularx}

	\label{tab:Results}
\end{table}

\subsection{Performance Evaluation of the Audio Content Based Features}

The Music Video Dataset was compiled to foster the development and evaluation of visual features that are able to capture the subtle relationship between the visual layer of music videos and the underlying music itself. In this sense the audio classification results presented in the upper part of Table \ref{tab:Results} (a) serve 
as a baseline for all consecutive evaluations to compare the performance of the visual features against the audio-only classification results. Results for MFCC were provided due to their popularity in music research domains \cite{siedenburg2016comparison}. Further Chroma features were included because they provide an abstract kind of harmonic description and are frequently used in different tasks such as audio fingerprinting \cite{miotto2008music} or synchronization of audio with music scores \cite{ewert2009high}. The psychoacoustic features were included due to their reported advantage in classification tasks \cite{lidy2005evaluation}.

The aggregation of the \textit{MVD-VIS} subset aimed at clearly defined and well differentiated classes. 
Classification accuracies provided in Table \ref{tab:Results} (a) show that these requirements have been met which can be observed by the high accuracies for the combined feature-sets \textit{RP-TRH-TSSD} (a10). The high  accuracies of the distinct feature-sets \textit{RP} (a4) and \textit{SSD} (a3) conclude that the selected classes are well differentiated by spectral and rhythmical characteristics - at least in terms that are captured by the corresponding feature sets. The intended overlaps of the classes of the \textit{MVD-MM} classes are observable as well. 
Analysis of the confusion matrices shows extensive mutual confusions of the classes \textit{Hard Rock}, \textit{Pop Rock} and \textit{Indie}. These confusions stretch out to \textit{Metal} and \textit{Country} for the combined \textit{MVD-MIX} dataset. 
The presented results are comparable to the evaluation provided by \cite{SCHINDLER_2012amr} where the same feature sets had been evaluated on the four de facto MIR music classification benchmark sets. Although (a10) provides best results there is no significant improvement (p \textless  0.05) over (a9) which has half the dimensions of (a10), thus (a9) will be used for further evaluations.

\subsection{Performance Evaluation of the low-level color based affective features}
\label{eval_vis_lowlevel}

This evaluation serves as a comprehensive bottom-up evaluation to investigate the performance of low-level image features in describing the music related semantic content of music videos. The evaluation is based on the seven color and illumination based image processing feature sets described in Chapter \ref{ch7:visual_features}. An initial evaluation of these low-level visual features had been provided in \ref{ch7:intro}, but only the performance of the combined visual feature space was evaluated. The performance results for their distinct color related image features are presented in Table \ref{tab:Results} (b). The results indicate that Color Names (CN), Wang Affective Factors (WAF) and Global Color Statistics (GCS) perform better in discriminating the different classes but are not reliable to describe music genres.  However, it was able to confirm the common stereotype of 'darkness' related to \textit{Heavy Metal} music \cite{farley2009demons}. Videos of the class \textit{Metal} contain significantly more black pixels (independent t-test, p \textless 0.05) than other classes of the \textit{MVD-MIX} subset and the second smallest brightness values (see Figure \ref{fig:mvd_vis_color_dist}). The lowest values on \textit{brightness} and \textit{colorfulness} for \textit{Opera} (see Figure \ref{fig:mvd_vis_color_dist}) are a strong indicator for the inferior lightning conditions at such venues and thus have no relation to the music itself.

\subsection{Performance Evaluation of the Visual Vocabulary} 
\label{eval_vis_voc}

This evaluation focuses on the visual concept detection based approach presented in Section \ref{ch8:visual_vocabulary}. The rational behind this evolution is to estimate if the high-level semantic concepts extracted from the frames of the music videos contain relevant information to discriminate the classes of the MVD. 
Part of the evaluation is to assess appropriate methods to aggregate the softmax scaled visual object detection results for each music video frame into a single feature vector. The resulting vector is the representative descriptor for the corresponding music video. 
This single vector representation is a requirement of the machine learning algorithms applied in the experiments and further abstracts from variations in length. 
Seven statistical moments (minimum, maximum, mean, median, standard deviation, variance, kurtosis, skewness) were calculated from the prediction results for each music video. The effectiveness of these moments as a feature in the experimental classification tasks was analyzed by testing all possible combinations. 
The best-performing results using \textit{Visual Concepts} are listed in Table \ref{tab:Results} (c). Results show high classification accuracies for visual vocabularies based on the \textit{ImageNet} model. The most relevant aggregations for this model are \textit{MAX}  ($v_{in}3$) and \textit{STD}  ($v_{in}2$). This can be explained by the type of feature response which is the probability of a certain concept of the vocabulary to be present in a frame. 
In that sense the maximum values of all object detection results of a music video (\textit{MAX}) describe which visual concepts had the highest prediction values and thus 'reliably' appeared in at least one frame of the video. This information appeared to be most discriminative with an accuracy of 73.15\% on the \textit{MVD-VIS} dataset ($v_{in}3$). The best performing combination ($v_{in}5$) reached 74.36\% which is not significantly better than ($v_{in}3$) (p \textless 0.05). The visual vocabulary results for the \textit{Imagenet} model improved the previously reported accuracy of 50.13\% ($v_{co}8$) using low-level visual features (see Chapter \ref{ch7:results}) by 24.23\%. This high accuracy supports the initial hypothesis (see Chapter \ref{ch1:research_questoins}) that music videos make use of easy identifiable visual concepts. The low improvement of the combination with color features ($vc1$ ) and ($vc5$) indicates that this information is provided by high semantic concepts (e.g. apparel, buildings, music instruments, vehicles, etc.).  This will be discussed further in the analysis of visual stereotypes in Section \ref{eval_visual_stereotypes}. Comparing the results of the visual vocabulary based approach with state-of-the-art audio features shows that the described visual approach outperforms the music features \textit{Chroma} and the de-facto standard feature \textit{MFCC}.

\begin{table}[t]
	\caption{Classification results for cross-genre music themes evaluation using Support Vector machines and 10-fold cross-validation. Values represent accuracies for the corresponding theme. Columns titled with one of the\textit{ MVD-}\{\textit{VIS,MM,MIX}\} subsets' names represent results where the theme had been added as additional class to the corresponding set. Columns marked with \textit{TH} represent results where only the four classes of the MVD-THEMES dataset have been used in the experiments. In this table the \textit{Audio-Only} results serve as baseline for the other experiments. Results exceeding the baseline are depicted in bold letters.}
	
	\begin{tabularx}{\columnwidth}{p{0.13\textwidth}|XXXX|XXXX|XXXX}
		
		\hline
		\rowcolor{gray!30}
		& \multicolumn{4}{c|}{\textbf{Audio-Only}} & \multicolumn{4}{c|}{\textbf{Visual-Only}} & \multicolumn{4}{c}{\textbf{Audio-Visual}} \\ 
		\hline
		& & & & & & & & & & & & \\ [-0.9ex]

		\multicolumn{1}{c|}{\textbf{Theme}} & \multicolumn{1}{X}{\textbf{VIS}} & \multicolumn{1}{X}{\textbf{MM}} & \multicolumn{1}{X}{\textbf{MIX}} & \multicolumn{1}{X|}{\textbf{TH}} & \multicolumn{1}{X}{\textbf{VIS}} & \multicolumn{1}{X}{\textbf{MM}} & \multicolumn{1}{X}{\textbf{MIX}} & \multicolumn{1}{X|}{\textbf{TH}} & \multicolumn{1}{X}{\textbf{VIS}} & \multicolumn{1}{X}{\textbf{MM}} & \multicolumn{1}{X}{\textbf{MIX}} & \multicolumn{1}{X}{\textbf{TH}} \\ 
		
		\hline
		& & & & & & & & & & & & \\ [-0.9ex]
		\textbf{\smaller Christmas}      & 67.6  &  36.7  &  29.5  &  52.9  &  \textbf{71.7}  &  \textbf{65.5}  &  \textbf{64.0}  &  \textbf{88.9}  &  \textbf{87.5}  &  \textbf{70.8}  &  \textbf{75.0}  &  \textbf{90.4}  \\
		\textbf{\smaller K-Pop}          & 86.0  &  65.4  &  68.6  &  86.0  &  \textbf{88.4}  &  \textbf{81.6}  &  \textbf{80.4}  &  \textbf{91.7}  &  \textbf{95.5}  &  \textbf{88.2}  &  \textbf{82.7}  &  \textbf{90.0}  \\
		\textbf{\smaller Protest Song}   & 50.0  &  21.7  &   7.7  &  47.5  &          23.7   &  \textbf{33.3}  &  \textbf{16.7}  &  \textbf{75.5}  &          44.4   &  \textbf{57.1}  &  \textbf{30.3}  &  \textbf{77.5}  \\
		\textbf{\smaller Broken Heart}   & 75.0  &  28.6  &  28.6  &  54.9  &          51.2   &          21.9   &          16.7  &   \textbf{70.2}  &          61.0   &  \textbf{31.9}  &          25.5   &  \textbf{68.6}  \\ [0.4ex]
		
		\hline
	\end{tabularx}
	
	\label{tab:ResultsChristmas}
\end{table}


\subsection{Performance Evaluation of Audio-Visual Combinations}
\label{eval_audiovisual}

The rational behind a multi-modal approach is to utilize information of different modalities to improve the performance for a dedicated task. In the previous sections the two modalities \textit{audio} and \textit{video} have been evaluated separately. This evaluation att\-empts to answer the question, if their combinations can improve  classification accuracy, using the \textit{audio} results (see Table \ref{tab:Results} (a)) as baseline. 
Again different combinations of audio features and visual vocabulary aggregations were evaluated in classification experiments. 
To reduce the number of required experiments, weak performing aggregations of the predicted semantic concpets of a music video, as identified in the previous task, have been skipped. Only \textit{MEAN}, \textit{MAX} and \textit{STD} aggregations over all video frames have been used. Feature-sets have been combined using an early fusion approach. Table \ref{tab:Results} e) shows selected best-performing audio-visual combinations for the \textit{SVM} classifier. The additional information can be harnessed well by the \textit{SVM} classifier where all results show noticeable and some remarkable improvements over the baseline. Despite the high integrity of the classes of the \textit{MVD-VIS} subset, an improvement of 2.94\% ($av1$) was accomplished over the best performing audio combination ($a10$). An even higher improvement was observed for the less differentiated set \textit{MVD-MM} (+6.84\%) and the bigger \textit{MVD-MIX} dataset (+10.82\%). The visual information ($v_{in}5$) showed outstanding improvements for the \textit{TSSD} ($av4$) audio feature set. 
The highest improvement  of 16.43\% for ($av4$) was observed for the bigger \textit{MVD-MIX} dataset. All mentioned improvements are significant (p \textless 0.05).

\subsection{Analysis of non-audible Music Themes }
\label{eval_themes}

\begin{table}
	
	\caption{Salient ILSVRC Synsets descendingly ranked by their minimal difference to other genres. 
	}
	\small
	\begin{tabularx}{\textwidth}{lllll}

		\hline
		\rowcolor{gray!30}
		
		\textbf{Country}    & \textbf{Dance}     & \textbf{Metal}       & \textbf{Opera}       & \textbf{Reggae}      \\ 
		
		\hline                                   
		
		1. cowboy hat       & 1. brassiere       & 1. spotlight         & 1. theater curtain   &  1. seashore coast   \\
		5. drumstick        & 3. maillot         & 2. electric-guitar   & 3. hoopskirt         &  2. academic gown    \\
		8. restaurant       & 4. lipstick        & 4. drumstick         & 5. stage             &  3. capuchin         \\
		9. tobacco shop     & 9. seashore coast  & 6. matchstick        & 11. flute            &  5. black stork      \\
		10. pickup truck    & 10. bikini         & 7. drum              & 19. harmonica        &  7. sunglasses       \\
		11. acoustic guitar & 15. sarong         & 8. barn spider       & 21. marimba          &  8. orangutan        \\
		13. violin fiddle   & 16. perfume        & 10. radiator         & 25. oboe             &  9. titi monkey      \\
		16. jeep landrover  & 17. trunks         & 12. chain            & 26. french horn      &  10. lakeshore       \\
		18. tractor trailer & 18. ice lolly      & 14. grand piano      & 27. panpipe          &  11. cliff drop      \\
		19. tow truck       & 19. pole           & 23. spider web       & 30. grand piano      &  17. elephant        \\
		21. minibus         & 20. bubble         & 24. nail             & 31. cello            &  23. steel drum      \\
		23. electric guitar & 30. miniskirt      & 28. brassiere        & 48. pipe organ       &  24. macaw           \\
		33. thresher        & 42. feather boa    & 37. loudspeaker      & 55. harp             &  25. coonhound       \\
		
		\hline
		
	\end{tabularx}
	
	\label{tab:salient_concepts}
\end{table}

This evaluation attempts to answer the question, if the presented multi-modal approach can improve the classification performance of cross-genre music themes such as \textit{Christmas}. The task corresponding to this evaluation refers to music tagging and is aligned to the MusiClef multi-modal music tagging task \cite{orio2012musiclef}. The evaluation was performed in two ways. In a general experiment the accuracy for discriminating the classes of the \textit{MVD-THEMES} dataset was evaluated. 
Further,  each theme was added separately to the datasets \textit{MVD}-\textit{VIS,MM,MIX} and the accuracy for discriminating this theme from the other classes was measured. This investigates how good the audio-visual approach can discriminate the music themes from the overlapping music genres of the dataset. The audio results (see Table \ref{tab:ResultsChristmas} \textit{``Audio-Only''}) again serve as baseline and clearly show low accuracy values for classifying non-audible concepts by audio content based features. Their classification accuracies for all themes, especially when combined with the  other \textit{MVD} sub-sets such as the \textit{MVD-MIX} dataset, are low. Except for \textit{Broken Heart} almost all visual vocabulary based approaches already performed better than the audio-only results such as \textit{Christmas} combined with \textit{MVD-MIX} (+34,5\%). More remarkable for the same combination is the observed improvement of 45.5\% using the introduced audio-visual approach. Similar high improvements were observed for the themes \textit{K-Pop} and \textit{Protest Song}. 
Only the theme \textit{Broken Heart} showed no general improvements by an audio-visual approach. Only its discriminability within the \textit{MVD-Themes} dataset was increased.


\subsection{Analysis of Visual Stereotypes}
\label{eval_visual_stereotypes}

The final evaluation is concerned with the analysis whether the extracted features are able to capture genre or thematically related visual stereotypes. The term frequencies of the \textit{ImageNet} Synsets for each class were calculated. For each concept of a class the largest minimal difference to the term frequencies of the other classes was calculated. This resulted in a list of the most salient visual concepts for each genre. Table \ref{tab:salient_concepts} provides an overview of these salient concepts for selected genres. For the discussion of visual stereotypes Synsets such as \textit{Abaya} were removed (see Figure \ref{fig:ch8:synset_examples}). These are concepts that are not well discriminated by the applied model. Manually inspecting series of video frames, which were misclassified as \textit{Abaya} revealed that the model tends to classify dark and blurred images, or images with larger dark regions, according this category. Although the frequency distribution of these ``mis-classifications'' has shown to be a discriminative feature in music video classification, it is difficult to derive a semantic relation to music and music related visual stereotypes.

The discussion of visual stereotypes is started with the genre \textit{Country}. For this genre the visual concepts \textit{cowboy hat} and utility vehicles such as the \textit{pickup truck} or a \textit{tractor trailer} are identified to be most salient. These observations correspond to an evaluation of perceived extra-musical associations with country music \cite{shevy2008music} and to aesthetic description provided in \cite{frith2005sound}. 
It was also possible to confirm the mentioned \textit{'movement towards the warm, orange tones that became the dominant 'look' of many contemporary country videos.'} \cite{frith2005sound}. The average values for \textit{red} and \textit{yellow} for \textit{Country} in Figure \ref{fig:mvd_vis_color_dist} are highest and second highest for all genres. 

\begin{figure*}[t]
	\centering
	\includegraphics[width=1.0\textwidth]{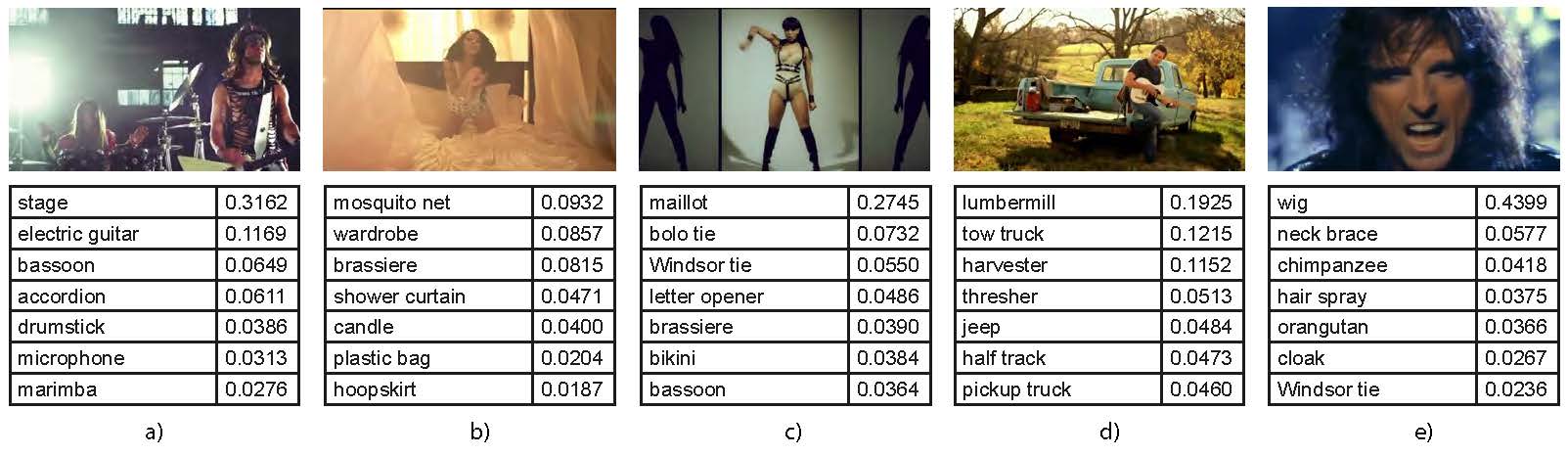}
	\caption{Example frames of music videos and their top-ranked synsets ordered by descending probability. Frames taken from videos of the genres a) \textit{Metal}, b,c) \textit{Dance}, d) \textit{Country} and e) \textit{Hard-Rock}}
	\label{fig:ch8:Example_Frames}
\end{figure*}

The second stereotype addresses the over-sexualization of contemporary popular \textit{Dance} music \cite{hall2012sexualization}. 
Figure \ref{fig:ch8:Example_Frames} b) and c) provide two examples of dance video frames. 
As shown in Table \ref{tab:salient_concepts}, the most salient concept for the \textit{MVD-VIS} category \textit{Dance} is \textit{Brassiere} (for training examples of the model see Figure \ref{fig:ch8:synset_examples}).
Also, eight further of the most salient Synsets are referred to low-covering clothes such as \textit{Maillot}, \textit{Bikini}, \textit{Trunks} or \textit{Miniskirts}. Closeups of body parts or people dancing on \textit{poles} are also among the top ranking Synsets. 
The accumulation of most salient Synsets which relate to images showing women in under- or swim-wear are a strong indication towards the proof of the stereotype of over-sexualization in contemporary \textit{Dance} music.

For Heavy \textit{Metal} music the common stereotype of \textit{darkness} was already confirmed in Section \ref{eval_vis_lowlevel}.  Videos of the class \textit{Metal} contain significantly more black pixels (independent t-test, p \textless 0.05) than other classes of the \textit{MVD-MIX} subset and the second smallest brightness values.
A further interesting result of the analysis is the salient Synset \textit{matchstick}. This is assigned by the applied model to video frames showing all kinds of fire. Fire, on the other hand is a typical element related to \textit{Metal} \cite{farley2009demons}. Further common concepts such as \textit{spotlight}, \textit{electric guitar}, \textit{drums} and \textit{loudspeaker} are performance related elements. \\

This analysis provides arguments towards answering research question \textit{RQ5} \textit{``Is it possible to verify applied methods within the production process?''}. Generally, it was not possible to identify literature explicatively stating that the identified visual stereotypes should be applied to music videos of the corresponding genres. As described in Chapter \ref{ch3:albumart_history}, however, marketing departments make use of easy identifiable symbols and fashion styles to visually relate new music acts and actors to their music genre and style. Further, there are salient visual objects which have already been addressed in music psychological studies. Thus, the correlation between these visual concepts and their associated music genre has been observed and documented. The quantitative analysis provided in this section verifies these observed associations and thus concludes that there are common visual patterns that are deliberately applied to music videos, at least for the analyzed genres \textit{Metal}, \textit{Dance} and \textit{Country}.

\subsection{General Discussions}

The evaluation and different analyses presented in this chapter indicate that music related visual information can be harnessed on the level of visual concepts. The discussion in Section \ref{eval_vis_voc} showed that frequencies and intensities of detected visual concepts differ between music genres as far as that they can be used to discriminate them by an accuracy of up to 75.36\% for the \textit{MVD-VIS} dataset and that this performance can be further increased though audio-visual combinations (as discussed in Section \ref{eval_audiovisual}). 
In Section \ref{eval_visual_stereotypes} it was shown that some of the most salient visual concepts are associated with observed music related visual stereotypes, as reported by music psychology literature. Figure \ref{fig:ch8:Example_Frames} illustrates the top-seven synsets of five example music video frames ranked by their estimated probability. 

Despite the proof of certain music related visual stereotypes, their occurrence in music video genres does not denote the genre of the music video. The discriminating information lies in the different frequencies of all of the detected concepts such as the concept \textit{Abaya} (see Figure \ref{fig:ch8:synset_examples}). For this concept the model was trained on images showing women fully veiled in black fabric. Part of the images showed these women as full persons in a natural scenery, others showed portraits or close-ups, which resulted in images consisting of mostly black pixels. In this evaluation this lead to the issue that the model labeled many low saturated and/or blurred video frames with \textit{Abaya}. Although these are clearly mis-classifications, the concept \textit{Abaya} is identified as a highly significant feature. In its right designation - ``low saturated or blurred frames, as they often result from fast camera motion'' - this concept corresponds to video features described in Chapter \ref{ch3:mv_overview} and Chapter \ref{ch9:intro} such as \textit{fast panning}, \textit{distortion overlays} or \textit{fade in/outs}.

The salient Synsets for \textit{Reggae} music exhibits many animal related concepts (see Table \ref{tab:salient_concepts}). This is also an artifact of the applied model. Reggae videos are often shot at seashores or in tropical landscapes. Because the set of visual concepts the model was trained to identify, does not contain concepts for landscapes, it assigns the labels for the corresponding visual content to those categories which show similar content in the training images. For the landscapes shown in \textit{Reggae} music videos, the model refers to the some of the 400 animal categories and especially to those which show animals in the corresponding environment such as forests. Figure \ref{fig:ch8:synset_examples} shows such a Synset example - images of \textit{Capuchin} monkeys sitting on branches in trees. The \textit{Capuchin} is also one of the most salient concepts detected in \textit{Reggae} music videos.

\begin{figure*}[t]
	\centering
	\includegraphics[width=0.9\textwidth]{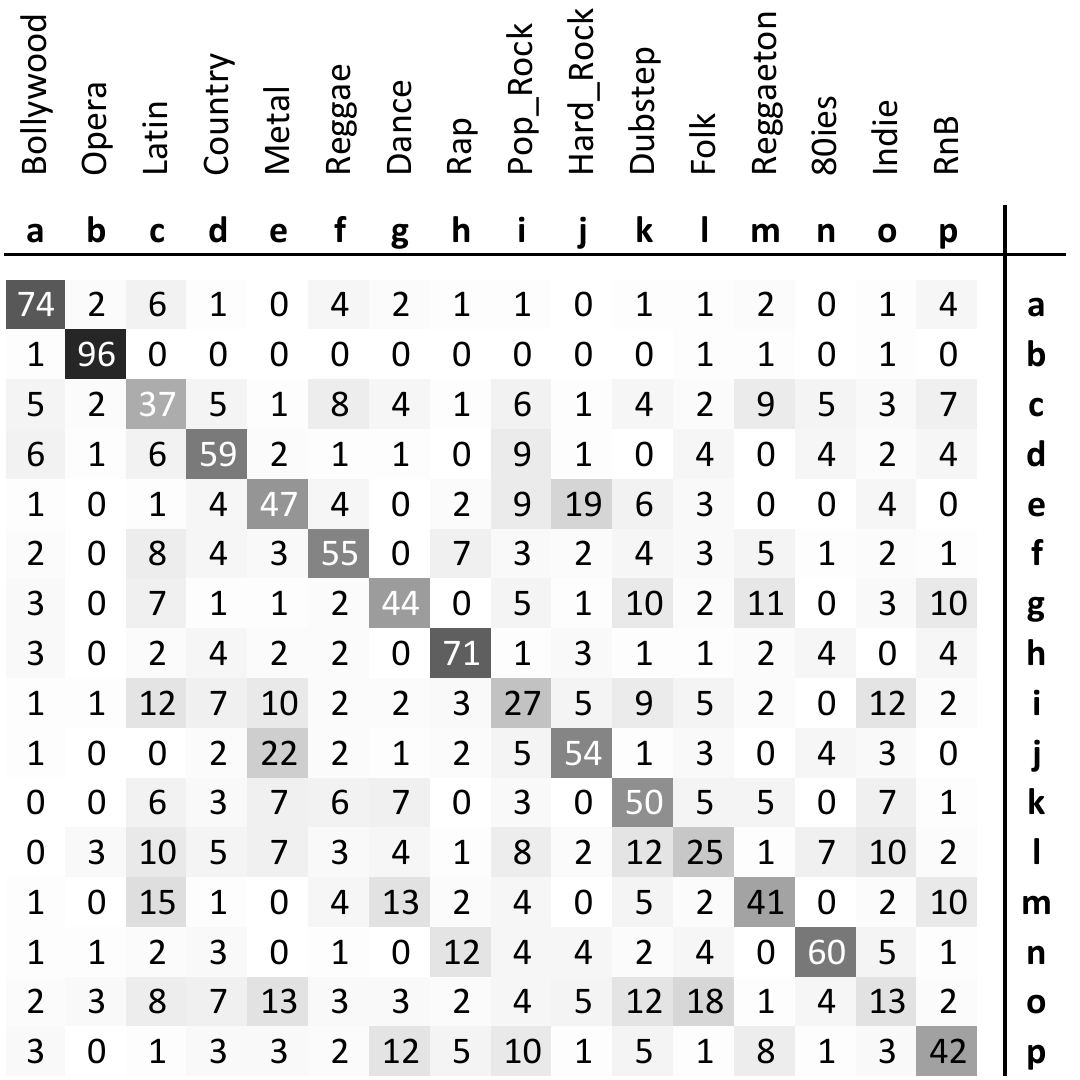}
	\caption{Confusion matrix of visual vocabulary based genre classification using best performing combination of mean+max aggregation and support vector machines on the MV-MIX dataset. Values describe how often a row-genre has been predicted as a column-genre. Due to the constant class size of 100 tracks, the diagonal elements represent the accuracies of 10-fold cross-validation for the given genre.}
	\label{tab:mv_mix_confusion_matrix}	
\end{figure*} 

\section{Summary}
\label{ch8:conclusions}

This chapter introduced semantic analysis of music video content. The focus of the presented research was on harnessing the information provided by the visual layer to approach the Music Information Retrieval problem space. Different feature-sets were analyzed in the context of different tasks. The results show that high-level concept detection approaches based on convolutional neural networks not only outperform traditional low-level image features as presented in Chapter \ref{ch7:intro}, but also are superior to audio-content based descriptors in semantic music tagging tasks. The introduced audio-visual approaches improve to audio-only baseline by up to 16.43\% for genre classification and up to 45.5\% for thematic music tagging tasks. The evaluation also indicates, that object vocabularies have the potential to capture the semantic and genre stereotypical information of music videos. Future work will focus on training specialized convolutional neural networks that include a wider range of music and genre related concepts.

\pagebreak

Using the Softmax scaled detection probabilities as extracted features and making them invariant towards the varying lengths of music videos already demonstrated the potential of harnessing the visual layer of music videos to improve the accuracy of music classification tasks. It is though expected that recent approaches based on deep neural networks would provide further improvements. Future work includes investigating if recurrent neural networks could be used to learn temporal progressions and relationships of the visual conception depending on their music genre. Neural networks further facilitate direct modeling of multiple modalities. A model which is trained on both - audio and visual input - can benefit from mutual semantic dependencies. This could yield higher classification accuracy.\\

\noindent
\textit{\smaller This approach and its corresponding evaluation presented in this chapter was published in \textit{Alexander Schindler and Andreas Rauber. ``Harnessing Music-Related Visual Stereotypes for Music Information Retrieval''} in the journal ACM Transactions on Intelligent Systems (TIST), special issue on ``Intelligent Music Systems and Applications'', volume 8, issue 2, January 2017, article no. 20, ACM New York, NY, USA \cite{Schindler_2016tist}. }


\chapter{Conclusions}

\epigraph{
	``The whole strenuous intellectual work of an industrious research worker would appear, after all, in vain and hopeless, if he were not occasionally through some striking facts to find that he had, at the end of all his criss-cross journeys, at last accomplished at least one step which was conclusively nearer the truth. ''} 
{--- Max Planck, Nobel Lecture in Physics, 1920}



This thesis is based on the observation that the genre of many music videos can be correctly guessed by only watching to their visual content without hearing the actual music and the derived hypothesis that there exist music related visual patterns. The validation of this hypothesis is approached from a Music Information Retrieval (MIR) perspective using data analytical methods to identify patterns or features and evaluate their correlation to semantic music-concepts such as genre, theme or performing artist. Because an analytical approach can only show degrees of correlations, Chapter \ref{ch3:intro} strove for identifying literature describing practices in music marketing and the intentions of using visual media.

\medskip
The review of the state of the art highlighted that audio-visual approaches are underrepresented in the MIR research domain, Especially audio-visual approaches to harness music related visual information. Only a few studies experimented with ensembles of acoustic and visual features extracted from album-arts to approach music emotion recognition and genre classification tasks. The literature review provided in Chapter \ref{ch2:intro} points out that recent advances in visual computing are capable of extracting abstract and subjective concepts such as movie genres. Because the production processes in music related visual media are based on traditional cinematic processes (but far more complex as described in Chapter \ref{ch3:mv_overview} and \ref{ch9:intro}) it is reasonable that these visual computing methods can also be applied to the MIR problem space.

\medskip
To verify that experimentally assessed relations between visual concepts and music genres/styles are not only based on statistical correlations, it was important to study the production processes. Chapter \ref{ch3:intro} extensively reviewed musicological literature to identify relevant descriptions of the application of visual media within the music marketing process. It was shown, that historically visual media has been used extensively to advertise new music acts, especially in music magazines and print media in general. Marketing units of music labels were responsible for making these new acts visually identifiable. Fashion and apparels were mentioned to be common means to achieve this goal. This is a strong indication that a visual language has been developed to distinguish different music styles and genres.

\medskip
The literature review in Chapter \ref{ch2:intro} reviled that audio-visual analysis of music videos has been lacking attention at the start of this dissertation and thus appropriate datasets were missing. The only dataset available was the \textit{DEAP} Dataset \cite{koelstra2012deap} for Emotion Analysis using EEG, Physiological and Video Signals. This set is intended to map into a three-dimensional emotional space, which makes it different from the classification tasks used in the evaluations. Further obstacles are the small size of only 40 music videos and the absence of discrete categorical labels.
Thus, the aim of Chapter \ref{ch4.2:intro} was to introduce datasets which are suitable for the development and evaluation of audio-visual approaches to MIR tasks. This was first approached in Section \ref{ch4.2:intro} by augmenting the Million Song Dataset (MSD) with additional genre labels and state-of-the-art audio features extracted from 30 to 60 second samples provided by the online music streaming service 7digital. These additions facilitated comprehensive experiments on the MSD and provided first large-scale benchmark results. 
The major contribution of Section \ref{ch4.1:intro} is the introduction of the \textit{Music Video Dataset (MVD)}. This is a carefully aggregated dataset to develop audio-visual approaches to MIR tasks. Especially the \textit{MVD-VIS} sub-set is ideal for identifying music-related visual patterns. This is based on the premise that the audio tracks within the classes have been selected by their acoustic similarity in terms of rhythmic, instrumentation, expression and composition. Thus, the inherent question is, if the songs are similar in terms acoustic properties, do they also share similar visual patterns in the video layer? This property of the dataset makes it easier to draw conclusions and makes them more comprehensible.
The \textit{MVD-MM} sub-set is intended to be used to analyze if visual patterns overlap in a similar fashion as acoustic features do in datasets with coarsely defined classes.
The \textit{MVD-Themes} subset was aggregated to evaluate if visual information can be used to solve tasks that are commonly difficult to approach by using audio content alone.\\

\medskip
Chapter \ref{ch5.1:intro} reports on experiments performed on the introduced datasets to provide general initial results and baseline results for the consecutive audio-visual experiments.
The first experiments were conducted on the extended MSD. This also provided some valuable experiences in large-scale MIR. For example, best performing approaches in 2011 utilized Support Vector Machines (SVM) as classifier in tasks such as genre or emotion classification. By default, the SVM is a quadratic function. Although linear approximations are available, the computational costs are still high, especially for the high dimensional music feature space. Facing these obstacles only low-dimensional featuresets could be evaluated in the initial experiments with lower than state-of-the-art results.
The experiments using Deep Neural Networks (DNN) have shown that these models have several advantages. First, they can be trained iteratively, batch by batch, which makes them scalable even with large input dimensions. Second, they have shown high accuracy values in Chapter \ref{ch5.1:MSD}. Because the SVM results presented in this Chapter have not been obtained using the best performing feature-set combination, it cannot be concluded that DNNs outperform SVMs with state-of-the-art audio features.\\
Chapter \ref{ch5.2:TEN} provided valuable insights into the proprietary feature-set provided with the MSD. These feature-sets were extracted and provided by the former company \textit{The Echonest}. The brief description of the features made them incomprehensible and unreliable for experiments. The empirical experiments presented in this chapter showed that these features are indeed semantically similar to the referenced music features. The evaluation of which mid-level features contribute most to tasks relevant in this thesis and how to appropriately aggregate, served as a benchmark for the experiments in the consecutive chapters.

\medskip
The most prevalent type of such information is the human face. As outlined in Section \ref{ch3:mv_overview} the general goal of music videos is to promote the performing artist. Consequently, the artist is shown as much as possible. Some artists, often well established ones, renounce the promotional advantage and create artful concept videos. Such videos were excluded from the experiments and only those where the performing artist is shown were included.

\textit{Harnessing facial information of the performing artist} was evaluated in Chapter \ref{ch6:intro}.  Audio content and visual based classifiers were combined using an ensemble classifier. The audio classifier used Temporal Echonest Features which were introduced in Section \ref{ch5.2:intro} to predict artist labels. The audio classifier's low precision of 37\% and a recall of 36\% outlined the problem of discriminating artists by using conventional audio and music features which attempt to capture acoustic or compositional characteristics, whereas the repertoire of an artist varies in instrumentation, style and genre. Also, the production methods change over time which results in different acoustic characteristics.
The approach was evaluated using state-of-the-art face recognition method based on a Local Binary Patterns (LBP) predictor. The two modalities - audio and video - were combined through bootstrap aggregation. For each modality 10 classifiers were created and trained on sub-samples of their according training-sets. The final prediction for a music video was calculated on the basis of weighted majority voting of the resulting 20 predictions. The proposed cross-modal approach showed that the initial audio content based baseline could be increased by 27\% through information extracted from the visual part of music videos.

The aim of this evaluation was to set a starting point into a field that has not yet received a lot of attention. It was shortly followed by the position paper \cite{Schindler2014dlfm} which introduced the advantages of an audio-visual approach to Music Information Retrieval. Although the application of face recognition demonstrated the potential of such an approach, the \textit{artist identification task} is of lesser relevance to the research field. Nevertheless, detecting faces in music related visual media has further relevant applications.

\medskip

An \textit{evaluation of low-level color based affective features} was presented in Chapter \ref{ch7:intro}. The comparative evaluation of visual features on their performance in music classification tasks focused on the color information of music videos. The hypothesis of this evaluation was, that common colloquial associations such as ``happy music is bright'' and ``sad and angry music is dark'' are reflected in the choice of colors of the corresponding music video. Further, this evaluation served as a comprehensive bottom-up evaluation to investigate the performance of low-level image features in describing the music related semantic content of music videos.
A set of diverging approaches based on psychological or perceptive models has been applied to extract different kinds of low-level semantic information. Seven color and illumination based image processing feature sets described in Section \ref{ch7:visual_features} including a descriptor capturing rhythmical changes in illumination was introduced. 

There were effects noticeable indicating that Color Names (CN), Wang Affective Factors (WAF) and Global Color Statistics (GCS) perform better in discriminating the different classes. Nevertheless, the general conclusion of this evaluation is, that none of the evaluated low-level visual features, nor any combination of them, is able to reliably predict music genres. Their low music related information value was further confirmed in Chapter \ref{ch8:intro} where no significant improvement was noticed by combining the low-level features with high-level semantic concepts such as apparel, buildings, music instruments, vehicles, etc.

\noindent
However it was able to confirm the common stereotype of 'darkness' related to \textit{Heavy Metal} music \cite{farley2009demons}. Videos of the class \textit{Metal} contain significantly more black pixels (independent t-test, p \textless 0.05) than other classes of the \textit{MVD-MIX} subset and the second smallest brightness values. The lowest values on \textit{brightness} and \textit{colorfulness} for \textit{Opera} are a strong indicator for the inferior lightning conditions at such venues and thus have no relation to the music itself. It was also possible to confirm the mentioned \textit{'movement towards the warm, orange tones that became the dominant 'look' of many contemporary country videos.'} \cite{frith2005sound}. The average values for \textit{red} and \textit{yellow} are significantly highest and second highest for this genre.

\medskip
The \textit{evaluation of high-level visual features} was presented in Chapter \ref{ch8:intro}. It focused on the performance of visual concept detection using a visual vocabulary based approach in music genre and cross-genre classification tasks. The results of the visual concept detection approach outperformed the accuracy of low-level visual features by 24.23\% on the same data-set. This high accuracy supports the initial hypothesis that music videos make use of easy identifiable visual concepts (see Research Question 1 in Chapter \ref{ch1:research_questoins}). An interesting observation was that the genre classification accuracy of the visual vocabulary based approach is significantly outperforming state-of-the-art low-level music features such as \textit{Chroma} and the de-facto standard feature \textit{MFCC}. This also supports the hypothesis that \textit{these visual patterns can be used to derive further music characteristics} (see Research Question RQ3 in Chapter \ref{ch1:research_questoins}).

\medskip

To answer \textit{if the information extracted from music related visual media improves the performance of current Music Information Retrieval tasks} a benchmark had to be assessed to compare the performance of visual-related features against. For this reason the Music Video Dataset (MVD) was compiled following the intentions and constraints described in Section \ref{ch4:dataset_creation}. State-of-the-art music features and classifiers were used to assess genre classification accuracies. Due to the data-set's characteristics, a very high accuracy is already achieved on the MVD-VIS sub-set. This was by intention, because the purpose of this well defined sub-set was to analyze if music videos of highly similar sounding songs from different artists use similar visual patterns. Nevertheless, despite this high degree of acoustic coherence of the \textit{MVD-VIS} subset, an improvement of 2.94\% was accomplished over the best performing audio combination through the combination of audio and visual information. An even higher improvement was observed for the less acoustically differentiated set \textit{MVD-MM} (+6.84\%) and the bigger \textit{MVD-MIX} dataset (+10.82\%). The visual information showed outstanding improvements, raning from 7.84\% to 16.43\%, over some well-known audio feature sets such as the Temporal Statistical Spectrum Descriptors (TSSD).

The largest improvement, however was noticed for the classification performance of non-audible or cross-genre music themes such as \textit{Christmas} or \textit{Protest Songs}. These themes often have only few musical connotations and are thus difficult to describe using the audio signal alone. This was clearly shown by the low audio-based accuracies. Using visual features, almost all visual vocabulary based approaches already performed better than the audio-only results such as \textit{Christmas} combined with \textit{MVD-MIX} (+34,5\%). More remarkable for the same combination the improvement of 45.5\% was observed. Similar high improvements were observed for the other themes.

\medskip 

\textit{Research Question 4 (RQ5) ``Is it possible to verify concepts within the production process?''} was intended to evaluate if the identified patterns are just capturing abstract visual patterns or if these visual features correspond real visual concepts that have been intentionally applied during the production of the music video. The conclusion confirms this assumption. This is based on the following arguments:

\begin{itemize}
	
	\item The \textit{MVD-VIS} datsaet has been assembled to maximize the acoustic coherence of its genre-related classes. In other words, the songs within a class sound highly similar. This is confirmed by the audio classification accuracy of 93.79\% presented in Table \ref{tab:ch7:Results} (see Chapter \ref{ch7:Audio_Features}). The intention was to analyze if the visual representation shows similar coherent visual patterns.
	
	\item The results presented in Table \ref{tab:Results} (see Chapter \ref{ch8:eval_results}) showed an accuracy of 74.36\%, using statistics of high-level visual concepts detected in video frames of the \textit{MVD-VIS} dataset. This is a strong indicator that there exist class discriminating sets of visual features.
	
	\item The literature search in Chapter \ref{ch3:intro} identified articles which described that visual accentuation is part of branding a new music act to make them immediately assignable to a music style or genre - at least in the pre-internet era. This can only be accomplished through patterns with high recognition value.
	
	\item The analysis in Chapter \ref{eval_visual_stereotypes} showed the most salient visual concepts for the genre-related classes of the \textit{MVD} dataset from which many are specific for the corresponding genre. Further, music psychological and musicological literature was identified which identified similar audio-visual correlations for certain analyzed objects.
	
\end{itemize}

Thus, by verifying that these visual accentuation was deliberately applied to a certain song from a certain genre also verifies that the extracted information is music and task related. On the other hand, it was also intended to evaluate if the presented approach can be utilized to quantify music-related visual stereotypes in music videos. This was approached by identifying genre or thematic related stereotypes. Through automatically identifying the most frequent salient visual concepts for each genre through the visual vocabulary approach presented in Chapter \ref{ch8:intro} it was possible to relate them to reported music related stereotypes. 

Visual stereotypes were identified for the genre \textit{Country} for which the \textit{cowboy hat} and utility vehicles such as the \textit{pickup truck} or a \textit{tractor trailer} are highly salient concepts. These observations correspond to an evaluation of perceived extramusical associations with country music \cite{shevy2008music} and to aesthetic description provided in \cite{frith2005sound}. The second stereotype addresses the over-sexualization of contemporary popular \textit{Dance} music \cite{hall2012sexualization}. Eight of the provided top ranking examples referred to revealing clothes, closeups of body parts or people dancing on \textit{poles}. For Heavy \textit{Metal} music the common stereotype of \textit{darkness} was confirmed using low-level visual features. Also \textit{fire} as a typical element related to \textit{Metal} \cite{farley2009demons} was identified using the visual concept based approach. For \textit{Opera} a high number of top-ranking classical instruments related concepts can be observed.

\section{Future Work}

The answered research questions, gained insights and conclusions set up a good starting point for further research to harness audio-visual correlations within the MIR domain.

\paragraph{Complex audio-visual modeling:}
Based on the new options provided by Deep Neural networks (DNN) it is easier to train integrated audio-visual models. Different to the two-phase approaches presented in this thesis, where first features are extracted from the different modalities and then fused either in the feature-space or by ensembles of classifiers, a multi-modal DNN can take visual and acoustic data as input to learn a common representation and directly predict results. Especially Recurrent Neural Networks (RNN) are expected to better utilize the sequential information provided by video frames and audio data.
This could lead to improved audio-visual representations which capture the correlations between the two modalities better and perform better in the specific tasks on provided datasets.

The current success of deep neural network based approaches in visual computing has also improved face recognition systems. Recent methods \cite{farfade2015multi} have overcome obstacles reported in Section \ref{ch6:obstacles}. Features such as headcount, position of a face, the area it takes within the video frame, the amount of time it is shown, etc. could contain semantically valuable information to distinguish for example solo artists from bands or studio settings from live performances. Analyzing facial expressions could contribute to music emotion recognition systems.

\paragraph{Cross-modal modeling:}
Similarly, DNNs can be used to learn mutual representation spaces. This has been demonstrated for visual and textual inputs \cite{yi2015shared} and still need more attention for visual and acoustic inputs. Recent advances linked symbolic to recorded music in a learned cross-modal representation for improved audio-sheet music correspondence \cite{dorfer2018learning,mueller2019cross}.
Merging acoustic and visual information into the same representation space facilitates a wide range of new options and tasks. For instance, it facilitates to use visual inputs to search directly for music or sounds (e.g. using the image of a bird to search for tweets, using the image at a party to search for adequate music). Such an approach has recently been reported on general video-datasets \cite{chen2017deep}. In this approach a cross-domain audio-visual representation is learned to retrieve a video sequence using an audio sequence as query and vice-versa.

Such an approach could be extended to the music domain to use visual inputs to query for music. This could be used to automatically recommend music for videos.
In a further step this representation space can be extended to further modalities such as the text domain. By learning relationships between visual concepts and acoustic patterns such as birds and tweets, these concepts can be linked to words or word embeddings from textual corpora. Thus, text based search in music content might be achieved without the need for text-based annotations.

%

\paragraph{Multi-task music representation learning:}
Designing a well defined feature-set requires detailed knowledge about the problem domain as well as the characteristics of the dataset. Various factors have to be considered. These can span from perceptual properties such as mood to stylistic and genre specific characteristics, but may also include or focus on musicological aspects such as epochs or mutual influences of composers. To describe these factors, many of them require different feature-sets. On the one hand these are difficult to extract and manage. On the other hand, such combinations often do not generalize or scale well.
Representation learning represents a promising alternative. Features and feature-sets are not composed but learned - usually using Deep Neural Networks (DNN) in a supervised fashion from a given ground-truth annotation. The obstacle of this approach is the requirement for large amounts of annotations - for each factor/characteristics that should be considered.
Annotating such large amounts of data is time and resource-intensive.

The visual layer of music videos provides a wide range of contextual information. 
State-of-the-art Convolutional Neural Networks (CNN) such as the InceptionResNetV2 architecture \cite{SzegedyIV16} already reach an accuracy of 80.1\% on the Top-1 Error and 95.1\% accuracy on the Top-5 Error. Visual concepts detected in the music video frames with high prediction confidence could serve as a substitution for manual annotations. Here the knowledge of the annotations from the visual model is transferred to the music domain. As outlined in this thesis, the frequency distribution of these visual concepts is discriminative for music genres. Using Deep Learning concepts such as triplet networks or mutli-task learning, the information from the visual domain can be harnessed to either transfer the visual semantics to the music representation, or to guide the learning process by the visual semantics. This could especially contribute to MIR tasks such singer and instrument recognition.


\begin{appendices}

\chapter{Publications}

\begin{itemize}
	
	\item Alexander Schindler and Peter Knees. Multi-Task Music Representation Learning from Multi-Label Embeddings. In Proceedings of the International Conference on Content-Based Multimedia Indexing (CBMI2019). Dublin, Ireland, 4-6 Sept 2019. \cite{schindler_knees:cbmi:2019}
	
	\item Anahid Jalali, Clemens Heistracher, Alexander Schindler, Bernhard Haslhofer, Tanja Nemeth, Robert Glawar, Wilfried Sihn, Peter De Boer. Predicting Time-to-Failure of Plasma Etching Equipment using Machine Learning. In Proceedings of the IEEE International Conference on Prognostics and Health Management (PHM2019), June 17-19, 2019, in San Francisco, USA. \cite{jalali2019predicting}
	
	\item Anahid N Jalali, Alexander Schindler, Bernhard Haslhofer. Understandable Deep Neural Networks for Predictive Maintenance in the Manufacturing Industry In ERCIM News, Number 116, Jan 2019. \cite{jalali2019understandable}
	
	\item Alexander Schindler and Andreas Rauber. On the unsolved problem of Shot Boundary Detection for Music Videos. In Proceedings of the 25th International Conference on MultiMedia Modeling (MMM2019), January 8-11, 2019, in Thessaloniki, Greece. \cite{schindler2019}
	
	\item Alexander Schindler, Andrew Lindley, David Schreiber, Martin Boyer and Thomas Philipp. Large Scale Audio-Visual Video Analytics Platform for Forensic Investigations of Terroristic Attacks. In Proceedings of the 25th International Conference on MultiMedia Modeling (MMM2019), January 8-11, 2019, in Thessaloniki, Greece. \cite{schindler2019large}
	
	\item Nemeth, Tanja, Fazel Ansari, Wilfried Sihn, Bernhard Haslhofer, and Alexander Schindler. PriMa-X: A reference model for realizing prescriptive maintenance and assessing its maturity enhanced by machine learning. Procedia CIRP 72 (2018): 1039-1044. \cite{nemeth2018prima}
	
	\item Alexander Schindler and Sven Schlarb. Contextualised Conversational Systems. In ERCIM News, Number 114, July 2018. \cite{schindler2018contextualised}
	
	\item Alexander Schindler, Thomas Lidy and Andreas Rauber. Multi-Temporal Resolution Convolutional Neural Networks for Acoustic Scene Classification. In Proceedings of the Detection and Classification of Acoustic Scenes and Events 2017 Workshop (DCASE2017), November 2017. \cite{schindler2018multi}
	
	\item Alexander Schindler, Thomas Lidy and Andreas Rauber. Multi-Temporal Resolution Convolutional Neural Networks for the DCASE Acoustic Scene Classification Task. Technical report, DCASE2017 Challenge, November 2017. \cite{schindlermulti}
	
	\item Botond Fazekas, Alexander Schindler, Thomas Lidy, Andreas Rauber. A multi-modal deep neural network approach to bird-song identification. LifeCLEF 2017 working notes, Dublin, Ireland \cite{fazeka2018multi} 
	
	\item Alexander Schindler, Thomas Lidy, Stefan Karner and Matthias Hecker. Fashion and Apparel Classification using Convolutional Neural Networks. In Proceedings of the 9th Forum Media Technology (FMT2017), St. Poelten, Austria, October 29, 2017. \cite{schindler2018fashion}
	
	\item Alexander Schindler and Andreas Rauber. Harnessing Music related Visual Stereotypes for Music Information Retrieval. ACM Transactions on Intelligent Systems and Technology (TIST) 8.2 (2016): 20 \cite{Schindler_2016tist}
	
	\item Alexander Schindler, Thomas Lidy, and Andreas Rauber. Comparing shallow versus deep neural network architectures for automatic music genre classification. In Proceedings of the 9th Forum Media Technology (FMT2016), St. Poelten, Austria, November 23 - November 24 2016. \cite{schindler2016comparing} 
	
	\item Alexander Schindler, Sergiu Gordea, and Harry van Biessum. The europeana sounds music information retrieval pilot. In Proceedings of the International Conference on Cultural Heritage (EuroMed2016), Lecture Notes in Computer Science, Cyprus, October 31 - November 5 2016. Springer. \cite{schindler2016europeana}
	
	\item Thomas Lidy and Alexander Schindler. CQT-based convolutional neural networks for audio scene classification. In Proceedings of the Detection and Classification of Acoustic Scenes and Events 2016 Workshop (DCASE2016), pages 60--64, September 2016. \cite{lidy2016cqt}
	
	\item Thomas Lidy and Alexander Schindler. CQT-based convolutional neural networks for audio scene classification and domestic audio tagging. Technical report, DCASE2016 Challenge, September 2016.
	
	\item Thomas Lidy and Alexander Schindler. Parallel convolutional neural networks for music genre and mood classification. Technical report, Music Information Retrieval Evaluation eXchange (MIREX 2016), August 2016. \cite{Lidy_Schindler_MIREX2016}
	
	\item Thomas Lidy, Alexander Schindler and Michela Magas. MusicBricks: Connecting Digital Creators to the Internet of Music Things. In ERCIM News, Number 101, April 2015. \cite{lidy2015musicbricks}
	
	\item Alexander Schindler and Andreas Rauber. An audio-visual approach to music genre classification through affective color features. In Proceedings of the 37th European Conference on Information Retrieval (ECIR'15), Vienna, Austria, March 29 - April 02 2015. \cite{schindler2015}
	
	\item Alexander Schindler. A picture is worth a thousand songs: Exploring visual aspects of music. In Proceedings of the 1st International Digital Libraries for Musicology workshop (DLfM 2014), London, UK, September 12 2014. \cite{Schindler2014dlfm}
	
	\item Roman Graf, Alexander Schindler, and Reinhold Huber-Moerk. A fuzzy logic based expert system for quality assurance of document image collections. In International Journal of Arts \& Sciences IJAS 2014 to appear, Valetta, Malta, March 2-6 2014. \cite{graf2014fuzzy}
	
	\item Roman Graf, Reinhold Huber-Moerk, Alexander Schindler, and Sven Schlarb. Duplicate detection approaches for quality assurance of document image collections. In Proceedings of the International ACM Conference on Management of Emergent Digital EcoSystems (MEDES'13) to appear, Neumuenster Abbey, Luxembourg, October 28-31 2013. \cite{graf2013duplicate}
	
	\item Alexander Schindler and Reinhold Huber-Moerk. Towards objective quality assessment in digital image collections. In Proceedings of the 2nd Workshop on Open Research Challenges in Digital Preservation (ORC'13) to appear, Lisbon, Portugal, September 6 2013. \cite{schindlertowards}
	
	\item Alexander Schindler and Andreas Rauber. A music video information retrieval approach to artist identification. In Proceedings of the 10th International Symposium on Computer Music Multidisciplinary Research (CMMR2013) to appear, Marseille, France, October 14-18 2013. \cite{SCHINDLER_2013CMMR}
	
	\item Reinhold Huber-Moerk and Alexander Schindler. Automatic classification of defect page content in scanned document collections. In Proceedings of the 8th International Symposium on Image and Signal Processing and Analysis (ISPA 2013) to appear, Trieste, Italy, September 4-6 2013. \cite{huber2013automatic}
	
	\item Sven Schlarb, Peter Cliff, Peter May, William Palmer, Matthias Hahn, Reinhold Huber-Moerk, Alexander Schindler, Rainer Schmidt, and Johan van der Knijff. Quality assured image file format migration in large digital object repositories. In Proceedings of the 10th International Conference on Digital Preservation (IPres2013) to appear, Lisbon, Portugal, September 2-5 2013. \cite{schlarb2013quality}
	
	\item Reinhold Huber-Moerk and Alexander Schindler. A keypoint based approach for content characterization in document collections. In Proceedings of the 9th International Symposium on Visual Computing (ISVC'13) to appear, Rethymnon, Crete, Greece, July 29-31 2013. \cite{huber2013image}
	
	\item Alexander Schindler and Andreas Rauber. Capturing the temporal domain in echonest features for improved classification effectiveness. In Adaptive Multimedia Retrieval, Lecture Notes in Computer Science, Copenhagen, Denmark, October 24-25 2012. Springer. \cite{SCHINDLER_2012amr}
	
	\item Andreas Rauber, Alexander Schindler, Nicu Sebe, Henning Mueller, Shara Monteleone, Yiannis Kompatsiaris, Spiros Nikolopoulos, Alexis Joly, and Henri Gouraud. Latest trends in multimedia search computing. In Nicu Sebe, editor, Latest Trends in Multimedia Search Computing (Media Search Cluster White Paper). European Commission - Infomation Society and Media, December 2012.
	
	\item Reinhold Huber-Moerk, Alexander Schindler, and Sven Schlarb. Duplicate detection for quality assurance of document image collections. In Proceedings of the 9th International Conference on Digital Preservation (IPres2012), Toronto, Canada, October 1-5 2012. \cite{huber2012duplicate}
	
	\item Roman Graf, Reinhold Huber-Moerk, and Alexander Schindler. An expert system for quality assurance of document image collections. In Proceedings of the International Conference on Cultural Heritage (EuroMed2012), Lecture Notes in Computer Science, Lemesos, Cyprus, October 29 - November 3 2012. Springer. \cite{graf2012expert}
	
	\item Alexander Schindler, Rudolf Mayer, and Andreas Rauber. Facilitating comprehensive benchmarking experiments on the million song dataset. In Proceedings of the 13th International Society for Music Information Retrieval Conference (ISMIR 2012), pages 469-474, Porto, Portugal, October 8-12 2012. \cite{schindler2012}
	
	\item Reinhold Huber-Moerk and Alexander Schindler. Quality assurance for document image collections in digital preservation. In Proceedings of the 14th International Conference on Advanced Concepts for Intelligent Vision Systems (ACIVS 2012), Lecture Notes in Computer Science, Brno, Czech Republic, September 4-7 2012. Springer. \cite{huber2012quality}
	
	\item Alexander Schindler. Million song dataset integration into the clubmixer framework. In Proceedings of the 12th International Society for Music Information Retrieval Conference (ISMIR 2011), Miami, USA, October 24-28 2011. \cite{schindlermillion}
	
	\item Roman Graf, Reinhold Huber-Moerk, and Alexander Schindler. Quality assurance for scalable braille web service using human interaction by a computer vision system. In Post-conference Proceedings of the International Conference on Integrated Information (IC-ININFO 2011), Kos Island, Greece, September 29 2011. \cite{grafquality}
	
	\item Alexander Schindler and Andreas Rauber. Clubmixer: A presentation platform for mir projects. In Marcin Detyniecki, Peter Knees, Andreas Nuernberger, Markus Schedl, and Sebastian Stober, editors, Adaptive Multimedia Retrieval. Context, Exploration and Fusion Adaptive Multimedia Retrieval. Context, Exploration and Fusion, volume 6817 of Lecture Notes in Computer Science, Linz, Austria, August 17-18 2010. Springer. \cite{schindler2010}
	
	\item Alexander Schindler. Quality of service driven workflows within the microsoft .net environment. Master's thesis, Vienna University of Technology, 2009.

\end{itemize}
\end{appendices}



\appendix

\bibliographystyle{plain}
\bibliography{references}

\includepdf[pages=-]{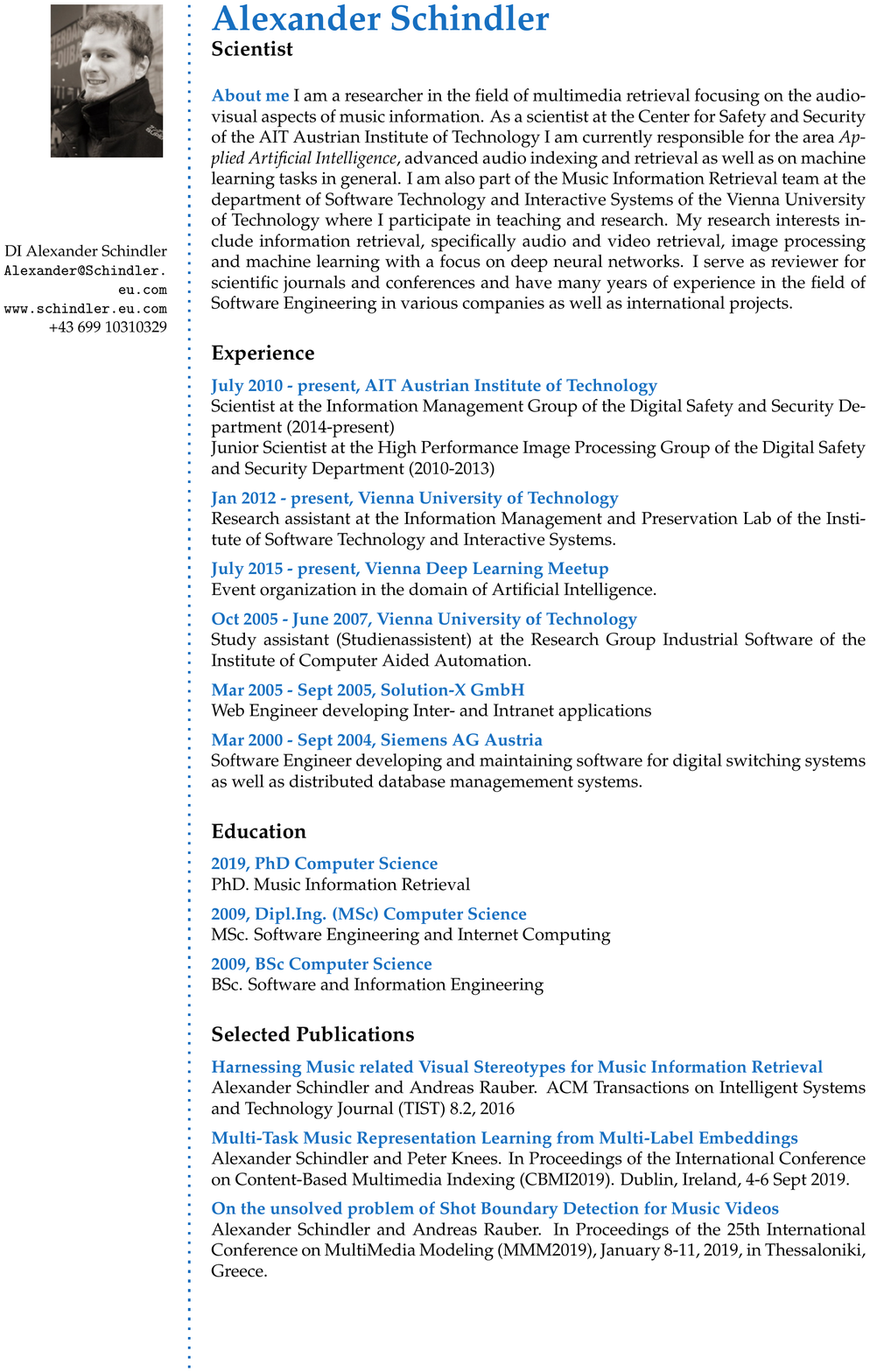}

\end{document}